\documentclass[pdflatex,sn-basic]{sn-jnl}

\usepackage{graphicx}%
\usepackage{multirow}%
\usepackage{amsmath,amssymb,amsfonts}%
\usepackage{amsthm}%
\usepackage{mathrsfs}%
\usepackage[title]{appendix}%
\usepackage{xcolor}%
\usepackage{textcomp}%
\usepackage{manyfoot}%
\usepackage{booktabs}%
\usepackage{algorithm}%
\usepackage{algorithmicx}%
\usepackage{algpseudocode}%
\usepackage{listings}%
\usepackage{comment}%
\usepackage{ulem}

\begin{document}

\title[Dark sector and dynamics in clusters]{Exploring the dark sector through galaxy dynamics in clusters}


\author*[1,2]{\fnm{Andrea} \sur{Biviano}}\email{andrea.biviano@inaf.it}
\author[3,1]{\fnm{Lorenzo} \sur{Pizzuti}}\email{lorenzo.pizzuti@unimib.it}

\author[4,5,6]{\fnm{Antonaldo} \sur{Diaferio}}\email{antonaldo.diaferio@unito.it}

\affil*[1]{\orgdiv{INAF-Osservatorio Astronomico di Trieste}, \orgaddress{\street{via G. B. Tiepolo 11}, \city{Trieste}, \postcode{I-34131}, \country{Italy}}}
\affil[2]{\orgdiv{IFPU-Institute for Fundamental Physics of the Universe}, \orgaddress{\street{via Beirut 2}, \city{Trieste}, \postcode{34014}, \country{Italy}}}

\affil[3]{\orgdiv{Dipartimento di Fisica G. Occhialini}, \orgname{Universit\`a degli Studi di Milano-Bicocca}, \orgaddress{\street{Piazza della Scienza 3}, \city{Milano}, \postcode{I-20126}, \state{Italy}}}

\affil[4]{\orgdiv{Dipartimento di Fisica}, \orgname{Universit\`a di Torino}, \orgaddress{\street{via P. Giuria 1}, \city{Torino}, \postcode{10125}, \country{Italy}}}
\affil[5]{\orgdiv{Istituto Nazionale di Fisica Nucleare}, \orgname{sezione di Torino}, \orgaddress{\street{via P. Giuria 1}, \city{Torino}, \postcode{10125}, \country{Italy}}}
\affil[6]{\orgdiv{Accademia delle Scienze di Torino}, \orgaddress{\street{via Maria Vittoria 3}, \city{Torino}, \postcode{10123}, \country{Italy}}}

\abstract{Galaxy clusters are among the most powerful laboratories to probe fundamental physics, from the formation of cosmic structures to the nature of dark matter, dark energy, and gravity. Realizing their full potential requires accurate and unbiased determinations of their mass profiles, a longstanding challenge in cluster cosmology.
In this context, the kinematics of member galaxies in clusters has become a key tool to reconstruct the mass distribution and to investigate the dynamical properties of clusters. The advent of large spectroscopic datasets has enabled the application of increasingly sophisticated models and robust techniques, which make kinematic analyses competitive with other probes of cluster dynamics.

In this review, we provide a comprehensive overview of cluster kinematics, covering the main methods to infer dynamical masses and mass profiles, their physical assumptions, and the impact of systematics such as deviation from dynamical equilibrium, triaxiality, and substructures. We discuss the complementarity with other techniques to investigate the structure of clusters and highlight recent applications to tests cosmological models. We focus in particular on the role of kinematical studies in constraining the dark sector (dark matter and dark energy) and in providing novel insights into the behavior of gravity at Mpc scales.

Finally, we outline current limitations and future prospects, emphasizing that the combination of improved modeling and forthcoming spectroscopic surveys will establish cluster kinematics as a fundamental probe in the era of precision cosmology.}




\maketitle
\tableofcontents

\section{Structure and aim of the review}
\label{sec:aims}
We are currently transitioning from an era of cluster detection to the realm of precision cluster physics. While recent photometric surveys have cataloged thousands of systems, the true cosmological capability of galaxy clusters lies in understanding their internal structure and mass distribution. Thanks to the dramatic increase in spectroscopic data for cluster galaxies, the kinematics of member galaxies in clusters is nowadays a very valuable instrument to unlock these prospects, offering a direct probe of the gravitational potential that is complementary to lensing and X-ray analyses.

This review is devoted to the analysis of galaxy cluster kinematics, with the goal of providing a coherent overview of the methods used to infer cluster mass profiles, and of the most relevant results obtained in recent years. We focus on the main techniques developed to perform kinematical studies, from velocity dispersion measurements, Jeans modeling, caustics, as well as recent developments based on machine learning (ML hereafter). For each class of methods, we discuss the key assumptions and the main sources of systematic uncertainty. In particular, we discuss what can be robustly inferred from the projected positions and line-of-sight (los, hereafter) velocity distribution of cluster galaxies, and what systematic effects must be controlled before these inferences can be used for constraining the properties of the dark sector and the nature of gravity. In fact, if on one hand, galaxy kinematics constrains the total cluster mass profiles, the cluster assembly history, and the orbital distribution of galaxies, on the other hand, when combined with lensing, X-ray, and SZ observations, it offers independent tests of the dark sector and of the gravitational law on Mpc scales. 

The structure of the review is as follows. Section~\ref{sec:introduction} provides a brief historical overview of kinematical studies of galaxy clusters. Section~\ref{sec:methods} presents the main techniques for dynamical mass reconstruction and phase-space analysis. Section~\ref{sec:results} summarises the principal results from the literature concerning kinematical tests of the dark sector and gravity. Finally, Section~\ref{sec:future} discusses 
future prospects for the cluster kinematical studies in light of the upcoming spectroscopic and multi-wavelength surveys and improved dynamical modeling.
 

\section{Introduction: A hundred years of cluster kinematics}
\label{sec:introduction}

Galaxy clusters are the most massive gravitationally bound structures in the Universe and represent a fundamental natural laboratory at the edge between modern astrophysics and cosmology. Within the current standard model of the hierarchical formation of cosmic structure, 
clusters trace the high peaks of the initial matter density fluctuations and grow via mergers and accretion over cosmic time (e.g., \citealt{Kravtsov_2012}). As such, they can be used to constrain the parameters of the cosmological model (e.g., \citealt{Haggar24,Fumagalli_2024}), providing an independent probe, when combined with other observables, to lift possible degeneracies. 
In particular, the abundance of clusters as a function of redshift is exquisitely sensitive to the growth of structures, while the cluster spatial distribution and correlations provide complementary information (e.g.,~\citealt{Allen_2011,Balaguera2014,Sartoris_2016}). Moreover, the concentration-mass relation, substructure content, and baryonic distribution in clusters are influenced by the underlying cosmology (e.g., \citealt{Giocoli_2012,Biviano_2017,Despali_2018}).

The internal structure of these objects further carries pivotal information. In the context of the $\Lambda$CDM (Lambda Cold Dark Matter) cosmological model, most of the matter content in clusters (roughly 80\%) appears in the form of collisionless, cold dark matter (DM) hereafter; the remaining budget is dominated by a diffuse ionized gas at high temperature, the intra-cluster medium (ICM) emitting in the X-ray band mostly via thermal bremsstrahlung ($\sim 10 -  15\%$ of the total mass). Stellar mass in galaxies represents only a tiny fraction of the mass, from 1 to 5\% (e.g.,~\citealt{BS06,Andreon_2010,Palmese_2020}). A significant fraction of the stellar mass is concentrated in the brightest cluster galaxy (BCG) sitting at the center of the gravitational potential and the surrounding intra-cluster light (ICL), i.e. diffuse stars not associated to any galaxy \citep[e.g.,][]{DeMaio20,Montenegro25}. While processes involving ordinary (baryonic) matter play a central role in the core of these structures, the overall dynamics is dominated by gravitational interaction, making clusters ideal laboratories for investigating deviations from General Relativity (GR, see e.g.,~\citealt{Sakstein_2016,Hammami17,Mitchell_2018,liu21,Boumechta24,Vogt24,Famaey25,Pizzuti25a}, and \citealt{Cataneo_2018} for a review), probing the properties of the dark sector (e.g.,~\citealt{Takada_2007,SartorisDM,Morandi_2016,Dai_2018,Biviano_2023}) and studying the evolution of galaxies in dense environments (e.g.,~\citealt{Boselli_2006,Wetzel2013,Muriel14,Lopez2022,Durret2022,cakr25}).

A fundamental quantity that encapsulates much of the physical information encoded in galaxy clusters is their total mass \citep{Pratt2019,Wu_2021,Doubrawa_2023,Sereno25,Ingoglia25,Iqbal25} and its radial distribution among the various components \citep{Crone_1997,Ettori_2013,Sartoris_2020,Pizzardo24,Pizzuti24c,Pizzuti25a,Sereno25b}. The mass profile governs the depth and shape of the gravitational potential well, regulates the thermodynamical state of the intracluster medium, controls the orbital structure of galaxies, and provides a direct link between clusters and the underlying cosmological model. Accurate determinations of cluster mass profiles are therefore essential not only for cosmological applications, such as cluster abundance and scaling relations \citep[e.g.,][]{Giodini_2013}, but also for understanding the assembly history of clusters, the nature of DM and its interplay with baryons (e.g.,~\citealt{Lagana13}), and the possible presence of new gravitational physics. 

Among the various observational probes of cluster mass, the kinematics of member galaxies occupies a special and long-standing role \citep[e.g.,][]{Zwicky1933,Smith1936,Rood+72,Carlberg_1997,Geller_1999,Biviano_2003,Mamon_2013,Ho_2019,Li2023,Sereno25}. The motion of galaxies within clusters is directly driven by the total gravitational potential $\Phi(\bf{r})$, and thus encodes dynamical information on the total matter distribution and its radial structure. By linking the observed positions and velocities of galaxies to the underlying potential, cluster kinematics provides a powerful tool to infer cluster mass profiles, the orbits of member galaxies, and to assess the dynamical state of clusters. Many kinematical methods for cluster analyses are sensitive to departures from equilibrium, projection effects, contamination of interlopers and substructure content; while on one side these effects constitute a limitation, biasing the constraints on total mass and mass distribution \citep[e.g.,][]{Saro_2013,Old_2017,Pizzuti2020syst}, on the other side they offer insight into the complex, non-linear processes that shape clusters over cosmic time \citep[e.g.][]{De_Luca_2021,Kimmig_2023}. 

\begin{figure}
         \centering
         \includegraphics[width=0.9\textwidth]{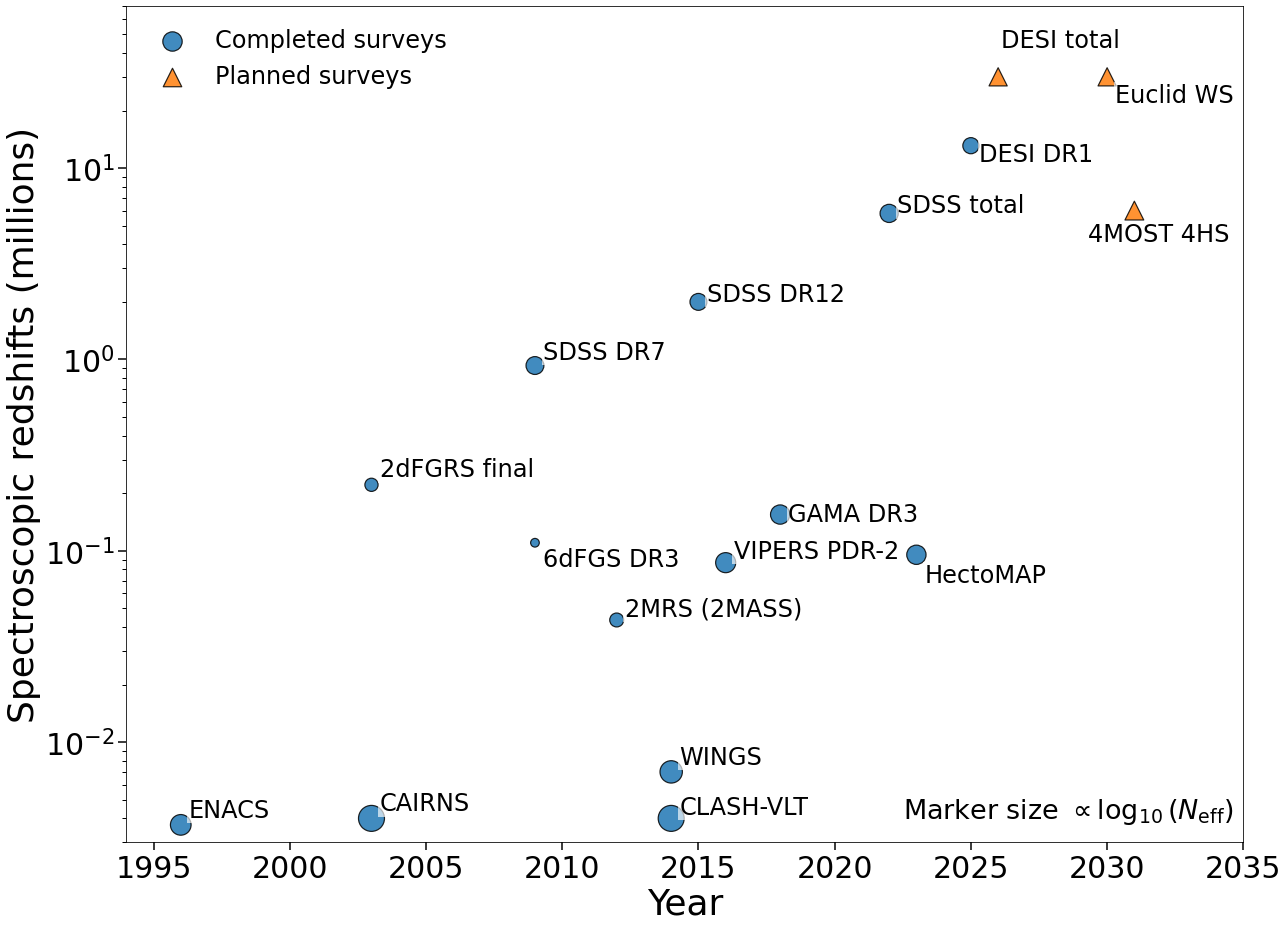}
         \caption{\label{fig:spectra} 
         Number of galaxies with measured spectroscopic redshift provided by past surveys, and estimated numbers for operating and planned surveys. Blue points mark surveys for which data have been already - or partially - released, while orange points represent the expectations from upcoming projects. We have also included spectroscopic surveys dedicated to clusters of galaxies, namely, ENACS, CAIRNS, WINGS, and
         CLASH-VLT (relevant references can be found in the main text). The size is proportional to the logarithm of the number of galaxies per cluster.} 
        \end{figure}

The use of galaxy motions to infer the mass content of clusters pre-dates other relevant techniques in the literature - such as gravitational lensing and hydrostatic equilibrium of the X-ray-emitting ICM --- and played a central role in the historical discovery of dark matter \citep{Zwicky1933,Zwicky1937,Smith1936}. Over the past century, cluster kinematics has evolved from simple velocity dispersion measurements to a sophisticated approach supported by large spectroscopic surveys, advanced statistical methods, and high-resolution numerical simulations. For decades, the efficiency of kinematical techniques has been limited by the low availability of spectroscopic data to reconstruct the velocity field of the member galaxies. However, with the advent of Stage III (e.g., the Sloan Digital Sky Survey, SDSS, \citealt{Alam_2021}, the Galaxy And Mass Assembly, GAMA, survey \citealt{Hopkins_2013})  and Stage IV (e.g., Dark Energy Spectroscopic Instrument, DESI, \citealt{DESI21}, 4-meter Multi-Object Spectroscopic Telescope, 4MOST, \citealt{4most19}) surveys, as well as of dedicated observational campaigns (e.g., 
The Cluster and Infall Region Nearby Survey, CAIRNS, \citealt{Rines+03}, Cluster Infall Regions in the SDSS, CIRS, \citealt{RinesDiaferio06}, the Wide-Field Nearby Galaxy-Cluster Survey, WINGS, \citealt{Cava+09}, the Hectospec cluster survey HeCS, \citealt{Rines+13}, the Cluster Lensing and Supernova Survey with Hubble follow-up, CLASH-VLT, \citealt{Rosati14}), an unprecedented amount of high-precision redshifts of galaxies has been collected with massive coverage from the center to the outskirt --- beyond the virial region --- of several systems. This improvement radically transformed kinematics of clusters into a competitive set of techniques for the study of galaxy clusters and their tight connection to cosmology. Figure~\ref{fig:spectra} shows an estimate of the number of spectroscopic redshifts of galaxies available (or planned) from releases of past, ongoing or upcoming surveys.

Nowadays, kinematics of member galaxies is not only a key tool for reconstructing cluster mass profiles, but also a powerful probe of the dark sector and a promising avenue for testing possible deviations from GR.


\subsection{The dawn of cluster kinematics}

The scientific investigation of galaxy clusters began long before they were recognized as physical systems. In the late eighteenth century, Charles Messier and William Herschel cataloged ``nebulae'', noting remarkable concentrations of such objects on the sky, most notably in the Virgo region. However, it was only in the early twentieth century, following the establishment of the extragalactic nature of galaxies through the work of Slipher and Hubble - based on the pioneering discovery of H. Leavitt - that clusters were understood as genuine assemblies of galaxies rather than features of the Milky Way.

A decisive turning point came in the 1930s, when \cite{Hubble31} estimated for the first time the velocity dispersion of galaxy clusters, noting for example that the velocity dispersion of the Coma cluster was larger than for other concentrations of galaxies, quoting a value of $\sim 700$ km/s. This result represented a preliminary indication of a velocity dispersion-richness relation later formalized by \cite{Bahcall81}. 

Fritz Zwicky used the estimate provided by \cite{Hubble31} for the Coma cluster and applied the virial theorem to obtain an estimate of its mass. This was the first kinematical approach to constrain the total matter distribution in systems of galaxies \citep{Zwicky1933}. Zwicky's analysis revealed that the visible mass was insufficient by more than an order of magnitude to gravitationally bind the system, leading to the conclusion that clusters must contain large amounts of unseen matter.
In fact, his estimates, adopting a modern value for the distance to Coma, implied mass-to-light ratios 
$\sim 300\text{--}500 \, M_\odot/L_\odot$,
suggesting that more than $\sim 90\%$ of the cluster mass was not accounted for by luminous matter alone. This discrepancy, which Zwicky famously attributed to the presence of {\it dunkle Materie}, represents the first compelling dynamical evidence for what is now known as DM on Mpc scales \citep{Zwicky1933,Zwicky1937}.

In the following decades, similar virial analyses were applied to other nearby clusters, confirming systematically large mass-to-light ratios and reinforcing the conclusion that galaxy clusters are dominated by unseen matter \citep[e.g.][]{Smith1936,Page1952,Abell1958}. However, progress in cluster kinematics remained slow for much of the mid-twentieth century, primarily due to the limited availability of spectroscopic data. Early studies typically relied on small galaxy samples.

The situation began to change in the 1970s, when improved spectroscopic capabilities enabled the construction of larger redshift samples for several clusters. The ubiquity of DM on large scales, complementing independent evidence from galaxy rotation curves, and the need for a dominant non-baryonic mass component, were confirmed by comparisons between the increasingly more accurate optically-derived virial masses, X-ray–derived masses, and cluster luminosities \citep{Bahcall1977,FormanJones1982}. By the late 1970s, clusters had thus emerged as key laboratories for testing gravitational physics and for studying the DM distribution in the Universe.

Early determinations of cluster virial masses implicitly assumed dynamical equilibrium and spherical symmetry. The increased number of redshifts for cluster members allowed to identify deviations of the
cluster velocity distributions from simple Gaussian shapes and showed that many systems exhibit multiple kinematic components \citep{Yahil1977,GellerBeers1982}, as originally suggested by \citet{vandenBergh61}. Such features provided early evidence that clusters are dynamically complex systems with several substructures, frequently affected by recent accretion and mergers. These findings challenged the simplistic view of clusters as relaxed, virialized objects and highlighted potential biases in virial mass estimates. The early evidence for the complex structure of clusters was strengthened by the analyses conducted in the following decade \citep{DresslerShectman1988,WestBothun1990}, that confirmed the emerging picture of clusters as continuously evolving systems within the hierarchical framework of structure formation.

The discovery of complex structure of clusters prompted investigations on the departure from dynamical equilibrium and its effects on kinematical mass estimates. Early numerical and analytical studies already pointed out that virial and Jeans-based methods can be significantly biased by subclustering, streaming motions, and mergers \citep[e.g.][]{White1976,Merritt1987}. Observational analyses combining positional and velocity information showed that dynamically young clusters --- often identified through substructure tests or non-Gaussian velocity distributions --- tend to exhibit inflated velocity dispersions and, consequently, overestimated virial masses \citep[e.g.][]{DresslerShectman1988,Bird94,Girardi_1997}. Moreover, several studies started to address the problem of interlopers --- foreground and background galaxies that fall in the projected radial range of a cluster --- as a possible source of systematic bias in cluster mass estimates based on kinematics (see, e.g. \citealt{Yahil1977,denHartogKatgert1996}). 

In parallel, theoretical work clarified the limitations of kinematic mass determinations based solely on line-of-sight velocities. In particular, the degeneracy between the total mass profile and the orbital anisotropy of cluster galaxies was formally identified \citep{BinneyMamon1982,Merritt1987}. Different combinations of gravitational potential and velocity anisotropy profiles can reproduce the same projected velocity dispersion profile, placing fundamental constraints on the information content of kinematic data. 

At the same time, increasingly detailed spectroscopic surveys revealed that cluster galaxy populations were not dynamically homogeneous. Observational studies uncovered clear evidence for morphological and kinematical segregation, with early-type galaxies, in particular the brightest ones, preferentially concentrated toward cluster centers and exhibiting lower velocity dispersions than late-type systems \citep[e.g.][]{Dressler1980,Sodre1989,Biviano+92,CollessDunn1996}. These findings were interpreted as signatures of environmental processes shaping the evolutionary path of galaxies in clusters, reflected in their distribution in projected phase-space \citep[p.p.s. hereafter, see, e.g.,][]{Biviano+02,Collister2005}. 

By the end of the twentieth century, galaxy kinematics had therefore evolved from Zwicky’s original virial argument into a mature, yet intrinsically complex, diagnostic of cluster mass, dynamical state, and assembly history. The advent of large spectroscopic programs such as the Canadian Network for Observational Cosmology (CNOC) survey \citep{Yee1996}, and the ESO Nearby Abell Cluster Survey \citep[ENACS,][]{Katgert+96}
played a pioneering role in this transformation. By stacking clusters and analyzing their average properties, many studies began to combine spatial and velocity information to constrain mass profiles and orbital structure \citep{Carlberg_1997,van_der_Marel_2000,Biviano2004}, bridging the gap between classical virial estimates and the fully resolved dynamical analyses of the modern era.

A more comprehensive historical overview of the first two centuries of cluster research, including the development of dynamical studies, is provided by \citet{Biviano2000}, who documents how clusters gradually emerged as key laboratories for studying dark matter, large-scale structure, and galaxy evolution.

\subsection{The rise of modern dynamical studies}
\label{sec:rise}
Beyond the steady growth of spectroscopic samples, the final years of the XX century and the early years of the XXI century were characterized by a conceptual broadening of cluster dynamical analyses. The focus shifted from establishing global mass estimates and identifying general dynamical properties to constructing self-consistent models of the internal structure of clusters. Rather than relying primarily on velocity dispersion profiles, increasing attention was devoted to the full p.p.s. --- that is, the projected positions of galaxies and their line-of-sight, rest-frame velocities  --- and the interplay between equilibrium and non-equilibrium processes.

In this context, dynamical models based on the moments of the distribution function in  phase space (collisionless Boltzmann and Jeans' equations) became more systematically implemented with parametric descriptions of both the mass profile and the velocity anisotropy, often informed by numerical simulations \citep[e.g.,][]{Cole1996}. The use of higher-order velocity moments and likelihood-based approaches helped to extract additional information from increasingly rich datasets \citep[e.g.][]{van_der_Marel_2000,LokasMamon2003,Katgert2004}. At the same time, alternative techniques that do not assume strict dynamical equilibrium gained prominence. \citet{DiaferioGeller97} and \citet{Diaferio1999} developed the Caustic method that allows the reconstruction of mass profiles from the escape-velocity envelope in the p.p.s., extending dynamical studies well beyond the virial radius and into the infall region.

\begin{figure}
         \centering
         \includegraphics[width=\textwidth]{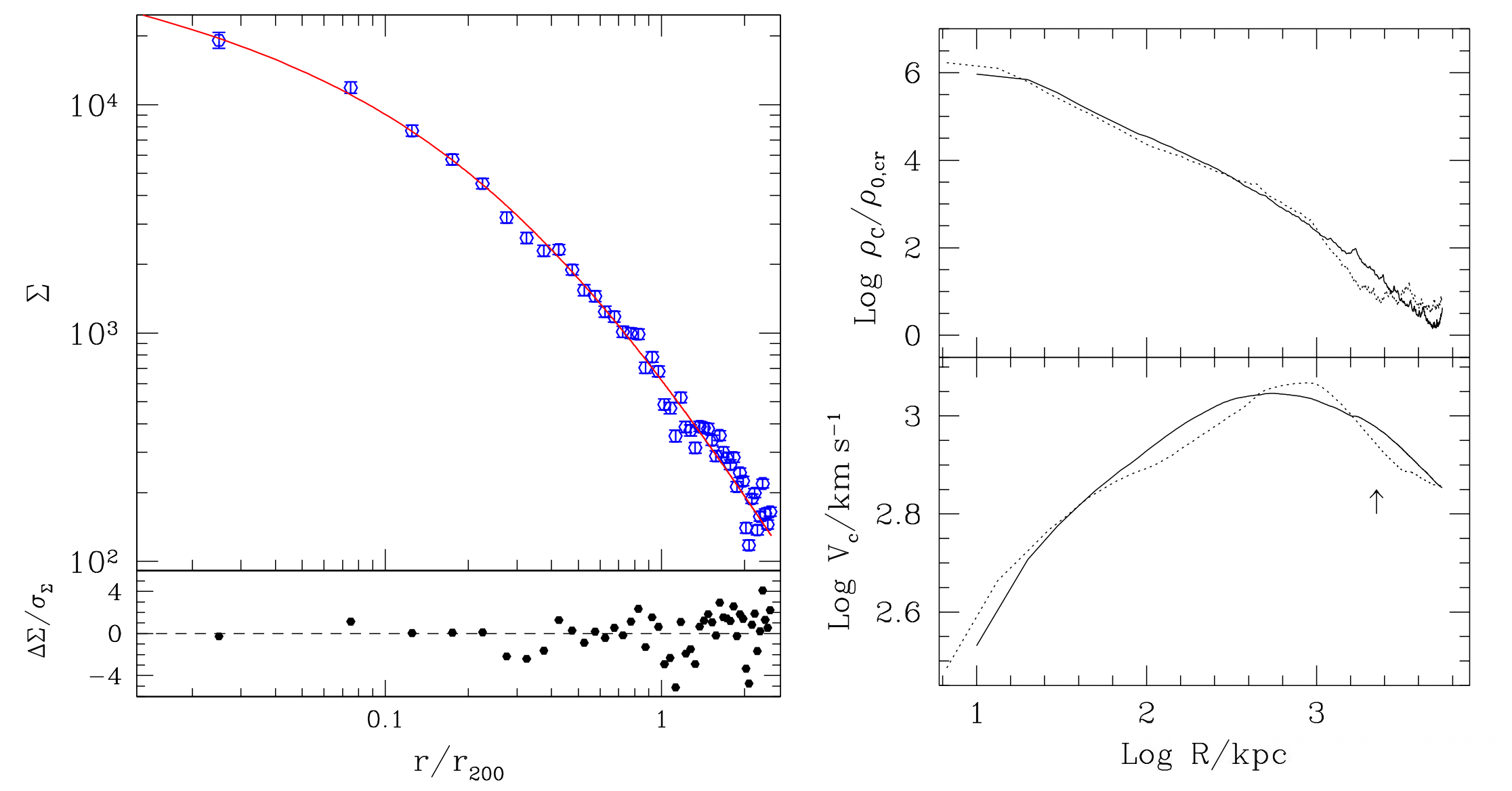}
         \caption{\label{fig:lin} Left panel: projected radial distribution of $\sim 6000$ galaxies in a stacked cluster, and the best-fit NFW profile (red solid line). Right panel: density (top) and circular velocity (bottom) of a simulated DM halo of $4.6\times 10^{14} \,{\rm M}_\odot$ at $z=0$ (solid curve) and at $z = 0.5$ (dotted curve).
         Images reproduced with permission from [left] \citet{Lin2004}, copyright by AAS; and from [right] \citet{Ghigna1998}, copyright by RAS.}
        \end{figure}

These methodological advances proceeded alongside a rapid progress in cosmological simulations. N-body CDM simulations developed in the 90's, {provided detailed expectations for the orbital structure and anisotropy of DM particles and subhalos \citep[e.g.][see the right panel of Fig.~\ref{fig:lin}]{Cole1996,Ghigna1998}. In particular, they predicted a nearly-universal halo density profile, that became known as the NFW profile \citep{Navarro1996,NFW97}: 
\begin{equation}
    \rho(r) = \frac{\rho_0}{(r/r_\text{s})(1+r/r_\text{s})^2}\,,
\end{equation}
where $\rho_0$ is the central density, and 
the scale radius $r_\text{s} \equiv r_{-2}$, is the radius at which the logarithmic slope of the density is $-2$. The central density can also be rephrased in terms of the radius $r_{200}$ (or the corresponding mass $M_{200}$), 
\begin{equation}
    \rho_0 = \frac{M_{200}}{4\pi r_\text{s}^3\left[\ln (1+c_{200})-c_{200}/(1+c_{200})\right]}\,,
\end{equation}
where $c_{200} = r_{200}/r_\text{s}$ is the concentration parameter, and $M_{\Delta} = 4\pi\,\rho_c(z)\, r^3_\Delta/3 $ is the mass at the radius $r_\Delta$ enclosing a spherical overdensity $\Delta$ times the critical density of the Universe $\rho_c(z)$ at redshift $z$.} 
Observational studies began to test whether galaxies in clusters trace the underlying DM distribution. Satellite galaxies in clusters were found to follow approximately a NFW number density profile \citep{Lin2004} (left panel of Fig.~\ref{fig:lin}).  

In parallel to these developments, simpler global estimators based on velocity dispersion measurements were extensively explored as practical mass proxies. Under the assumption of virial equilibrium, the total mass of a cluster, $M$, is expected to scale with the galaxy velocity dispersion, $\sigma_v$, following a relation that naturally emerges from self-similar models of halo formation,
 $M \propto \sigma_v^3 / H(z)$, where $H(z)$ is the Hubble constant at the cluster mean redshift, $z$, {and $M$ denotes the mass within a radius enclosing a specific overdensity, as will be detailed in Sect.~\ref{sec:scaling}}. This expectation has been quantitatively confirmed and calibrated in numerical simulations, which revealed a tight correlation between halo $M$ and DM $\sigma_v$, with a slope close to the virial prediction and a small intrinsic scatter \citep{Evrard2008}. Numerical simulations incorporating baryonic physics and subhalo populations further refined this picture, highlighting the impact of tracer selection and galaxy formation processes on the normalization and slope of the relation as a function of redshift \citep[e.g.][]{Munari2013,Saro_2013,Armitage+19a}.

The increased complexity of kinematical analyses also brought the impact of systematic effects into a sharper focus. The treatment of interlopers became a central concern (see e.g. \citealt{Mamon2010}) leading to the development of refined phase-space selection algorithms calibrated on simulations \citep[e.g.][]{Lokas2006,Wojtak2007,Mamon_2013, SD13}. 

At the same time, the advancement in cosmological simulations allowed to quantify systematic effects induced by a disturbed cluster morphology or departures from dynamical equilibrium with an increasing level of detail in controlled frameworks. The analysis of samples of mock clusters demonstrated that projection effects and deviations from equilibrium introduce non-negligible biases and scatter in the recovered mass profiles and velocity anisotropy, even when large spectroscopic samples are available \citep[e.g.][]{vanHaarlem1993,Biviano2006,Wojtak2007}. These studies clearly indicated that the dynamical state of clusters and their assembly history are in tight connection to the reliability of kinematical mass measurements. Recent works have also shown that, if not properly accounted for, the (non-) equilibrium state of clusters is a substantial source of bias also when stacked samples are considered \citep{Cai2025}. All such findings motivated the systematic calibration of dynamical methods against realistic mock catalogues, and the development of robust criteria to classify the dynamical state of a cluster \citep[e.g.,][]{Knebe2000,Old_2017,Ferragamo2020,Pizzuti2020syst,Zhang22,Benavides23,Cerini2023}. 

In the last decade, cluster kinematics has thus entered a genuinely modern phase: no longer limited to global $\sigma_v$ measurements, it has became a multidimensional diagnostic which combines phase-space information, theoretical modeling and expectations from cosmological simulations, with modern, increasingly rigorous statistical and numerical approaches \citep[e.g.][]{Ho_2019,Shi_2024}. This transformation set the stage for the precision era of cluster kinematical studies; the information extracted from large spectroscopic samples can now be translated into precise constraints on cluster mass profiles with interesting cosmological implications.

\subsection{Cluster kinematics and the dark sector}

Reconstructing cluster mass distribution from galaxy kinematics has long been a primary motivation for dynamical studies. Mass profiles inferred from galaxy motions can be directly compared with those obtained from X-ray observations and gravitational lensing, providing critical cross-checks and helping to identify and disentangle systematic uncertainties arising in different mass probes \citep{Pratt2019}. 

The scientific scope of cluster kinematics has expanded significantly beyond its original goal, over the past two decades. In particular, cluster kinematics is increasingly recognized as a probe of fundamental physics. The motions of galaxies respond to the gravitational potential in which they orbit, making them sensitive to both the distribution of (visible and dark) matter and the laws of gravity \citep{Schmidt_2010,Gronke2016,Pizzuti17}. In modified gravity (MG hereafter) theories, galaxies may experience additional forces or altered dynamical relations, potentially leading to discrepancies between dynamical and lensing mass estimates (e.g. \citealt{Pizzuti26} and references therein). As a result, cluster kinematics offers a complementary approach to testing gravity on Mpc scales, where screening mechanisms may operate differently than in the Solar System \citep{Burrage_2018,Brax_2021}.

At the same time, the dynamics of galaxies in clusters provides insight into the growth of structure and the influence of dark energy (DE hereafter), especially in the cluster outskirts \citep{Merafina14}. Detailed kinematical reconstructions of cluster masses further allow investigating the properties and nature of DM \citep{Serra2011a,SartorisDM,Biviano_2023}. 
Moreover, the assembly history of clusters shapes the orbital structure of their galaxy populations, producing characteristic trends in the velocity anisotropy profile which depend on the galaxy population \citep[e.g.][]{Lau2010,Mamon2019,Maraboli+26,Pedratti26}. These features are tightly connected to the dynamical state of the cluster and to its recent merger history, linking kinematics of clusters to the evolution of the large-scale structure \citep{Miyatake2025}.

\subsection{Beyond equilibrium: kinematics in the outskirts of clusters}
One of the most significant conceptual developments with respect to traditional methods, has been the growing emphasis on the outer regions of clusters. While the inner regions are often close to a dynamically stable configuration, cluster outskirts are dominated by infall, mergers, and ongoing accretion from the cosmic web
\citep[e.g.,][]{Haines+18,Malavasi2020}.
In these regions, traditional equilibrium-based methods become unreliable, but galaxy kinematics remains informative when interpreted in projected phase space and in combination with other approaches \citep{Diaferio2005,Umetsu25}.

The identification of caustics - sharp features in the galaxy p.p.s. distribution \citep{Geller_1999,Diaferio1999} has provided a way to estimate cluster mass profiles well beyond the virial radius without assuming equilibrium {\citep[e.g.,][]{Geller_1999,Rines+00,Biviano_2003,Rines+03,RinesDiaferio06,Rines+13,Guennou+14,Sohn2019ApJ...871..129S,Pizzardo21}}. This technique has also provided the possibility of a direct measurement of the mass accretion rate of clusters \citep{DeBoni+16, Pizzardo21, Pizzardo22, Pizzardo23} and has opened a new observational window onto cluster assembly and mass accretion \citep[e.g.,][]{
Walker2019,Pizzardo23}, independently of the dynamical state. 

The outskirts of clusters are also the perfect environment where the connections between kinematics, cosmology, and gravity become particularly relevant. Features such as the splashback radius \citep[e.g.,][]{Adhikari2014} link galaxy dynamics to the recent accretion history of clusters, while providing physically motivated definitions of cluster boundaries that go beyond traditional overdensity radii \citep{DK14,DMKM17,Bianconi+21}. The orbits of galaxies further show typical features at large scales \citep{Abdullah2025}, where the balance between outgoing and infalling material provides specific features which may depend non-trivially on the cosmological model. At the same time, beyond the virial radius the density rapidly decreases, marking a transition region between the screened and unscreened regimes of many theoretically viable MG models; these regions thus appear to be ideal for detecting possible signatures of departures from GR \citep{Adhikari_2016}.

Overall, such a multifaceted picture demonstrates that kinematical studies of galaxy clusters keep climbing 
a steep rising path: with a solid and broad methodological framework, sustained by an increasingly robust control of systematics, and an increasing amount of high-quality spectroscopic data,
cluster kinematics is reaching 
the maturity required to deliver a wide range of results, spanning from astrophysical processes to cosmological constraints. The versatility of the approaches proposed in the last decades --- which will be discussed in the next section ---  is now becoming optimal to fully exploit the forthcoming generation of large spectroscopic datasets.

\section{Kinematics and mass profiles: from the virial theorem to Machine Learning approaches}
\label{sec:methods}

In this Section, we review the basic aspects and theoretical motivations of the methods used to reconstruct galaxy cluster mass profiles based on kinematics. The basic observables in all these methods are the projected spatial distribution of cluster galaxies and their los velocities. Methods that require direct distance measures for cluster galaxies, 
such as the Action model of \citet{TM04} or the orbital roulette method of \citet{BL04} are not discussed in this review, as three-dimensional (3D hereafter) distances for cluster galaxies are not currently measurable with the needed accuracy, except for the very nearby clusters, Virgo and Fornax, so these methods are not of general use.

We start presenting simple global estimators, such as the virial theorem, eventually progressing toward the description of modern approaches based on more sophisticated phase-space analyses and ML techniques.

\subsection{The virial theorem}\label{sec:virial}

The earliest dynamical estimates of cluster masses \citep{Zwicky1933,Zwicky1937} are based on the virial theorem, which relates the total kinetic $K$ and potential energies $U$ of an isolated gravitationally bound system in equilibrium:
\begin{equation}\label{eq:virial}
2K + U = 0.
\end{equation}
The total kinetic energy of a self-gravitating system of $N$ galaxies can be written as
\begin{equation}\label{eq:kin}
K = \frac{1}{2} \sum_{i=1}^{N} m_i \langle v^2 \rangle\,,
\end{equation}
where $m_i$ is the mass of a test galaxy, and $\langle v^2 \rangle$ is the mean quadratic velocity of the galaxies. Note that observations typically access only the los component of the velocities. If galaxy orbits are close to an isotropic distribution, or if most of the radial extent of the cluster is covered observationally and the cluster is approximately spherical \citep{TheWhite1986,Girardi98}, we can write
$\langle v^2 \rangle\, \simeq  3\,\langle v^2_\text{los} \rangle = 3\,\sigma_\mathrm{los}^2$, where $\sigma_\mathrm{los}^2$ is the los velocity dispersion, and the last equality comes from the condition that velocities are computed in the cluster rest frame, i.e. $\langle v \rangle = 0$. 

If each galaxy is at a (3D) distance ${\bf r}_i$ from the cluster center, the total  potential energy can be written in terms of the potential as 
\begin{equation}\label{eq:pot}
    U = - \sum_i{\bf r}_i\cdot m_i\nabla\Phi({\bf r}_i)\,\sim - G\frac{\mathcal{M}_\mathrm{tr}M}{R}\,,
\end{equation}
with $\mathcal{M}_\mathrm{tr} = \sum_i m_i$ being the total mass of the galaxies, tracers of the gravitational potential, and $R$ is the typical radius of a cluster within which the virial theorem holds, also called the virial radius.

Combining  eqs.~\eqref{eq:kin}, \eqref{eq:pot} into eq.~\eqref{eq:virial}, we obtain the classical virial mass estimator:
\begin{equation}
M \sim \frac{3 \, \sigma_\mathrm{los}^2 \, R}{G}\, .
\end{equation}
if all the masses $m_i$ of the individual galaxies are comparable. 

More refined formulations \citep[e.g.][]{Limber1960,Heisler85} are based on the projected virial theorem, which explicitly accounts for the fact that only projected positions and los velocity are observable. In this framework, the mass estimator can be written as
\begin{equation}
M_\mathrm{v} = \frac{3\pi}{2G} \sigma_\mathrm{los}^2 R_\mathrm{PV},
\end{equation}
where $R_\mathrm{PV}$ is the projected virial radius, defined as the inverse harmonic mean of the projected pairwise separations $R_{ij}$:
\begin{equation}
R_\mathrm{PV} = \frac{N(N-1)}{\sum_{i<j} R_{ij}^{-1}}.
\end{equation}

In practical applications, the virial theorem has been modified to account for the fact that galaxy clusters are observed within a finite aperture, rather than as fully enclosed systems. In this case, an additional surface pressure term arises \citep{TheWhite1986,AE11}, leading to
\begin{equation}
2K + U = S,
\end{equation}
where $S$ depends on the density and velocity dispersion of the tracers at the system boundary. Neglecting this term generally leads to an overestimation of the cluster mass
\citep{MP16}. Corrections for this effect have been discussed in detail in the context of galaxy clusters \citep[e.g.,][]{Girardi98,Biviano2006}, and typically amount to $\sim 10$--$30\%$ depending on the radial extent of the data and the orbital structure of the tracer population. In \citet{AE11}, expressions of the virial theorem inclusive of the surface pressure term are presented for a variety of mass profiles.

The main problem for the application of the virial theorem to clusters of galaxies is that it assumes that galaxies, the tracers of the potential, have a similar spatial distribution as the total mass. This is generally known as the "mass traces light" assumption.
As shown by \citet{Smith80} and \citet{Merritt1987}, if galaxies and total matter do not share similar distributions, the virial theorem can lead to substantial under- or over-estimates of the cluster mass. Other methods have established that red, passive galaxies have a distribution similar to that of the cluster mass \citep[e.g.,][]{Gray_2002,Biviano_2003}, so they are the most appropriate population of cluster galaxies to be used as tracers for the virial mass estimator.

The virial mass estimator has been extensively applied to galaxy clusters in early and intermediate stages of spectroscopic surveys \citep{WojtakLokas2007,Evrard2008, Old_2015}, thanks to its simplicity and minimal observational requirements. 
\cite{Abdullah_2020} have shown that this approach can still be effectively implemented in modern large spectroscopic datasets, provided that accurate membership selection and appropriate corrections for boundary terms are included. In particular, they combined advanced method for interloper rejection with virial mass estimates corrected for the surface pressure term, allowing for a consistent determination of cluster dynamical properties across a wide range of systems. Some works have further extended its application to non-standard frameworks \citep{Harko_2007,LopezC22}. 

\begin{figure}
         \centering
         \includegraphics[width=\columnwidth]{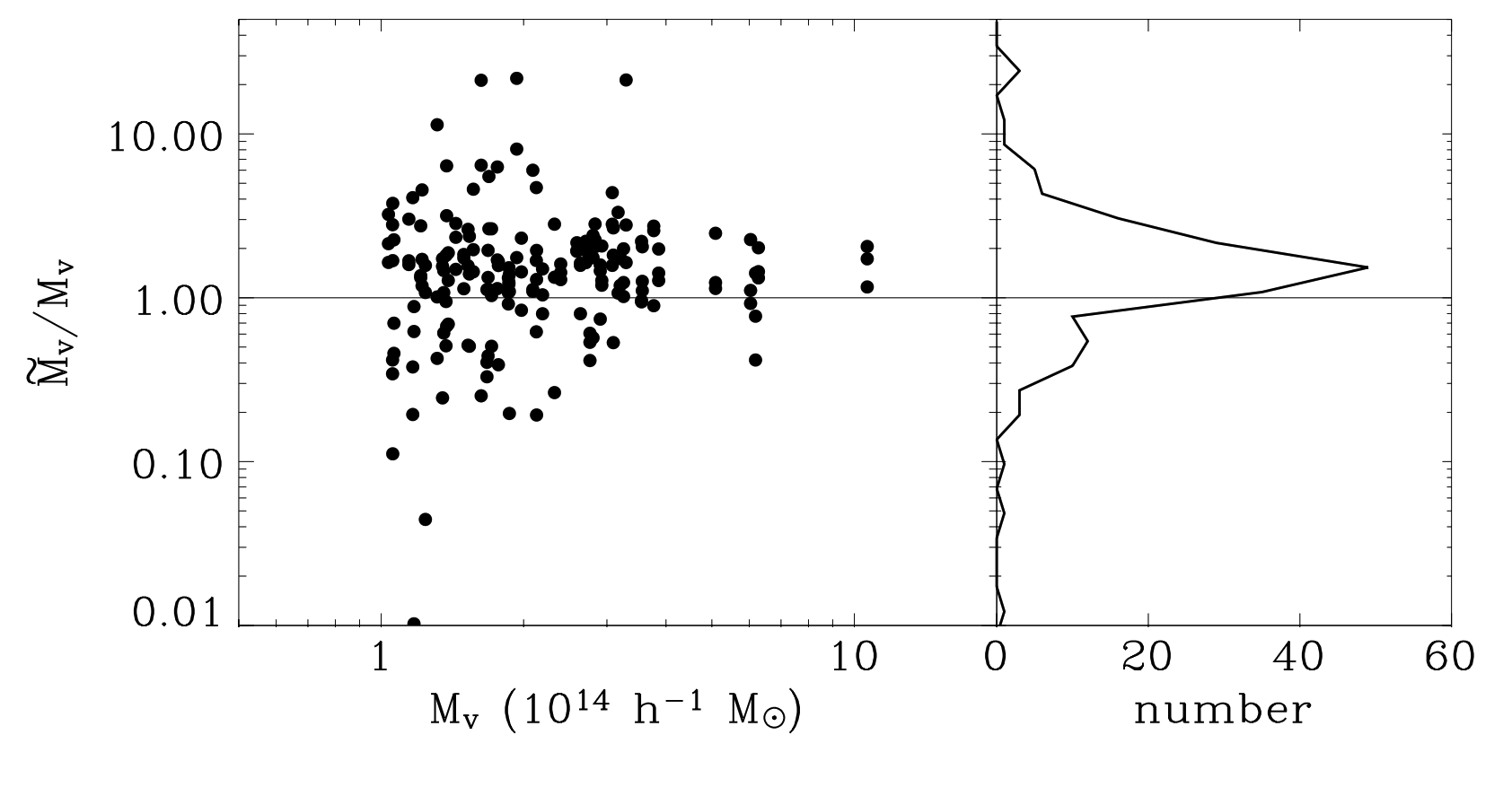}
         \caption{\label{fig:massVir}  Left panel: ratio between the virial and true mass, $\tilde{M}_v/M_v$ vs. the true mass, $M_v$, of simulated cluster-size halos. 
         Right panel: distribution of $\tilde{M}_v/M_v$. The virial masses are estimated considering all the galaxies within the projected radius $1.5\,h^{-1}$ Mpc. 
         Image reproduced with permission from \citet{Biviano2006}, copyright by ESO.}
    \end{figure}

Despite its simplicity, tests based on numerical simulations and mock galaxy catalogues have shown that the overall performance of virial mass estimates is affected by a substantial intrinsic scatter and bias, reflecting the combined impact of several effects not captured by the simple virial framework, beyond the specific correction associated with the surface pressure term  \citep[e.g.,][]{Biviano2006,Old_2015}. As an example, Fig.~\ref{fig:massVir}, adapted from \cite{Biviano2006}, shows the comparison between the masses estimated using the (surface-pressure-term corrected) virial theorem $\tilde{M}_\text{v}$ and the true cluster masses ${M}_\text{v}$ for a set of clusters extracted from the cosmological hydrodynamical simulations of \cite{Borgani04}. The distribution exhibits a positive bias of $\sim 40\%$ in the ratio $\tilde{M}_\text{v}/M_\text{v}$, with a scatter of 0.85. This is mostly due to the presence of interlopers, and \citet{Biviano2006} showed that if the virial mass is estimated excluding tracers that entered the system only recently, the mass bias is strongly reduced, as most interlopers are removed from the sample. 

An alternative to the virial mass estimator that enjoyed some popularity in the past is the Projected Mass estimator \citep{Heisler85}, 
\begin{equation}
M_{PM}=\frac{32}{\pi G N} \Sigma_i (v_i - \overline{v})^2 R_i,
\label{eq:pm}
\end{equation}
where the sum is over the $N$ cluster galaxies with cluster-centric distance $R_i$ and los velocities $v_i$. This equation is valid for isotropic orbits; a more general formulation for different orbital distributions of cluster galaxies can be found in \citet{Perea+90}.
This method was shown to be more sensitive to the presence of interlopers than the virial theorem \citep{Perea+90}, and for this reason it is not widely used anymore, except for the detection of the interlopers themselves \citep[][see Sect.~\ref{sec:interlopers}]{WL07b}.

\subsection{The mass--velocity dispersion scaling relation}
\label{sec:scaling}
 While the virial theorem provides a baseline for cluster mass estimation, its intrinsic limitations have motivated the development of alternative approaches and --- more generally --- the use of statistical mass proxies. In particular, the connection between
 the $\sigma_v$ of cluster members, $M$, and global observational properties of galaxy clusters naturally leads to the formulation of specific scaling relations \citep[e.g.,][]{Biviano2006,Evrard2008,Andreon2010b,Munari2013,SE15}, which are now extensively used in the analysis of large cluster samples and cosmological surveys \citep{Barahona2022,Pandya2024Examining}.  
 
Under simple assumptions, one expects $M \propto \sigma_v^3$, as predicted by the self-similar model of gravitational collapse \citep{Kaiser1986}, in which galaxy clusters form through scale-free gravitational processes and are therefore expected to follow simple power-law scaling relations --- modulo additional dependencies on the cluster structure, redshift and dynamical state. A tight $M-\sigma_v$ relation is expected for a wide range of mass profiles, given the condition of dynamical equilibrium, as defined by the Jeans eq.~\eqref{eq:Jeans} below
\citep{Diemer+13}.

Cosmological simulations have been extensively used to calibrate such relation, showing that the halo $\sigma_v$ and its $M$ are connected by a tight power-law with relatively small intrinsic scatter under controlled conditions. In particular, \cite{Evrard2008} performed a detailed study of simulations with different cold dark matter cosmologies, physics and resolutions to find that the 3D
velocity dispersion of DM particles within the virial radius can be written as
\begin{equation}\label{eq:Evrard}
    \sigma_\mathrm{DM}(M,z) = \sigma_\mathrm{DM,15}\left(\frac{h(z)M_{200}}{10^{15}\,{\rm M_\odot}}\right)^\alpha,
\end{equation}
where the exponent and normalization show only a percent-level variation depending on the simulation detail, $\alpha = 0.336  \pm  0.003$, $\sigma_\mathrm{DM,15} = 1083 \pm 4$ km/s. In the above equation, $h(z)=H(z)/100$ is the reduced Hubble parameter at redshift $z$.

\cite{Saro_2013} extended this framework by investigating the $M-\sigma_v$ relation for galaxies in realistic observational conditions using the semi-analytic models of \cite{DeLucia2007} applied to cosmological simulations. Starting from eq.~\eqref{eq:Evrard} they calibrated the relation between the $\sigma_\mathrm{los}$ of halo galaxies and the halo $M$, finding a slope consistent with the self-similar expectation ($1/\alpha \simeq 2.9$) and a normalization in agreement with \cite{Evrard2008}'s result at the few percent level. However, they also showed that the same systematics affecting virial theorem-based mass estimates (e.g., triaxiality, interlopers, and the selection of galaxy tracers; see Sect.~\ref{sec:syst}) introduce significant scatter and potential biases, especially when only a few galaxies are used for the estimate of $\sigma_\mathrm{los}$.

A step forward in understanding the impact of baryonic physics and tracer selection on the $M-\sigma_v$ relation was made by \cite{Munari2013}, who analyzed a suite of cosmological simulations including different implementations of gas physics and feedback processes. They showed that while DM particles follow a scaling relation fully consistent with the self-similar expectation, $\sigma_\mathrm{DM} \propto M^{1/3}$, subhalos and simulated galaxies exhibit systematically steeper relations and higher normalizations. This effect reflects the presence of a velocity bias at the $\sim 10\%$ level, arising from dynamical processes such as dynamical friction (i.e. a gravitational drag on a massive body moving in a cloud of lighter objects)  and tidal disruption, which selectively affect galaxy tracers \citep[see also][]{Anbajagane2022}. As a consequence, the calibration of the $M-\sigma_v$ relation sensitively depends on the adopted tracer population and on the modeling of baryonic physics.
Nonetheless, this simple analytical scaling has been extensively used in the literature to estimate cluster masses \citep[e.g.,][]{Mu_oz_Rodr_guez_2022,Sif_n_2025,Rinaldi2026}, providing a powerful and versatile tool for cosmological applications.

       \begin{figure}
         \centering
         \includegraphics[width=0.7\columnwidth]{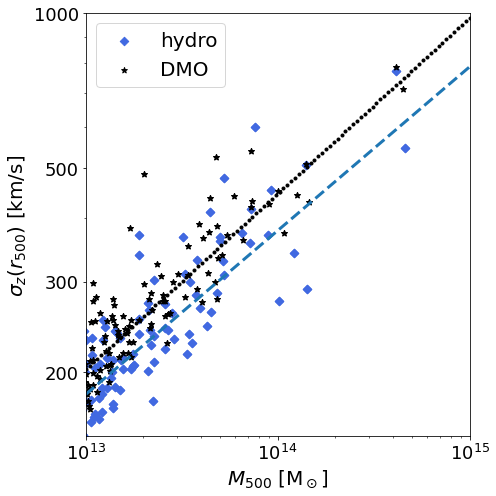}
         \caption{\label{fig:DIANOGA_scaling}  Example of scaling relations for $\sigma_{{\rm los}}$ computed at $r_{500}$ (labelled $\sigma_z$ in the figure),
         as a function of the mass $M_{500}$, from a cosmological $\Lambda$CDM simulation of the DIANOGA-SIDM suite \citep{Ragagnin2024}. The black stars and blue diamonds refer to the DM-only and hydrodynamical cases, respectively, with the best-fit scaliong relations indicated by the dotteed black and dashed blue lines, respectively. Reproduced with permission, copyright by ESO.}
    \end{figure}
    
For illustrative purposes, Fig.~\ref{fig:DIANOGA_scaling} presents an example of the scaling between the mass, $M_{500}$, and $\sigma_{{\rm los}}$, computed at the corresponding radius $r_{500}$, estimated for halos extracted from a $\Lambda$CDM simulation at $z=0.46$ assuming a DM only (black) and hydrodynamical (blue) setup. The snapshot comes from the DIANOGA-SIDM cosmological simulations suite of \cite{Ragagnin2024}. In the hydrodynamical run, $\sigma_\mathrm{los}$ are computed by considering all the particles (baryons and DM). While the slope of the scaling relation is consistent between the two cases ($\alpha = 0.34, 0.31$ for the DM-only and the hydrodynamical setups, respectively), the normalization and scatter differ, with a larger spread found when baryons are included.

\begin{figure}[ht]
         \centering
         \includegraphics[width=0.8\columnwidth]{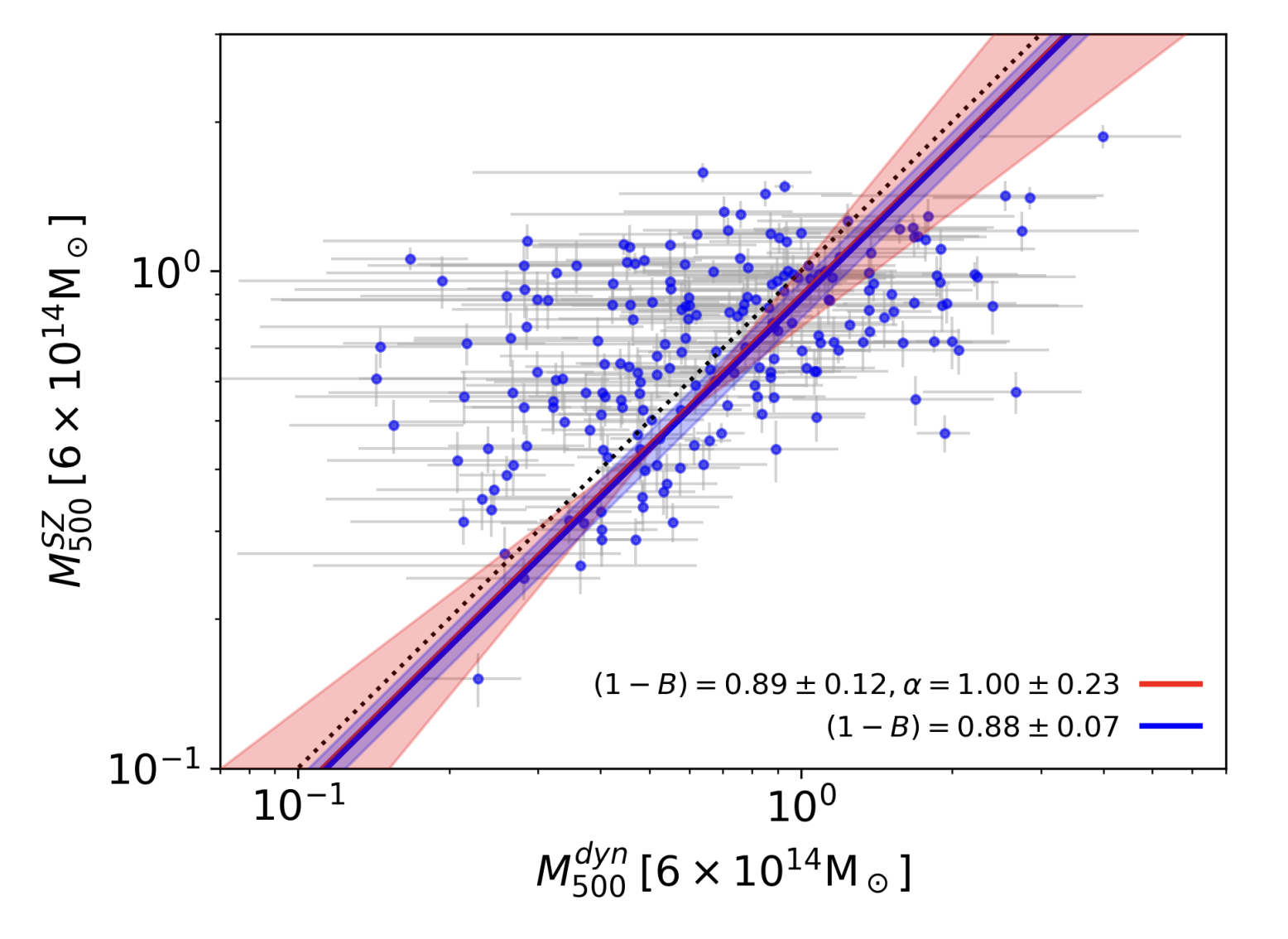}
         \caption{\label{fig:Ferra}  Comparison between dynamical and SZ-based masses of the cluster sample analysed by \cite{Ferragamo2021Velocity}. Blue and red lines represent the scaling relation between the two mass estimates obtained by fixing its slope to unity, and by letting the slope be a free parameter, respectively. The corresponding bands are the 68\% confidence regions and the black dotted line marks the identity relation. Reproduced with permission, copyright by ESO.}
    \end{figure}

Following the work of \cite{Munari2013}, subsequent studies have focused on refining the $M-\sigma_v$ relation by explicitly quantifying the impact of galaxy velocity bias and by extending its calibration to realistic observational regimes. More generally, the use of $\sigma_{{\rm los}}$ as a mass proxy has been framed within the broader context of cluster scaling relations and their cosmological applications \citep[e.g.,][]{SE15}. In particular, large spectroscopic surveys have enabled the use of $\sigma_{{\rm los}}$ as a statistical mass proxy across sizable cluster samples. For instance, \cite{Ferragamo2021Velocity} analyzed a sample of Planck-selected clusters with spectroscopic follow-up, deriving dynamical masses through scaling relations based on $\sigma_{{\rm los}}$. They then compared these dynamical mass estimates to the masses obtained from the Sunyaev--Zel'dovich (SZ) effect on the Cosmic Microwave Background spectrum (Fig.~\ref{fig:Ferra}). They found that the SZ-based masses are systematically lower than the dynamical estimates, corresponding to a mass bias of $(1 - b) \simeq 0.83$, consistent with the level of bias commonly inferred in SZ cluster cosmology analyses.

A major step forward in the interpretation of the $M-\sigma_v$ relation was achieved by \cite{Anbajagane2022}, who calibrated the galaxy velocity bias at nearly percent-level precision using a suite of cosmological simulations. 
Their results strengthened and extended \cite{Munari2013}'s results: the normalization, slope, and scatter of the $M-\sigma_v$ relation not only depend on halo mass and redshift, but also on the selection of galaxy tracers, as quantified through the adopted stellar-mass threshold. In particular, different tracer selections lead to systematic variations in the measured velocity dispersion, reflecting the impact of dynamical processes such as dynamical friction \citep{Adhikari_2016}, which preferentially affect the most massive galaxies.

The increasing availability of large and homogeneous datasets of clusters has further made it possible to investigate the role of the dynamical state in shaping kinematic scaling relations. 
In particular, confirming earlier indications based on numerical simulations \citep{Biviano2006},
\cite{Damsted2023} showed that clusters exhibiting significant velocity substructure tend to have enhanced $\sigma_{{\rm los}}$ at fixed richness, as well as a substantially larger intrinsic scatter, indicating that departures from dynamical equilibrium contribute significantly to the observed variance of the $M-\sigma_v$ relation. 

The theoretical $M-\sigma_{{\rm v}}$ relations
of \citet{Evrard2008} and/or \citet{Munari2013} were supported by the comparison of observed $\sigma_{{\rm los}}$ with mass estimates obtained from SZ data \citep{Rines2016}
and gravitational lensing  
\citep{SE15,Sereno25}, for clusters at $z \lesssim 0.9$. At higher $z$ ($0.9 \lesssim z \lesssim 1.4$),
\cite{Abdulshafy2025Dynamical} analyzed a sample of 14 clusters from the GOGREEN and GCLASS surveys \citep{Balogh+17}, and derived dynamical masses through the virial mass estimator combined with an advanced membership assignment technique (see Sect.~\ref{sec:syst}). They obtained a scaling relation 
in good agreement with simulation-based calibrations such as those of \cite{Evrard2008} and \cite{Munari2013}. 
These results highlight the potential of $\sigma_\mathrm{los}$ to infer masses even at high redshift. However, the normalization and scatter of the relation remain sensitive to the adopted membership selection and to the specific dynamical mass estimator, as well as to the limited size of current high-$z$ spectroscopic samples.

Compared to other commonly used mass-scaling relations, $\sigma_v$ offers a complementary perspective. X-ray and SZ-based relations, such as $Y_\mathrm{X}$--$M$ or $Y_\mathrm{SZ}$--$M$, are known to exhibit relatively low intrinsic scatter and are among the most robust mass proxies. Here, $Y_\mathrm{X}$ is defined as the product of the gas mass and temperature, $Y_\mathrm{X} \equiv M_\mathrm{gas}\,T_X$, tracing the total thermal energy of the intracluster medium \citep{Kravtsov2006,Short_2010}, while $Y_\mathrm{SZ}$ is the integrated Compton-$y$ parameter, proportional to the los integral of the electron pressure, and provides a direct measure of the thermal energy content of the ICM \citep{Arnaud2010}. These observables are found to correlate tightly with $M$, with relatively small intrinsic scatter \citep[e.g.,][]{Angulo2012,Pratt2019}. However, the intrinsic scatter of the $M-\sigma_v$ relation is also quite small, $\sim 15$\%, according to \citet{SE15} and \citet{Seppi+25}. 

X-ray and SZ-based scaling relations rely on assumptions about the thermodynamical state of the ICM, in particular hydrostatic equilibrium, and are therefore sensitive to non-thermal pressure support and departures from equilibrium \citep[e.g.,][]{Nelson2014,Eckert2019}. Analyses based on large cluster samples have further refined these relations and quantified their scatter, selection effects, and dependence on cluster properties, confirming their overall robustness while highlighting residual systematics related to gas physics and dynamical state \citep[e.g.,][]{Bocquet2019,Salvati2019,Adam_2024}. 

Weak-lensing measurements provide a relatively direct probe of the underlying gravitational potential, without relying on assumptions about the dynamical or thermodynamical state of the cluster. For this reason, weak-lensing masses have played a key role in establishing relations between halo mass and purely photometric observables, such as the optical richness, which can be efficiently measured in large imaging surveys \citep[e.g.,][]{Rozo2010,Simet2017,McClintock2019}. Nevertheless, weak-lensing measurements are affected by projection effects from large-scale structure along the los, as well as by shape noise and other sources of statistical uncertainty, leading to significant scatter for individual systems \citep[e.g.,][]{Hoekstra2013,Umetsu20,Grandis2021,Giocoli25}. 

Interesting correlations with cluster masses have also emerged from radio observations, which probes the non-thermal component of the intracluster medium, traced by synchrotron emission from relativistic electrons in the presence of $\mu$G-level magnetic fields. In particular, diffuse radio halos have been shown to follow a scaling relation of their total radio power $P_\nu$ with the cluster mass. These relations are now well established over a broad frequency range and reflect the connection between gravitational energy released during mergers and the generation of turbulence that re-accelerates relativistic particles \citep[e.g.,][]{Cassano2007,Cuciti2023}. Analyses based on homogeneous samples combining LOFAR observations and X-ray-selected clusters have further extended these studies, showing that the $P_\nu$--$M$ relation can be recovered with relatively small scatter once the dependence on mass and redshift is properly accounted for \citep{Balboni2025}. At the same time, these works highlight a significant dependence on the dynamical state of the system, with merging clusters hosting brighter and more extended radio halos, while more relaxed systems tend to be radio-quiet or lie below the main correlation. This behaviour reflects the transient nature of the non-thermal emission and its strong link to cluster assembly processes.
Unlike X-ray and SZ observables, radio emission does not trace the bulk thermal energy of the intracluster medium, but rather the interplay between turbulence, cosmic rays, and magnetic fields. As a consequence, radio-based scaling relations generally exhibit a more complex dependence on cluster properties, including merger history and magnetic field strength. Nevertheless, they provide a unique probe of the non-thermal energy budget of clusters, offering complementary constraints on cluster masses and on the physical processes driving structure formation.

In contrast, the $M-\sigma_v$ relation  is largely insensitive to gas
physics, unlike thermodynamical observables such as X-ray or SZ quantities, or non-thermal tracers such as radio emission, and can be efficiently applied to large spectroscopic datasets. Comparison of cluster masses derived from kinematics and from thermodynamic observables can be used to assess the departure from full thermalization of the intra-cluster gas, and estimate the hydrostatic mass bias \citep[e.g.][]{Andreon+17,FBC17,Lovisari+20}.
While the estimate of $\sigma_v$ is sensitive to projection effects just as the gravitational lensing mass estimates, rich groups projected along a cluster line-of-sight can be detected by their location in the p.p.s., and removed from the analysis in the mass estimates based on galaxy kinematics \citep[see, e.g., the lensing vs. kinematic mass estimate comparison in][Fig.~8 in that paper]{Biviano_2023}.


Since the $M-\sigma_v$ relation is not independent of the dynamical state of the system and the properties of the tracer population, it is most effectively used in a statistical framework, where its combination with other observables, that are subject to different systematics,
enables a more complete and robust characterization of cluster masses and of the underlying physical processes driving their assembly.

\subsection{Dynamical estimates based on the projected phase-space distribution}

Velocity dispersions offer only a partial view of the whole kinematical information that can be extracted from galaxy clusters. Indeed, more sophisticated techniques to infer cluster mass profiles rely on the combination of positions and velocities of cluster member galaxies. As mentioned in Sect.~\ref{sec:virial}, only projected positions and line-of-sight velocities are accessible to the observer, i.e., only half of the full six-dimensional phase-space $(\bf{x},\bf{v})$. As a consequence, the reconstruction of the underlying dynamical structure requires the adoption of physically motivated assumptions to deal with the missing information.

A widely adopted framework is to treat galaxy clusters as collisionless systems in dynamical equilibrium, in which galaxies act as tracers of the underlying gravitational potential. Under these assumptions, the phase-space distribution function evolves according to the collisionless Boltzmann equation, which provides the theoretical basis for most dynamical modelling techniques we discuss in this review.

The validity of the collisionless approximation can be justified by estimating the two-body relaxation time of a cluster. For a system of $N$ particles with characteristic crossing time $t_{\mathrm{cross}} \sim R / \sigma_v$, the relaxation time can be written as
\citep{BT87}
\begin{equation}
t_{\mathrm{relax}} \approx \frac{0.1 \, N}{\ln N} \, t_{\mathrm{cross}}.
\end{equation}
In the case of galaxy clusters, typical values are $N \sim 10^3$ galaxies, $R \sim 2\,\mathrm{Mpc}$, and $\sigma_v \sim 10^3\,\mathrm{km\,s^{-1}}$, yielding a crossing time $t_{\mathrm{cross}} \sim 2\,\mathrm{Gyr}$. This leads to a relaxation time of $\sim 30$ Gyr, larger than the age of the Universe. Since most of a cluster mass is not in galaxies, but in the DM component, a more realistic estimate for $t_{{\rm relax}}$ is ten times larger \citep{Sarazin86}.
This result implies that encounters between galaxies are too rare to efficiently redistribute energy and drive the system toward collisional equilibrium. In this regime, the internal kinematics of galaxy clusters is well described by a smooth gravitational potential, with galaxies behaving as collisionless tracers. Collective gravitational effects dominate over discrete encounters, providing a solid ground for the use of the collisionless Boltzmann equation and its moment equations (Jeans equations) to perform kinematical analysis, as detailed below. Note, however, that clusters are complex systems, and often in a disturbed dynamical state \citep[e.g.][]{Roberts_2019,Bilton_2020,Wen2024}. As already mentioned, non-relaxed clusters introduce significant bias in kinematical mass estimates, and this will be particularly important when using kinematics to test theories beyond the standard cosmological model (e.g., \citealt{Pizzuti2020syst}, see Sect.~\ref{sec:syst}).

\subsubsection{Dynamics of collisionless systems}
\label{sec:dyn}
A self-gravitating system of N collisionless particles governed by a total potential $\Phi(\bf{x}, t)$ can be described in terms of the number of objects $\text{d}N$ which, at a given time $t$, are in a 3D
position between $\bf{x}$ and $ \bf{x}+\text{d}^3\bf{x}$, and are characterized by velocities between $\bf{v}$ and  and $\bf{v} + \text{d}^3\bf{v}$:
\begin{equation}
   \text{d}N = f(\bf{x}, \bf{v}, t)\text{d}^3\bf{x}~\text{d}^3\bf{v},
\end{equation}
where $f > 0$ is called distribution function and it represents the density of objects in phase-space. The dynamical quantities of the system are identified by the six-dimensional vector $w = (\bf{x},\bf{v})$ and its time derivative $\dot{w} = (\bf{v},-\nabla\Phi)$, where the total gravitational potential is related to the mass density profile through the Poisson equation:
\begin{equation}
    \nabla^2\Phi = 4\pi G\rho(\bf{x})\,.
\end{equation}
If the collisions are negligible, the flow of points in the phase-space is characterized by a
regular "drift" motion, and the distribution function satisfies the Vlasov or collisionless Boltzmann equation,
\begin{equation}\label{eq:Vlasov}
    \frac{D}{\text{d}t}f(\bf{x}, \bf{v}, t) = 0\,,
\end{equation}
where the convective derivative $D/\text{d}t = \partial_t + \bf{v}\cdot\nabla$ describes the conservation of the density of points in phase space as seen by an observer comoving with the flow.

Integrating the Vlasov equation over velocities yields the continuity equation
\begin{equation}\label{eq:cont}
    \frac{\partial \nu(\bf{x},t)}{\partial t} +  \nabla\cdot(\langle \bf{v}\rangle\nu) = 0\,,
\end{equation}
where we have defined the number density and the mean velocity of the particles as 
\begin{equation}
    \nu(\bf{x},t) = \int\text{d}^3\bf{v}\,f(\bf{x},\bf{v},t)\,,\qquad \langle v_i(\bf{x},t)\rangle = \frac{1}{\nu}\int\text{d}^3\bf{v}\,v_if(\bf{x},\bf{v},t)\,.
\end{equation}

The first moment of eq.~\eqref{eq:Vlasov} is obtained by multiplying it by $v_i$ and integrating again in velocity space,
\begin{equation} \label{eq:firstMoment}
    \frac{\partial (\nu \langle{v_j}\rangle)}{\partial t} + \frac{\partial (\nu\langle{v_i v_j}\rangle)}{\partial x_i} + \nu\frac{\partial \Phi}{\partial x_j} = 0,
\end{equation}
where
\begin{equation}
    \langle{v_i v_j}\rangle = \frac{1}{\nu} \int v_i v_j f(\bf{x},\bf{v},t) \text{d}^3 \bf{v}\,.
\end{equation}
The Jeans equation \citep{Jeans19} can be obtained by combining eq.~\eqref{eq:cont} and eq.~\eqref{eq:firstMoment} as 
\begin{equation}
\frac{\partial (\nu \, \langle{v_i}\rangle)}{\partial t}
+ \frac{\partial (\nu \, \langle{v_i}\rangle \langle{v_j}\rangle)}{\partial x_j}
+ \frac{\partial (\nu \, \sigma^2_{ij})}{\partial x_j}  +\nu \, \frac{\partial \Phi}{\partial x_i}
= 0\,,
\end{equation}
where we have introduced the velocity dispersion tensor $\sigma_{ij} = \langle{(v_i -\langle v_i\rangle)(v_j -\langle v_j\rangle)}\rangle$.
If the system has reached dynamical equilibrium, time derivatives can be neglected, and the dynamics is fully specified by the gradients of the gravitational potential for a given structure of the velocity field.

Note that in principle there are no relations connecting the six independent components of $\sigma^2_{ij}$ to the number density $\nu$. One can imagine to compute higher moments of the Vlasov equation, giving rise to a hierarchy of relations that need to be closed somehow. This closure can be achieved either by truncating the series at a given order, or by making specific assumptions on the velocity tensor.

A particularly relevant and commonly applied simplification is obtained by assuming that the system is spherically symmetric, non-rotating, and stationary. In this case, all quantities depend only on the radial coordinate $r$, and the mean streaming motions vanish, that is $\langle v_r \rangle = \langle v_\theta \rangle = \langle v_\phi \rangle = 0$ \citep[e.g.,][]{Falco2013b}. Under these assumptions, the Jeans equation reduces to a single scalar equation \citep[e.g.,][]{BT87}:
\begin{equation}
\frac{\mathrm{d} (\nu \sigma_r^2)}{\mathrm{d} r} + \frac{2\beta(r)}{r} \nu \sigma_r^2 = -\nu \frac{\mathrm{d} \Phi}{\mathrm{d} r}\,,
\end{equation}
where $\sigma_r$ is the radial velocity dispersion, and we have introduced the velocity anisotropy profile
\begin{equation}
\beta(r) = 1 - \frac{\sigma_\theta^2 + \sigma_\phi^2}{2\sigma_r^2}\,,
\end{equation}
where $\sigma_\theta$ and $\sigma_\phi$ are the two tangential components of the velocity dispersion tensor, usually assumed to be identical.
The function $\beta(r)$ encodes fundamental information about the orbital structure of the system \citep[e.g.,][]{Biviano2004,Wojtak2009}: $\beta = 0$ corresponds to isotropic orbits, $\beta > 0$ to radially biased orbits, and $\beta < 0$ to tangentially dominated motions. The anisotropy profile profile further traces the dynamical state of a galaxy cluster, providing hints about the presence of merging events \citep{Hou2009,Biviano_2026}, and it has been shown to depend on the properties and morphology of galaxies in dense environments \citep[e.g.,][]{Biviano1997,Mamon2019,Maraboli+26,Pedratti26}.

By expressing the gravitational potential in terms of the enclosed mass $M(r)$ through $\mathrm{d}\Phi/\mathrm{d}r = GM(r)/r^2$, the Jeans equation can be rewritten as:
\begin{equation}
\frac{\mathrm{d} (\nu \sigma_r^2)}{\mathrm{d} r} + \frac{2\beta(r)}{r} \nu \sigma_r^2 = -\nu \frac{G M(r)}{r^2}\,,
\end{equation}
or, equivalently,
\begin{equation}
G M(r) = - r \sigma_r^2(r) \left[ \frac{d \ln \nu}{d \ln r} + \frac{d \ln \sigma_r^2}{d \ln r} + 2 \beta(r) \right].  
\label{eq:Jeans}
\end{equation}

Integration of this equation leads to the virial theorem presented in Sect.~\ref{sec:virial} \citep[see, e.g.,][]{AE11}. Equation~(\ref{eq:Jeans}) represents a key anchor for dynamical mass modeling in astrophysics, but its use is limited by the fact that only projected quantities are available to the observer. Indeed, it is generally not possible to infer independently $M(r), \sigma^2_r(r)$ and $\beta(r)$; direct measurements of the velocity anisotropy are very challenging, since tangential velocities are difficult to observe \citep[e.g.][]{Molnar_2003,Abdullah2013}.

\subsubsection{Solving the Jeans equation}
\label{sec:jeanssolution}
From a formal point of view, the Jeans equation can be solved for the radial velocity dispersion profile $\sigma_r(r)$ given a tracer number density $\nu(r)$, a total mass profile $M(r)$, and a velocity anisotropy profile $\beta(r)$. The general solution can be written as a first-order integral,
\begin{equation}\label{eq:Jeansol}
\sigma_r^2(r) = \frac{1}{\nu(r)} \int_r^\infty \nu(s)\,\frac{G M(s)}{s^2}
\exp\left[2\int_r^s \frac{\beta(t)}{t}\,\text{d}t\right] \text{d}s,
\end{equation}
which highlights the non-local nature of the problem: the kinematics at a given radius depends on the global mass and anisotropy structure of the system. The integral of $\beta(t)/t$ can be solved analytically for a broad variety of physically-motivated parametric anisotropy models (see e.g., \citealt{Mamon_2013}). An interesting feature of eq.~\eqref{eq:Jeansol} is its independence of the normalisation of the number density $\nu(r)$ --- i.e., the total number of tracers used to infer the velocity dispersion.

The connection between the intrinsic and observed (projected) kinematics is established through the projection of $\sigma_r(r)$ along the line of sight, yielding the projected velocity dispersion profile
\begin{equation}
\Sigma(R)\sigma_\mathrm{los}^2(R) = 2 \int_R^\infty 
\nu(r)\,\sigma_r^2(r)\left(1 - \beta(r)\frac{R^2}{r^2}\right)
\frac{r\,\text{d}r}{\sqrt{r^2 - R^2}},
\end{equation}
which is a classical Abel-type integral. Above, $\Sigma(R)$ is the projected surface number density profile of the tracers. Inverting this relation is, in principle, possible: given $\Sigma(R)$ and $\sigma_\mathrm{los}(R)$, one can recover $\nu(r)$ or $\sigma_r(r)$ through Abel inversion techniques. The tracer density can be obtained from the projected density as
\begin{equation}
\nu(r) = -\frac{1}{\pi} \int_r^\infty \frac{d\Sigma}{dR}\frac{\text{d}R}{\sqrt{R^2 - r^2}}.
\end{equation}
The inversion is considerably more complicated for the velocity dispersion tensor; because of the mass-anisotropy degeneracy, it cannot be performed without knowledge of either $M(r)$ 
\citep{BinneyMamon1982,Solanes90,Dejonghe1992,Diakogiannis+14a,Diakogiannis+19} or $\beta(r)$ \citep{MB10,Wolf10}.
In practice, these inversions are highly sensitive to noise, binning, and extrapolation assumptions, which limits their applicability to high-quality datasets \citep[see, e.g.,][]{Biviano_2013,Annunziatella2016}.
For this reason, modern analyses generally adopt forward-modelling approaches, where parametric forms of $M(r)$ and $\beta(r)$ are assumed and projected quantities are directly compared to the data. 

A widely used implementation is the Jeans Anisotropic Modelling (JAM), originally developed for stellar systems \citep{Cappellari2008} and later extended to larger scales. In its most general formulation, JAM solves the Jeans equations beyond the spherically symmetric assumption, for an axisymmetric ($\partial/\partial \phi = 0$) and steady state ($\partial/\partial t = 0$) configuration. In cylindrical coordinates $(R,z,\phi)$, the Jeans equation can be re-casted as a system of two equations:
\begin{equation}\label{eq:jeans1}
\frac{\partial (\nu \sigma_R^2)}{\partial R} + \frac{\partial (\nu \sigma_{Rz}^2)}{\partial z} + \nu \left( \frac{\sigma_R^2 - \sigma_\phi^2}{R} \right) = - \nu \frac{\partial \Phi}{\partial R},
\end{equation}
\begin{equation} \label{eq:jeans2}
\frac{\partial (\nu \sigma_{Rz}^2)}{\partial R} + \frac{\partial (\nu \sigma_z^2)}{\partial z} + \nu \frac{\sigma_{Rz}^2}{R} = - \nu \frac{\partial \Phi}{\partial z},
\end{equation}
where $\sigma_{Rz}^2 \equiv \langle v_R v_z \rangle$ represents the mixed velocity moment, encoding correlations between radial and vertical motions. In general, the Jeans equations \eqref{eq:jeans1},\eqref{eq:jeans2} form a coupled system through $\sigma_{Rz}^2$. However, in the JAM framework it is commonly assumed that the velocity ellipsoid is aligned with the cylindrical coordinate system ($\sigma_{Rz}=0$), and that the velocity anisotropy in the meridional plane, 
\begin{equation}
\beta_z = 1 - \frac{\sigma_z^2}{\sigma_R^2},
\end{equation}
is constant. Both the tracer density $\nu(r)$ and the gravitational potential $\Phi(r)$ are then modelled through flexible parametric decompositions, often based on Multi-Gaussian Expansions (MGE), which allow for an efficient deprojection of the Jeans equations. This approach makes it possible to explore high-dimensional parameter spaces in a fast and efficient forward-modeling framework.

Complementary to JAM, alternative forward-modelling techniques exploit the full distribution of the tracers in the p.p.s.
If the number of tracers is large enough $N\sim \mathcal{O}(100)$, one possibility is to connect the observed p.p.s. of galaxies to the intrinsic phase space density $f(E,L)$, expressed in terms of total energy $E$ and angular momentum $L$ of the system. This connection can be achieved by integrating with respect to the Cartesian components of the velocity along the polar coordinates $(R,\theta)$ in the plane of the sky, \citep[see, e.g.,][]{Dejonghe1992}
\begin{eqnarray}\label{eq:gdens}
g(R,v_{z}) 
&=&
2\int_R^\infty
\frac{r\,d r}{(r^2-R^2)^{1/2}}
\int_{-\infty}^{+\infty} 
\int_{-\infty}^{+\infty} 
f(E,L) \,d v_\theta \,d v_R \,.
\end{eqnarray} 
The gravitational potential of the system is then related to $f(E,L)$ through the Poisson equation. $f(E,L)$ is not known \textit{a priori}, and some assumptions need to be made. \cite{MG04} used a scale-free, separable function of $E$ and $L$, 
\begin{equation}\label{eq:fel}
f(E,L) = f_0 \epsilon^{\alpha-1/2} L^{-2 \beta}, 
\end{equation}
to estimate the mass density profile of 41 nearby galaxy systems. Note that, the energy component in eq.~\eqref{eq:fel} can be determined (typically numerically) once the total mass profile is assumed.
Alternatively, to select the functional form of $f(E,L)$ one can rely on halos extracted from cosmological simulations \citep[e.g.,][]{Wojtak2007,Wojtak2009}.
This method has been employed by e.g., \cite{Wojtak2010} to estimate the total mass profiles of 41 nearby clusters.

Another example is the \textsc{MAMPOSSt} (Modelling of Anisotropy and Mass Profiles of Observed Spherical Systems) method of \cite{Mamon_2013}, which solves the Jeans equation in the p.p.s. by jointly fitting parametric forms for the gravitational potential, the tracer number density, and the orbital velocity anisotropy. Assuming spherical symmetry and a Gaussian ansatz for the 3D distribution of velocities $f(v_r,v_\theta,v_\phi)$, for a given set of models the algorithm first computes $\sigma_r$, eq.~\eqref{eq:Jeansol}. The radial component is then projected along the line of sight to obtain the observable 
\begin{equation}
\sigma_{{\rm los}}^2(R,r) = \left[1 - \beta(r)\frac{R^2}{r^2}\right] \sigma_r^2(r),
\end{equation}
where $R$ and $r$ are the projected and 3D radius, respectively.

Under the Gaussian assumption for the 3D velocity distribution, the probability density of observing a galaxy at projected radius $R$ with line-of-sight velocity $v_{{\rm los}}$, eq.~\eqref{eq:gdens}, is given by
\begin{equation}
g(R,v_{{\rm los}}) = \sqrt{\frac{2}{\pi}} \int_R^\infty 
\frac{\nu(r)}{\sigma_{{\rm los}}(R,r)} 
\exp\left[-\frac{v_{{\rm los}}^2}{2\sigma_{{\rm los}}^2(R,r)}\right]
\frac{r\,\text{d}r}{\sqrt{r^2 - R^2}},
\end{equation}
which represents the los integral of the phase-space distribution. The corresponding probability density at fixed projected radius is obtained as:
\begin{equation}
    q\, (R, v_{{\rm los}}) = \frac{2\pi R \, g\, (R, v_{{\rm los}})}{N_\text{proj}(R_\text{max})-N_\text{proj}(R_\text{min})} \, ,
\end{equation}
where $N_\text{proj}(\,R\,)$ is the integral of the surface density $\Sigma(R)$, providing the cumulative number of galaxies at $R$. $R_\text{min}$ and $R_\text{max}$ are the minimum and maximum projected radii in the p.p.s, respectively. The log-likelihood for a set of parameters $\Theta$ of the model is then constructed as the sum of $\ln q(R_i,v_{z,i}|\Theta)$ over all galaxies, allowing one to constrain simultaneously $M(r)$ and $\beta(r)$ without binning the data.

The key advantage of methods like \textsc{MAMPOSSt} is that they consider all the information of the p.p.s., rather than cumulative moments of the velocity field, leading to tight constraints on the cluster mass profile when the number of galaxies is $N \gtrsim 100$. Note, however, that the \textsc{MAMPOSSt} technique requires a specific form of the 3D velocity distribution --- on top of the assumptions of spherical symmetry and dynamical equilibrium --- and it further considers negligible streaming motions. While such requirements can in principle introduce systematics in the kinematical analysis, several tests using halos extracted from cosmological simulations have consistently shown that the method performs robustly with a reasonable small scatter in recovering the mass and anisotropy parameters, even for non-Gaussian velocity fields  \citep{Mamon_2013,Old+14,Old_2015,Aguirre2021,Read2021}.

\paragraph{The Jeans swindle}
On a final note, it is worth recalling that the Jeans equation is formally derived under the assumption of a steady-state system embedded in a well-defined gravitational potential. In a cosmological context, however, this assumption raises a subtle issue known as the {\it Jeans swindle}, namely the apparent inconsistency arising from neglecting the contribution of the homogeneous background density when solving Poisson’s equation. 

If the total density is decomposed in the sum of a halo contribution and a background contribution,
\begin{equation}
\rho(r) = \rho_h(r) + \rho_\mathrm{bg},
\end{equation}
the inclusion of the background density in the standard Jeans solution leads to a divergent term to the velocity dispersion when integrating to large radii \citep{Falco2013}. Indeed, the background mass grows as $M_\mathrm{bg}(r) \propto r^3$, causing the integral solution of the Jeans equation to explode.

The standard approach consists in neglecting $\rho_\mathrm{bg}$ when computing the gravitational potential, effectively assuming that only density fluctuations contribute to the dynamics. This prescription, traditionally referred to as the Jeans swindle, has long been regarded as a mathematically inconsistent but empirically successful approximation. However, \citet{Falco2013} showed that this procedure can be rigorously justified once the cosmological context is properly taken into account. In particular, when the Jeans equation is generalized to include the effects of the expanding Universe, an  additional term arises,
\begin{equation}
S(r,t) = \langle v_r\rangle \frac{\partial \langle v_r\rangle}{\partial r} + \frac{\partial \langle v_r\rangle}{\partial t},
\end{equation}
which accounts for the presence of a non-zero mean radial velocity field $\langle v_r\rangle$. At large radii, where the velocity field is dominated by the Hubble expansion, $v_r \simeq H r$, and the additional term exactly cancels the contribution of the background density (and of the cosmological constant) to the gravitational potential. 
As a consequence, the divergent term disappears, and the Jeans equation reduces to the standard form where only the halo mass contributes to the dynamics. 

In this sense, the Jeans swindle is not an ad hoc assumption, but rather the result of a consistent treatment of the system within an expanding Universe. This result also clarifies the domain of validity of the standard Jeans formalism. While it provides an accurate description within the virialized region, where streaming motions are negligible, a more general treatment is required at larger radii. In particular, the inclusion of infall motions leads to a generalized Jeans equation that remains valid beyond the virial radius and can reproduce the velocity dispersion profile out to several virial radii \citep{Falco2013b}.

\subsubsection{Applications of the Jeans analysis to galaxy clusters}

The spherical Jeans formalism has been extensively applied to a wide range of astrophysical systems, including globular clusters \citep[e.g.,][]{Henault2019}, elliptical galaxies \citep[e.g.,][]{LokasMamon2003,Agnello2014}, and galaxy clusters. In the context of clusters, it provides a powerful tool to infer total mass profiles \citep[e.g.,][]{Li2023} and further test cosmological models, especially when combined with independent probes such as gravitational lensing observations \citep[e.g.,][]{Pizzuti22}.

\paragraph{Mass-profile reconstruction}
In the context of galaxy clusters, the applications of JAM have demonstrated its potential for dynamical mass reconstruction. In particular, \citet{Shi_2024} employed JAM to analyze the dynamics of mock clusters from the TNG300 simulation, using satellite galaxies as tracers (typically $\sim 150$ per cluster), and showed that the method can recover virial masses with small bias ($\lesssim 0.03$ dex) and a scatter of $\sim 0.18$ dex. Interestingly, the mass enclosed within the half-mass radius of the tracers, is found to be more tightly constrained than $M_{200}$ when the inclination angle between the intrinsic and observed frame is considered a free parameter. These findings also highlight the robustness of JAM against several observational systematics, including moderate contamination from interlopers and spectroscopic incompleteness, which were shown to have a limited impact on the recovered velocity dispersion profiles. However, residual biases can still arise from departures from dynamical equilibrium, projection effects, and deviations from the assumed axisymmetric geometry, especially in dynamically active systems.

\begin{figure}[ht]
         \centering
         \includegraphics[width=.5\columnwidth]{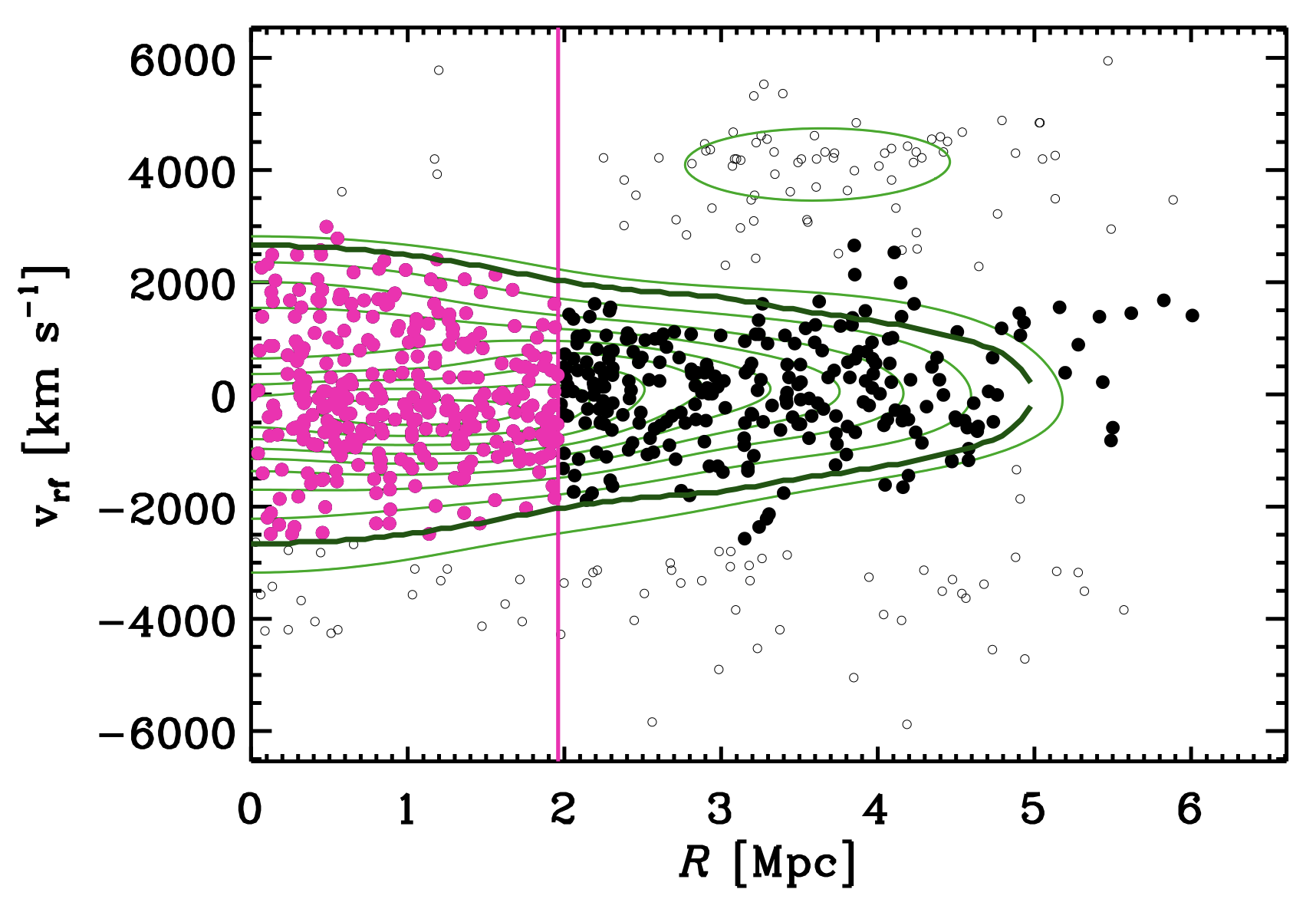}
        \includegraphics[width=.49\columnwidth]{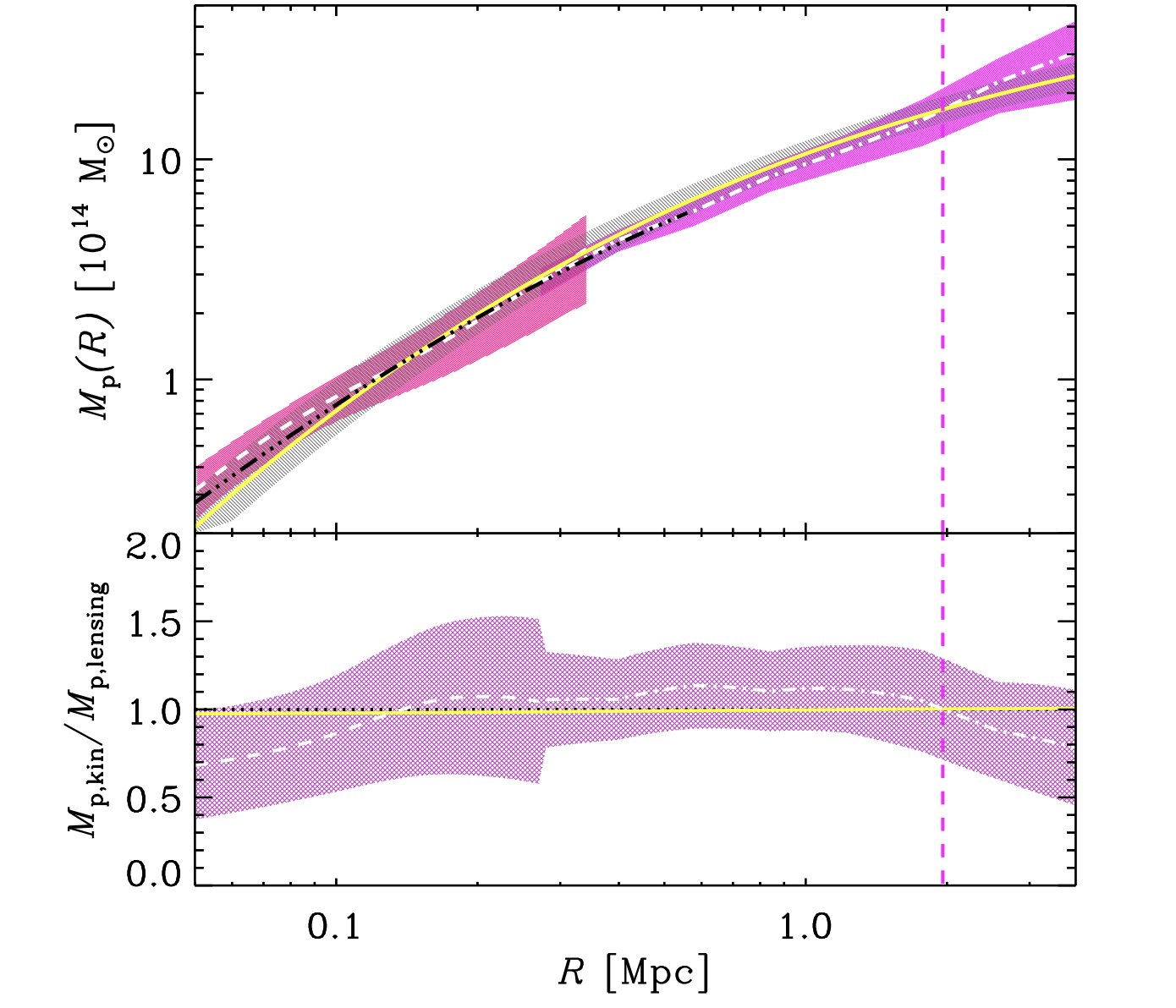}
         \caption{\label{fig:phaseMACS}  Left panel: p.p.s. distribution ($v_{{\rm los}}$ is labelled as $v_\mathrm{rf}$) of the galaxy cluster MACS J1206.2-0847 obtained from the spectroscopic data collected within the CLASH-VLT collaboration \citep{Rosati14}. Filled circles represent the identified cluster members; the pink points indicates galaxies within $R= r_{200}^{L}$, with $r_{200}^{L}$ being the value of the virial radius estimated by the strong+weak lensing analysis of \cite{Umetsu12}. The trumpet-shaped curves show the caustics in the p.p.s. Right panel: total projected mass profile of the same cluster obtained from \textsc{MAMPOSSt} (gray band and yellow curve) compared to the strong (pink) and weak (violet) lensing profile of \cite{Umetsu12}. While \textsc{MAMPOSSt} provides the 3D mass profile this has been projected for comparison with the lensing (projected) mass profile.  
         All the filled regions indicate the $1\,\sigma$ uncertainties. The bottom plot shows the relative difference and the vertical solid and dashed lines identify the position of $r_{200}^{L}$. Images reproduced with permission, copyright by ESO.}
    \end{figure}

\textsc{MAMPOSSt} has been extensively applied to reconstruct masses and velocity anisotropy profiles of single and stacked galaxy clusters \citep[e.g.][]{Biviano_2013,Biviano_2017,Biviano_2021,Balestra2016,Verdugo2016,Capasso_2019,Mamon2019,Pizzuti25b}, as well as to infer the properties of DM when combined with information from gravitational lensing \citep{SartorisDM}, or with the stellar kinematics of the BCG \citep{Sartoris_2020,Biviano_2023}. Figure~\ref{fig:phaseMACS} shows an example of p.p.s (left) and of the mass profile reconstructed by the \textsc{MAMPOSSt} analysis of \cite{Biviano_2013} (right) of MACS J1206.2-0847, a massive galaxy cluster at $z=0.44$, for which high precision spectroscopic data have been collected within the CLASH-VLT program \citep{Rosati14}. A total of 330 member galaxies have been used in the \textsc{MAMPOSSt} fit, leading to an inferred mass distribution in excellent agreement with independent strong and weak lensing estimate by \cite{Umetsu12}.

The  \textsc{MAMPOSSt} method has been further employed jointly with lensing analyses to test the nature of gravity \citep{Pizzuti2016,Pizzuti17}. In this regard, a specific extension of \textsc{MAMPOSSt} to MG has been developed in the form of the \textsc{MG-MAMPOSSt} pipeline \citep{Pizzuti2021}, where the gravitational potential entering the Jeans equation is generalized to account for a wide variety of models beyond $\Lambda$CDM. 
The \textsc{MG-MAMPOSSt} approach represents a powerful flexible framework for probing the gravity and dark sector at the Mpc scale \citep{Pizzuti22,Pizzuti25a,Pizzuti26}, as detailed in Sect.~\ref{sec:results}.

\paragraph{Velocity anisotropy profiles}
As already mentioned, the velocity anisotropy is a key quantity in kinematical analyses of galaxy clusters. To infer its profile, several techniques have been employed. One possibility is to assume a parametric profile for $\beta(r)$ and jointly fit it along with the mass distribution of the system with a Maximum Likelihood approach \citep[e.g.,][]{Wojtak2009,Mamon_2013,Li2023}. However, the solutions for $\beta(r)$ and $M(r)$ are degenerate, since increasing $\beta(r)$ and decreasing $M(r)$, or vice-versa, would lead to the same radial velocity dispersion  --- the so-called mass--anisotropy degeneracy. 
On the other hand, non parametric \citep{BinneyMamon1982,Host2009} and analytical \citep{Stark2019} methods can be used, but those generally requires additional assumptions and information on the mass distribution (see, e.g. \citealt{Biviano_2013, Maraboli+26}). 

\begin{figure}[ht]
         \centering
         \includegraphics[width=\columnwidth]{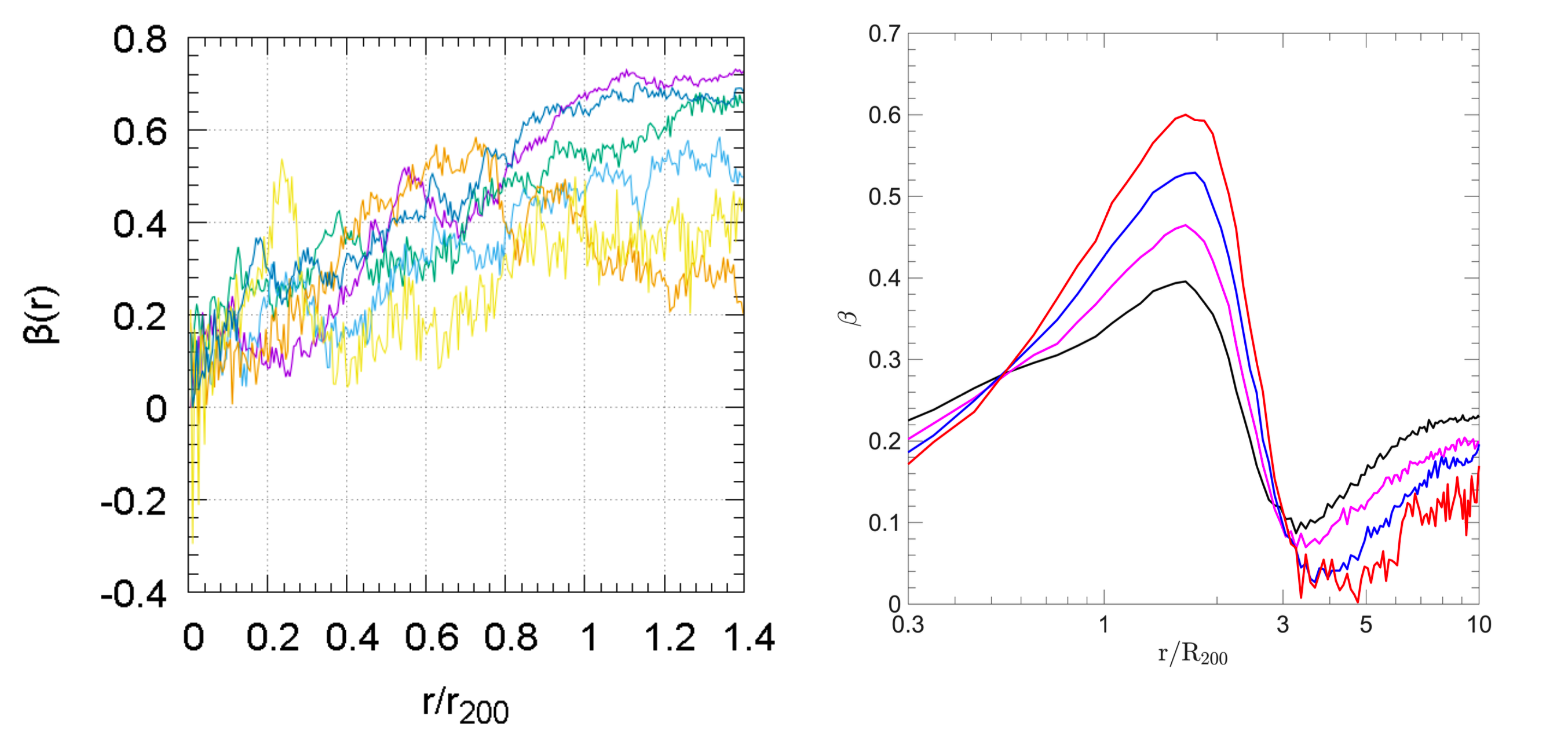}
         \caption{\label{fig:anisEx}  Left panel: velocity anisotropy profiles $\beta(r)$ as a function of the normalized radius $r/r_{200}$ of six cluster-size DM halos ($M_{200} \in[2 \times 10^{14}, \,1.7 \times 10^{15}]\, {\rm M}_\odot$) from a $\Lambda$CDM cosmological simulation at $z=0$. Credits: \cite{PizzutiThesis}. Right panel:  stacked anisotropy profiles of simulated clusters in four mass bins (from $M \ge 10^{13.9}\,\text{M}_\odot/h$ to $M \ge 10^{14.9}\,\text{M}_\odot/h$) at $z=0$ out to several virial radii, reproduced from Figure 3 of \cite{Abdullah2025}. The velocity anisotropy decreases towards isotropic orbits at around $\sim 3\,r_{200}$. Images reproduced with permission, copyright by AAS.}
    \end{figure}

Indications on how the anisotropy behaves as a function of the distance from the cluster centre can be obtained from the analysis of mock systems: cosmological simulations showed that $\beta(r)$ is typically an increasing function of the radius, with more isotropic orbits in the core and more radial in the outskirts (see e.g. \citealt{Hansen_2006,Mamon2010}). 
$\beta(r)$ reaches a peak at $r \sim 1.5$--$2\,r_{200}$, then it declines toward $r \sim 3$--$4\,r_{200}$, where the coexistence of galaxies on first infall and backsplash populations leads to significant orbital mixing and an enhancement of the tangential velocity components \citep[][see the right panel of Fig.~\ref{fig:anisEx}]{Diaferio1999,Abdullah2025}. At even larger radii, $\beta(r)$ increases again as the dynamics becomes dominated by galaxies undergoing their first infall, and by the Hubble flow.

The radial orbits in the external region are therefore interpreted as reflecting the memory of the infall of galaxies under the influence of the cluster gravitational potential \citep[e.g.][]{Lotz2019}, while several environmental processes are thought to produce an isotropisation of the orbits near the centre \citep[e.g.][]{Abdullah2025}. This scenario is at least partially falsified when cluster triaxiality is taken into account. In fact, the central isotropy of galaxy orbits in clusters is a result of averaging $\beta$ in spherical shells. In the inner regions, the local velocity ellipsoids are strongly aligned with the major axis
of the halo, resulting in $\beta \approx 0.35$  along the major axis, and $\beta \lesssim 0$ perpendicular to it \citep{Wojtak13}.

The left panel of Fig.~\ref{fig:anisEx}, shows the anisotropy profile of six massive dark matter halos ($M_{200} \in[2 \times 10^{14}, \,1.7 \times 10^{15}]\, {\rm M}_\odot$) from a cosmological $\Lambda$CDM simulation carried out with the GADGET-3 code \citep{Springel_2005,Bonafede_2011}, where the increasing radial trend can be visually appreciated. This trend has also been observed in real clusters and confirmed by several studies based on kinematical data \citep[e.g.,][]{NK96, Biviano2004,Lemze_2009,Mamon2019,Barrena24,Pizzuti25b,Valk2025}{, and up to $z \sim 1$ \citep{Biviano2009,Biviano_2021}. There are indications that the orbits of galaxies in clusters tend to become more radially elongated at higher $z$ \citep{Biviano2009,Biviano_2026}. The strongest indication in this direction comes from the detailed study of nine, very well sampled clusters at $0.2 \lesssim z \lesssim 0.5$ by \citet{Biviano_2026}. They found significantly more radial orbits than found for clusters at lower $z$ by \citet{Wojtak2010} and \citet{Li2023}. \citet{Biviano_2026} also found consistency between the observed, more radial, $\beta(r)$ of their medium-$z$ clusters and the average $\beta(r)$ of simulated halos that were matched in $z$ and mass to the observed sample.}

The velocity anisotropy of member galaxies {does not only depend on $z$; it may also be} linked to the morphological and physical properties of the galaxies themselves, with different populations following on average different orbits \citep{Biviano2004,Boselli_2006,Biviano2009, Adhikari2018,Mamon2019,Pedratti26}. Early-type galaxies (ellipticals and lenticulars), which preferentially reside in the dense inner regions of clusters, are typically found to have nearly isotropic velocity distributions, consistent with a long residence time and efficient dynamical processing within the cluster potential. In contrast, late-type (star-forming) galaxies tend to exhibit more radially anisotropic orbits, especially at large radii, due to their more recent accretion from the field and infall along preferential directions. However, opposite trends have also been observed in some clusters, with early-type galaxies on more radial orbits than late-type galaxies \citep{Aguerri2017,Mercurio2021}.

Refined analyses based on stacked samples with high-quality spectroscopic data
provide more insight into this topic. In particular, the study of the CLASH-VLT sample by \citet{Maraboli+26}, based on the same sample of clusters studied by \citet{Biviano_2026}, indicates that, despite the clear segregation of galaxy populations in phase space, their velocity anisotropy profiles are similar within uncertainties. All populations — including red, blue, high- and low-mass galaxies — follow a common radial trend, with progressively more radial orbits toward the outskirts. Only mild differences are observed, mainly at large radii, where red and blue galaxies exhibit small deviations, although their significance depends sensitively on the assumed mass, number density, and anisotropy models. An interesting feature emerges when adopting non-parametric approaches based on the inversion of the Jeans equation, which reveal a non-monotonic behavior of the anisotropy profile: a local maximum in $\beta(r)$ at intermediate radii ($\sim 200$--$300$ kpc), followed by a dip before the profile resumes its increase toward the outskirts. This ``bump'' suggests that the orbital structure of cluster galaxies may retain imprints of the complex interplay between relaxation processes and ongoing accretion, pointing to deviations from simple, monotonic prescriptions of $\beta(r)$.

In conclusion, the orbital differences of different galaxy populations are related to the history of their assembly process into the clusters, modulated by the transformation and destruction processes operated by the hostile cluster environment \citep[e.g.][]{Iannuzzi2012,Bialas_2015,Annunziatella2016,Lotz2019,Lokas2020,Boselli_2022,Xie_2025}.

\subsubsection{Estimating masses in the cluster outskirts}\label{sec:outskirts}
\paragraph{The caustic method}
The \textit{caustic} method \citep{DiaferioGeller97,Diaferio1999,Serra+11} has been developed to estimate cluster mass profiles well beyond the virial region, where the assumption of dynamical equilibrium breaks down and traditional techniques such as the virial theorem or Jeans analysis are no longer valid. This regime corresponds to the infall region of clusters, where matter is still accreting and the dynamics are intrinsically non-linear and time-dependent \citep{Diaferio1999,Diaferio2009umei.confE..42D,Geller2011AJ....142..133G}. 

Similarly to other techniques, also this one makes use of the distribution of galaxies in the p.p.s. As shown in the left plot of Fig.~\ref{fig:phaseMACS}, galaxy clusters exhibit a characteristic trumpet-shaped pattern in p.p.s., whose sharp edges --- known as caustics --- trace the escape velocity profile of the system. This structure arises from the combination of gravitationally bound and infalling galaxies and is a direct imprint of the underlying gravitational potential \citep{DiaferioGeller97}. As shown in numerical simulations, the amplitude of the caustics is closely related to the escape velocity, and thus to the cluster mass profile \citep{DiaferioGeller97, Diaferio1999,Serra+11}. 

Operationally, the method identifies the caustics as regions of steep gradients in the galaxy density distribution in the p.p.s. along the velocity axis (solid curves in the left plot of Fig.~\ref{fig:phaseMACS}). The caustic amplitude ${\cal A}(R)$ is then used as a proxy for the escape velocity, allowing the cumulative mass profile to be estimated through
\begin{equation}
G \, [M(r)-M(r_0)] = \int_{r_0}^r {\cal A}^2(x)\,{\cal F}_{\beta}(x)\,dx \, ,
\label{eq:cau}
\end{equation}
where ${\cal F}_{\beta}(x)$ is a function that depends on both the gravitational potential and the velocity anisotropy profile $\beta(r)$. In practice, ${\cal F}_{\beta}$ is often approximated as a constant, motivated by numerical simulations that show a relatively weak radial dependence outside the virial region. This approximation is known to break down in the inner regions, leading to a systematic bias in the mass estimate if applied within $\lesssim r_{200}$ \citep{Serra+11,Logan22}. However, it provides a robust description in the outskirts, where the method is specifically designed to operate.

The approximately constant value of ${\cal F}_{\beta}$ has been estimated using numerical simulations, but it mostly varies according to the specific algorithm adopted to estimate the $\sigma_{{\rm los}}$ of cluster galaxies \citep{Serra+11}.\footnote{The algorithm to locate the caustics in the p.p.s. is based on the arrangement of the galaxies on a binary tree \citep{Diaferio1999}. The velocity dispersion of the galaxies on the main branch of the tree sets the caustic location and determines the calibration of ${\cal F}_{\beta}$. The velocity dispersion thus  depends on the criterion adopted to identify the galaxies on the main branch of the tree and can differ from the $\sigma_{{\rm los}}$ derived with other traditional methods. These latter methods are usually limited to the galaxies within the cluster central region, unlike the galaxies on the main branch of the tree \citep{Serra+11}. } 
\citet{Diaferio1999} adopted a value ${\cal F}_{\beta} \simeq 0.5$ 
that was used in several studies \citep[e.g.,][]{Rines+00,Rines+03,Rines+13}. Later, \citet{Serra+11}  and, separately, \citet{Gifford2013} adopted a higher value, ${\cal F}_{\beta} \simeq 0.65-0.7$, and other investigations, based on large hydrodynamical simulations, adopt a lower normalization, ${\cal F}_{\beta} \simeq 0.4$ with a mild scatter \citep{Pizzardo2023A&A...675A..56P}. This wide range of values of ${\cal F}_{\beta}$ present in the literature is not a consequence of an alleged difficulty of its estimate, but it simply mirrors the different choices adopted in the N-body simulations for the calibration procedure: the major role is played by the algorithm for the estimate of the cluster galaxies $\sigma_{{\rm los}}$, the chosen radial range of the mass profile, and the number of the velocity tracers (DM particles or simulated galaxies) in the p.p.s. \citep{Serra+11}.  When estimating the mass profiles of real clusters in their outer regions with the caustic technique, it is thus crucial to apply exactly the same procedure that was used with the N-body simulations for the calibration of ${\cal F}_{\beta}$.  

\citet{GM13} have proposed to adopt the NFW $M(r)$ model and observational determinations of $\beta(r)$ to go beyond the constant ${\cal F}_{\beta}$ assumption. We emphasize however that this proposal of Gifford and collaborators originates from their goal to use the caustic method to measure the cluster mass only within $r_{200}$. On the contrary, the relevant and specific original purpose of the caustic method is to measure the cluster mass profile beyond $r_{200}$. 

Indeed, a key strength of the caustic technique is that it does not rely on dynamical equilibrium assumptions, so it is  suited for probing cluster mass profiles out to several virial radii, where other dynamical methods fail but where important information about cluster growth and mass accretion is encoded \citep{Diaferio1999,DeBoni+16,Pizzardo22}. 
Another practical advantage of the method is that it does not require an explicit parametric modeling of the tracer density profile or of the velocity anisotropy --- although the latter is hidden in the function ${\cal F}_{\beta}(r)$ --- as needed in Jeans-based analyses. This simplification 
comes however at the cost of increased statistical scatter, partly due to the variance in $\beta(r)$ across different clusters \citep{Biviano_2026}.

Tests on numerical simulations show that the caustic method can recover cluster mass profiles with a typical accuracy of $\sim 20$--$30\%$ over a wide radial range, although larger scatter can occur depending on the data quality and sampling \citep[e.g.,][]{Diaferio1999,Serra+11,Pizzardo23}. The main source of uncertainty is the assumption of spherical symmetry and thus projection effects: the caustic technique does estimate different mass profiles for the same cluster observed along different los \citep{Serra+11, Svensmark+15}. The above mentioned effect of the variance in $\beta(r)$ plays a substantially minor role. 
The performance of the caustic method also depends on the number of spectroscopic tracers \citep{Serra+11,Logan22}. Reliable identification of the escape-velocity edge typically requires several tens to $\sim 200$ galaxies per cluster \citep{Serra+11}. In the regime of sparse sampling, stacking techniques can be employed to recover average mass profiles with significantly reduced scatter \citep[e.g.,][]{Rines+03, RinesDiaferio06, Rines+13}, enabling applications to large cluster samples in current and future surveys \cite[e.g.,][]{Gifford2017, Pizzardo24}. Alternatively, a statistical correction can be applied to recover the intrinsic escape-velocity edge \citep{Halenka+22,Rodriguez+24}.

Finally, since eq.~\eqref{eq:cau} provides a differential estimate of the mass profile, the caustic method can be combined with other, more precise, mass estimates in the inner regions, obtained for example by gravitational lensing \citep{Umetsu25},
or by techniques that use galaxies kinematics in a dynamical equilibrium configuration \citep[e.g., \texttt{MAMPOSSt}, see][]{Biviano_2013}. This hybrid approach leverages the strengths of different methods that have different strengths in different cluster regions, 
and provides a reliable reconstruction of the cluster mass profile from the core to the infall region \citep{Biviano_2003,Biviano_2013}.

The caustic method has been widely applied to observational datasets, providing mass profiles for individual clusters and large samples \citep[e.g.][]{Geller_1999,Rines+00,Biviano_2003,Rines+03,RinesDiaferio06,Alpaslan+12,Rines+13,Guennou+14,Maughan2016,Sohn2019ApJ...871..129S,Pizzardo21}. The caustic technique has been shown to provide cluster mass estimates in good agreement with the virial estimates \citep[][see Sect.~\ref{sec:virial}]{Rines+03,RinesDiaferio06} and with the masses derived from the Jeans equation \citep{FBC17}. More importantly, the caustic and weak lensing mass profiles extending to the infall region  agree within the uncertainties in clusters where the two probes, both independent of the dynamical state of the cluster, can be applied \citep{Diaferio2005,Geller2013ApJ...764...58G,Geller2014}.
{ On the other hand, comparisons between caustic and X-ray cluster mass profiles might show some disagreement \citep[e.g.][]{Sereno2015MNRAS.450.3649S, Ettori2019A&A...621A..39E, Logan22}. These discrepancies are mostly due to the application of the caustic method to the very central region of the cluster, for which the method is not designed. X-ray mass estimates are usually limited within $r_{500}$, sometimes within $r_{200}$. The caustic method was conceived to provide reliable estimates beyond $r_{200}$ \citep{Diaferio1999, Serra+11}, so possible discrepancies are not surprising.\footnote{{The misunderstanding of considering and using the caustic method as a simple  mass estimator within $r_{200}$ rather than an estimator of the mass beyond $r_{200}$ can also lead to the incorrect conclusion that the method is not independent of the dynamical state of the cluster \citep[e.g.][]{Monteiro22}.}} In addition, the X-ray estimate usually assumes hydrostatic equilibrium, unlike the caustic method: if the intracluster medium is out of equilibrium, the X-ray mass is likely to be incorrect. Indeed, in some clusters where  spectroscopic,  X-ray and gravitational lensing data are available, the X-ray mass disagrees with the caustic and lensing masses because of the lack of hydrostatic equilibrium of the intracluster medium \citep{Diaferio2005}.}
With the advent of next-generation spectroscopic surveys, the role of the caustic technique is expected to become increasingly important for studies of the cluster outskirts and mass accretion \citep{DeBoni+16}, and for testing GR on Mpc scales \citep[e.g,][]{Pizzardo24,Butt24}.

\paragraph{Other methods}
A technique strictly related to the caustic method is the {\it Fair Galaxies} mass estimation of \citet{Cupani+08,Cupani+10}. Similarly to the caustic technique, the Fair Galaxies method is not well suited to estimate the mass in the inner cluster regions, while it provides a cluster mass estimate in the outer, out-of-equilibrium, regions.
This method is based on the spherical collapse model and it exploits the non-equilibrium, infall region of a cluster, for estimating its mass, given that the infall velocity is related to the mass over-density. Since the full kinematics of cluster galaxies is not directly observable, the method relies on numerical simulations to identify a region in p.p.s. in which galaxies are likely to have their los velocity equal to their radial infall velocity.

\citet{Cupani+10} compared the mass estimates of nine clusters obtained with the Fair Galaxies technique, with those obtained by \citet{RinesDiaferio06} via the caustic technique. The two estimates are consistent within $\sim 1.5 \, \sigma$ for seven clusters, but the uncertainties are on average significantly larger for the Fair Galaxies method than for the caustic one. 

Another method that aims to use galaxies in the cluster infall region to determine cluster masses is that of \citet{Falco+14}. The mean infall velocity of structures outside the cluster depends on the cluster mass, according to a relation that is determined from simulations. Similarly to \citet{Cupani+10}, \citet{Falco+14} seek for suitable tracers of the infall pattern, to break projection effects. They argue that these tracers may be found in bound systems with a small velocity dispersion, lying on a preferential direction in 3D space, and forming filament-like structures in the projected velocity space. Application of this method to the Coma cluster of galaxies yielded a value in agreement with previous estimates with comparable error bars.

Like \citet{Falco+14}, also the method of \citet{Hamabata+19} exploits the dependence of the infall velocities of galaxies in the cluster outskirts on the cluster mass. They model the los velocity distribution in the cluster outskirts and find that the shape of this distribution shows a mild dependence on the cluster mass. They estimate that they can reach a 15\% accuracy on the mean mass of a cluster sample which comprises a total of $\sim 10^5$ cluster galaxies. 

\subsection{Towards Machine Learning approaches}
\label{Sec:ML}
The increasing volume and complexity of spectroscopic and photometric data from current and upcoming surveys naturally motivate the adoption of ML techniques as complementary tools to traditional dynamical analyses. Unlike Jeans-based methods, which start from physical assumptions and well-motivated equations (e.g., the Vlasov equation), ML approaches are generally data-driven and aim at learning the mapping between observables and underlying physical quantities directly from simulations or labeled datasets.

A first concrete application of machine learning to cluster kinematical mass estimate was presented by \cite{Ntampaka2015,Ntampaka2016_ML}, who explored how to go beyond the standard $M-\sigma_v$ scaling relation by exploiting the full information content of the los velocity distribution. Using mock cluster catalogues extracted from $N$-body simulations and projected along multiple los, these works introduced a framework in which each cluster is represented by the empirical distribution of galaxy los velocities, and supervised learning algorithms are trained to directly map this distribution to the underlying halo mass.

In this approach, the full velocity probability distribution function (PDF hereafter) is used as input, rather than being reduced to a finite set of summary statistics. This approach allows the model to capture non-Gaussian features of the phase-space distribution --- such as asymmetries, heavy tails, or multi-component structures --- which are difficult to model analytically and are often associated with mergers, projection effects, and ongoing accretion.

Within this framework, it was shown that a significant fraction of the scatter in dynamical mass estimates is driven by variations in the shape of the velocity distribution itself. Higher-order moments of the velocity PDF, such as the kurtosis, already correlate with the residuals of the $M-\sigma_v$ relation and can be used as simple corrections to reduce the scatter. More generally, however, ML methods that operate on the full velocity distribution lead to a substantial improvement in mass predictions compared to traditional scaling relations, demonstrating that a large amount of dynamical information is encoded in the detailed structure of the velocity PDF.

More recently, deep learning approaches have been developed to operate directly on projected phase-space representations. Convolutional neural networks (CNNs), for instance, can treat the distribution of galaxies in p.p.s. as an image, where each pixel encodes the number density of tracers. In practice, the input data are constructed by transforming discrete galaxy measurements into continuous phase-space distributions, typically using kernel density estimators, and sampling them on a regular grid to produce fixed-size images suitable for neural network architectures. The network is trained on simulated cluster catalogues, where the true halo mass is known, to learn a mapping between these phase-space ``images'' and the underlying mass. Through successive convolutional layers, CNNs extract hierarchical features from the data, identifying patterns such as the extent of the velocity envelope, asymmetries, substructures, and distortions induced by projection effects. In this sense, these methods can be viewed as a data-driven generalization of classical phase-space techniques, exploiting the full morphology of the galaxy distribution rather than relying on summary statistics such as $\sigma_v$. 

Early applications demonstrated that ML models can substantially improve
dynamical mass reconstruction under controlled simulated conditions.
\citet{Ho_2019} applied one- and two-dimensional CNNs to the distribution
of galaxy los velocities and to the joint $(R,v_{{\rm los}})$ distribution,
respectively. Their models achieved a log-normal scatter as low as
$0.132$ dex, more than a factor of two smaller than that obtained with the
standard $M$--$\sigma_v$ relation. In a complementary approach,
\citet{Armitage2019} trained several supervised regression algorithms on
combinations of spectroscopic and other cluster observables extracted from
the \textsc{MACSIS} hydrodynamical simulations. For their simulated
spectroscopic sample, the scatter in the predicted-to-true mass ratio
decreased from $0.130$ dex for the $\sigma_v$--mass relation to
$0.031$ dex when multiple observables were included. These results
demonstrate that a significant information gain can be achieved with ML, although the quoted
precision was obtained when the training and test samples were generated
within closely matched simulation frameworks.
       \begin{figure}[ht]
         \centering
         \includegraphics[width=.99\linewidth]{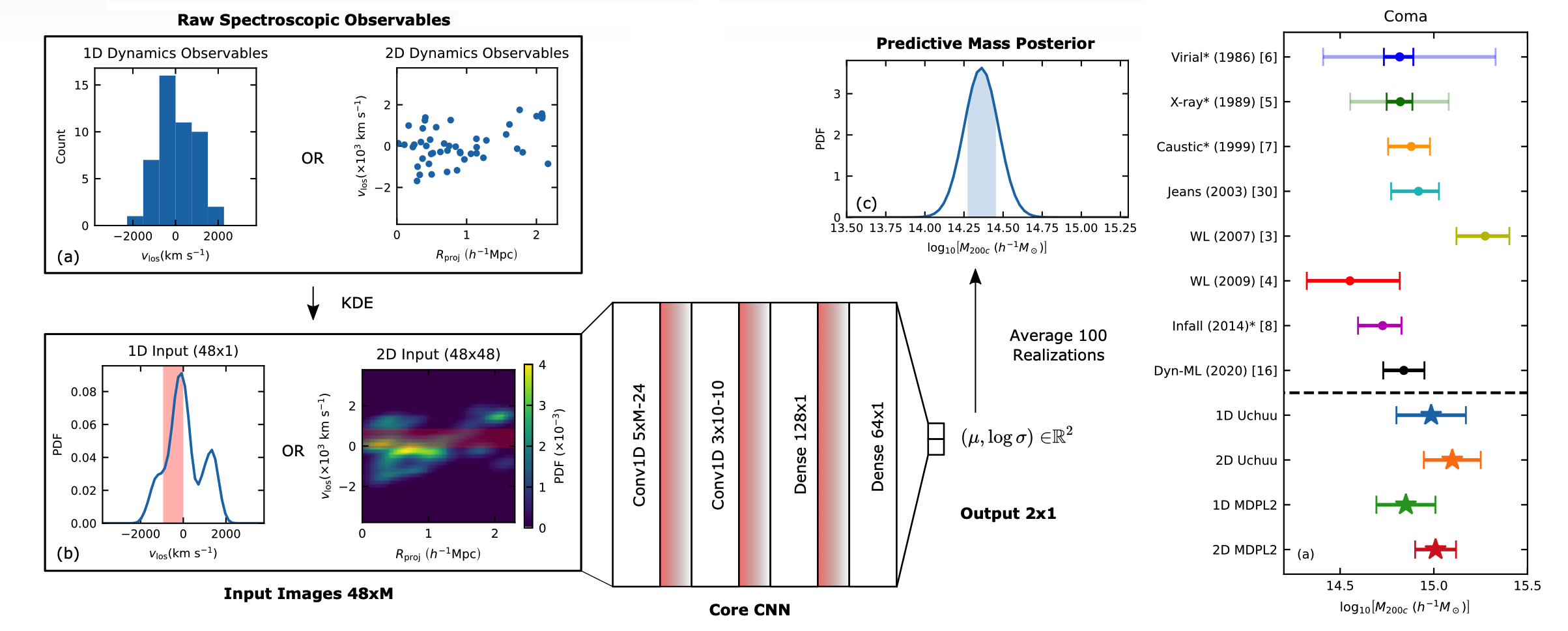}
        
         \caption{\label{fig:testHo}  Left panel: Machine-learning workflow for dynamical cluster mass inference. (a) Dynamical observables derived from galaxy data, in terms of los velocities and projected radii. (b) Convolutional neural network architecture used to extract features from phase-space distributions. (c) Example of the resulting predictive posterior on the cluster mass. Right panel: resulting estimates of $M_{200}$ (four bottom stars with errorbars), along with their 16th to 84th percentile confidence intervals, compared to other results from the literature. Figures adapted from Figs. 1 and 3 of \cite{Ho2022_ML}, copyright by the author(s).}
    \end{figure}

Further developments have focused on incorporating a rigorous probabilistic framework within deep learning models. In particular, \citet{Ho2021,Ho2022_ML} introduced approximate Bayesian CNNs for cluster mass estimation, enabling the reconstruction of full predictive posteriors rather than single-point estimates. In these models, the network predicts a full posterior distribution for the cluster mass rather than a single-point estimate.  In practice, the output encodes both a mean estimate and an associated uncertainty, capturing the intrinsic scatter of the dynamical observables as well as the error related to the mapping learned by the neural network itself, arising from the finite and imperfect nature of the training data. This error is typically estimated by evaluating the network multiple times under different realizations of its internal parameters (e.g., via Monte Carlo dropout), effectively approximating the marginalization over model uncertainties (left plot of Fig.~\ref{fig:testHo}). \citet{Ho2022_ML} applied this framework to the Coma cluster, obtaining
$M_{200}=10^{15.10\pm0.15}\,h^{-1}\,{\rm M_\odot}$, consistent with
previous measurements, and tested the robustness of the result across
multiple simulated training catalogues (right panel of
Fig.~\ref{fig:testHo}).

\begin{figure}[ht]
         \centering
         \includegraphics[width=.99\linewidth]{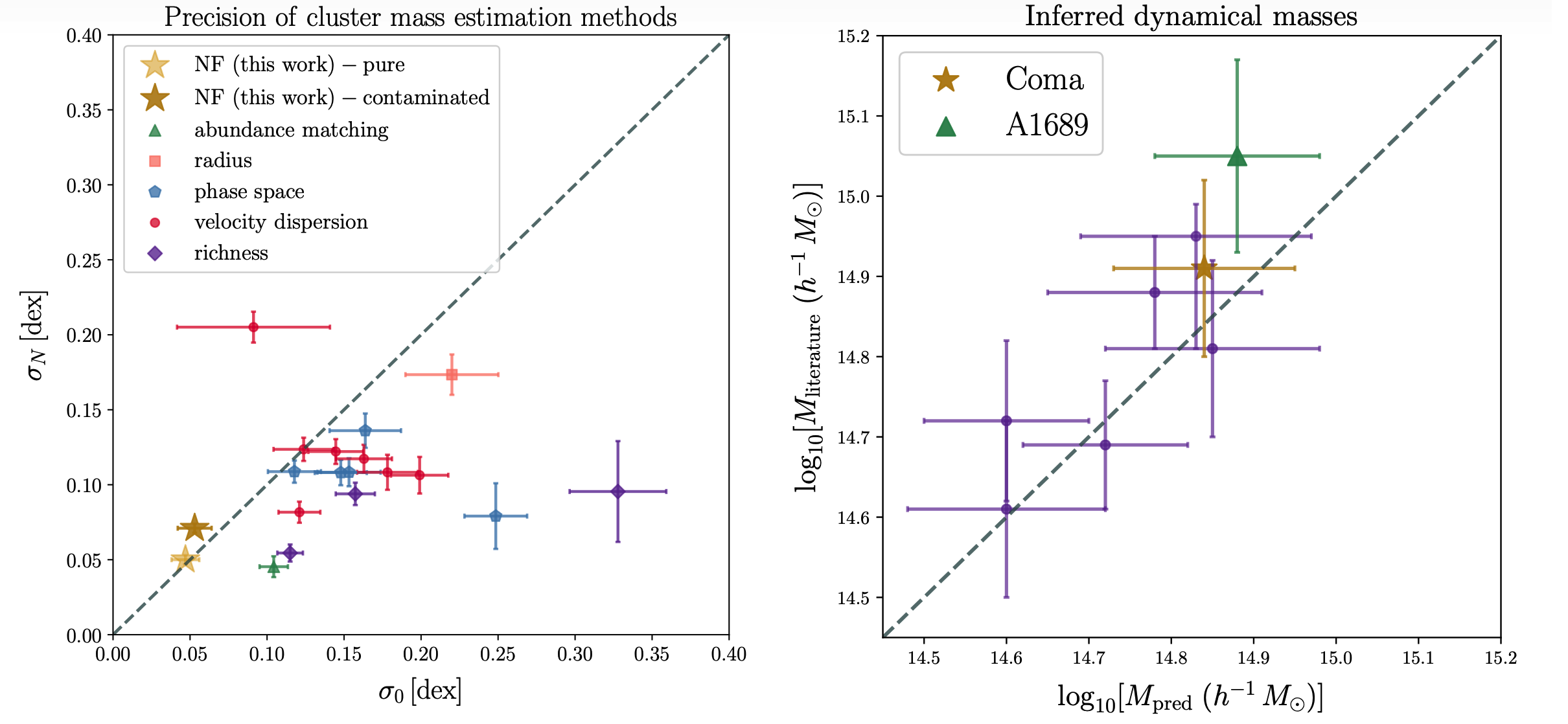}
        
         \caption{\label{fig:Kodicomparison}  Left panel: Fig.~6 of \cite{Kodi_2020}. Performance of the normalizing-flow mass estimator compared to standard cluster mass estimators based on richness, $\sigma_{{\rm los}}$, galaxy distributions, and abundance matching. The uncertainties are decomposed into a richness-dependent scatter $\sigma_N$ (for clusters with 100 members) and a richness-independent systematic term $\sigma_0$, defined in eq.~(8) of  \cite{Kodi_2020}. Right panel: comparison of the Neural Flow mass estimates and constraints from the literature. The golden star and the green triangle indicates the Coma cluster and the cluster Abell 1689. Adapted from Fig.~8 of \cite{Kodi_2020}. In both figures, uncertainties are $1\,\sigma$. Images reproduced with permission, copyright by RAS.}
    \end{figure}

Further developments have explored alternative probabilistic frameworks that go beyond approximate Bayesian neural networks. In particular, \citet{Kodi_2020} introduced a normalizing-flow approach to model the full p.p.s., enabling a flexible and explicit reconstruction of the underlying probability density without relying on restrictive parametric assumptions. By learning an invertible mapping between a simple latent distribution and the complex observed p.p.s. structure, this method provides a powerful way to capture non-Gaussian features and correlations in the data, which are difficult to model within traditional dynamical approaches. 
They applied the Neural Flow (NF) model to a sample of eight real galaxy clusters, showing that it can infer cluster dynamical masses in remarkable agreement with other estimates in the literature. The performance of their approach and the comparison with the mass estimates in the literature of their group sample are displayed in the left and right plots of Fig.~\ref{fig:Kodicomparison}. Closely related to this method, \citet{Kodi_2021} proposed another approach which relies on simulation-based inference (SBI, also referred to as ``likelihood-free inference''). In this framework, neural density estimators are trained on simulations to directly learn the posterior distribution of model parameters, bypassing the need for an explicit analytical likelihood \citep[see, e.g.,][]{Alsing2019,Lemos23}. 

A direct comparison between ML and conventional likelihood-based inference
was  presented by \citet{Mundow2025}. They inferred the cluster
virial mass directly from the redshift-space distribution of tracers within
and around clusters, without first identifying members or removing
interlopers. The mock catalogues were extracted from MultiDark--Planck 2
(MDPL2), a DM-only cosmological $N$-body simulation of a
$(1\,h^{-1}\,{\rm Gpc})^3$ volume adopting a Planck cosmology, with subhalos
and nearby distinct halos used as proxies for galaxies. Using maps of the
tracer number density in redshift space, the autoencoder CNN achieved an rms scatter of $0.10$ dex in the virial mass, compared with $0.16$ dex for
the maximum-likelihood estimator.

This comparison requires some care because the two estimators have
different conditional properties. The maximum-likelihood estimator is
constructed to be asymptotically unbiased at fixed true mass, whereas a
network trained by minimizing the mean squared error approximates the
posterior mean $\langle\log M \mid{\rm data}\rangle$ and is therefore
calibrated at fixed inferred mass. The smaller CNN scatter cannot thus be
interpreted solely as additional dynamical information extracted by ML,
since it also depends on the mass distribution encoded in the training
sample. When mean peculiar velocities were used as input, the scatter was
$0.12$ dex for tracers binned in redshift space and $0.16$ dex when they
were binned according to noisy observed distances, for which inhomogeneous
Malmquist bias becomes severe. The resulting $0.04$ dex degradation
provides a direct measure of the sensitivity of the reconstruction to the
observational representation adopted in the mock catalogues.

ML approaches have also started to extend beyond basic cluster mass reconstruction to the inference of cosmological parameters from ensemble cluster properties. \citet{Qiu_2024} developed a framework that combines multiple cluster observables --- $\sigma_\mathrm{los}$, properties of the ICM, and stellar content --- within a supervised learning scheme trained on multi-cosmology hydrodynamical simulations. The analysis demonstrates that the complex correlations among cluster properties can be efficiently mapped into constraints on cosmological parameters (e.g., $\Omega_\mathrm{m}$ and $\sigma_8$), achieving competitive precision with respect to traditional approaches. Instead of relying on a simplified one-to-one mass proxy, the method of \cite{Qiu_2024} makes use of the full multi-dimensional structure of cluster scaling relations. 

Other applications of ML to the kinematics of clusters concern the analysis of the dynamical state and the identification of substructures in phase space. \cite{Henriksen2024} employed a clustering algorithm (k-means) to the combined spatial and kinematical distribution of galaxies in the cluster Abell 2146, using projected positions and line-of-sight velocities as input features. In their approach, galaxies are partitioned into a set of distinct substructures by minimizing the distance from cluster centroids in the 3D space $(x,y,v_z)$. The optimal number of substructures is determined using standard clustering diagnostics, such as the silhouette score and the elbow method, leading to a decomposition of the cluster into multiple dynamically coherent components. \cite{Henriksen2024} showed how the procedure can be used to reconstruct radial mass profiles under the assumption that the identified components are gravitationally bound but not virialized.

Overall, the main limitation of these approaches is not the flexibility of the network architecture, but the transfer of a relation learned from
simulations to real observations. Performance on a held-out test set
drawn from the same simulation and mock-generation pipeline does not, by
itself, demonstrate that the estimator is unbiased when applied to survey
data. Approximate Bayesian neural networks and simulation-based inference can quantify the uncertainty conditional on the adopted simulations, but they do not automatically account for an incorrect galaxy--halo
connection, baryonic prescription, cosmology, or observational selection
function. This simulation-to-data mismatch therefore represents a source
of systematic uncertainty distinct from the statistical and network
uncertainties returned by the model.

Constructing suitable training sets is particularly demanding because
the mock catalogues must be sufficiently large to sample halo mass,
redshift, dynamical state, environment, and projection, while also
reproducing the properties of the observations. Hydrodynamical
simulations model galaxy formation more directly, but their limited
volumes generally provide fewer massive clusters and independent survey
light cones. A practical alternative is to use large-volume
dark-matter-only simulations populated with galaxies through a
semi-analytic model. Such catalogues can reproduce survey geometry, flux
limits, redshift selection, foreground and background galaxies, and
projection effects at comparatively low computational cost.

This strategy was implemented by \citet{Kodi_2021}, who populated the
$1\,h^{-1}\,{\rm Gpc}$ MDPL2 simulation with \textsc{SAG}
(Semi-Analytic Galaxies), a semi-analytic galaxy-formation
model that assigns baryonic and observable properties to galaxies
evolving within the DM halo merger trees. From this, they constructed an SDSS-like mock catalogue. The
mock included an apparent-magnitude limit, redshifts
accounting for the Hubble flow, random observer distances, multiple los, and interloper contamination. The \textsc{SAG} model was selected
in part because of its treatment of orphan satellites, which improves the
description of galaxies in cluster cores. Nevertheless, a semi-analytic
catalogue does not eliminate simulation-to-data mismatch: its satellite
abundance, radial distribution, luminosity function, orphan treatment,
and galaxy velocity bias may still differ from those of real clusters.

Robust applications therefore require the complete survey selection to
be forward-modelled, with validation across different (semi-analytic and
hydrodynamical) prescriptions, and tests on observations with independent
mass constraints. The mass and redshift distributions of the training
set must also be controlled, because a supervised estimator learns a
conditional relation that depends on the training prior. In particular,
it can exhibit regression towards the mean and biased predictions near
the boundaries of the mass range represented in the training catalogue.

\subsection{Kinematics and systematics: a multi-layered challenge}
\label{sec:syst}
The dynamical methods discussed in the previous sections extract cluster masses from different summaries of the same distribution of observables, that is, projected positions and line-of-sight velocities.
As we have mentioned several times, each technique is prone to different kinds of systematic uncertainties, which strongly depend on the  method employed and the assumption made. For instance, a $\sigma_\mathrm{los}$ estimate
contaminated by interlopers, directly biases virial and scaling-relation
masses; an incorrect tracer density or anisotropy model affects Jeans-based inference; an ill-defined phase-space boundary propagates into caustic masses; and departures from equilibrium alter all estimators that interpret galaxy motions as tracers of a stationary potential. The relevant question is therefore not only whether interlopers, substructures, or non-relaxed dynamics are present, but how
they enter a given estimator and how they can be mitigated. 

\begin{table}[ht]
\centering
\small
\setlength{\tabcolsep}{4pt}
\caption{Main sources of systematic uncertainty in cluster kinematics, their typical impact on the Virial, Jeans, Caustic and Machine Learning (V, J, C, ML) mass estimates, and commonly adopted mitigation strategies.
}
\label{tab:systematics}
\begin{tabular}{lcc}
\toprule
Effect & Mass bias & Methods \\
$\rightarrow$ Mitigation       &           & concerned  \\
\toprule

{Spatial and velocity bias} & order of magnitude & V \\
$\rightarrow$ use early-type galaxies & & \\[0.1cm]

$\cal{F}_{\beta}$ uncertainty & {$\sim 30$\%} & {C} \\
$\rightarrow$ rely on simulations & & \\[0.1cm]

Interlopers & $\sim 10$--$30\%$ & VJC \\
$\rightarrow$ p.p.s. cleaning & & \\[0.1cm]

Incomplete sampling & Model-dependent & VJ \\
$\rightarrow$ radial completeness correction & & \\[0.1cm]

Substructures & $\sim 10$--$50\%$ (up to $\sim 2\times$) & VJ \\
$\rightarrow$ identification and removal & & \\[0.1cm]

Non-equilibrium & $\sim 20\%$ (larger in outskirts) & VJ \\
$\rightarrow$ selection criteria & & \\[0.1cm]

Triaxiality & Up to $\sim 50\%$ & VJC \\
$\rightarrow$ stacking in p.p.s. & & \\[0.1cm]

Rotation & $\sim 25\%$ (in 1/4 clusters) & VJC \\
$\rightarrow$ identification and $\sigma_{{\rm los}}$ correction & & \\[0.1cm]

Simulation-to-data mismatch & Model- and survey-dependent & ML \\ $\rightarrow$ survey-matched mocks & & \\
and cross-simulation tests & & \\ [0.1cm]

\bottomrule
\end{tabular}
\end{table}

In Table~\ref{tab:systematics} we report a summary of the main source of systematic uncertainties affecting the mass estimates in the kinematical analysis of galaxy clusters. Some uncertainties are specific to some methods, and have been discussed in the relevant sections (the light-traces-mass assumption in the virial theorem, Sect.~\ref{sec:virial}, and the $\cal{F}_{\beta}$ term uncertainty in the Caustic method, Sect.~\ref{sec:outskirts}), others affect all methods. ML-based mass estimators are generally exposed to the same observational and dynamical effects because these alter the p.p.s. distribution used as input. However, the mass biases quoted in
Table~\ref{tab:systematics} for the V, J, and C methods cannot be
transferred directly to ML estimators: their magnitude and even
their sign depend on whether the corresponding effects, and their
correlations, are adequately represented in the training
catalogue. The additional systematic specific to ML is therefore
the mismatch between the simulations used for training and the
survey data to which the estimator is applied.

A quantitative example is provided again by \citet{Kodi_2021}. For their
SDSS-like mock catalogue, the CNN mass estimates had a
mean residual $\log (M_{\rm true}/M_{\rm pred})=0.04$ dex and a scatter $\sigma_{\epsilon}=0.16$ dex. Interlopers caused pronounced mass overestimates for some
low-mass systems, whereas an extreme test mimicking the loss of
spectroscopic targets in cluster cores due to fibre collisions
shifted the mean predicted mass by only $-0.017$ dex. As discussed
in Sect.~\ref{Sec:ML}, \citet{Mundow2025} found that replacing
redshift-space coordinates with noisy distance estimates affected by inhomogeneous Malmquist bias increased the rms scatter from $0.12$ to $0.16$ dex. Note that the systematic error budget of ML-based dynamical estimators is still being characterised and remains strongly dependent on the 
simulations used for training and on the target survey. For the purposes of this review, we therefore limit the discussion to the examples above,
rather than attempting to assign a general systematic uncertainty to
ML-based mass estimates.

In general, systematic effects in dynamical analysis are not independent and often act simultaneously, making their calibration and mitigation particularly challenging. A robust dynamical analysis therefore requires a careful, self-consistent treatment of all the systematics at the same time, ideally integrating cosmological simulations to complementary multi-wavelength observations. In the following we discuss the different systematics in turn, reviewing the main findings and the strategies implemented to reduce the impact of such systematics  in kinematical mass modeling.

\subsubsection{Observational uncertainties}
The uncertainties in the galaxy redshift estimates inflate the observational estimate of the cluster $\sigma_v$, but if they are known, their inflating effect can be corrected for, as shown by \citet{DdZdT80}:
\begin{eqnarray}
\sigma_c = (\sigma_o^2 - \delta_v^2)^{1/2} \\
\mathrm{with} \; \delta_v \equiv c \, \delta_z /(1+z_c), \nonumber
\end{eqnarray}
where $\sigma_c$ and $\sigma_o$ are the corrected and observed velocity dispersion, $z_c$ is the cluster redshift, and  $\delta_z$ is the average error on the cluster galaxy redshifts.

In general, modern spectroscopic surveys achieve uncertainties $\lesssim 150$~km~s$^{-1}$ for the rest-frame velocities of cluster members \citep[e.g.,][]{Rines+03,Biviano_2013,Biviano_2021} much below the typical velocity dispersion of clusters. Since the velocity errors add in quadrature to the cluster intrinsic $\sigma_{{\rm los}}$, small uncertainties in galaxy redshifts can be neglected. Notable exceptions occur when we are dealing with groups, rather than clusters of galaxies, characterized by small $\sigma_v$, or when the redshifts are obtained via very low resolution (typically slitless) spectroscopy, such as in the Euclid survey \citep{Euclid_LeBrun}.

Another observational uncertainty that needs to be considered is miscentering, that is the spatial offset between an observationally chosen cluster center and the true minimum of the gravitational potential well. \citet{Adami98} used a very limited sample of nearby clusters to estimate that optical and X-ray determinations of cluster centres are in agreement within $\sim 60 \pm 30$~kpc, on average. This preliminary estimate was confirmed by later studies. Based on a sample of 129 X-ray emitting groups, \citet{George2012} found that BCGs near X-ray centroids trace the cluster centre to $\lesssim 75$ kpc accuracy, on average; the X-ray peak and the centroid of the cluster galaxy distribution are less accurate centre estimators. Centering is however less accurate for those 30\% of groups that have multiple bright or massive galaxies.  \citet{Zhang2019} compared the optical and X-ray centres of 145 optically-identified clusters, and found that the two centre determinations are similar, within $\sim 50$ kpc (at the 1 $\sigma$ confidence level) in 60-80\% of the cases. \citet{Seppi2023} studied a sample of 87 massive X-ray emitting clusters and estimated an average offset between the X-ray and optically identified centres of $\simeq 76 \pm 30$ kpc.

Miscentering by $\lesssim 100$ kpc is not expected to have strong impacts on cluster dynamical $M(r)$ estimates, since these are in general based on samples of galaxies that reach to much larger distances, and not many of them occupy the inner $\sim 100$ kpc, a region dominated by the BCG \citep[e.g.,][]{Zhang26b}. An example of the robustness of the dynamical Jeans analysis vs. miscentering is provided by \citet{Mamon2019}, whose results show very little dependence on whether the BCG or the X-ray peaks are chosen as centres for their sample of 54 clusters. However, these authors only considered regular clusters, and miscentering is expected to be stronger in irregular clusters \citep{George2012}. In this respect, a more critical test is provided by the Jeans analysis of the Coma cluster by \citet{Pedratti26}. The Coma cluster is a classical example of a cluster where there is not a single dominating BCG, but two. \citet{Pedratti26} showed that selecting either of the two BCGs as the cluster centre, did not change significantly the \textsc{MAMPOSSt}-inferred cluster $M(r)$.

Given the typical scale of the cluster centering uncertainty, it might be argued that it is impossible to constrain the inner slope, $\gamma$, of $M(r)$, by galaxy dynamics. In fact, \citet{Sartoris_2020} have shown that, to constrain $\gamma$, using only galaxies as tracers of the potential is ineffective, and it is necessary to include the BCG stellar kinematics in the Jeans analysis (see Sect.~\ref{sec:kinDM}). Still, these constraints are valid only if the BCG is indeed located at the cluster centre.

\subsubsection{The selection of the tracers of the gravitational potential}
\paragraph{Interlopers}
\label{sec:interlopers}
The identification of cluster members is the first and most critical step in any kinematical analysis. Interlopers can significantly bias measurements of $\sigma_{{\rm los}}$ and phase-space reconstructions. Interlopers can be defined in two ways: (1) galaxies that are at a projected radius $R$ from the cluster centre, but have a 3D distance from the centre $r \gg R$;  (2) galaxies that are unbound to the cluster but are not well separated from the cluster members in p.p.s.. More specific definitions have been given in the literature, for instance, \citet{Wojtak2007} defined as interlopers those galaxies that are located beyond twice the cluster virial radius in 3D distance, with a velocity with respect to the cluster that exceeds the escape velocity.

The impact of interlopers on kinematic mass estimates has been extensively quantified using mock galaxy catalogues extracted from cosmological simulations. 
\citet{Mamon2010} showed that projection effects alone lead to a substantial fraction of contaminants even after standard velocity cuts, with interlopers accounting for $\sim 20\%$ of galaxies within the virial radius and becoming dominant in the outer regions. \cite{Wojtak2018} further demonstrated that residual contamination in membership selection systematically propagate into dynamical mass estimates, leading to  
an overestimate of cluster masses. However, interloper removal techniques (described below) can mitigate the effect of contaminants; interlopers can even lead to 
an under-estimate of the cluster $\sigma_v$ \citep{Cen97,Biviano2006}, since most interlopers that remain after an initial selection of cluster members in velocity space, have in fact low velocities compared to the velocity distribution of cluster members \citep{Mamon2010}. More generally, tests based on realistic mock observations indicate that imperfect membership selection is one of the dominant contributors to the intrinsic scatter of dynamical mass estimators \citep[e.g.,][]{Old_2015,Ntampaka2015}.
In addition, interlopers blur the edges of the p.p.s. distribution, directly affecting methods based on escape velocity profiles, such as the caustic technique. 
This effect is particularly important in the cluster outskirts, where the transition between bound and infalling galaxies is gradual and projection effects are more severe. 
Numerical simulations show that uncertainties in the identification of the caustic amplitude contribute to the typical $\sim 20$--$30\%$ scatter in mass estimation \citep{Diaferio1999,Gifford2013,Pizzardo23}. 

A variety of techniques have been developed to mitigate interloper contamination. Classical methods include velocity clipping \citep{Yahil1977,Zabludoff_1993}, and the more sophisticated shifting gapper \citep[\texttt{SG},][]{Fadda1996,Girardi98}. In the \texttt{SG} method, galaxies are first grouped in radial bins, of typical size $\sim 0.4$ Mpc/h, or of a size large enough to contain at least 15 galaxies. Within each bin, the los velocities are sorted, and gaps between consecutive velocities are computed.
Large gaps (above a given threshold, typically $\sim 1000$ km s$^{-1}$) are interpreted as separations between dynamically distinct populations. Galaxies separated from the main body by such gaps are rejected as interlopers. This procedure is iterated outward in radius, by shifting the radial bins along the distance to the cluster centre, 
allowing the velocity selection to adapt to the radial dependence of $\sigma_{{\rm los}}$. A similar approach is employed in the peak+gap (\texttt{P+G}) method \citep{Girardi1993,Biviano_2013}, which performs an initial identification of the main velocity peak via an adaptive kernel density estimation, before running the \texttt{SG} method.

Another approach which directly analyzes the galaxy distribution in p.p.s. is provided by the \texttt{CLUMPS} algorithm \citep{Biviano_2021}. The method estimates the density of galaxies in the $(R, v_{{\rm los}})$ plane by binning the data and smoothing the resulting distribution with a Gaussian kernel. For each radial bin, the main peak in velocity space is identified, and the surrounding minima define the velocity boundaries of the cluster population. Galaxies lying outside these boundaries are rejected as interlopers. 

Another p.p.s.-based method for interloper removal is the \texttt{GalWeight} algorithm of \cite{Abdullah_2018}, which assigns to each galaxy a weight according to its position in $(R,v_{{\rm los}})$. 
The total weight is written as the product of a dynamical weight and a phase-space density weight,
\begin{equation}
W_\mathrm{tot}(R,v_{{\rm los}})=W_\mathrm{dy}(R,v_{{\rm los}})\,W_\mathrm{ph}(R,v_{{\rm los}}).
\end{equation}
The dynamical term is constructed from the projected number density profile, the velocity dispersion profile, and the radial distribution of galaxies, while $W_\mathrm{ph}$ is estimated with an adaptive kernel density method in the $(R,v_z)$ plane. 
An optimal weight contour is then selected to separate members from interlopers, with a final radial cut chosen as either the virial or turnaround radius. 
Tests on simulations show high completeness, $\simeq0.99$ within the virial radius, with contamination fractions of $\simeq12$\%; at larger radii the completeness remains high but the contamination increases, especially in the infall region. 

\citet{SD13} investigate two alternative methods to identify  the cluster members defined as the gravitational bound galaxies: the first method uses the location of the caustics in the p.p.s.; the second method adopts a binary hierarchical tree, based on the
projected galaxy pairwise energy.   In the former method, the cluster members are simply identified with the galaxies within the upper and lower caustics. In the latter method, all the galaxies in the cluster's field of view are arranged into a hierarchical binary tree. This hierarchy is built by calculating the projected pairwise binding energy ($E_{ij}$) between galaxies, using the equation 
\begin{equation}
E_{ij} = -G \frac{m_i m_j}{R_p} + \frac{1}{2} \frac{m_i m_j}{m_i + m_j} \Pi^2,
\end{equation}
where $R_p$ is the pair projected separation, $\Pi$ is the line-of-sight velocity difference and $m_i, m_j$
are the two galaxy masses. Next, the algorithm determines where to cut the binary tree to separate candidate cluster members from the unlinked background. By measuring $\sigma_{{\rm los}}$ as one descends from the root down the main branch, a distinct, nearly isothermal "plateau" is identified. Cutting the tree at the node starting this plateau filters out obvious interlopers and yields a core group of members. Tests on simulated clusters indicate that the caustic location method returns a completeness fraction $f_c=0.95\pm 0.03$ at $3r_{200}$ and an interloper fraction that increases from $f_i=0.02^{+0.05}_{-0.02}$ at $r_{200}$
to $f_i=0.08^{+0.11}_{-0.05}$ at $3\,r_{200}$; the binary tree method returns $f_c=1.00^{+0.00}_{-0.05}$ at $3 \, r_{200}$, $f_i = 0.03^{+0.05}_{-0.02}$ at $r_{200}$, and $f_i = 0.13^{+0.10}_{-0.06}$ at $3r_{200}$. The advantage of these methods, and particularly the caustic location method, is that they do not need to assume dynamical equilibrium and naturally extend the member identification into the infall region \citep{Sohn2019ApJ...871..129S,Sohn2020ApJ...891..129S,Sohn2020ApJ...902...17S}

All the methods described so far do not make any assumptions about the cluster mass profile, but they depend on several, rather {\it ad hoc} parameters (such as the size of the velocity gap), that need to be calibrated with numerical simulations \citep[as done in, e.g.,][]{Biviano_2026}. Alternative approaches try to establish the presence of interlopers by relying on mass or mass profile estimates, often by iteration. 

The simplest of these methods relies on the comparison of the virial and projected mass estimates (see Sect.~\ref{sec:virial}), relying on the different sensitivities of the two mass estimates on the presence of interlopers \citep{Perea+90,WL07b}. Another method that relies on the virial mass estimate, is that of \citet{denHartogKatgert1996}. In this method, galaxies are rejected as interlopers if their velocities exceed the maximum velocity allowed for a bound particle at the galaxy radial distance from the cluster centre. 

Similar in spirit to the method of \citet{denHartogKatgert1996}, but more sophisticated, is the \texttt{Clean} algorithm of \cite{Mamon_2013}. \texttt{Clean} assumes parametric models for $M(r)$ and $\beta(r)$,
using a scaling relation from an initial estimate of $\sigma_\mathrm{los}$ to obtain a first guess value for $M_{200}$, and from this, $c_{200}$ using a relation from cosmological numerical studies. Adopting the NFW profile, 
$\sigma_{{\rm los}}(R)$ is then derived through the Jeans equation. Galaxies are then rejected if their rest-frame velocity satisfies
$|v_{z}(R)| > 2.7 \, \sigma_\mathrm{los}(R)$.
The procedure is iterative, since both $\sigma_\mathrm{los}(R)$ and the mass estimate are updated using the progressively cleaned sample. This method therefore links interloper rejection to an explicit dynamical model.
The \texttt{Clean} method has been  applied with some modification, to $\sim 100$ clusters detected by the Planck satellite through the SZ effect 
\citep{Sereno25}. The cluster masses obtained with the \texttt{Clean} membership selection were found to be in agreement with weak lensing mass estimates to within 8\%, and to have a systematic error of only $\sim 5$\% when estimated with $\gtrsim 50$ cluster members. 

Overall, the above described methods typically achieve a compromise between completeness ($\sim 80$--$95\%$) and purity ($\sim 80$--$95\%$), but no technique can simultaneously maximize both. In particular, aggressive rejection schemes reduce contamination at the cost of removing galaxies in the infall regions, which are dynamically informative but difficult to classify. This issue becomes increasingly important at large radii, where there is a gradual transition between virialized and infalling populations. 

In this respect, a significant improvement might come from ML approaches. In particular, \citet{Farid+22} introduced the \texttt{C}$^2$\texttt{-GaMe} classification algorithm, based on a suite of ML models that differentiates galaxies into orbiting, infalling, and background (interloper) populations, using p.p.s.\ information as input. Their method identifies 95\% of the orbiting galaxies, and 50\% of the infalling galaxies, at the expense of a 10\% false positive rate. The novelty of the method is that it is able to distinguish these two classes of cluster members, which are characterized by different p.p.s. and orbital distributions. Distinguishing the two populations might be important for mass determinations based on the virial theorem or the $M-\sigma$ scaling relation (see Sect.~\ref{sec:virial},\ref{sec:scaling}).

\begin{figure}[ht]
    \centering
    \includegraphics[width=0.6\linewidth]{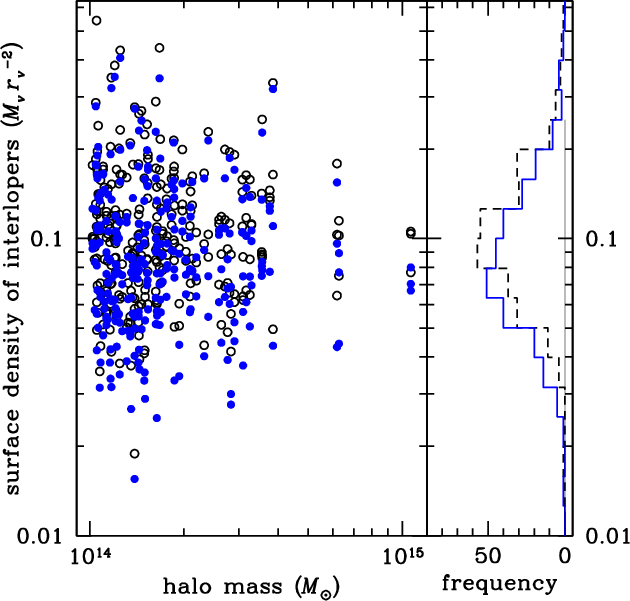}
    \caption{Mean interloper surface density (in virial units) versus halo mass (at an overdensity of 200 critical) for 93 halos viewed along three orthogonal directions. Black (respectively blue) circles and histogram are for interlopers before (respectively after) a 2.7 $\sigma_v$ clipping. Note the large variance across different halos. Image reproduced with permission from \citet{Mamon2010}, copyright by ESO. \label{fig:interlopers}}
\end{figure}

Rather than trying to separate cluster members from interlopers, an alternative approach is to include both populations in the dynamical modeling. This is the only approach that allows to remove the low velocity interlopers, and not only those that occupy the external regions of the cluster p.p.s..
While \cite{Prada2003} modeled the velocity distribution of interlopers with a constant surface density in velocity space \citep[see also][]{Wojtak2007}, 
\cite{Tomooka2020} modeled the velocity distribution of interlopers with a radial-dependent Gaussian. Using  cosmological numerical simulations, \cite{Mamon2010} showed that the velocity distributions of interlopers is in fact a combination of a Gaussian and a flat components, with a density that does not depend on radial distance from the cluster center. They also showed that the surface density of interlopers in cluster virial units has a large variance across different clusters (see Fig.~\ref{fig:interlopers}). This feature makes it very difficult to constrain the interloper fraction in any individual cluster, so that the statistical modeling of the interlopers can only be achieved by the stacking of many clusters. However, stacking requires at least approximate estimates of individual cluster masses anyway, as we discuss below.

Stacking can be done by defining bins in mass from weak lensing estimates, or in mass proxies, such as optical richness or SZ signal. A composite p.p.s. is obtained by combining all galaxies from different clusters within each bin \citep{Pizzuti25b}. A different stacking strategy consists in avoiding binning by using normalized cluster-centric distances and rest-frame velocities, using, respectively, approximate estimates of the individual cluster $r_{200}$ and $v_{200}$ for the normalization \citep[e.g.][]{van_der_Marel_2000,Biviano_2003,Katgert2004,Mamon2019}. If $r_{200}, v_{200}$ are derived from kinematics, interlopers must be rejected before stacking, thereby invalidating the possibility of a joint cluster members and interlopers dynamical modeling. Independent estimates of $r_{200}, v_{200}$ from, e.g., weak lensing, or X-ray, can be used instead, or in combination with estimates from kinematics \citep{Mamon2019}.
This stacking strategy also implicitly assumes a degree of universality in the cluster dynamical structure, but
numerical simulations have shown that the variance in the cluster internal structure and the uncertainties in the individual cluster reference frames (usually defined by the position of the BCG in p.p.s.)
can induce a moderate amount of bias in the mass estimates from kinematics of stacked cluster samples \citep{Cai2025}.

\paragraph{Incompleteness}
Still related to member selection, the incompleteness of spectroscopic samples is an additional and often less-emphasized source of systematic uncertainty. In real observations, selection effects such as fiber collisions, slit positioning constraints, or magnitude limits introduce a spatially-dependent incompleteness in the tracer population. 
If not properly accounted for, this affects the estimate of the projected surface number density profile $\Sigma(R)$, and therefore propagates into dynamical analyses that rely on the full p.p.s. distribution.

Numerical and observational studies have shown that while spatial incompleteness can bias the spatial distribution of tracers, it does not significantly affect the velocity distributions, since the observational selection function is almost entirely
independent of the rest-frame cluster galaxy velocities \citep[e.g.,][]{Biviano2006,Biviano_2013}, except when only the brightest (hence most massive) galaxies are selected, since their velocities might have been affected by dynamical friction \citep{Biviano+92,Stein97,Ferragamo2020}.

For this reason, dynamical methods based primarily on velocity information (like, e.g., the $M-\sigma_v$ relation) tend to be more robust against moderate levels of spatial incompleteness, than methods that rely also on the galaxy spatial distribution (like, e.g., the virial theorem). However, since $\sigma_v$ is a function of radius, the global estimate of a cluster $\sigma_v$ can be mildly affected by a radial selection function. Moreover, sparse sampling in velocity space tend to cut the tails of the velocity distribution, and therefore bias $\sigma_\mathrm{los}$ estimates low \citep{Wojtak2018}.

\paragraph{Different cluster galaxy populations}
Selecting cluster members on the basis of their properties can lead to potential bias in the cluster mass estimates, because cluster galaxies with different properties can have different spatial and velocity distributions \citep[e.g,][]{Tammann72,Capelato1980,Dressler1980,Sodre1989,Biviano+02}.
This selection is particularly relevant when cluster galaxies are extracted from spectroscopic field surveys that are not tailored for the population of cluster galaxies, and focus on the presence of emission-lines in the galaxy spectra for a reliable estimate of the galaxy redshifts \citep[e.g.][]{EC_Mellier_2025}. Emission-lines are typically present in the spectra of blue/star-forming/late-type galaxies (LTGs hereafter), but they are faint, or not present, in red/quiescent/early-type galaxies (ETGs hereafter) that are the most abundant population in clusters. 

LTGs are a believed to be a younger population of cluster members than ETGs, and tend to occupy the more external regions of the cluster p.p.s. \citep[e.g.][]{MD77,Biviano+02}. The different p.p.s. distribution of LTGs and ETGs is well established in low-redshift clusters \citep[e.g.,][]{Cava_2017}, and it appears to be present also in $z \gtrsim 1$ clusters \citep{Biviano2016}.
If in dynamical equilibrium in the cluster potential, application of the Jeans equation to, separately, the two populations of cluster galaxies, should lead to the same result for $M(r)$, and (possibly) different $\beta(r)$ \citep{Carlberg1997}. On the other hand, the different $\sigma_\mathrm{los}$ and number density profiles of LTGs and ETGs can lead to different mass estimates through the $M-\sigma_v$ scaling relation and application of the virial theorem \citep[][see also Sect.~\ref{sec:virial}]{Carlberg1997,Biviano1997,Biviano2006,Bayliss2017}.

Another observational selection that can potentially induce a bias in the cluster mass estimates is that on luminosity. Spatial luminosity segregation is known to be present in clusters of galaxies, brighter (more massive) galaxies being more centrally concentrated than fainter (less massive) ones \citep{Capelato1980,Adami1998}, although the importance of this effect is still matter of investigation and it may depend on the cluster mass \citep{Kim2020}. Failure to account for this effect may bias the virial mass estimate. 
Selection in galaxy luminosity can also affect the cluster $\sigma_v$ estimate, but only when the sample is restricted to the $\lesssim 10$ brightest members, which typically display a $\sim 10$\% negative bias, consequence of dynamical friction and the fact that they entered the cluster environment in the early stages \citep{Biviano+92,Adami1998,Goto2005,Old2013,Wu+13,Bayliss2017}. The work of \citet{Armitage+18} confirms that velocity bias is not a major issue when estimating cluster masses from kinematic methods. 

\subsubsection{Non-rotating spherical systems in dynamical equilibrium?}

Most kinematical methods to infer cluster masses rely on the assumption that the systems are in a steady state. However, clusters are evolving systems, and deviations from dynamical equilibrium are common, especially in massive systems and at intermediate to high redshift. In the standard model of structure formation, galaxy clusters assemble hierarchically through the continuous accretion of smaller halos and galaxy groups, as well as through major mergers \citep[e.g.,][]{WCS2010}. As a consequence, they are generically neither relaxed nor structurally simple systems. Instead, they exhibit a complex internal structure characterized by clumps (substructures) and intrinsically triaxial mass distributions. Clusters hosting a significant level of substructures are far from equilibrium, and their galaxy populations have not yet undergone complete phase mixing \citep[e.g.,][]{Cava_2017}. Moreover, the accretion of groups into clusters from the surrounding cosmic web, does not occur on perfectly radial orbits, and this can eventually ingenerate rotation in the merged system. 

\paragraph{Substructures}
\label{sec:sub}
Substructures are a natural outcome of the hierarchical formation scenario \citep[e.g.][]{Yu2018} and are commonly observed in both simulations and real systems. They correspond to dynamically distinct components, physically associated with the cluster, such as infalling groups or remnants of recent mergers \citep[e.g.,][]{GellerBeers1982,DresslerShectman1988,Stein97,GB02,Girardi_2015}. The presence of substructures can have a direct impact on kinematical analysis, primarily by introducing coherent velocity patterns and deviations from Gaussianity in the velocity distribution. Numerical simulations and observational studies have shown that substructures can bias the $\sigma_v$ and mass estimates \citep[e.g.,][]{Bird1995,Pinkney1996,RL11,Old_2017,Tucker+20}. Even more affected than $\sigma_v$ are the higher moments of the velocity distribution, like the kurtosis, which are used to break the mass-velocity anisotropy degeneracy in the Jeans equation \citep{SLM04}. The effect of substructures is particularly strong when they correspond to infalling or merging groups observed along the line of sightlos. 
If not identified, these groups widen the observed velocity distribution, which is then interpreted as characteristic of a single, dynamically "hotter" system \citep[e.g.][]{Takizawa2010,Aguirre2021}.

Observationally, substructures are often identified in the p.p.s. as localized overdensities  or as distinct peaks in the velocity distribution \citep{DresslerShectman1988,ZF93,Boschin2006,Hou2009,Yu_2015,Yu2016ApJ...831..156Y,Benavides23,Girardi2024} and they may be traced differently by galaxies of different types and luminosities \citep[e.g.][]{Biviano_1996,Girardi_2015,Costa2025}. 
Simple diagnostics of the dynamical state of a cluster
are the positional offset between the X-ray centroid and the BCG, and the magnitude difference between the BCG and the second most luminous galaxy \citep{Ramella+07,Lopes2018}. Other diagnostics
can be obtained from the global statistical properties of galaxy velocities. In dynamically relaxed systems, the los velocity distribution is expected to be close to a Gaussian, as a consequence of virial equilibrium. Deviations from Gaussianity can therefore be used as indicators of departures from equilibrium. Several common statistical tests have been adopted for this purpose; for instance the Anderson--Darling, Kolmogorov--Smirnov, and Shapiro--Wilk tests, as well as measurements of skewness and kurtosis of the distribution \citep[e.g.,][]{Einasto12,Ribeiro2013,Roberts_2019}. In particular, the Anderson--Darling  metric has been shown to perform very well as a criterion to distinguish between relaxed and un-relaxed clusters \citep{Hou2009,Pizzuti2020syst}; it assigns a larger weight to the tails of the distribution, effectively enhancing its sensitivity to deviations at high and low velocities. This diagnostic is particularly relevant in the context of galaxy clusters, where infalling groups or merger remnants often populate the high-velocity wings of the distribution, producing excess kurtosis or asymmetric tails \citep{BFG90,ZF93}.

\begin{figure}[ht]
    \centering
   \includegraphics[width=0.35\linewidth]{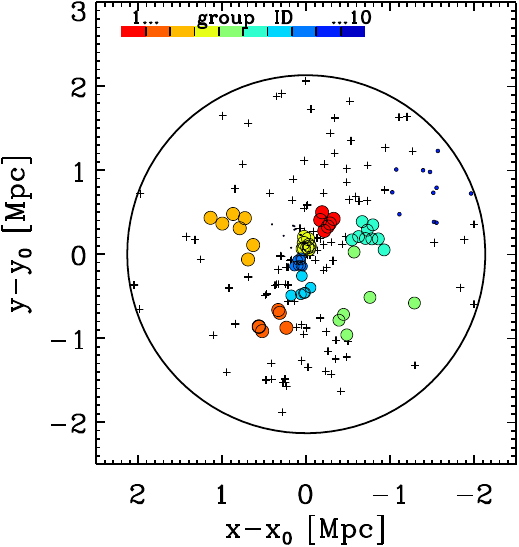}\hspace{1em}
    \includegraphics[width=0.6\linewidth]{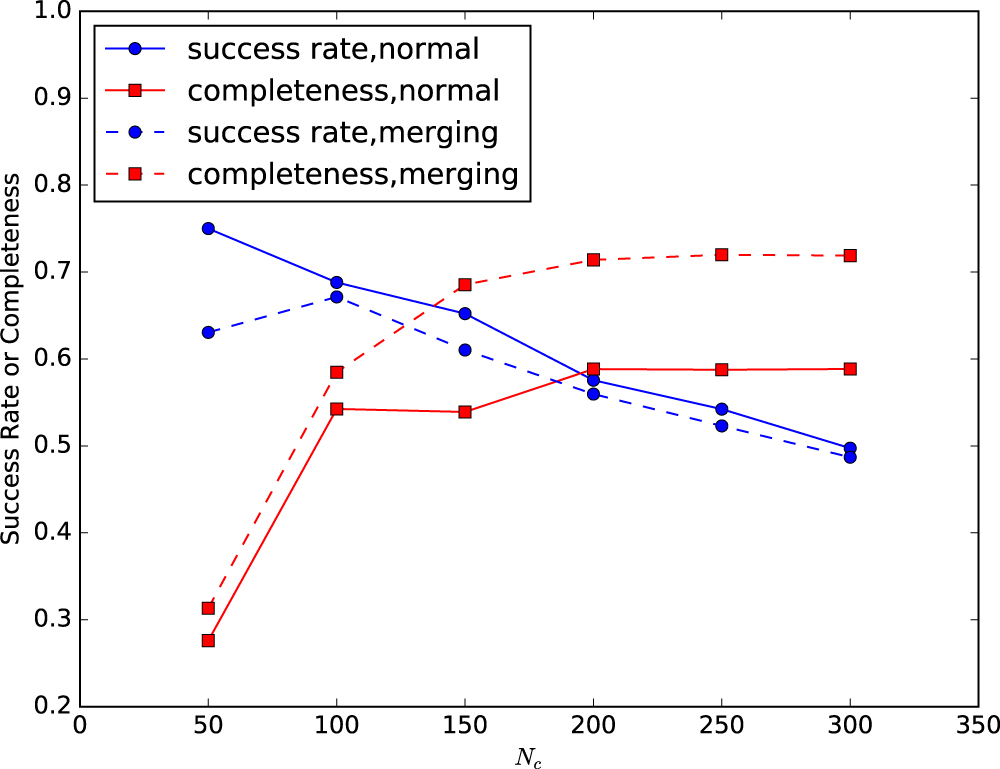}
    \caption{Left panel: substructures in the Bullet cluster identified by the \texttt{DS+} method. From \citet{Benavides23}, image reproduced with permission, copyright by ESO.
    Right panel: Success rate (blue dots) and completeness (red squares) for merging (solid lines) and normal (dashed lines) mock clusters, for the identification of substructures
    with the \texttt{Blooming Tree} algorithm. From \citet{Yu2018}, image reproduced with permission, copyright by AAS. \label{fig:subs}}
\end{figure}

More sophisticated methods to characterize a cluster dynamical state use the combined information coming from the spatial and velocity distributions of galaxies. The most classical method of this class is that of
\citet{DresslerShectman1988}, which quantifies the probability of a cluster to contain significant subclusters, by estimating local deviations of the mean velocity and $\sigma_\mathrm{los}$ from the global cluster values. This method is particularly sensitive to localized kinematic sub-clumps and has been widely applied in spectroscopic surveys \citep[e.g.,][]{Einasto12,deCarvalho17,Barrena24}. 
The \texttt{DS+} method of \cite{Benavides23} extended \citet{DresslerShectman1988}'s method by considering substructures of different multiplicities, and by separately considering deviations in mean velocity and $\sigma_\mathrm{los}$, and it allows the identification of individual subclusters and their members. \texttt{DS+} reaches 
a completeness of $\sim 80\%$ and a purity of $\gtrsim 60\%$ in the recovery of substructures in
simulated clusters. 

The method was applied to the Bullet cluster \citep[described in Sect.~\ref{sec:modgrav};][]{Benavides23}. Besides the well known substructure that represents the most recent group collision with the cluster \citep{Barrena+02}, several other substructures were identified, that trace the past cluster accretion history and can account for the cluster elongated shape (see the left panel of Fig.~\ref{fig:subs}). Including these substructures in the lens model of the Bullet cluster improved the fit to the multiple image positions \citep{Rihtarsic+26}. Moreover, 
\citet{Benavides23}'s mass estimate of the main colliding group, based on the \texttt{DS+} method, was confirmed by \citet{Cho+26}'s gravitational lensing analysis.

Another method for substructure detection is the \texttt{Blooming Tree} algorithm \citep{Yu2018}. This method reconstructs the hierarchical binding structure of galaxy systems by exploiting pairwise binding energies, and it allows the identification of substructures across a wide range of scales. Tests on numerical simulations indicate that the method achieves performances comparable to \texttt{DS+} (see the right panel of Fig.~\ref{fig:subs}). The advantage of the \texttt{Blooming Tree} algorithm is that it can also be applied for the identification of the structure surrounding the cluster, including filaments \citep{Yu2025ApJ...991..220Y}, and it can even be extended to investigate the hierarchical structure of open star clusters \citep{Yu2020ApJ...899..144Y}.

The \texttt{Galaxy Cluster Mixture Model} of \citet{Tucker+20} models the observed distribution of galaxies in a cluster field as a combination of a main
cluster and up to four subclusters, each characterized by a NFW projected surface density, and a Gaussian velocity distribution. The method then defines the probability of any galaxy to belong to the main cluster or any subcluster. To determine the number of subclusters effectively present in the system, 
the method considers the change in the Bayes factor obtained with a different number of subclusters in the model. Application of this method to the cluster Abell~267 showed that neglecting the presence of substructures leads to an overestimate of the cluster mass by $\sim 20$\%.

\paragraph{Triaxiality}
The topic of substructures is related to the deviation of clusters from sphericity, since both reflects the anisotropic process of hierarchical assembly of clusters from the cosmic web
\citep[e.g.,][]{Colberg1999MNRAS.308..593C, Malavasi2020,Musso2021,Gouin2021}. 
In addition to substructures, the intrinsic shape of galaxy clusters represents a fundamental source of systematic uncertainty. DM halos are not spherical but triaxial, with a general tendency toward prolate configurations, as predicted by cosmological simulations \citep{Saro_2013,Limousin_2013} and confirmed by multi-wavelength observations \citep{Sereno2006,Morandi_2011,Kim_2024}. The degree of triaxiality depends on the halo mass, assembly history, and accretion rate, and is particularly pronounced in the outskirts where ongoing accretion along filaments is significant.

Triaxiality affects all projected observables \citep[e.g.,][]{Wojtak2013,Stapelberg_2022,Zhang_2023}. In the context of kinematical analyses, the orientation of the cluster with respect to the los can significantly bias $\sigma_\mathrm{los}$ measurements and inferred mass profiles. Prolate clusters aligned along the los appear more concentrated and exhibit enhanced $\sigma_\mathrm{los}$, while the opposite occurs for systems elongated in the plane of the sky \citep{Skielbo2012,Wojtak2013,Pizzuti2020syst}. These projection effects are further degenerate with the velocity anisotropy, and can therefore mimic or amplify the mass--anisotropy degeneracy in Jeans-based analyses.
Moreover, triaxiality also affects caustic mass estimates, since they too are derived under the spherical symmetry assumption \citep{Serra+11, Svensmark+15}.

Direct measurements of cluster triaxiality are difficult, requiring lengthy observations in X-ray, and at submm and optical wavelengths \citep{Sereno2006,Sereno2018,Morandi_2011}. The few determinations of cluster axis ratios provided a distribution of values that is consistent with expectations for massive halos in a $\Lambda$CDM cosmology \citep{Sereno2018}. ML techniques can potentially improve the current observational situation and at least allow for a clear detection of the most problematic cases, that is, clusters with their main axis oriented along the line-of-sight \citep{Delgado2026}. An alternative strategy to the problem of triaxiality is to smooth out the systematics by stacking clusters in p.p.s. \citep[e.g.,][]{Bilton2018,Mamon2019,Maraboli+26}. As shown by \citet{van_der_Marel_2000}, stacking provides a simple and effective way to reduce the impact of triaxiality by averaging over different cluster orientations. 

\paragraph{Rotation}
One of the usual assumptions made in the solution of the Jeans equation is that clusters do not rotate (see Sect.~\ref{sec:dyn}). Gradients in the mean velocity of clusters as a function of radius have been found in several clusters \citep{Biviano_1996,Girardi+96} and interpreted as evidence of the presence of substructures or unidentified structures along the los to the cluster. Indeed, disentangling this scenario from one of rotation is difficult. Groups are unlikely to merge with clusters along perfectly radial orbits \citep[see, e.g., the NGC~4839 group in the Coma cluster,][]{Biviano_1996,Lyskova+19}. If not identified and removed as substructures (see Sect.~\ref{sec:sub}), the presence of these groups on partially tangential orbits may be mis-interpreted as a global rotational motion of the whole cluster. For this reason, searches for rotation in clusters of galaxies tried to exclude substrctured clusters from their analyses.

\citet{HL07} searched for rotation in 899 nearby clusters and found only 12 tentative rotating clusters. On the other hand, \citet{MP17} found a higher fraction of rotating clusters, $\sim 1/4$. Not accounting for the kinetic energy component in the rotational motion of rotating clusters, causes a mass overestimate of 20-30\% \citep{MP17}. \citet{Ferrami+23}
found evidence for rotation in the central region of two massive clusters, for which they estimated a rotational velocity of $\sim 0.15 \, \sigma_{{\rm los}}$. Their work was extended to other 17 clusters by \citet{Castellani+25}, who detected rotation in eight of them, with even larger rotational velocities, $v_{{\rm rot}} \gtrsim 0.15 \, \sigma_{{\rm los}}$. Evidence of rotational motion, with comparable velocity to that found in optical studies, has also been discovered by X-ray analyses of the ICM of some clusters \citep{Bartalesi+25,Bartalesi+26}.

\section{Testing gravity and the dark sector with kinematics}
\label{sec:results}

From the picture outlined in the previous sections, it emerges that galaxy clusters occupy a privileged role for investigating the nature of the dark components of the universe and of the gravitational interaction. First, as the endpoint of the structure formation process, cluster shapes and abundances are strongly dependent on the cosmological model --- which, in turn, is connected to the density budget and properties of DM and DE.  
Second, galaxy clusters probe gravitational potentials on Mpc scales, where both the internal dynamics of collapsed (virialized) halos and the transition to the infall region can be observed. 

Several extensions of the $\Lambda$CDM paradigm introduce new degrees of freedom which may produce observable signatures exactly in the regimes probed with galaxy clusters.  As such, the kinematical techniques discussed in Sect.~\ref{sec:methods} provide interesting insights on the physics of galaxy clusters in a complementary way to X-ray, SZ, and lensing observations. While X-ray and SZ analyses probe the thermodynamic state of the ICM, and gravitational lensing measures the projected mass distribution through the propagation of null geodesics in curved spacetimes, galaxy velocities respond to the force law governing the motion of non-relativistic tracers. The p.p.s. distribution of cluster members therefore encodes information not only on the mass profile and velocity anisotropy, but also on possible departures from the standard CDM paradigm and the GR description of gravity.

\subsection{Kinematics as probe for DM}
\label{sec:kinDM}
\subsubsection{The inner slope of the mass density profile}
Historically, testing the dark sector with the kinematics of galaxies in clusters contributed to the foundation of modern cosmology, providing the first evidence of DM thanks to the pioneering analyses of \cite{Zwicky1933,Zwicky1937,Smith1936}.
In the standard CDM paradigm, DM is assumed to be collisionless and pressureless, and the structure of galaxy clusters is determined primarily by gravitational collapse and hierarchical assembly of DM halos. \cite{Navarro1996} showed that in cosmological N-body simulations the density of virialised haloes tends to follow a nearly-universal distribution over a broad range of masses and scales, the so-called NFW profile (see Sect.~\ref{sec:rise}).
The NFW model has been widely adopted in literature as a good description for the total mass of both simulated and real galaxy clusters \citep{Geller_1999,Rines+00,Rines+03,Rines+13,RinesDiaferio06,Biviano_2013,Balmes14,Mantz_2016,Umetsu_2016,Pizzuti25b}. 
The simplicity of this parametrization, together with the fact that its main properties (such as the expression of the enclosed mass and projected quantities) can be expressed in closed analytical form, makes the NFW profile a powerful framework for studying gravitationally bound astrophysical systems and for comparing observations with numerical predictions. 

\begin{figure}[ht]
         \centering
         \includegraphics[width=.5\linewidth]{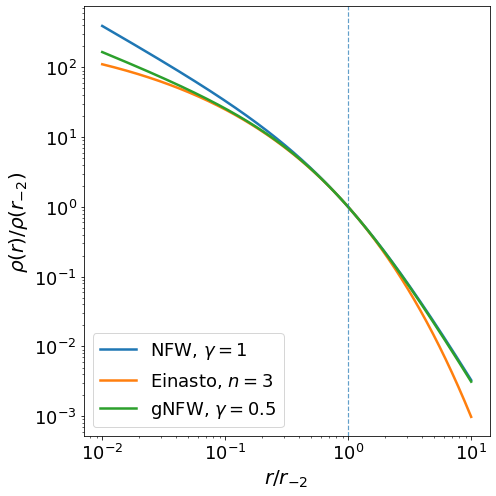}
         \caption{\label{fig:profExample} Density profiles for different models with different inner slopes. The vertical dashed line indicates the normalised radius $r/r_{-2}=1$. While the exponent $\gamma$ of the gNFW only affects the central region, changing the Einasto shape parameter $n$ modifies both the inner and the outer regions.}
    \end{figure}

However, the universality of the NFW profile has been questioned by increasingly accurate simulations. In particular, the inner logarithmic slope of the density profile does not appear to be strictly fixed, but may vary continuously with radius, becoming progressively shallower towards the centre \citep[e.g.,][]{Stadel2009}. 
On the observational point of view, analyses at both galactic \citep{Delpopolo09,deblok2010,Dehghani_2020} and cluster scales \citep{Laporte12,Newman_2013,Biviano_2023} have shown that the inner density profiles of DM haloes can significantly deviate from the standard NFW expectation, often indicating core-like structures, although the interpretation remains widely debated and sensitive to several assumptions \citep[see e.g.,][]{Delpopolo09,Peirani_2017,He_2020}.
This discussion has motivated the introduction of more flexible models with additional parameters. Popular choices are simple generalisations of the NFW (gNFW, hereafter), where the central slope is controlled by an exponent $\gamma \in [0,2]$ \citep[e.g.,][]{Zhao96,Wyithe_gNFW_2001}:
\begin{equation} \label{eq:gNFW_profile}
    \rho_\mathrm{gNFW}(x) = \frac{\rho_0}{(r/r_\text{s})^\gamma (1 + r/r_\text{s})^{3 - \gamma}}\,,
\end{equation}
or models like the Einasto (\citealt{Einasto65}, see also \citealt{Baes_2022}) profile:
\begin{equation}
\rho(r) = \rho_{-2} \, \exp \left\{ - 2 n \left[ \left( \frac{r}{r_{-2}} \right)^{1/n} - 1 \right] \right\} \,,
\end{equation}
where $\rho_{-2}$ is the central density at $r = r_{-2}$, and the shape parameter $n$ controls the radial variation of the logarithmic slope, and typically lies in the range $1/n \sim 0.1$--$0.3$ for cluster-sized haloes in numerical simulations. In Fig.~\ref{fig:profExample} we show an example of how the NFW, Einasto and gNFW normalised density profiles $\rho(r)/\rho(r_{-2})$ behave as functions of $r/r_{-2}$ Note that in the gNFW model, the scale radius is connected to $r_{-2}$ by the relation $r_\text{s} = r_{-2}/(2-\gamma)$.

Departures from the NFW profile of the pure CDM scenario are expected once additional physical processes beyond collisionless gravitational collapse are taken into account. In particular, the inner regions of DM haloes are shaped by a complex interplay between baryonic physics and the (possible) intrinsic properties of the DM particle. Baryons can alter the central density profile in non-trivial and often competing ways. The accumulation of cooled gas and the growth of the BCG deepen the central potential well, generally leading to a contraction of the DM distribution and a steepening of the inner slope \citep[e.g.,][]{Gnedin2004,Peirani_2017}. Conversely, mechanisms that inject energy into the DM component can counteract this effect. For instance, dynamical friction exerted by infalling satellites transfers orbital energy to the halo, potentially flattening the central cusp \citep{Delpopolo09,Boldrini2021}. Additional energy input can arise from feedback processes: while supernova-driven outflows might be particularly relevant in low-mass systems \citep{Governato2010}, active galactic nuclei are expected to dominate in massive clusters, where they may reduce the central density and partially erase cusps \citep{Martizzi2012}. The overall impact of these mechanisms depends strongly on the mass scale and assembly history of the system, and remains a matter of active investigation in hydrodynamical simulations \citep[e.g.,][]{Schaller2015,He_2020,Del_Popolo_2021}.

Not only baryonic processes, also the nature of DM can lead to deviations from the NFW profile. A variety of non-standard scenarios --- including warm, fuzzy, decaying, or self-interacting dark matter (SIDM) --- predict a suppression of the central density relative to the CDM expectation, often leading to the formation of cored profiles \citep[e.g.,][]{Maccio_2012,Ragagnin2024,Pozo2025}. 
Among these alternatives, SIDM provides a well-defined physical mechanism to modify the inner structure of haloes. In such models, DM particles scatter elastically with a cross section per unit mass $\sigma/m$, leading to heat conduction within the halo. In regions where the scattering rate is sufficiently high, this process redistributes energy from the outer, hotter regions to the colder central parts, effectively isotropizing the velocity distribution and reducing the central phase-space density. As a result, the steep cusps predicted by collisionless CDM are transformed into approximately isothermal cores \citep[e.g.,][]{Spergel2000,Robertson2018,Tran_2026}.

The impact of self-interactions at a given radius is controlled by the
local scattering rate per particle,
$\Gamma(r)=\rho(r)\langle(\sigma/m)v_{\rm rel}\rangle$, where
$v_{\rm rel}$ is the relative velocity of the interacting particles.
A useful dimensionless measure is the accumulated number of scatterings,
$N_{\rm scat}(r)\simeq\Gamma(r)\,t_{\rm age}$. The radius $r_1$ at which
$N_{\rm scat}(r_1)=1$ approximately separates an inner region whose
density and velocity structure can be modified by self-interactions from
an outer region that remains close to the collisionless-CDM prediction
\citep{Kaplinghat2016}.

For velocity-independent cross sections in the range
$\sigma/m=0.1$--$1~\mathrm{cm^2\,g^{-1}}$, DM-only simulations
find characteristic core radii increasing from
$r_{\rm core}\sim0.1\,r_{-2}$ to $\sim0.7\,r_{-2}$, where $r_{-2}$
denotes the scale radius of the corresponding collisionless halo
\citep{Rocha2013}. The wider region affected by heat redistribution can
extend up to $r_1$, which is at most of order $r_{-2}$ for these models.
Thus, the main structural modifications are expected approximately
within $r/r_{-2}\lesssim0.1$--$1$, with the precise transition depending
on the cross section, halo age and concentration. In cluster cores,
baryons can further alter the SIDM density profile, so these values
should not be considered as universal
\citep{Robertson2018}.

Since the characteristic collision velocity ranges from tens of
$\mathrm{km\,s^{-1}}$ in dwarf galaxies to about
$10^3~\mathrm{km\,s^{-1}}$ in clusters, observations on different halo
scales probe the cross section at different relative velocities.
The smaller cross sections generally allowed on cluster scales compared
with those invoked to produce sizeable galaxy cores have therefore
motivated velocity-dependent models in which the effective cross section
decreases with increasing relative velocity
\citep{Kaplinghat2016,Tulin2018}.

Because both baryonic effects and DM physics can leave comparable imprints on $\rho(r)$, accurately determining the shape of the density profile as a function of cluster mass, redshift, and dynamical status,
is essential to disentangle their respective roles. To probe the cluster mass density profile over several decades in radius, from the central regions dominated by the BCG to the outskirts, a combination of multiple observational techniques is required. In this framework, kinematical analyses play once again a crucial role by providing information on the dynamical mass distribution, complementing lensing/X-ray measurements and direct observations of the baryonic profiles.

\subsubsection{The DM equation of state}
Beyond constraints on the shape of the DM density profile or on the SIDM cross section, the combination of kinematical and lensing information can also be used to test a more basic assumption of the CDM paradigm, namely that DM behaves as an effectively pressureless component. In the standard picture, the DM equation-of-state parameter,
\begin{equation} \label{eq:eosDM}
    w_\mathrm{DM} \equiv \frac{p_\mathrm{DM}}{\rho_\mathrm{DM}c^2},
\end{equation}
is assumed to be negligibly small. While this hypothesis is usually built into dynamical modeling, it can be tested by comparing probes that respond differently to the gravitational field.  Following the relativistic formalism proposed by \cite{Faber_2006}, 
\cite{Serra2011a} applied the comparison between kinematic and lensing mass
profiles to test whether DM behaves as an effectively pressure-less
component on cluster scales. The basic idea is that, in the weak-field limit,
galaxy motions and light deflection probe different combinations of the metric
potentials. The metric of a galaxy cluster can be described as a linear perturbation of the Friedmann--Lemaître--Robertson--Walker (FLRW) spacetime. In the conformal Newtonian gauge, assuming a flat background, it takes the form
\begin{equation}
    \text{d}s^2 = -(1+2\Phi/c^2)\text{d}t^2 + a^2(t)(1-2\Psi/c^2)\text{d}\bf{x}^{\,2},
    \label{eq:ds2}
\end{equation}
where $\Phi$ and $\Psi$ are the two scalar gravitational potentials; non-relativistic tracers respond primarily to the time-time potential $\Phi$, whereas photons depend on the so-called Weyl potential, $\Phi_\mathrm{lens} =(\Phi+\Psi)/2$. The kinematic mass profile, inferred from the Jeans equation or from the caustic technique, is sensitive to the potential governing the motion of non-relativistic tracers, whereas gravitational lensing constrains the curvature potential that determines photon trajectories. In GR, the relation between these two potentials depends not only on the density of the gravitating component, but
also on its pressure terms. Therefore, comparing the two mass profiles provides a way to reconstruct an effective density ($\rho_{{\rm DM}}$) and pressure ($p_{{\rm DM}}$) for the dark component and to test the pressureless-CDM assumption. Since a bound system such as a galaxy cluster need not be described by an isotropic perfect fluid, the radial and tangential pressures ($p_r$ and $p_t$, respectively) can differ. It is therefore useful to define an effective equation-of-state parameter from the directional average of the pressure tensor, which changes the definition of eq.~\eqref{eq:eosDM} as
\begin{equation}
    w_\mathrm{DM}(r) =
    \frac{p_r(r)+2p_t(r)}
         {3\,\rho(r)c^2}.
\end{equation}

\subsubsection{Observational constraints}

A first example of kinematical approaches to reconstruct the DM distribution in clusters was provided by the analysis of the Coma cluster by \citet{LokasMamon2003}. In their work, the total mass profile is modeled as the sum of the stellar, ICM, and DM components, with the latter described by a gNFW profile. To overcome the mass-anisotropy degeneracy of the Jeans equation (see 
Sect.~\ref{sec:jeanssolution}), \citet{LokasMamon2003} considered higher-order velocity moments, in particular the projected kurtosis profile, which provides additional sensitivity to the orbital structure of the system. They obtained a DM distribution consistent with profiles predicted by cosmological simulations over a wide radial range ($\gtrsim 0.03\,r_\mathrm{vir}$); however, the results showed a strong degeneracy between the inner slope and the concentration of the halo. In particular, models with different central behaviours (cuspy versus cored) can provide equally good fits to the data when coupled with appropriate values of the concentration parameter, illustrating the intrinsic limitations of kinematical constraints in the absence of additional information from the innermost regions. At the same time, the analysis indicates that DM dominates the mass budget at all radii, contributing $\sim 85\%$ of the total mass within the virial radius.

\begin{figure}[ht]
         \centering
         \includegraphics[width=\linewidth]{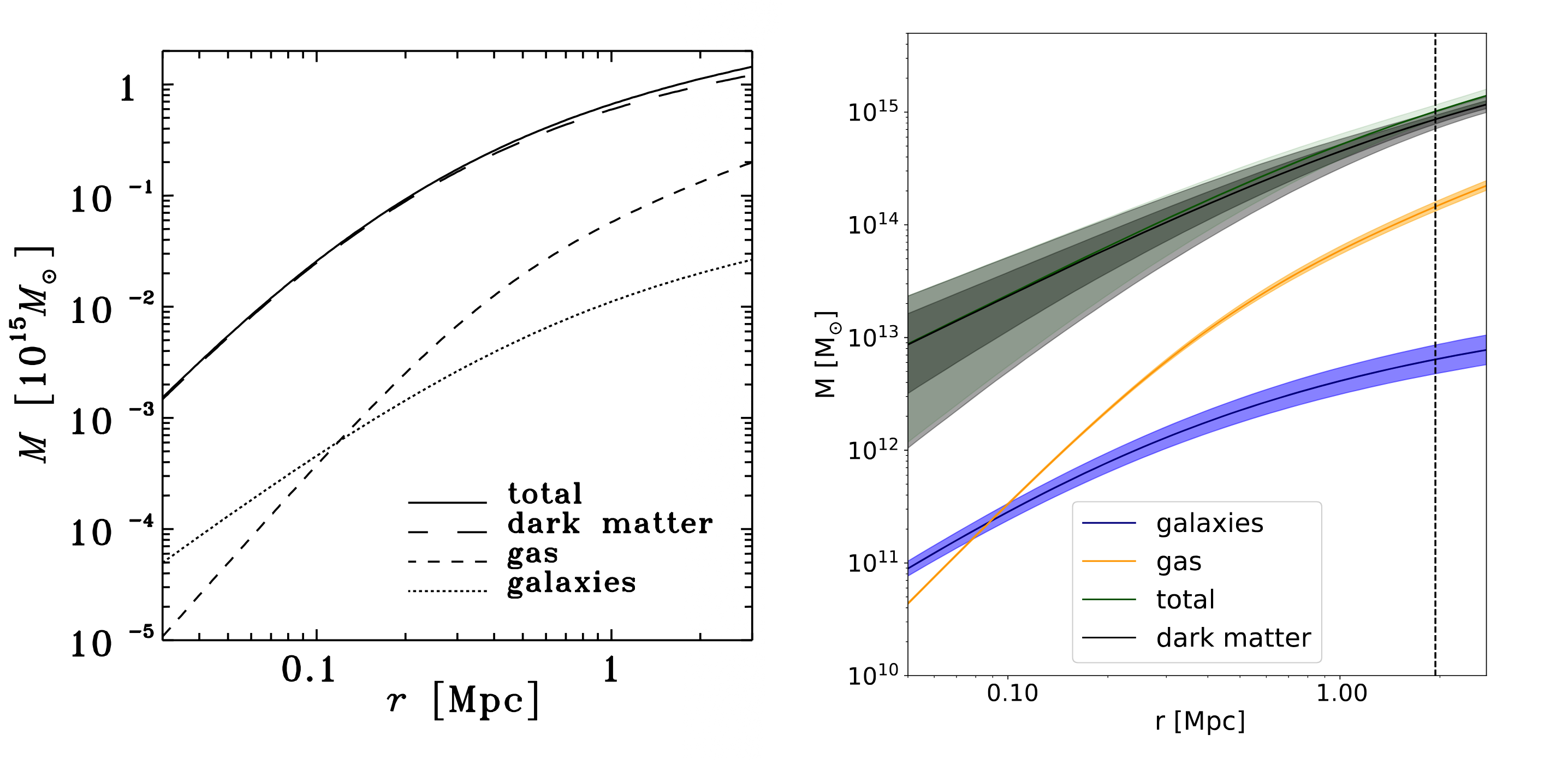}
         \caption{\label{fig:ComaExample} Multi-component mass profiles of the Coma cluster obtained by \cite{LokasMamon2003} (left panel), and \cite{Pedratti26} (right panel); in the latter, the grey inner and outer shaded regions indicate the 68\% and 95\% confidence region of the DM profile, respectively, while the orange and blue bands represent the 68\% limits of the
gas and galaxies mass profiles. The DM virial radius $r_{200} = 1.94$ Mpc is marked by the vertical dashed line. Left panel: copyright by RAS. Right panel: copyright by ESO.}
    \end{figure}

Analyses based on larger spectroscopic samples have confirmed this picture. \citet{Pedratti26} performed a detailed kinematic reconstruction of the Coma cluster, using an unprecedented spectroscopic dataset from DESI, and the \texttt{MG-MAMPOSSt} technique, jointly modeling $M(r)$ and $\beta(r)$. By explicitly separating the baryonic (stellar, gas), and DM components, they obtained tight constraints on the overall mass distribution, while confirming the degeneracy between the scale radius $r_\text{s}$ and the inner slope $\gamma$ of the gNFW model.
In Fig.~\ref{fig:ComaExample} we compare the DM profiles of the Coma cluster obtained by \citet[][left panel]{LokasMamon2003} and \citet[][right panel]{Pedratti26}. Despite the substantial differences in the quality of the spectroscopic datasets adopted, the reconstructed DM distributions are in remarkable agreement (within the uncertainties) over the radial range probed by galaxy kinematics. 

Taken together, these two studies provide a prototypical example of how p.p.s. analyses can be used to infer the DM budget and distribution in clusters. At the same time, they also make clear that  the combination with complementary probes is fundamental to break the degeneracies when testing deviations from the standard CDM paradigm. In this sense, a step forward was made by \cite{Sartoris_2020} and \cite{Biviano_2023}, who solved the Jeans eqaution by including in the \texttt{MAMPOSSt} analysis an additional kinematical dataset describing the BCG stellar velocity dispersion. Note that it is perfectly fine to use the Jeans equation, which applies to collisionless systems, to constrain deviations from the CDM paradigm, and eventually constrain the DM self-interacting cross-section, as long as the tracer of the gravitational potential can be treated as a collisionless fluid. This is indeed the case for galaxies in clusters, where mergers are suppressed by the cluster high velocity dispersion, and for stars in the BCG, another powerful tracer of the inner cluster $M(r)$.

By combining the distribution of cluster galaxies in p.p.s. and the stellar kinematics of the BCG \citet{Sartoris_2020} and \citet{Biviano_2023} were able to reach the constraining power needed to perform a precise reconstruction of the diffuse DM profile, $\rho_{{\rm DM}}(r)$, over three order of magnitudes in radius, from $\sim r_{200}$ down to $\sim 1$ kpc (see also e.g. \citealt{Geller2014}). Since the 
BCGs of both clusters are located at their cluster centres, as indicated by the comparison with the centres inferred from the X-ray and strong lensing analyses \citep{Umetsu12,Caminha16}, the inner slopes of the two cluster $\rho_{{\rm DM}}(r)$'s can provide useful insights into the nature of DM.

\cite{Sartoris_2020} applied this approach to Abell S1063, a massive cluster at $z=0.35$ observed as part of the CLASH and Frontier Fields programmes, and extensively followed-up spectroscopically within the CLASH-VLT campaign. The CLASH-VLT dataset, obtained mainly with the VIMOS spectrograph at the VLT and complemented in the core by MUSE integral-field spectroscopy, provided a large sample of spectroscopic members together with a direct measurement of the BCG stellar velocity dispersion profile. The total mass profile was decomposed into four components: the stellar mass of the BCG, the stellar mass in the other cluster galaxies, the ICM mass, and the diffuse DM component modeled by a gNFW profile. The two kinematical tracers probe complementary radial regimes: the BCG stellar velocity dispersion constrains the inner potential, at $\lesssim 50$ kpc, while the p.p.s. distribution of cluster members fixes the mass profile at the Mpc scale, and the two datasets togetther allow to break the degeneracy between the gNFW scale radius and the inner slope, constrained with high precision, $\gamma_\mathrm{DM} = 0.99 \pm 0.04$, and
in excellent agreement with the NFW expectation. The resulting total mass profile was also found to be consistent with independent strong+weak lensing and X-ray hydrostatic reconstructions, although the cluster is known to be not fully dynamically relaxed \citep{Mercurio2021}.

The same methodology was subsequently applied by \cite{Biviano_2023} to MACS J1206.2$-$0847 (MACS J1206 hereafter), another CLASH-VLT cluster for which previous analyses had suggested a much shallower, nearly cored DM distribution \citep[see e.g.,][]{Sand2004,LBJ22}. The discrepancy is most likely due to the improved precision of the spectroscopic data describing the BCG kinematics available to \citet{Biviano_2023} compared to those used by \citet{Sand2004}.
The inferred inner slope from the combined analysis of the BCG stellar velocity dispersion and the galaxy p.p.s., is incompatible with a core, $\gamma_\mathrm{DM} = 0.7^{+0.2}_{-0.1}$, but shallower than the slope in Abell S1063. This value is however not in tension with expectations from hydrodynamical $\Lambda$CDM simulations, once the intrinsic scatter of cluster-scale haloes and the decreasing trend of $\gamma_\mathrm{DM}$ with halo mass are taken into account \citep{He_2020}.

The comparison between the two clusters is particularly informative. Since Abell S1063 and MACS J1206 were analysed with the same dynamical methodology and with data of comparable quality, the difference in their best-fitting values of $\gamma_\mathrm{DM}$ suggests an intrinsic cluster-to-cluster variance in the inner structure of massive DM haloes, as also suggested by the gravitational lensing analysis of \citet{Newman2013a}. Such variance can arise from differences in the assembly history, the relative role of baryonic processes in the central regions and the cluster dynamical state.

An approach similar to those of \cite{Sartoris_2020} and \cite{Biviano_2023} has been employed in studies combining the stellar kinematics of the BCG with gravitational lensing information \citep{Newman_2013,Newman2013a}. In particular, \cite{Sagunski2021} considered a sample of relaxed galaxy groups and clusters with available strong-lensing constraints and spatially resolved stellar kinematics of the BCGs to specifically constrain SIDM models. Their modelling was based on a semi-analytic Jeans description of SIDM haloes, in which the inner region affected by self-interactions is treated as approximately isothermal, while the outer halo is matched to a collisionless NFW-like profile. The transition radius is set by the condition that DM particles have scattered on average once over the lifetime of the system. For massive clusters, they found an upper limit for the cross section $\sigma/m < 0.35~\mathrm{cm^2\,g^{-1}}$,
while for galaxy groups they obtained a larger value, $\sigma/m  = 0.5 \pm 0.2~\mathrm{cm^2\,g^{-1}}$.
The comparison between groups and clusters is important because the typical scattering velocity increases with halo mass. Their results therefore point towards velocity-dependent SIDM models in which the effective cross section decreases at high relative velocities.

\cite{ODonnell_2026} applied a related, but more detailed, strategy to the strong-lensing cluster MACS J0138.0$-$2155. They constructed a self-consistent model of a single cluster by combining strong-lensing constraints from multiple images with a two-dimensional stellar velocity dispersion map of the BCG measured with MUSE. 
By measuring the cluster density profile over two orders of magnitude in radius, the authors constrained the SIDM interaction cross-section,
$    \sigma/m < 0.613~\mathrm{cm^2\,g^{-1}}
    \; (95\%~\mathrm{C.L.})$
at an interaction velocity $\langle v_\mathrm{pair} \rangle < 2090~\mathrm{km\,s^{-1}}$. The lensing-only and kinematics-only reconstructions were found to be mutually consistent, while their combination produced a tighter upper limit on the SIDM cross section. This analysis illustrates the same physical principle mentioned above: lensing fixes the projected mass distribution on the scales where multiple images form, whereas the stellar kinematics of the central galaxy probes the inner potential where the distinction between a cuspy and a cored DM profile is most relevant.

The constraints on the nature of DM offered by the strong lensing analyses coupled with the BCG stellar kinematics are stronger than those obtained by the joint analysis of the BCG stellar and cluster galaxy kinematics. However, it has been argued by \citet{He_2020} that $\gamma_\mathrm{DM}$ can be biased by incorrect estimates of $r_\text{s}$. The cluster $M(r)$ must therefore be estimated with good accuracy not only near the cluster centre but also at large radii, where strong lensing is ineffective, while cluster galaxies are efficient tracers of the gravitational potential.

The combination of gravitational lensing and cluster galaxy kinematics has been also used to constrain the DM equation of state $w_\mathrm{DM}$
\citep{Serra2011a,SartorisDM}. In particular,
using simulated clusters and observational data for two real clusters, \cite{Serra2011a} found values of $w_\mathrm{DM}$ consistent with the pressureless CDM expectation within the uncertainties. This idea was later applied by \cite{SartorisDM} to the CLASH-VLT cluster MACS J1206, using \texttt{MAMPOSSt} to reconstruct the dynamical mass profile from the p.p.s. distribution of member galaxies, as in \cite{Biviano_2013}, and comparing it with the high-precision strong+weak gravitational-lensing analysis of \cite{Umetsu12}. Also in this case, the inferred equation-of-state parameter was found to be consistent with $w_\mathrm{DM}=0$, supporting the pressureless nature of DM on cluster scales.

While very powerful, this approach inherits the systematics affecting both sides of the comparison. On the dynamical side, the kinematic mass profile is biased by the effects discussed in Sect.~\ref{sec:syst}: deviation from spherical symmetry, dynamical equilibrium (except for the caustic method adopted by \citealt{Serra2011a}), substructure, and membership selection. Lensing, on the other hand, does not require the cluster to be dynamically relaxed, but it is not free from modelling uncertainties: triaxiality, orientation bias, correlated structures along the line of sight, and the treatment of the baryonic mass in the cluster core can all bias the deprojection from projected lensing observables to a 3D mass profile (see e.g., \citealt{Grandis2021,Giocoli25}). Since the DM equation of state is inferred precisely from the mismatch between lensing and dynamical mass profiles, these systematics must be controlled before interpreting any deviation from $w_\mathrm{DM}=0$ as evidence for non-standard DM physics.

\subsection{Kinematics to probe the nature of gravity}
\label{sec:modgrav}
\subsubsection{Beyond GR}
GR is by far one of the most successful theories in the history of science, able to provide a self-consistent description of the gravitational interaction across several scales and in a large variety of regimes. Its predictions have been tested with great precision by an unlimited amount of independent probes, from Solar System constraints \citep[e.g.,][]{Bertotti2003,Touboul17}, to astrophysical analyses \citep[e.g.,][]{Collett_2018,Jyoti2019,EHT_2019,Liu_2022}, and gravitational waves detection (e.g.,\citealt{Abbott_2019,Krishnendu_2021}, see also \citealt{Baker_2015}). However, while many cosmological-scale tests of GR exist in the literature \citep[e.g.,][]{Planck2016,Ishak_2018,8r4r-qg6t}, they are still less precise than the astrophysical counterparts, living room for possible extensions of the current gravitational framework. Interestingly, already in the years following Einstein's formulation, alternative geometric theories were explored, mostly from a formal and theoretical perspective \citep[e.g.,][]{Eddington1923}. Later, higher-order curvature terms in the gravitational action were investigated in connection with the ultraviolet behaviour of gravity and quantum corrections in regimes of very large curvature, such as the early Universe or the vicinity of classical singularities \citep[e.g.,][]{Stelle1977}. In this context, modifications of GR are motivated by the expectation that the Einstein-Hilbert action should be regarded as an effective low-energy description to be supplemented by additional operators at sufficiently high energies. 

A different picture emerged after the discovery of the late-time accelerated expansion of the Universe \citep{Riess1998,Perlmutter99}. In the context of the $\Lambda$CDM model, the acceleration is explained by the cosmological constant $\Lambda$ in Einstein's GR equations, which can be further interpreted as a fluid component, DE, with negative pressure and constant density: $p = w\rho c^2$, $w = -1$. While DE should account for $\sim$70\% of the cosmological energy budget \citep[e.g.,][]{Planck2018}, this description is not free from conceptual difficulties; for instance, the observed value of $\Lambda$ is extremely small compared to the natural vacuum-energy scale expected from quantum-field-theory arguments, leading to the well-known cosmological-constant problem \citep[e.g.,][]{Weinberg89}, while string theory adds another challenge, due to the difficulty in obtaining stable de Sitter vacuum, essential for $\Lambda$CDM (e.g, \citealt{Cicoli22}).
In addition, tensions between the predictions of the $\Lambda$CDM model and empirical data, as well as discrepancies among independent datasets have been emerging in recent years. Examples include the Hubble tension (e.g., \citealt{H0DN2026}) and the $S_8$ tension 
(e.g., \citealt{Abbott_2022, PhysRevD.110.123508}) although the latter substantially weakened with the final KiDS-Legacy cosmic-shear
analysis, which obtained $S_8=0.82 \pm 0.02$, a value consistent with Planck at $0.7 \, \sigma$ \citep{Wright2025}.
Moreover, recent studies combining DESI and CMB data seem to favor a time-dependent equation of state for DE (e.g., \citealt{Adame_2025}).

The increasingly complex observational landscape has motivated the exploration of alternatives to the standard cosmological framework. Over the last 20 years, numerous theoretical models have been proposed to alleviate the aforementioned tensions; those can be broadly classified in two main groups. The first retains GR as the theory of gravity but replaces the cosmological constant with a dynamical DE component, such as a scalar field with a peculiar equation of state \citep[e.g.,][]{YOO_2012,kjpb-r698}. The second class include models that modify the gravitational sector itself. MG theories mimic the background evolution of the $\Lambda$CDM scenario by adding new features in the gravitational interaction \citep[see e.g.,][for detailed reviews]{Ishak_2018,Cantata21}. In recent years, a broad landscape of MG models has been developed, differing both in the degrees of freedom added to GR and in the scale at which deviations from the Einstein equations become relevant. Popular examples include scalar--tensor theories, such as Brans--Dicke-like models, $f(R)$ gravity, Galileon and Horndeski theories, as well as their extensions to degenerate higher-order scalar--tensor (DHOST) theories. Other classes include vector--tensor and tensor--vector--scalar models, massive and bimetric gravity, higher-dimensional scenarios, and non-local modifications of gravity \citep[e.g.,][]{Carroll2004,DeFelice2010,Clifton2012,Joyce2016}. 

\begin{figure}[ht]
    \centering
    \includegraphics[width=0.6\linewidth]{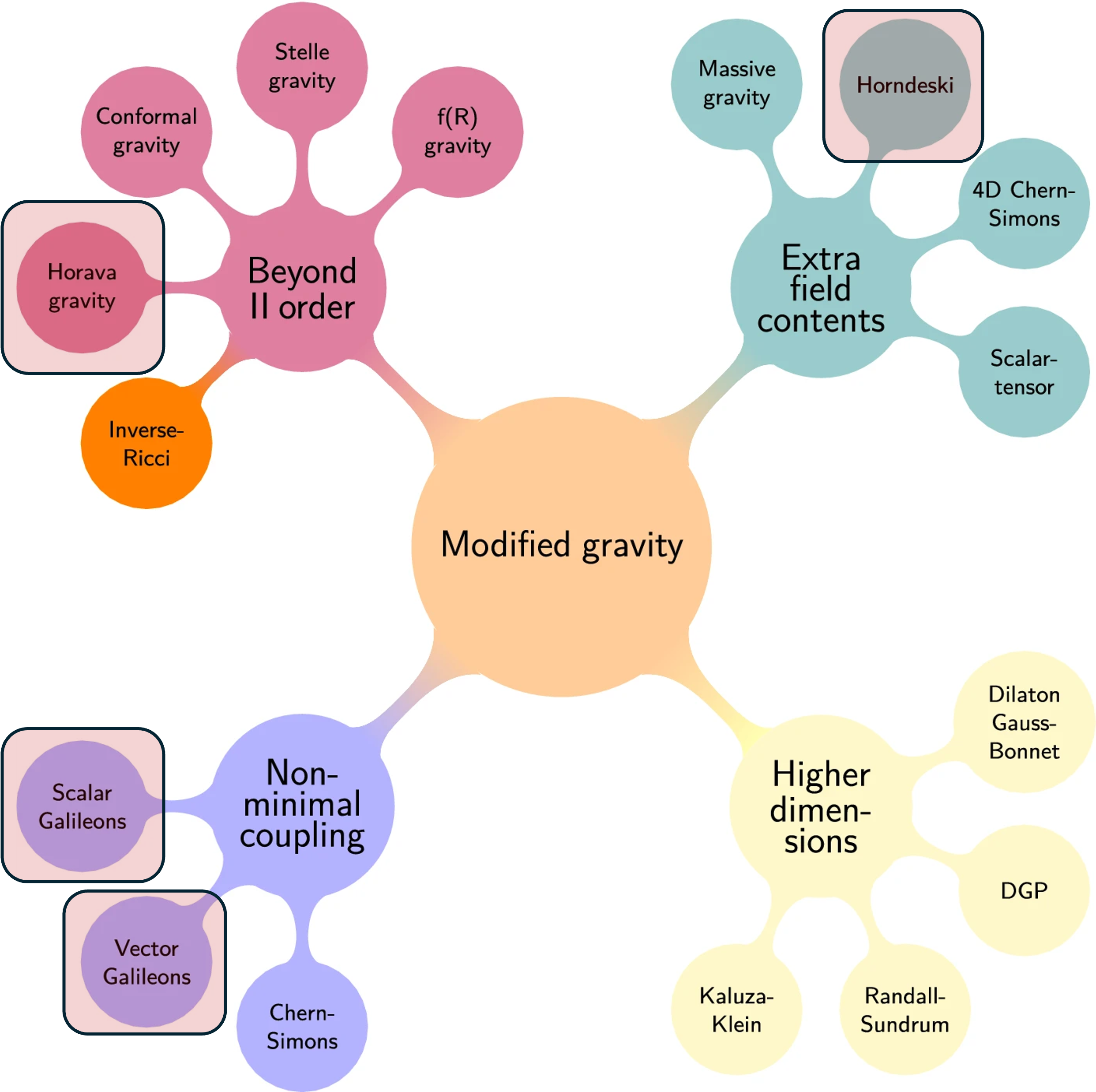}
    \caption{
    \label{fig:MG_overview}
    Schematic overview of different classes of MG theories, classified according to the type of modification introduced with respect to GR. Nodes outlined with a box identify some of the classes whose viable parameter space was strongly reduced by the luminal propagation speed inferred from GW170817/GRB~170817A \citep{Baker_2017}. Image adapted with permission from \cite{Shankaranarayanan2022}, copyright by the author(s).
    }
\end{figure}

A schematic overview of some of the main classes of modified-gravity theories is shown in Fig.~\ref{fig:MG_overview} from \cite{Shankaranarayanan2022}. The classification is based on the type of modification introduced with respect to GR, such as additional fields, higher-order curvature terms, extra dimensions, or non-minimal coupling.
The landscape represented in Fig.~\ref{fig:MG_overview} was substantially reduced by the
multimessenger observation of GW170817 and GRB~170817A, which constrained
the propagation speed of gravitational waves to agree with that of light
to approximately one part in $10^{15}$ \citep{Abbott_2019,Baker_2017}.
Among the theories shown in the figure, this result most strongly affects
Horndeski gravity and the scalar- and vector-Galileon sectors. In the
absence of finely tuned cancellations, luminal propagation significantly restricts the
Horndeski action, excluding models such as the quartic
and quintic scalar Galileons as late-time cosmological models, while others (e.g., the
cubic Galileon and $f(R)$ gravity) remain compatible. Analogous restrictions
apply to generalized-Proca or vector-Galileon interactions, while the
tensor-speed parameter of Ho\v{r}ava gravity is constrained to lie extremely
close to its GR value. Cosmologically active scalar--Gauss--Bonnet
couplings are also strongly restricted. These bounds reduce the viable
parameter space of many broad classes displayed in
Fig.~\ref{fig:MG_overview}; for example, massive gravity remains viable for a sufficiently
small graviton mass, while standard Chern--Simons gravity satisfies the
propagation-speed constraint.

The boundary between extended DE or DM models and MG is not always sharp; following the definition of \cite{Joyce2016}, one may broadly refer to MG as any model in which the Strong Equivalence Principle of GR is violated. However, whether by changing the gravitational sector or by adding new fields to the energy budget of the Universe, extensions of $\Lambda$CDM generally modify the formation, evolution, and assembly of cosmic structures, leaving in principle observable imprints.

Galaxy clusters are particularly valuable in this respect. Modifications of gravity can affect the spatial distribution and abundance of galaxy clusters \citep[e.g.,][]{Ferraro_2011,Cataneo2015,Vogt24}, while also altering mass--observable relations \citep{Mitchell_2021} or structural properties \citep{Lombriser_2012}. Moreover, imprints of new interactions in the gravitational sector can potentially emerge from independent probes of the mass distribution, by comparing how non-relativistic tracers (gas and galaxies) and photons respond to the gravitational field (\citealt{Terukina_2014, Sakstein_2016,Pizzuti2016}; see also \citealt{Cataneo_2018} for a review).

Note that, to account for the stringent constraints on GR in the local Universe, viable extended theories of gravity in cosmological context should implement a screening mechanism to effectively hide the modification at small scales and in dense regions \citep{Brax_2021,Koyama2016,Burrage_2018}. The way in which the screening takes place provides a distinct phenomenology, resulting in specific signatures on the inferred mass distribution of galaxy clusters \citep[e.g.,][]{Pizzuti22}. In this regard, the kinematics of member galaxies may exhibit peculiar features in different regions of the p.p.s. if gravity is modified. Indeed, while for many MG models the internal region of clusters is expected to be fully screened, departures from GR still affect the dynamics in the outskirts \citep[e.g.,][]{Pizzuti2024b,Butt24}. 

In the last decade, several studies have employed kinematical mass profile reconstructions of clusters to test gravity and constrain specific MG models, either as a single probe \citep[e.g.,][]{Rosselli_2023,Pizzuti25a} or in combination with other observables \citep{Gronke2016}. However, general p.p.s. analyses are affected by rather strong degeneracies among the degrees of freedom arising in MG and the parameters of the mass profile \citep{Pizzuti17}.  To make this generic degeneracy explicit, we use \textit{Chameleon gravity} below as a worked phenomenological example. Its 
formulation provides a transparent illustration of how an additional
degree of freedom enters the dynamical mass and affects the matter distribution in non-trivial ways.

Chameleon gravity \citep{Chameleon} is a class of scalar-tensor theories where a new dynamical field $\varphi(t,\bf{x})$ modifies the geodesics of non-relativistic particles (e.g., galaxies) by generating a fifth force which depends on the local matter density $\rho(x)$. In the quasi-static limit --- when time derivatives can be neglected --- the equation of motion of the field reads
\begin{equation}
\label{eq:campophi}
\nabla^2 \varphi
= \frac{\partial V(\varphi)}{\partial \varphi}
+ \frac{Q}{M_\mathrm{Pl}c^2}
\sum_j \rho_j\, e^{Q\varphi/(M_\mathrm{Pl}c^2)},
\end{equation}
where $Q$ represents the coupling with matter, and the sum runs over all matter species with density $\rho_j$. In principle $Q$ can also be different for each component (dark or baryonic), i.e., the field has a specific coupling with different species. $M_\mathrm{Pl} = (8\pi G)^{-1/2}$ is the reduced Planck mass and $c$ is the speed of light. Depending on the shape of the field potential $V(\varphi)$, eq.~\eqref{eq:campophi} can be customized such that gradients are negligible when $\rho_j$ is large, and the field is screened. However, when the density drops the field starts to change and its gradient provides a fifth force modifying the Newtonian potential. In spherical symmetry and weak field limit, the usual Poisson equation becomes
\begin{equation}
\label{eq:gradient}
\frac{\mathrm{d}\Phi}{\mathrm{d}r}
=
\frac{G M(r)}{r^2}
+
\frac{Q}{M_\mathrm{Pl}}
\frac{\mathrm{d}\phi}{\mathrm{d}r} = \frac{G}{r^2}\left[ M(r) + M_\mathrm{eff}\right],
\end{equation}
where we have defined the effective mass associated with the scalar
field,
\begin{equation}
M_\mathrm{eff}(r)
=
\frac{Q}{G}
\frac{r^2}{M_\mathrm{Pl}}
\frac{\mathrm{d}\phi}{\mathrm{d}r}.
\end{equation}
From the structure of eq.~\eqref{eq:gradient} it is clear that any probe of the gravitational potential $\Phi$ --- such as a p.p.s. analysis of galaxy kinematics or hydrostatic equilibrium X-ray measurements --- will be affected by a degeneracy between the "true" mass $M(r)$ and the effective contribution of the field. In fact, the same depth of the potential can be achieved by either having a more massive system in standard gravity or turning on the chameleon field. This degeneracy would prevent any stringent constraints on such models with single mass profile reconstructions at cluster scales if no additional information is provided. Interestingly, a nice feature of Chameleon gravity is that the null geodesics are not affected by the fifth force. In other words, inferring masses using photon propagation as tracers (gravitational lensing) is sensitive to the true mass only; thus, a combination of relativistic and non-relativistic analyses can break the degeneracy and allow to place bounds on the magnitude and shape of the fifth force \citep{Wilcox2015,Boumechta24,Pizzuti26b}. 

The multi-probe approach is not a specific strategy to constrain Chameleon gravity only, but rather a fundamental ingredient for any cluster mass-based test of gravity. 
Galaxies, being non-relativistic tracers, respond primarily to the time-time potential $\Phi$, whereas photons depend on the Weyl potential, $\Phi_\mathrm{lens} =(\Phi+\Psi)/2$ (see eq.~\eqref{eq:ds2}). In GR, in the absence of significant anisotropic stress, the two potentials
are equal, up to second order corrections. If gravity departs from GR, this equality is in general not true; MG models can change the relation between matter fluctuations, $\Phi$ and $\Psi$, giving rise to modified Poisson equations. In comoving Fourier space, they can be written as (e.g.~\citealt{Planck2016}),
\begin{align}
k^2 \Psi &=
-4\pi G a^2 \,\mu(k,a)\,\bar{\rho}_m \,\Delta, \\
k^2(\Phi+\Psi) &=
-8\pi G a^2 \,\Sigma(k,a)\,\bar{\rho}_m \,\Delta .
\end{align}
In this notation $\bar{\rho}_m$ is the average matter density at time $a$ and  $\Delta$ the gauge-invariant comoving density contrast; $\mu(k,a)$ controls the motion of non-relativistic objects, while $\Sigma(k,a)$ controls light deflection.
In modified gravity, generally  $\mu\ne\Sigma\ne1$ and a gravitational slip $\eta(k,a)= \Phi/\Psi \ne 1$ appears. Determinations of cluster masses with kinematics of galaxies (or via hydrostatic equilibrium of the ICM) is therefore the natural probe for the time-time potential $\Phi$, while lensing provides an independent measurement of the relativistic potential (or better, the sum $\Psi+\Phi$). In this context, as for the DM equation-of-state test discussed in
Sect.~\ref{sec:kinDM}, a discrepancy between the mass profiles inferred
from galaxy kinematics or ICM hydrostatics and those reconstructed from lensing should be regarded as a phenomenological null-test, and not as unambiguous evidence for modified gravity \citep{Schmidt_2010}. The
functions $\mu$, $\Sigma$, and $\eta$ quantify this inconsistency, but do
not by themselves identify its physical origin. Indeed, a gravitational
slip may also be generated by anisotropic stress or non-zero effective
pressure in the DM distribution. Its interpretation is further
complicated by the possibility that the additional degrees of freedom
couple differently to dark and baryonic matter
\citep[e.g.,][]{Hammami17}, whereas probe-dependent systematics can mimic
the same departure from unity by biasing the reconstructed mass profiles.

One route to break this degeneracy is to ``over-constrain" the
gravitational potentials with independent tracers. As mentioned above, in metric theories
with universal matter coupling, cluster galaxies and the ICM respond to
the same dynamical potential but are affected by different astrophysical
systematics. Agreement between their inferred mass profiles, combined
with a common offset from lensing, would therefore favour a common
physical origin over a tracer-specific bias. Conversely, a residual
difference between galaxy and gas dynamics, once their respective
systematics are controlled, could indicate a non-universal coupling of
the additional degrees of freedom to baryonic and dark matter
\citep{Hammami17,Gronke2016}.

Further discrimination comes from the radial shape of the signal
and its dependence on cluster properties. Since screening mechanisms
rely on nonlinear field interactions, the resulting
$M_{\rm dyn}(r)/M_{\rm lens}(r)$ need not be constant and may exhibit
model-dependent variations with radius, halo mass, environment, and
redshift \citep{Joyce2016,Falck2015,Pizzuti22b}. Likewise, the alternative
theories we will mention below predict specific
relations between the baryonic distribution, dynamics, and lensing,
which must be tested jointly and consistently across systems.
Identifying the underlying theory therefore requires model-specific
predictions for all available tracers; neither constant $\eta, \mu,\Sigma$ nor a
single global mass offset is sufficient.

\begin{figure}[ht]
    \centering
    \includegraphics[width=0.7\linewidth]{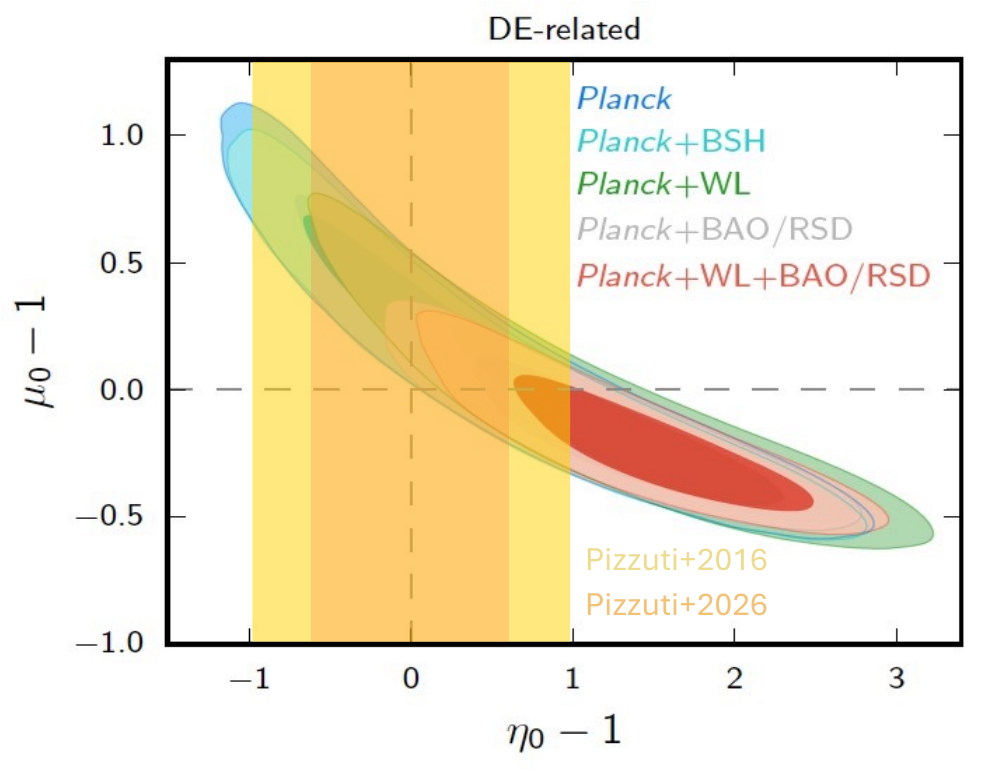}
    \caption{
    \label{fig:eta}
    Constraints on $\mu_0=\mu(z=0)$ vs.
    $\eta_0=\eta(z=0)$, obtained from the joint analysis of \cite{Planck2016}, copyright by ESO, combining different probes and assuming no dependence on the scale. The  lighter and darker vertical bands show the $2\sigma$-bounds found by \cite{Pizzuti2016} and \cite{Pizzuti26} (copyright by the author(s)) by using lensing and kinematics mass profile reconstructions of galaxy clusters. While the former refers to $\eta(z=0.44,r=1.96\,\text{Mpc})$, the latter is obtained by the joint analysis of nine clusters in the redshift range $z\in[0.1,0.5]$, assuming a negligible time evolution. Images reproduced with permission.
    }
\end{figure}

\cite{Terukina_2014} first applied the methodology by combining X-ray and weak lensing data of the Coma cluster to constrain the parameter space of Chameleon gravity. Later, \cite{Pizzuti2016} analysed the cluster MACS J1206 using the same combination of high precision kinematics and lensing data as in \cite{SartorisDM} to constrain the radial profile of the gravitational slip $\eta(r)$. They found  $\eta(1.96 \,\text{Mpc}) = 1.0\pm 0.9$ ($2\,\sigma$ including systematics), and no radial dependence, tightening previous constraints obtained combining CMB data, expansion probes and growth rate probes \citep{Planck2016}. \cite{Pizzuti26} extended their analysis to a sample of nine clusters from the CLASH-VLT collaboration, and found $\eta(1.0\, \text{Mpc}) = 0.9^{+0.5}_{-0.4}$ ($2\, \sigma$). These results highlight the potential of this method when high precision imaging and spectroscopic data are available to reconstruct cluster mass profiles. In Fig.~\ref{fig:eta} we compare the constraints in the $\eta$--$\mu$ diagram obtained by \cite{Planck2016}, and by \cite{Pizzuti2016} and \cite{Pizzuti26}.

Forecast conducted on mock halos  showed that, in the ideal case, with a few dozen clusters, percent-level constraints on $\eta$ can be achieved, sufficient to discriminate among a wide variety of viable models \citep{Pizzuti19}. However, such accuracy requires a rigorous control of systematic uncertainties which may translate a bias in the determination of the mass profiles into a spurious detection of MG. The study of cluster-size halos from $\Lambda$CDM N-body cosmological simulations in \cite{Pizzuti2020syst} demonstrated that using kinematic mass reconstructions based on Jeans analysis to constrain gravity would provide a fake indication of MG for roughly 70\% of systems.  Nonetheless, this percentage can be significantly reduced by a factor of three or more (down to $\lesssim 20\%$) if only regular clusters are selected, based on indicators of their dynamical state (e.g., the Anderson-Darling coefficient, see Sect.~\ref{sec:syst}). 

Beyond general tests of gravity, which do not assume a specific underlying theory, several studies focused on constraining specific classes of MG models using p.p.s. analyses in combination with other mass probes. For instance, \citet{Gronke2016} used cosmological hydrodynamical simulations of screened modified-gravity models, namely $f(R)$ and Symmetron scenarios, to quantify how cluster mass estimates inferred from lensing, galaxy/DM kinematics, and X-ray thermal observables differ from each other. In $f(R)$, the Einstein--Hilbert action is modified by replacing the Ricci scalar $R$ with a generic function, built such that it can reproduce the accelerated expansion of the Universe via modified Einstein equations \citep[e.g.][]{HuSawicki07}. They are equivalent to Chameleon models where the derivative $f_R =\partial f/\partial R$ acts as the new field (the scalaron) mediating a fifth force. Symmetron models are another class of scalar--tensor theories in which the coupling between the field and matter depends on the local environment. At high density, the effective potential has a symmetry-restoring minimum at vanishing field value, so that the coupling to matter is suppressed and GR is recovered. At low density, the symmetry is spontaneously broken, the field acquires a non-zero expectation value, and a fifth force can operate on cosmological and cluster-outskirt scales \citep[e.g.,][]{Hinterbichler2010,Brax2011}.

\citet{Gronke2016} showed that the ratio between kinematic and lensing masses provides a competitive probe of fifth forces on cluster scales, while mass estimates based on the temperature of the ICM are more difficult to interpret because non-thermal pressure support is degenerate with the imprint of MG. Similarly, \cite{Pizzuti17,Pizzuti22} focused on $f(R)$ gravity and DHOST models, respectively, to constrain their parameter space by joint lensing and Jeans-based kinematics analyses of the two CLASH-VLT clusters, MACS J1206 and Abell S1063. DHOST models extend Horndeski gravity through higher-order
field equations subject to degeneracy conditions that remove the
additional Ostrogradsky degree of freedom. In the subclasses relevant
to cluster tests, Vainshtein screening can be partially broken inside
matter, modifying the gravitational potentials through terms that
depend on the radial derivatives of the mass profile
\citep[e.g.,][]{Crisostomi2018}. While confirming again the strength of the multi-probe approach, with bounds competitive with the state-of-art cosmological tests of those models, the results pointed towards a slight indication of MG in the case of Abell S1063. This result seems to contradict the results of \cite{Sartoris_2020}, who found a good agreement between lensing and kinematic mass estimates. However, the application of the Anderson-Darling diagnostic revealed that the cluster is not dynamically relaxed, as also shown by \cite{Mercurio2021}. This outcome highlighted a subtle but crucial point: when searching for small effects such as those expected from viable modification of gravity, even tiny deviations from equilibrium --- which provide negligible effects in standard gravity --- may become an important source of bias.

Other approaches have focused on the stacked p.p.s.\ and infall kinematics around clusters, exploiting the fact that MG effects may become more visible outside the fully screened inner regions. This regime is particularly relevant for theories relying on specific kind of screening mechanism, the
Vainshtein screening. The Dvali--Gabadadze--Porrati (DGP) model is a braneworld construction in which gravity crosses over from four-dimensional to five-dimensional behaviour beyond a scale $r_c$
\citep{Dvali_2000}. Around massive sources, the additional scalar
brane-bending mode is suppressed within the Vainshtein radius by its
nonlinear derivative interactions. Galileon theories generalise this
effective scalar sector of DGP and employ the same screening
mechanism \citep{Nicolis2009,Koyama2016}. Consequently, deviations
from GR can be strongly suppressed within haloes while remaining
observable through velocities in their infall regions.

\cite{Lam2012,Lam2013} used the los velocity distribution of galaxies around simulated stacked clusters, in combination with weak-lensing mass information, to probe $f(R)$ and DGP-like \citep{Dvali_2000} models  on scales of a few to tens of Mpc. They found that in $f(R)$ gravity $\sigma_\mathrm{los}(R)$ can be enhanced by $\sim 10$--$30\%$, while the DGP framework produces either enhanced or suppressed velocities depending on the sub-class of models (branch) considered. In particular, they showed that a spectroscopic survey covering around $\sim 2000\,{\rm deg}^2$ could provide enough information to build stacked clusters to significantly improve constraints on $f(R)$ and DGP parameters, with sensitivity to the intermediate, weakly non-linear regime between virial cluster scales and large-scale redshift-space distortions. In a similar spirit, \citet{Zu2014} used the redshift-space cluster--galaxy cross-correlation function to model galaxy infall kinematics around clusters, showing that the velocity field in the infall region carries detectable signatures of $f(R)$ and Galileon gravity.

\citet{Shirasaki2021} measured the stacked p.p.s. of galaxies around 23 clusters from the Local Cluster Substructure Survey \citep[LoCuSS,][]{Smith+10}, using a combination of spectroscopic information and weak-lensing mass estimates. In their halo-model, the observable is the distribution of los velocities of cluster--galaxy pairs as a function of projected separation, including contributions from satellite galaxies, interlopers, and large-scale pairwise motions. This approach allowed them to infer the average $\sigma_v$ of galaxies in massive clusters and to compare it with the virial scaling expected from DM-only simulations in GR.
Interpreting possible signatures of modified gravity in terms of a modified Poisson equation with an effective gravitational constant $G_\mathrm{eff}$, they obtained
    $0.88 < G_\mathrm{eff}/G < 1.29$ at 68\%.
They also translated this constraint into bounds on $f(R)$ and DGP models, finding  $\log |f_R(z=0.2)|<-4.57$ assuming a Hu--Sawicki parametrisation of $f(R)$ \citep{HuSawicki07}, and 
$r_c>253\,h^{-1}{\rm Mpc}$ for the typical normal branch of DGP gravity. 

A related use of galaxy kinematics in the cluster outskirts is provided by the caustic technique \citep[e.g.,][]{Pizzardo24}. \citet{Butt24} proposed a joint analysis of hydrostatic and caustic mass estimates for the clusters A2029 and A2142 from the X-COP sample \citep{Eckert2017}, with the aim of constraining Chameleon and DHOST models. In this framework, the caustic method traces the escape-velocity profile from the p.p.s. distribution of galaxies and extends the dynamical information to radii where screening effects may be weaker. By calibrating the radial dependence of the caustic filling factor $F_\beta(r)$ with the hydrostatic reconstruction, instead of assuming it to be constant, they where able to significantly improve the constraints on the MG models. In the specific DHOST case, the joint analysis tightened the constraints on one of the free parameters by about a factor of two for A2029, while for A2142 it reduced a mild $\sim2\sigma$ deviation from GR to a result consistent with GR at $\lesssim1\sigma$. 
This analysis also provided an important cautionary example: even if the assumption of equlibrium is no more required in the caustic analysis, incorrect assumptions on $F_\beta(r)$ can be a further source of bias, and they can generate artificially tight constraints, or even produce spurious indications of MG.

\subsubsection{Gravitational redshift}
Another way of testing GR using kinematical data of cluster member galaxies is given by the gravitational redshift effect (GRS). Photons emitted by galaxies located deep
inside the cluster potential well, lose energy while ``climbing out" of the
potential, producing a small shift in the observed spectrum. In the weak-field limit this effect is proportional to the difference in gravitational potential,
    $z_\mathrm{g} \simeq \Delta \Phi/c^2$.
For a typical galaxy cluster, the corresponding velocity shift is only 
$\sim 10~{\rm km\,s^{-1}}$, i.e. about two orders of magnitude smaller than the typical scale of the velocity distribution of member galaxies. As such, it can safely be ignored in the error budget for a cluster $M(r)$ determination, but, for the same reason, it is very difficult to detect. In fact it can only be detected statistically, by stacking large samples of clusters and measuring the mean offset of the los velocity distribution as a function of projected radius \citep{Cappi1995,Wojtak2011}.

\citet{Rosselli_2023} used the GRS of cluster member galaxies as a direct test of GR on Mpc scales. Using SDSS DR16 spectroscopic galaxies, they constructed a sample of 3058 clusters at $z<0.5$, with masses in the range $10^{14}$--$10^{15}\,M_\odot$, and $\simeq 4.9\times10^4$ associated member galaxies. Differently from previous analyses, they did not assume the cluster centre to coincide with the BCG alone, but estimated it from the average angular position and redshift of the galaxies closest to the BCG, to provide a more stable estimate of the center of the gravitational potential well. The mean value of the los velocity offset between the cluster centre and all the cluster galaxies,
\begin{equation}
    \Delta =
    c\left[\ln(1+z_\mathrm{obs})-\ln(1+z_\mathrm{cen})\right],
\end{equation}
measured after stacking the cluster-galaxy p.p.s. and correcting for foreground/background contamination, encodes the GRS which may differ depending on the given theory of gravity.
\citet{Rosselli_2023} compared the measured signal with the predictions of GR, a strong-field $f(R)$ model, and a self-accelerating DGP model, also including the main relativistic and selection effects that can shift the mean at the level of a few ${\rm km\,s^{-1}}$, namely the transverse Doppler effect, light-cone effects, and surface-brightness modulation \citep{Kaiser13}.
They detected a clear negative signal, increasingly negative with projected radius, and measured an integrated value of
$\bar{\Delta}_\mathrm{int} =
    -11 \pm 3~{\rm km\,s^{-1}}$,
in agreement with the GR expectation for clusters in the same mass range as the observed sample. Here $\bar{\Delta}_\mathrm{int}$ denotes the mean velocity offset measured from the
full stacked p.p.s., including all member galaxies within a projected radius of $R<4 \, r_{500}$. By fitting the radial profile with a phenomenological rescaling of the gravitational acceleration, $G\rightarrow \alpha \, G$, they obtained
$
    \alpha = 0.86 \pm 0.25 ,
$
consistent with GR and with the self-accelerating DGP model within the uncertainties, while marginally disfavouring the strong-field $f(R)$ model considered in the analysis.

While this method appears to be very powerful, providing a complementary use of cluster kinematics with respect to standard dynamical mass reconstructions, it remains sensitive to many of the systematics we have already discussed, and which demand future careful analyses. The definition of the cluster centre, the contamination of interlopers, and the modelling of the small relativistic corrections could affect the GRS signal and thus generating again biased tests of gravity.

\subsubsection{MOdified Newtonian Dynamics (and friends)}
Cluster kinematics can be used to test more radical alternatives to the standard dark sector. Several models have been proposed not only to explain cosmic acceleration through a modification of gravity, but also to reduce, or even remove, the need for DM. The best known example is MOdified Newtonian Dynamics (MOND), originally introduced by \citet{Milgrom1983}, in which the Newtonian relation between acceleration and gravitational field is modified below a characteristic acceleration scale $a_0$. While MOND is remarkably successful in describing galaxy rotation curves and the radial acceleration relation, galaxy clusters have long represented one of its main challanges: even after applying the MOND prescription, the observed baryonic mass in stars and hot gas is generally insufficient to account for the full dynamical mass, leaving a residual missing-mass problem \citep[e.g.,][]{Sanders2003,Famaey2013,Kelleher2024}. A recent reassessment of this issue, presented by \cite{Zhang2026}, suggests that part of the residual missing mass in MOND clusters may be linked to an underestimation of the stellar and remnant mass components, although the conclusion depends on the assumed stellar population model, the spatial distribution of remnants, the ICL fraction, and the reliability of the hydrostatic mass estimates.

The work of 
\cite{Li2023} reinforced the issue of the MOND missing mass in clusters. These authors used the spherical Jeans equation and cluster $\sigma_\mathrm{los}$ to reconstruct dynamical mass profiles for HIFLUGCS clusters down to accelerations $\lesssim 10^{-11}\,{\rm m\,s^{-2}}$. By comparing the resulting dynamical accelerations with the baryonic accelerations inferred from gas and stellar profiles, they found that rich X-ray clusters deviate from the galaxy-calibrated radial acceleration relation, reinforcing the known MOND residual missing-mass problem in clusters, especially at intermediate radii. At larger radii the data may approach the galaxy relation, but the conclusion is limited by extrapolations of the gas mass and by the uncertain dynamical state of cluster outskirts. 

Spectroscopic surveys of cluster galaxies extending to the cluster outer regions, where the X-ray surface brightness is below current detection limits, are important to map the baryonic matter surrounding the cluster. Mapping this matter distribution is crucial to estimate the external field effect, which is a specific feature of some Lagrangian formulations of MOND, like AQUAL and QUMOND \citep[e.g.][]{Milgrom2025PhRvD.111j4033M}, and thus to properly describe the cluster galaxy velocity field with the observed baryonic mass alone. 
With this information, we can compare the mass estimates inferred from spectroscopic surveys of cluster galaxies, X-ray data and gravitational lensing. These mass estimates depend differently on the two gravitational potentials of the space-time metric of a relativistic theory \citep[e.g.][]{Milgrom2022PhRvD.106h4010M}. Therefore the discrepancies or consistency of these mass estimates are crucial to discriminate among the rich family of the relativistic extensions of MOND \citep{Skordis2021PhRvL.127p1302S}. 

A challenging 
test for MOND is given by the ``Bullet cluster'' phenomenology \citep{CGM04}. Following a collision of this cluster with a group, the intra-cluster gas has been displaced relative to both the galaxies
\citep{Barrena+02} and the gravitational potential, as traced by gravitational lensing \citep{Clowe+06b}. Since most of the baryonic mass is in the intra-cluster gas, in the absence of DM the gravitational lensing signal would be expected to trace the location of the baryons, while it traces the location of the galaxies. The coincidence of the galaxy distribution with the gravitational lensing signal is considered direct proof for the existence of an almost collisionless DM component. However, a recent analysis seems to reconcile MOND with the Bullet phenomenology on the basis that the gravitational lensing signal is actually generated by the mass surface density which is larger for galaxies than for the diffuse intra-cluster medium \citep{Hernandez26}. In addition, the amount of baryonic matter in the Bullet cluster, like other nearby clusters as mentioned above, might be $\sim 50$\% larger than previously thought \citep{Zhang2026PhRvD.114b3001Z}.

A striking feature of massive interacting clusters, like the Bullet cluster or El Gordo \citep{Menanteau2012ApJ...748....7M}, is the large relative velocity  $\gtrsim 3000$~km~s$^{-1}$ of  the main cluster substructures. These velocities are  rather unlikely in the standard $\Lambda$CDM cosmology, but expected in MONDian cosmologies \citep[e.g.][]{Kraljic2015JCAP...04..050K, Asencio2023ApJ...954..162A, Katz2013ApJ...772...10K}. These collision velocities are derived from the identification of shocks in the X-ray maps \citep[e.g.][]{Markevitch2004ApJ...606..819M} or from N-body modelling \citep[e.g.][]{Donnert2014MNRAS.438.1971D}. However, extensive spectroscopic surveys of a large sample of interacting clusters can provide crucial additional information on the velocity field of the galaxies in the system \citep{Yu2025ApJ...991..220Y} and thus provide stringent constraints on the models of the formation of cosmic structure and on the law of gravity that rules it.

Other theoretical frameworks follow a phenomenological motivation similar to MOND but implement the modification differently. In Scalar--Tensor--Vector Gravity, often referred to as MOdified Gravity (MOG), the gravitational interaction is supplemented by scalar fields and a massive vector field, leading to an effective gravitational coupling and a finite-range fifth-force contribution that can mimic DM on galactic and cluster scales \citep{Moffat2006,Brownstein2006}. Refracted Gravity (RG), instead, modifies the Poisson equation by introducing an effective gravitational permittivity $\epsilon(\rho)$ that depends on the local matter density. In high-density regions the theory recovers the Newtonian limit, whereas in low-density environments the gravitational field is effectively enhanced, producing DM-like phenomenology without the need of a halo of DM particles \citep{Matsakos2016,Cesare2020, Cesare2022A&A...657A.133C}. RG  was first proposed as a phenomenological model; an attempt to formulate it as a consistent covariant theory incorporating in a single scalar field the effects of both DM and DE at large scales is also available \citep{Sanna2023, Gervani2026JCAP...06..038G}.

Studying galaxy clusters properties in these alternative theories is therefore a mandatory step to connect the regime where the models are often most successful, namely galaxies, and the cosmological scales where the full dark sector is usually required even if gravity is modified. In this context, kinematics analyses can be used to trace the gravitational potential predicted by the baryonic components in frameworks without DM under the modified force law. An example is provided by the study of \citet{Pizzuti25a}, who tested RG in the clusters MACS J1206 and Abell S1063 by combining the p.p.s. distribution of cluster members with the stellar velocity-dispersion profile of the BCG through the \texttt{MG-MAMPOSSt} code \citep{Pizzuti2021}. They found that RG can reproduce the observed kinematics of the two clusters with a quality comparable to Newtonian gravity, although the latter is mildly preferred; however, the two clusters require mutually inconsistent values of the supposedly universal RG parameters, and those values are also in tension with constraints obtained on other astrophysical scales. While the results can be interpreted as an efficient way to rule out RG, they rise again the delicate issue of systematics. In particular, RG is extremely sensitive to deviation from a regular round shape of the objects, and the inconsistencies among the model parameters may be a direct consequence of small departures from spherical symmetry and dynamical relaxation.

\hspace{0.3cm}

In summary, gravity tests with galaxy clusters are becoming increasingly important in the current theoretical and observational landscape, where cosmological and astrophysical tensions are sharpened by the growing amount and quality of available data, and the physical nature of the dark sector remains unknown. In this context, cluster kinematics provides a rapidly evolving set of tools to probe the gravitational interaction from the innermost regions, dominated by the BCG, to the outskirts and infall regions, where several MG scenarios predict potentially detectable signatures. The studies reviewed in this section show that combining kinematical information with complementary probes (e.g., lensing, X-ray, or SZ measurements) allows the broad zoo of alternatives to $\Lambda$CDM to be constrained more robustly than with any single probe alone. However, the interpretation of possible non-standard signatures requires caution. Detecting small deviations from GR, or from the standard dark-sector picture, demands not only high-quality data, but also a careful treatment of the systematics that can bias the inferred gravitational potential. In the coming years, large spectroscopic surveys will make these tests statistically much more powerful; the limiting factor will increasingly shift from data volume to the accuracy with which astrophysical and observational uncertainties can be controlled.

\section{Conclusions: a bright future for cluster kinematics}
\label{sec:future}
The study of the dynamics of galaxies in clusters has followed a remarkable evolution in the last two decades. What started as a simple application of the virial theorem to a handful of redshifts has become a broad set of techniques capable of jointly reconstructing mass and velocity anisotropy profiles, connecting the kinematical properties to the assembly history of the cluster and its galaxies, as well as constraining the dark sector and possible departures from the standard gravitational framework. This rising trajectory mirrors the observational progress of the field: from the first velocity dispersions of nearby clusters, to hundreds or thousands of spectroscopic members in individual systems, and now to large samples of clusters where p.p.s. analyses can be performed statistically. The central message of this review is that cluster kinematics is nowadays much more than a classical mass estimator. 
Galaxy velocities not only quantify the depth of the potential well; they also encode the degree to which a cluster departs from an idealized, relaxed system, turning p.p.s. into a powerful diagnostic of many physical processes from kpc to Mpc scales, and across the cosmic history at late times.

A first lesson we draw from the methods discussed in this review is that there are several different ``dynamical mass'' definitions for a cluster, depending on the assumptions made. Virial estimators, $\sigma_v$ scaling relations, Jeans analyses, caustics, and ML estimators all exploit the same basic observables --- projected positions and line-of-sight velocities --- but use the available information in different ways and rely on different physical assumptions. Simple global estimators remain useful for large samples, especially when calibrated on simulations and corrected for selection effects, but they cannot fully describe the radial structure of the potential or the orbital distribution of the tracers. On the other hand, forward modelling of p.p.s. provides a much richer description, allowing one to infer mass and anisotropy profiles jointly, but at the price of stronger assumptions about equilibrium, symmetry, tracer density, and the parametric form of the gravitational potential. The caustic method extends the reach of kinematics into the infall region, where equilibrium cannot be assumed, but requires a careful treatment of the escape-velocity edge, sampling, projection effects, and the anisotropy-dependent filling factor. Finally, ML and simulation-based approaches add a complementary path: they can learn complex mappings between p.p.s. structure and cluster properties, but their reliability depends on how well the training simulations reproduce the relevant galaxy populations, selection functions, and --- as always --- observational systematics.

The second lesson that emerges from our overview is that the main limitation of cluster kinematics is not the lack of information in galaxy velocities, but the difficulty of isolating the physical signal from the layers of assumptions and observational complexity that shape the p.p.s.. We have seen that effects introduced by contamination from interlopers, departures from dynamical equilibrium, and irregular morphology cannot anymore be considered secondary details; with the increase of the statistical power, they become a fundamental part of the problem itself, producing biases that may be of the same order of (or even larger than) the constraints on the interesting physical quantities. They determine which galaxies are actually tracing the potential, which regions can be analysed with equilibrium models, and which observables are most sensitive to recent accretion or projection effects.
What makes the issue of systematics particularly relevant is the fact that the same features that complicate mass reconstruction also contain physical information. Substructures trace the hierarchical assembly of clusters; velocity segregation encodes a connection between galaxy orbits and environmental transformation; the outskirts preserve signatures of infall, splashback, and anisotropic accretion from the cosmic web. A mature use of cluster kinematics therefore requires a change of perspective: systematics should not only be minimized, but also modeled and, whenever possible, turned into diagnostics of the dynamical state of the system.

This point is even more important if we aim to use clusters as laboratories for precision cosmology. Cluster abundance studies require accurate mass calibration, and galaxy velocities offer a tool that is physically independent of lensing, X-ray, and SZ observables. However, the interpretation of kinematical masses must be tied to the selection of tracers and to the dynamical state of the systems. A cluster $\sigma_\mathrm{los}$ measured from passive, central, well-virialized galaxies is not equivalent to that measured from a population rich in blue, recently accreted, star-forming galaxies. Similarly, a stacked p.p.s. built from clusters with different orientations, accretion rates, and selection functions does not automatically represent a single equilibrium object. Future cosmological applications will therefore require dynamical likelihoods that include all the aforementioned effects as part of the inference --- not as external corrections. In this sense, the next step requires a synergy between the collection of more high-precision redshifts, and the development of new frameworks of data analysis that are sufficiently realistic to propagate observational and astrophysical uncertainties into the final mass calibration. In this regard, recent progress in artificial intelligence and deep learning, together with increasingly realistic cosmological simulations, offers a promising route forward. However, while these tools can build inference schemes that learn directly the full complex information in the p.p.s., there will always be the need of physically-motivated dynamical models (such as distribution-function techniques) that complement ML approaches to make a transparent connection with the underlying physics.

A third, fundamental lesson is that cluster kinematics is uniquely powerful when combined with independent probes of the mass distribution. Gravitational lensing maps the projected mass without relying on dynamical equilibrium; X-ray and SZ observations probe the thermodynamical state of the intracluster gas; galaxy kinematics measures the response of non-relativistic (collisionless) tracers to the gravitational potential. The comparison among these probes is therefore not only a way to reduce statistical uncertainties, it is a physical consistency test. Agreement between lensing and kinematics supports the interpretation of galaxies as reliable markers of the potential and constrains the orbital distribution of galaxies breaking the mass-anisotropy degeneracy. Disagreement may reveal non-equilibrium dynamics and the presence of dominant systematics effects, or, in more ambitious applications, a departure from the gravitational framework assumed in the modelling. The most informative analyses are therefore those where the different probes are not simply compared \textit{a posteriori}, but combined in a common likelihood or hierarchical framework, with each observable retaining its own physical assumptions and systematics.

This complementarity is central for tests of the dark sector and gravity. In the standard collisionless CDM picture, the global properties of a cluster (e.g. the total mass, concentration, orbital anisotropy) are connected to the assembly history of the halo. Kinematics can test these relations directly, especially when extended from the BCG-dominated inner regions to the virial radius and beyond. Joint analyses of BCG stellar kinematics, galaxy dynamics, lensing, and X-ray data can probe the inner logarithmic slope of the DM profile, the relative contribution of baryons and DM, and the consistency of cluster mass profiles with CDM expectations. At larger radii, caustics and infall modelling provide access to the transition between the virialized halo and the surrounding cosmic web, where the mass accretion rate, splashback features, and environmental dependence of cluster growth become relevant.

For MG and DE models, the role of kinematics is even more specific. Galaxy motions respond to the potential governing non-relativistic dynamics, whereas lensing depends on the relativistic combination of the metric potentials. This feature makes the combination of lensing and kinematics a natural test of the gravitational slip $\eta$, a general measurements of departures from GR, as well as of specific MG models exhibiting fifth forces at cluster scales. In many alternative theories the inner regions of massive clusters may be screened while the outskirts remain partially sensitive to deviations from GR. A purely dynamical mass reconstruction can be degenerate with changes in the true mass profile, but the degeneracy can be broken by comparing it with lensing, which is affected differently by the additional degrees of freedom. The current constraints are still mostly limited by the sample size and the control of systematics, but they demonstrate that cluster kinematics can offer tests of gravity in environments less explored by other probes and complementary to Solar-System experiments, cosmological large-scale structure, and gravitational waves.

Looking ahead, the field is entering a quantitatively new regime. Ongoing and planned wide-area spectroscopic surveys such as DESI \citep{DESIDR1}, 4MOST 4HS \citep{4HS} and CHANCES \citep{CHANCES}, WEAVE WCN, WWFCS, and the Cosmological Clusters Survey \citep{WEAVE}, the Subaru PFS Cosmology Survey \citep{PFS}, MOONS MOONRISE \citep{MOONRISE}, together with imaging and lensing surveys such as Rubin/LSST \citep{LSST}, and Roman HLWAS \citep{ROMAN}, and the spectroscopic and lensing Euclid surveys \citep{EC_Mellier_2025} will provide unprecedented samples of clusters with spectroscopic tracers over large radial ranges. The cluster gravitational potential will be probed using several tracers, galaxies of different spectral and morphological type, color, mass, and star-formation rate, but possibly also planetary nebulae \citep[e.g.][]{Spiniello+18}, globular clusters \citep[e.g.][]{Chaturvedi+22}, or ICL \citep[e.g.][]{Atek+25}. Studying cluster kinematics will allow to connect mass calibration to galaxy evolution, as the same p.p.s. that constrains the gravitational potential, also contains information on the
accretion history and evolution of the different cluster galaxy populations.

At the same time of these large spectroscopic surveys, eROSITA \citep{eROSITA}, Athena \citep{Athena}, CMB-S4 \citep{CMB-S4}, the Simons Observatory \citep{Simons}, and other X-ray and SZ facilities will improve the characterization of the intracluster medium and the thermodynamical state of large cluster samples. The combination of these datasets with those coming from the optical surveys will enable a shift from detailed studies of individual benchmark clusters to population-level dynamical
studies, where mass, redshift, environment, morphology, and dynamical state can be
controlled simultaneously. In addition, the same phase-space diagram that constrains the gravitational potential, also contains information on the accretion history and evolution of the different cluster galaxy populations.

This observational progress will demand equally significant progress in modeling. First, dynamical analyses should move toward hierarchical frameworks in which individual clusters are modeled jointly, while allowing
concentration, anisotropy, substructure, and velocity bias to vary from cluster to cluster according to population-level distributions. Such frameworks would constrain population trends and intrinsic scatter while retaining cluster-specific information, allowing sparsely sampled systems
to benefit from the joint analysis without diluting the information
provided by well-sampled clusters. Second, the boundary between equilibrium and non-equilibrium dynamics should be treated more flexibly. Standard Jeans analyses remain appropriate inside the virialized region of relaxed systems, but the infall region requires models that include mean radial streaming, Hubble expansion, and the transition between orbiting and infalling populations. Generalized Jeans formalisms, caustic-based approaches, and models of the full los velocity distribution in the outskirts provide promising paths in this direction. Third, simulation-based inference and ML methods should be integrated with physical modeling rather than treated as black boxes. Their strength lies in handling complex selection functions, high-dimensional observables, and non-Gaussian phase-space features; their weakness lies in their dependence on the realism of the simulations used for training. Robust applications will therefore require extensive validation across independent hydrodynamical simulations, semi-analytic galaxy models, and mock survey realizations.

The future of the field will also depend on the construction of clean and reproducible dynamical datasets. Membership selection, aperture definitions, centering choices, redshift quality flags, completeness masks, and adopted radial scalings often differ from study to study, making comparisons difficult. Large surveys will make it possible to standardize these choices, but only if the dynamical catalogues are built with explicit selection functions and publicly documented algorithms. This condition is particularly important for ML applications and for comparisons with simulations, where the same observable definitions must be applied to real and mock data. A robust dynamical pipeline for the next decade should therefore include not only redshift measurements and member catalogues, but also probabilistic membership assignments, completeness weights, substructure diagnostics, dynamical-state indicators, and covariance estimates for derived profiles.
 
The next generation of spectroscopic data coupled with improved theoretical modeling, will allow
kinematical studies to provide increasingly accurate and precise cluster mass and mass profile determinations, to trace the cluster assembly history, to test for different DM candidates, probe theories beyond GR, but also, in combination with other data, probe the physics of the cluster components. In doing so, the field can transform one of the oldest arguments for DM into one of the most precise laboratories for testing the nature of DM, DE, and gravity on cosmological scales. 

\vspace{0.5cm}

\textit{``Clusters of galaxies are the largest clearly recognizable structurally organized units of matter in the universe. They represent the fundamental cells of the cosmos. It is within these grand aggregates that the laws of gravitation, thermodynamics, and cosmic evolution must be tested on a grand scale, for they contain the total history of cosmic matter written in their distribution and motions''} \citep{Zwicky1957}.

\backmatter

\bmhead{Acknowledgements}
We thank two anonymous referees for their useful suggestions and comments. LP acknowledges support from the Italian Ministry for Research and University (MUR) under Grant ‘Progetto Dipartimenti di Eccellenza 2023–2027’ (BiCoQ).
AD acknowledges partial support from the grant InDark of the Italian National Institute of
Nuclear Physics (INFN). 
This research has made use of NASA’s Astrophysics Data System Bibliographic Services.

\section*{Declarations}

\bmhead{Conflict of interest} The authors declares no conflict of interest.

\phantomsection
\addcontentsline{toc}{section}{References}
\bibliography{sn-bibliography}

@article{Takada_2007,
   title={Probing dark energy with cluster counts and cosmic shear power spectra: including the full covariance},
   volume={9},
   ISSN={1367-2630},
   DOI={10.1088/1367-2630/9/12/446},
   number={12},
   journal={\njp},
   publisher={IOP Publishing},
   author={Takada, Masahiro and Bridle, Sarah},
   year={2007},
   month=dec,
   pages={446–446}
}

@ARTICLE{Durret2022,
       author = {{Durret}, F. and {Degott}, L. and {Lobo}, C. and {Ebeling}, H. and {Jauzac}, M. and {Tam}, S.-I.},
        title = "{Ram pressure stripping in the z {\ensuremath{\sim}} 0.5 galaxy cluster MS 0451.6-0305}",
      journal = {\aap},
         year = 2022,
        month = jun,
       volume = {662},
          eid = {A84},
        pages = {A84},
          doi = {10.1051/0004-6361/202142983},
archivePrefix = {arXiv},
       eprint = {2204.07445},
 primaryClass = {astro-ph.CO},
       adsurl = {https://ui.adsabs.harvard.edu/abs/2022A&A...662A..84D}
}

@article{Abell1958,
  author       = {Abell, G.~O.},
  title        = {The Distribution of Rich Clusters of Galaxies},
  journal      = {\apjs},
  volume       = {3},
  pages        = {211},
  year         = {1958},
  doi          = {10.1086/190036}
}

@article{Bahcall1977,
  author       = {Bahcall, N.~A.},
  title        = {Clusters of galaxies and the large-scale structure of the universe},
  journal      = {\araa},
  volume       = {15},
  pages        = {505--540},
  year         = {1977},
  doi          = {10.1146/annurev.aa.15.090177.002445}
}

@article{FormanJones1982,
  author       = {Forman, W. and Jones, C.},
  title        = {X-ray observations of clusters of galaxies},
  journal      = {\araa},
  volume       = {20},
  pages        = {547--598},
  year         = {1982},
  doi          = {10.1146/annurev.aa.20.090182.002555}
}

@article{DresslerShectman1988,
  author       = {Dressler, A. and Shectman, S.~A.},
  title        = {Evidence for substructure in rich clusters of galaxies},
  journal      = {Astronomical Journal},
  volume       = {95},
  pages        = {985--995},
  year         = {1988},
  doi          = {10.1086/114802}
}

@article{WestBothun1990,
  author       = {West, M.~J. and Bothun, G.~D.},
  title        = {Substructure in rich clusters of galaxies},
  journal      = {\apj},
  volume       = {350},
  pages        = {36--47},
  year         = {1990},
  doi          = {10.1086/168357}
}

@ARTICLE{BinneyMamon1982,
       author = {{Binney}, J. and {Mamon}, G.~A.},
        title = "{M/L and velocity anisotropy from observations of spherical galaxies, or must M87 have a massive black hole ?}",
      journal = {\mnras},
         year = 1982,
        month = jul,
       volume = {200},
        pages = {361-375},
          doi = {10.1093/mnras/200.2.361},
       adsurl = {https://ui.adsabs.harvard.edu/abs/1982MNRAS.200..361B}
}

@article{Merritt1987,
  author       = {Merritt, D.},
  title        = {Spherical stellar systems with spheroidal velocity distributions},
  journal      = {\apj},
  volume       = {313},
  pages        = {121--135},
  year         = {1987},
  doi          = {10.1086/164948}
}

@ARTICLE{Biviano_2013,
       author = {{Biviano}, A. and {Rosati}, P. and {Balestra}, I. and {Mercurio}, A. and {Girardi}, M. and {Nonino}, M. and {Grillo}, C. and {Scodeggio}, M. and {Lemze}, D. and {Kelson}, D. and {Umetsu}, K. and {Postman}, M. and {Zitrin}, A. and {Czoske}, O. and {Ettori}, S. and {Fritz}, A. and {Lombardi}, M. and {Maier}, C. and {Medezinski}, E. and {Mei}, S. and {Presotto}, V. and {Strazzullo}, V. and {Tozzi}, P. and {Ziegler}, B. and {Annunziatella}, M. and {Bartelmann}, M. and {Benitez}, N. and {Bradley}, L. and {Brescia}, M. and {Broadhurst}, T. and {Coe}, D. and {Demarco}, R. and {Donahue}, M. and {Ford}, H. and {Gobat}, R. and {Graves}, G. and {Koekemoer}, A. and {Kuchner}, U. and {Melchior}, P. and {Meneghetti}, M. and {Merten}, J. and {Moustakas}, L. and {Munari}, E. and {Reg{\H{o}}s}, E. and {Sartoris}, B. and {Seitz}, S. and {Zheng}, W.},
        title = "{CLASH-VLT: The mass, velocity-anisotropy, and pseudo-phase-space density profiles of the z = 0.44 galaxy cluster MACS J1206.2-0847}",
      journal = {\aap},
         year = 2013,
        month = oct,
       volume = {558},
          eid = {A1},
        pages = {A1},
          doi = {10.1051/0004-6361/201321955},
archivePrefix = {arXiv},
       eprint = {1307.5867},
 primaryClass = {astro-ph.CO},
       adsurl = {https://ui.adsabs.harvard.edu/abs/2013A&A...558A...1B}
}

@article{GellerBeers1982,
  author       = {Geller, M.~J. and Beers, T.~C.},
  title        = {Substructure in clusters of galaxies},
  journal      = {\pasp},
  volume       = {94},
  pages        = {421--435},
  year         = {1982},
  doi          = {10.1086/131028}
}

@article{Yahil1977,
  author       = {Yahil, A. and Vidal, N.~V.},
  title        = "{Velocity distribution and mass of the Coma cluster}",
  journal      = {\apj},
  volume       = {214},
  pages        = {347--357},
  year         = {1977},
  doi          = {10.1086/155267}
}

@ARTICLE{Lopez2022,
       author = {{L{\'o}pez-Guti{\'e}rrez}, M.~M. and {Bravo-Alfaro}, H. and {van Gorkom}, J.~H. and {Caretta}, C.~A. and {Durret}, F. and {N{\'u}{\~n}ez-Beltr{\'a}n}, L.~M. and {Jaff{\'e}}, Y.~L. and {Hirschmann}, M. and {P{\'e}rez-Mill{\'a}n}, D.},
        title = "{Environmental cluster effects and galaxy evolution: The H I properties of the Abell clusters A85/A496/A2670}",
      journal = {\mnras},
         year = 2022,
        month = nov,
       volume = {517},
       number = {1},
        pages = {1218-1241},
          doi = {10.1093/mnras/stac2526},
archivePrefix = {arXiv},
       eprint = {2209.00764},
 primaryClass = {astro-ph.GA},
       adsurl = {https://ui.adsabs.harvard.edu/abs/2022MNRAS.517.1218L}
}

@ARTICLE{Cole1996,
       author = {{Cole}, Shaun and {Lacey}, Cedric},
        title = "{The structure of dark matter haloes in hierarchical clustering models}",
      journal = {\mnras},
         year = 1996,
        month = jul,
       volume = {281},
        pages = {716},
          doi = {10.1093/mnras/281.2.716},
archivePrefix = {arXiv},
       eprint = {astro-ph/9510147},
 primaryClass = {astro-ph},
       adsurl = {https://ui.adsabs.harvard.edu/abs/1996MNRAS.281..716C}
}

@ARTICLE{Lokas2006,
       author = {{{\L}okas}, E.~L. and {Prada}, F. and {Wojtak}, R. and {Moles}, M. and {Gottl{\"o}ber}, S.},
        title = "{The complex velocity distribution of galaxies in Abell 1689: implications for mass modelling}",
      journal = {\mnras},
         year = 2006,
        month = feb,
       volume = {366},
       number = {1},
        pages = {L26-L30},
          doi = {10.1111/j.1745-3933.2005.00125.x},
archivePrefix = {arXiv},
       eprint = {astro-ph/0507508},
 primaryClass = {astro-ph},
       adsurl = {https://ui.adsabs.harvard.edu/abs/2006MNRAS.366L..26L}
}

@ARTICLE{Biviano2006,
       author = {{Biviano}, A. and {Murante}, G. and {Borgani}, S. and {Diaferio}, A. and {Dolag}, K. and {Girardi}, M.},
        title = "{On the efficiency and reliability of cluster mass estimates based on member galaxies}",
      journal = {\aap},
         year = 2006,
        month = sep,
       volume = {456},
       number = {1},
        pages = {23-36},
          doi = {10.1051/0004-6361:20064918},
archivePrefix = {arXiv},
       eprint = {astro-ph/0605151},
 primaryClass = {astro-ph},
       adsurl = {https://ui.adsabs.harvard.edu/abs/2006A&A...456...23B}
}

@ARTICLE{Evrard2008,
       author = {{Evrard}, A.~E. and {Bialek}, J. and {Busha}, M. and {White}, M. and {Habib}, S. and {Heitmann}, K. and {Warren}, M. and {Rasia}, E. and {Tormen}, G. and {Moscardini}, L. and {Power}, C. and {Jenkins}, A.~R. and {Gao}, L. and {Frenk}, C.~S. and {Springel}, V. and {White}, S.~D.~M. and {Diemand}, J.},
        title = "{Virial Scaling of Massive Dark Matter Halos: Why Clusters Prefer a High Normalization Cosmology}",
      journal = {\apj},
         year = 2008,
        month = jan,
       volume = {672},
       number = {1},
        pages = {122-137},
          doi = {10.1086/521616},
archivePrefix = {arXiv},
       eprint = {astro-ph/0702241},
 primaryClass = {astro-ph},
       adsurl = {https://ui.adsabs.harvard.edu/abs/2008ApJ...672..122E}
}

@ARTICLE{SartorisDM,
       author = {{Sartoris}, Barbara and {Biviano}, Andrea and {Rosati}, Piero and {Borgani}, Stefano and {Umetsu}, Keiichi and {Bartelmann}, Matthias and {Girardi}, Marisa and {Grillo}, Claudio and {Lemze}, Doron and {Zitrin}, Adi and {Balestra}, Italo and {Mercurio}, Amata and {Nonino}, Mario and {Postman}, Marc and {Czakon}, Nicole and {Bradley}, Larry and {Broadhurst}, Tom and {Coe}, Dan and {Medezinski}, Elinor and {Melchior}, Peter and {Meneghetti}, Massimo and {Merten}, Julian and {Annunziatella}, Marianna and {Benitez}, Narciso and {Czoske}, Oliver and {Donahue}, Megan and {Ettori}, Stefano and {Ford}, Holland and {Fritz}, Alexander and {Kelson}, Dan and {Koekemoer}, Anton and {Kuchner}, Ulrike and {Lombardi}, Marco and {Maier}, Christian and {Moustakas}, Leonidas A. and {Munari}, Emiliano and {Presotto}, Valentina and {Scodeggio}, Marco and {Seitz}, Stella and {Tozzi}, Paolo and {Zheng}, Wei and {Ziegler}, Bodo},
        title = "{CLASH-VLT: Constraints on the Dark Matter Equation of State from Accurate Measurements of Galaxy Cluster Mass Profiles}",
      journal = {\apjl},
         year = 2014,
        month = mar,
       volume = {783},
       number = {1},
          eid = {L11},
        pages = {L11},
          doi = {10.1088/2041-8205/783/1/L11},
archivePrefix = {arXiv},
       eprint = {1401.5800},
 primaryClass = {astro-ph.CO},
       adsurl = {https://ui.adsabs.harvard.edu/abs/2014ApJ...783L..11S}
}

@ARTICLE{Pizzuti17,
       author = {{Pizzuti}, L. and {Sartoris}, B. and {Amendola}, L. and {Borgani}, S. and {Biviano}, A. and {Umetsu}, K. and {Mercurio}, A. and {Rosati}, P. and {Balestra}, I. and {Caminha}, G.~B. and {Girardi}, M. and {Grillo}, C. and {Nonino}, M.},
        title = "{CLASH-VLT: constraints on $f(R)$ gravity models with galaxy clusters using lensing and kinematic analyses}",
      journal = {\jcap},
         year = 2017,
        month = jul,
       volume = {2017},
       number = {7},
          eid = {023},
        pages = {023},
          doi = {10.1088/1475-7516/2017/07/023},
archivePrefix = {arXiv},
       eprint = {1705.05179},
 primaryClass = {astro-ph.CO},
       adsurl = {https://ui.adsabs.harvard.edu/abs/2017JCAP...07..023P}
}

@ARTICLE{Mamon2019,
       author = {{Mamon}, G.~A. and {Cava}, A. and {Biviano}, A. and {Moretti}, A. and {Poggianti}, B. and {Bettoni}, D.},
        title = "{Structural and dynamical modeling of WINGS clusters. II. The orbital anisotropies of elliptical, spiral, and lenticular galaxies}",
      journal = {\aap},
         year = 2019,
        month = nov,
       volume = {631},
          eid = {A131},
        pages = {A131},
          doi = {10.1051/0004-6361/201935081},
archivePrefix = {arXiv},
       eprint = {1901.06393},
 primaryClass = {astro-ph.GA},
       adsurl = {https://ui.adsabs.harvard.edu/abs/2019A&A...631A.131M}
}

@ARTICLE{Pedratti26,
       author = {{Pedratti}, S. and {Pizzuti}, L. and {Fossati}, M. and {Biviano}, A. and {Boselli}, A. and {Ragagnin}, A. and {Carlin}, A.},
        title = "{Novel insights into the Coma cluster kinematics with DESI: I. Linking mass profile, orbital anisotropy, and galaxy populations}",
      journal = {\aap},
         year = 2026,
        month = jun,
       volume = {710},
          eid = {A218},
        pages = {A218},
          doi = {10.1051/0004-6361/202659068},
       adsurl = {https://ui.adsabs.harvard.edu/abs/2026A&A...710A.218P}
}

@ARTICLE{Pizzuti26,
       author = {{Pizzuti}, L. and {Biviano}, A. and {Umetsu}, K. and {Agostoni}, E. and {Autorino}, A. and {Pombo}, A.~M. and {Mercurio}, A. and {D'Addona}, M.},
        title = "{CLASH-VLT: Constraining deviation from GR with the mass profiles of nine massive galaxy clusters}",
      journal = {\jcap},
         year = 2026,
        month = mar,
       volume = {2026},
       number = {3},
          eid = {022},
        pages = {022},
          doi = {10.1088/1475-7516/2026/03/022},
archivePrefix = {arXiv},
       eprint = {2509.16317},
 primaryClass = {astro-ph.CO},
       adsurl = {https://ui.adsabs.harvard.edu/abs/2026JCAP...03..022P}
}

@article{Burrage_2018,
   title={Tests of chameleon gravity},
   volume={21},
   ISSN={1433-8351},
   DOI={10.1007/s41114-018-0011-x},
   journal={Living Rev. Relativ.},
   publisher={Springer Science and Business Media LLC},
   author={Burrage, Clare and Sakstein, Jeremy},
   year={2018},
   month=mar }

@article{Brax_2021,
   title={Testing Screened Modified Gravity},
   volume={8},
   ISSN={2218-1997},
   DOI={10.3390/universe8010011},
   number={1},
   journal={Universe},
   publisher={MDPI AG},
   author={Brax, Philippe and Casas, Santiago and Desmond, Harry and Elder, Benjamin},
   year={2021},
   month=dec,
   pages={11}
}

@article{Schmidt_2010,
   title={Dynamical masses in modified gravity},
   volume={81},
   ISSN={1550-2368},
   DOI={10.1103/physrevd.81.103002},
   number={10},
   journal={\prd},
   publisher={American Physical Society (APS)},
   author={Schmidt, Fabian},
   year={2010},
   month=may }

@ARTICLE{Gronke2016,
       author = {{Gronke}, M. and {Hammami}, A. and {Mota}, D.~F. and {Winther}, H.~A.},
        title = "{Estimates of cluster masses in screened modified gravity}",
      journal = {\aap},
         year = 2016,
        month = nov,
       volume = {595},
          eid = {A78},
        pages = {A78},
          doi = {10.1051/0004-6361/201628644},
archivePrefix = {arXiv},
       eprint = {1609.02937},
 primaryClass = {astro-ph.CO},
       adsurl = {https://ui.adsabs.harvard.edu/abs/2016A&A...595A..78G}
}

@ARTICLE{Merafina14,
       author = {{Merafina}, M. and {Bisnovatyi-Kogan}, G.~S. and {Donnari}, M.},
        title = "{Galaxy clusters in presence of dark energy: a kinetic approach}",
      journal = {\aap},
         year = 2014,
        month = aug,
       volume = {568},
          eid = {A93},
        pages = {A93},
          doi = {10.1051/0004-6361/201424062},
archivePrefix = {arXiv},
       eprint = {1404.7744},
 primaryClass = {astro-ph.CO},
       adsurl = {https://ui.adsabs.harvard.edu/abs/2014A&A...568A..93M}
}

@ARTICLE{Munari2013,
       author = {{Munari}, E. and {Biviano}, A. and {Borgani}, S. and {Murante}, G. and {Fabjan}, D.},
        title = "{The relation between velocity dispersion and mass in simulated clusters of galaxies: dependence on the tracer and the baryonic physics}",
      journal = {\mnras},
         year = 2013,
        month = apr,
       volume = {430},
       number = {4},
        pages = {2638-2649},
          doi = {10.1093/mnras/stt049},
archivePrefix = {arXiv},
       eprint = {1301.1682},
 primaryClass = {astro-ph.CO},
       adsurl = {https://ui.adsabs.harvard.edu/abs/2013MNRAS.430.2638M}
}

@ARTICLE{Armitage2019,
       author = {{Armitage}, Thomas J. and {Kay}, Scott T. and {Barnes}, David J.},
        title = "{An application of machine learning techniques to galaxy cluster mass estimation using the MACSIS simulations}",
      journal = {\mnras},
         year = 2019,
        month = apr,
       volume = {484},
       number = {2},
        pages = {1526-1537},
          doi = {10.1093/mnras/stz039},
archivePrefix = {arXiv},
       eprint = {1810.08430},
 primaryClass = {astro-ph.CO},
       adsurl = {https://ui.adsabs.harvard.edu/abs/2019MNRAS.484.1526A}
}

@ARTICLE{Cerini2023,
       author = {{Cerini}, Giulia and {Cappelluti}, Nico and {Natarajan}, Priyamvada},
        title = "{New Metrics to Probe the Dynamical State of Galaxy Clusters}",
      journal = {\apj},
         year = 2023,
        month = mar,
       volume = {945},
       number = {2},
          eid = {152},
        pages = {152},
          doi = {10.3847/1538-4357/acbccb},
archivePrefix = {arXiv},
       eprint = {2209.06831},
 primaryClass = {astro-ph.CO},
       adsurl = {https://ui.adsabs.harvard.edu/abs/2023ApJ...945..152C}
}

@ARTICLE{Benavides23,
       author = {{Benavides}, Jos{\'e} A. and {Biviano}, Andrea and {Abadi}, Mario G.},
        title = "{DS+: A method for the identification of cluster substructures}",
      journal = {\aap},
         year = 2023,
        month = jan,
       volume = {669},
          eid = {A147},
        pages = {A147},
          doi = {10.1051/0004-6361/202245422},
archivePrefix = {arXiv},
       eprint = {2212.00040},
 primaryClass = {astro-ph.GA},
       adsurl = {https://ui.adsabs.harvard.edu/abs/2023A&A...669A.147B}
}

@ARTICLE{Zhang22,
       author = {{Zhang}, Bowei and {Cui}, Weiguang and {Wang}, Yuhuan and {Dave}, Romeel and {De Petris}, Marco},
        title = "{The Three Hundred: cluster dynamical states and relaxation period}",
      journal = {\mnras},
         year = 2022,
        month = oct,
       volume = {516},
       number = {1},
        pages = {26-38},
          doi = {10.1093/mnras/stac2171},
archivePrefix = {arXiv},
       eprint = {2112.01909},
 primaryClass = {astro-ph.CO},
       adsurl = {https://ui.adsabs.harvard.edu/abs/2022MNRAS.516...26Z}
}

@ARTICLE{Ferragamo2020,
       author = {{Ferragamo}, A. and {Rubi{\~n}o-Mart{\'\i}n}, J.~A. and {Betancort-Rijo}, J. and {Munari}, E. and {Sartoris}, B. and {Barrena}, R.},
        title = "{Biases in galaxy cluster velocity dispersion and mass estimates in the small N$_{gal}$ regime}",
      journal = {\aap},
         year = 2020,
        month = sep,
       volume = {641},
          eid = {A41},
        pages = {A41},
          doi = {10.1051/0004-6361/201834837},
archivePrefix = {arXiv},
       eprint = {2006.05949},
 primaryClass = {astro-ph.CO},
       adsurl = {https://ui.adsabs.harvard.edu/abs/2020A&A...641A..41F}
}

@ARTICLE{Knebe2000,
       author = {{Knebe}, Alexander and {M{\"u}ller}, Volker},
        title = "{Quantifying substructure in galaxy clusters}",
      journal = {\aap},
         year = 2000,
        month = feb,
       volume = {354},
        pages = {761-766},
          doi = {10.48550/arXiv.astro-ph/9912534},
archivePrefix = {arXiv},
       eprint = {astro-ph/9912534},
 primaryClass = {astro-ph},
       adsurl = {https://ui.adsabs.harvard.edu/abs/2000A&A...354..761K}
}

@ARTICLE{vanHaarlem1993,
       author = {{van Haarlem}, Michiel and {van de Weygaert}, Rien},
        title = "{Velocity Fields and Alignments of Clusters in Gravitational Instability Scenarios}",
      journal = {\apj},
         year = 1993,
        month = dec,
       volume = {418},
        pages = {544},
          doi = {10.1086/173416},
       adsurl = {https://ui.adsabs.harvard.edu/abs/1993ApJ...418..544V}
}

@ARTICLE{White1976,
       author = {{White}, S.~D.~M.},
        title = "{The dynamics of rich clusters of galaxies.}",
      journal = {\mnras},
         year = 1976,
        month = dec,
       volume = {177},
        pages = {717-733},
          doi = {10.1093/mnras/177.3.717},
       adsurl = {https://ui.adsabs.harvard.edu/abs/1976MNRAS.177..717W}
}

@article{Girardi_1997,
doi = {10.1086/304113},
year = {1997},
month = {jun},
publisher = {},
volume = {482},
number = {1},
pages = {41},
author = {Girardi, M. and Escalera, E. and Fadda, D. and Giuricin, G. and Mardirossian, F. and Mezzetti, M.},
title = {Optical Substructures in 48 Galaxy Clusters: New Insights from a Multiscale Analysis},
journal = {\apj}
}

@ARTICLE{Hou2009,
       author = {{Hou}, Annie and {Parker}, Laura C. and {Harris}, William E. and {Wilman}, David J.},
        title = "{Statistical Tools for Classifying Galaxy Group Dynamics}",
      journal = {\apj},
         year = 2009,
        month = sep,
       volume = {702},
       number = {2},
        pages = {1199-1210},
          doi = {10.1088/0004-637X/702/2/1199},
archivePrefix = {arXiv},
       eprint = {0908.0938},
 primaryClass = {astro-ph.GA},
       adsurl = {https://ui.adsabs.harvard.edu/abs/2009ApJ...702.1199H}
}

@ARTICLE{Girardi2024,
       author = {{Girardi}, M. and {Boschin}, W. and {Mercurio}, A. and {Nocerino}, N. and {Nonino}, M. and {Rosati}, P. and {Biviano}, A. and {Demarco}, R. and {Grillo}, C. and {Sartoris}, B. and {Tozzi}, P. and {Vanzella}, E.},
        title = "{CLASH-VLT: Galaxy cluster MACS J0329--0211 and its surroundings using galaxies as kinematic tracers}",
      journal = {\aap},
         year = 2024,
        month = dec,
       volume = {692},
          eid = {A175},
        pages = {A175},
          doi = {10.1051/0004-6361/202451286},
archivePrefix = {arXiv},
       eprint = {2410.20133},
 primaryClass = {astro-ph.CO},
       adsurl = {https://ui.adsabs.harvard.edu/abs/2024A&A...692A.175G}
}

@ARTICLE{Boschin2006,
       author = {{Boschin}, W. and {Girardi}, M. and {Spolaor}, M. and {Barrena}, R.},
        title = "{Internal dynamics of the radio halo cluster Abell 2744}",
      journal = {\aap},
         year = 2006,
        month = apr,
       volume = {449},
       number = {2},
        pages = {461-474},
          doi = {10.1051/0004-6361:20054408},
archivePrefix = {arXiv},
       eprint = {astro-ph/0603192},
 primaryClass = {astro-ph},
       adsurl = {https://ui.adsabs.harvard.edu/abs/2006A&A...449..461B}
}

@ARTICLE{Yu_2015,
       author = {{Yu}, Heng and {Serra}, Ana Laura and {Diaferio}, Antonaldo and {Baldi}, Marco},
        title = "{Identification of Galaxy Cluster Substructures with the Caustic Method}",
      journal = {\apj},
         year = 2015,
        month = sep,
       volume = {810},
       number = {1},
          eid = {37},
        pages = {37},
          doi = {10.1088/0004-637X/810/1/37},
archivePrefix = {arXiv},
       eprint = {1503.08823},
 primaryClass = {astro-ph.CO},
       adsurl = {https://ui.adsabs.harvard.edu/abs/2015ApJ...810...37Y}
}

@article{Yu2018,
   title={Blooming Trees: Substructures and Surrounding Groups of Galaxy Clusters},
   volume={860},
   ISSN={1538-4357},
   DOI={10.3847/1538-4357/aac263},
   number={2},
   journal={\apj},
   publisher={American Astronomical Society},
   author={Yu, Heng and Diaferio, Antonaldo and Serra, Ana Laura and Baldi, Marco},
   year={2018},
   month=jun,
   pages={118}
}

@ARTICLE{Cava_2017,
       author = {{Cava}, A. and {Biviano}, A. and {Mamon}, G.~A. and {Varela}, J. and {Bettoni}, D. and {D'Onofrio}, M. and {Fasano}, G. and {Fritz}, J. and {Moles}, M. and {Moretti}, A. and {Poggianti}, B.},
        title = "{Structural and dynamical modeling of WINGS clusters. I. The distribution of cluster galaxies of different morphological classes within regular and irregular clusters}",
      journal = {\aap},
         year = 2017,
        month = oct,
       volume = {606},
          eid = {A108},
        pages = {A108},
          doi = {10.1051/0004-6361/201730785},
archivePrefix = {arXiv},
       eprint = {1708.08541},
 primaryClass = {astro-ph.GA},
       adsurl = {https://ui.adsabs.harvard.edu/abs/2017A&A...606A.108C}
}

@ARTICLE{Girardi_2015,
       author = {{Girardi}, M. and {Mercurio}, A. and {Balestra}, I. and {Nonino}, M. and {Biviano}, A. and {Grillo}, C. and {Rosati}, P. and {Annunziatella}, M. and {Demarco}, R. and {Fritz}, A. and {Gobat}, R. and {Lemze}, D. and {Presotto}, V. and {Scodeggio}, M. and {Tozzi}, P. and {Bartosch Caminha}, G. and {Brescia}, M. and {Coe}, D. and {Kelson}, D. and {Koekemoer}, A. and {Lombardi}, M. and {Medezinski}, E. and {Postman}, M. and {Sartoris}, B. and {Umetsu}, K. and {Zitrin}, A. and {Boschin}, W. and {Czoske}, O. and {De Lucia}, G. and {Kuchner}, U. and {Maier}, C. and {Meneghetti}, M. and {Monaco}, P. and {Monna}, A. and {Munari}, E. and {Seitz}, S. and {Verdugo}, M. and {Ziegler}, B.},
        title = "{CLASH-VLT: Substructure in the galaxy cluster MACS J1206.2-0847 from kinematics of galaxy populations}",
      journal = {\aap},
         year = 2015,
        month = jul,
       volume = {579},
          eid = {A4},
        pages = {A4},
          doi = {10.1051/0004-6361/201425599},
archivePrefix = {arXiv},
       eprint = {1503.05607},
 primaryClass = {astro-ph.CO},
       adsurl = {https://ui.adsabs.harvard.edu/abs/2015A&A...579A...4G}
}

@ARTICLE{Bird1995,
       author = {{Bird}, Christina M.},
        title = "{The Effects of Substructure on Galaxy Cluster Mass Determinations}",
      journal = {\apjl},
         year = 1995,
        month = jun,
       volume = {445},
        pages = {L81},
          doi = {10.1086/187895},
archivePrefix = {arXiv},
       eprint = {astro-ph/9503038},
 primaryClass = {astro-ph},
       adsurl = {https://ui.adsabs.harvard.edu/abs/1995ApJ...445L..81B}
}

@ARTICLE{Bird94,
       author = {{Bird}, Christina M.},
        title = "{Substructure in Clusters and Central Galaxy Peculiar Velocities}",
      journal = {\aj},
         year = 1994,
        month = may,
       volume = {107},
        pages = {1637},
          doi = {10.1086/116973},
archivePrefix = {arXiv},
       eprint = {astro-ph/9402058},
 primaryClass = {astro-ph},
       adsurl = {https://ui.adsabs.harvard.edu/abs/1994AJ....107.1637B}
}

@ARTICLE{denHartogKatgert1996,
       author = {{den Hartog}, R. and {Katgert}, P.},
        title = "{On the dynamics of the cores of galaxy clusters.}",
      journal = {\mnras},
         year = 1996,
        month = mar,
       volume = {279},
       number = {2},
        pages = {349-388},
          doi = {10.1093/mnras/279.2.349},
       adsurl = {https://ui.adsabs.harvard.edu/abs/1996MNRAS.279..349D}
}

@ARTICLE{Wojtak2009,
       author = {{Wojtak}, Rados{\l}aw and {{\L}okas}, Ewa L. and {Mamon}, Gary A. and {Gottl{\"o}ber}, Stefan},
        title = "{The mass and anisotropy profiles of galaxy clusters from the projected phase-space density: testing the method on simulated data}",
      journal = {\mnras},
         year = 2009,
        month = oct,
       volume = {399},
       number = {2},
        pages = {812-821},
          doi = {10.1111/j.1365-2966.2009.15312.x},
archivePrefix = {arXiv},
       eprint = {0906.5071},
 primaryClass = {astro-ph.CO},
       adsurl = {https://ui.adsabs.harvard.edu/abs/2009MNRAS.399..812W}
}

@article{Molnar_2003,
   title={Determining Tangential Peculiar Velocities of Clusters of Galaxies Using Gravitational Lensing},
   volume={586},
   ISSN={1538-4357},
   DOI={10.1086/346071},
   number={2},
   journal={\apj},
   publisher={American Astronomical Society},
   author={Molnar, Sandor M. and Birkinshaw, Mark},
   year={2003},
   month=apr,
   pages={731–734}
}

@BOOK{Jeans19,
       author = {{Jeans}, James Hopwood},
        title = "{Problems of cosmogony and stellar dynamics}",
         year = 1919,
    publisher = {Cambridge University Press},
       adsurl = {https://ui.adsabs.harvard.edu/abs/1919pcsd.book.....J}
}

@ARTICLE{Pinkney1996,
       author = {{Pinkney}, Jason and {Roettiger}, Kurt and {Burns}, Jack O. and {Bird}, Christina M.},
        title = "{Evaluation of Statistical Tests for Substructure in Clusters of Galaxies}",
      journal = {\apjs},
         year = 1996,
        month = may,
       volume = {104},
        pages = {1},
          doi = {10.1086/192290},
       adsurl = {https://ui.adsabs.harvard.edu/abs/1996ApJS..104....1P}
}

@ARTICLE{Abdullah_2018,
       author = {{Abdullah}, Mohamed H. and {Wilson}, Gillian and {Klypin}, Anatoly},
        title = "{GalWeight: A New and Effective Weighting Technique for Determining Galaxy Cluster and Group Membership}",
      journal = {\apj},
         year = 2018,
        month = jul,
       volume = {861},
       number = {1},
          eid = {22},
        pages = {22},
          doi = {10.3847/1538-4357/aac5db},
archivePrefix = {arXiv},
       eprint = {1805.06479},
 primaryClass = {astro-ph.CO},
       adsurl = {https://ui.adsabs.harvard.edu/abs/2018ApJ...861...22A}
}

@ARTICLE{Costa2025,
       author = {{Costa}, Alisson P. and {Ribeiro}, Andr{\'e} L.~B. and {de Morais Neto}, Flavio R. and {dos Santos Junior}, Juarez},
        title = "{Unveiling the Dynamics in Galaxy Clusters: The Hidden Role of Low-Luminosity Galaxies in Coma}",
      journal = {Universe},
         year = 2025,
        month = mar,
       volume = {11},
       number = {3},
          eid = {82},
        pages = {82},
          doi = {10.3390/universe11030082},
archivePrefix = {arXiv},
       eprint = {2503.03667},
 primaryClass = {astro-ph.GA},
       adsurl = {https://ui.adsabs.harvard.edu/abs/2025Univ...11...82C}
}

@ARTICLE{Wojtak2010,
       author = {{Wojtak}, Rados{\l}aw and {{\L}okas}, Ewa L.},
        title = "{Mass profiles and galaxy orbits in nearby galaxy clusters from the analysis of the projected phase space}",
      journal = {\mnras},
         year = 2010,
        month = nov,
       volume = {408},
       number = {4},
        pages = {2442-2456},
          doi = {10.1111/j.1365-2966.2010.17297.x},
archivePrefix = {arXiv},
       eprint = {1004.3771},
 primaryClass = {astro-ph.CO},
       adsurl = {https://ui.adsabs.harvard.edu/abs/2010MNRAS.408.2442W}
}

@ARTICLE{Fadda1996,
       author = {{Fadda}, D. and {Girardi}, M. and {Giuricin}, G. and {Mardirossian}, F. and {Mezzetti}, M.},
        title = "{The Observational Distribution of Internal Velocity Dispersions in Nearby Galaxy Clusters}",
      journal = {\apj},
         year = 1996,
        month = dec,
       volume = {473},
        pages = {670},
          doi = {10.1086/178180},
archivePrefix = {arXiv},
       eprint = {astro-ph/9606098},
 primaryClass = {astro-ph},
       adsurl = {https://ui.adsabs.harvard.edu/abs/1996ApJ...473..670F}
}

@ARTICLE{Girardi1993,
       author = {{Girardi}, M. and {Biviano}, A. and {Giuricin}, G. and {Mardirossian}, F. and {Mezzetti}, M.},
        title = "{Velocity Dispersions in Galaxy Clusters}",
      journal = {\apj},
         year = 1993,
        month = feb,
       volume = {404},
        pages = {38},
          doi = {10.1086/172256},
       adsurl = {https://ui.adsabs.harvard.edu/abs/1993ApJ...404...38G}
}

@ARTICLE{Wojtak2007,
       author = {{Wojtak}, R. and {{\L}okas}, E.~L. and {Mamon}, G.~A. and {Gottl{\"o}ber}, S. and {Prada}, F. and {Moles}, M.},
        title = "{Interloper treatment in dynamical modelling of galaxy clusters}",
      journal = {\aap},
         year = 2007,
        month = may,
       volume = {466},
       number = {2},
        pages = {437-449},
          doi = {10.1051/0004-6361:20066813},
archivePrefix = {arXiv},
       eprint = {astro-ph/0606579},
 primaryClass = {astro-ph},
       adsurl = {https://ui.adsabs.harvard.edu/abs/2007A&A...466..437W}
}

@ARTICLE{Mamon2010,
       author = {{Mamon}, G.~A. and {Biviano}, A. and {Murante}, G.},
        title = "{The universal distribution of halo interlopers in projected phase space. Bias in galaxy cluster concentration and velocity anisotropy?}",
      journal = {\aap},
         year = 2010,
        month = sep,
       volume = {520},
          eid = {A30},
        pages = {A30},
          doi = {10.1051/0004-6361/200913948},
archivePrefix = {arXiv},
       eprint = {1003.0033},
 primaryClass = {astro-ph.CO},
       adsurl = {https://ui.adsabs.harvard.edu/abs/2010A&A...520A..30M}
}

@ARTICLE{Navarro1996,
       author = {{Navarro}, Julio F. and {Frenk}, Carlos S. and {White}, Simon D.~M.},
        title = "{The Structure of Cold Dark Matter Halos}",
      journal = {\apj},
         year = 1996,
        month = may,
       volume = {462},
        pages = {563},
          doi = {10.1086/177173},
archivePrefix = {arXiv},
       eprint = {astro-ph/9508025},
 primaryClass = {astro-ph},
       adsurl = {https://ui.adsabs.harvard.edu/abs/1996ApJ...462..563N}
}

@article{Ghigna1998,
    author = {Ghigna, Sebastiano and Moore, Ben and Governato, Fabio and Lake, George and Quinn, Thomas and Stadel, Joachim},
    title = {Dark matter haloes within clusters},
    journal = {\mnras},
    volume = {300},
    number = {1},
    pages = {146-162},
    year = {1998},
    month = {10},
    issn = {0035-8711},
    doi = {10.1046/j.1365-8711.1998.01918.x}
}

@ARTICLE{Diaferio1999,
       author = {{Diaferio}, Antonaldo},
        title = "{Mass estimation in the outer regions of galaxy clusters}",
      journal = {\mnras},
         year = 1999,
        month = nov,
       volume = {309},
       number = {3},
        pages = {610-622},
          doi = {10.1046/j.1365-8711.1999.02864.x},
archivePrefix = {arXiv},
       eprint = {astro-ph/9906331},
 primaryClass = {astro-ph},
       adsurl = {https://ui.adsabs.harvard.edu/abs/1999MNRAS.309..610D}
}

@ARTICLE{LokasMamon2003,
       author = {{{\L}okas}, Ewa L. and {Mamon}, Gary A.},
        title = "{Dark matter distribution in the Coma cluster from galaxy kinematics: breaking the mass-anisotropy degeneracy}",
      journal = {\mnras},
         year = 2003,
        month = aug,
       volume = {343},
       number = {2},
        pages = {401-412},
          doi = {10.1046/j.1365-8711.2003.06684.x},
archivePrefix = {arXiv},
       eprint = {astro-ph/0302461},
 primaryClass = {astro-ph},
       adsurl = {https://ui.adsabs.harvard.edu/abs/2003MNRAS.343..401L}
}

@ARTICLE{cakr25,
       author = {{{\c{C}}ak{\i}r}, O{\u{g}}uzhan and {Owers}, Matt and {Kimmig}, Lucas and {Nulsen}, Paul and {Pak}, Mina and {Quattropani}, Gabriella and {Couch}, Warrick},
        title = "{The impact of cluster mergers on galaxy properties}",
      journal = {\pasa},
         year = 2025,
        month = nov,
       volume = {42},
          eid = {e163},
        pages = {e163},
          doi = {10.1017/pasa.2025.10121},
archivePrefix = {arXiv},
       eprint = {2506.04799},
 primaryClass = {astro-ph.GA},
       adsurl = {https://ui.adsabs.harvard.edu/abs/2025PASA...42..163C}
}

@ARTICLE{Pizzuti22,
       author = {{Pizzuti}, Lorenzo and {Saltas}, Ippocratis D. and {Umetsu}, Keiichi and {Sartoris}, Barbara},
        title = "{Probing vainsthein-screening gravity with galaxy clusters using internal kinematics and strong and weak lensing}",
      journal = {\mnras},
         year = 2022,
        month = may,
       volume = {512},
       number = {3},
        pages = {4280-4290},
          doi = {10.1093/mnras/stac746},
archivePrefix = {arXiv},
       eprint = {2112.12139},
 primaryClass = {astro-ph.CO},
       adsurl = {https://ui.adsabs.harvard.edu/abs/2022MNRAS.512.4280P}
}

@article{Agnello2014,
    author = {Agnello, A. and Evans, N. W. and Romanowsky, A. J.},
    title = "{Dynamical models of elliptical galaxies – I. Simple methods}",
    journal = {\mnras},
    volume = {442},
    number = {4},
    pages = {3284-3298},
    year = {2014},
    month = {08},
    issn = {0035-8711},
    doi = {10.1093/mnras/stu959}
}

@article{Henault2019,
    author = {Hénault-Brunet, V and Gieles, M and Sollima, A and Watkins, L L and Zocchi, A and Claydon, I and Pancino, E and Baumgardt, H},
    title = "{Mass modelling globular clusters in the GAIA era: a method comparison using mock data from an N-body simulation of M4}",
    journal = {\mnras},
    volume = {483},
    number = {1},
    pages = {1400-1425},
    year = {2019},
    month = {02},
    issn = {0035-8711},
    doi = {10.1093/mnras/sty3187}
}

@ARTICLE{Barrena24,
       author = {{Barrena}, R. and {Pizzuti}, L. and {Chon}, G. and {B{\"o}hringer}, H.},
        title = "{Unveiling the shape: A multi-wavelength analysis of the galaxy clusters Abell 76 and Abell 1307}",
      journal = {\aap},
         year = 2024,
        month = nov,
       volume = {691},
          eid = {A135},
        pages = {A135},
          doi = {10.1051/0004-6361/202451144},
archivePrefix = {arXiv},
       eprint = {2409.16981},
 primaryClass = {astro-ph.CO},
       adsurl = {https://ui.adsabs.harvard.edu/abs/2024A&A...691A.135B}
}

@phdthesis{PizzutiThesis,
    author = "Pizzuti, Lorenzo",
    title = "{Nature of gravity from the mass profiles of galaxy clusters}",
    school = "Trieste University",
    year = "2019",
    url = {https://hdl.handle.net/11368/2936435}
}

@article{Bialas_2015,
   title={On the occurrence of galaxy harassment},
   volume={576},
   ISSN={1432-0746},
   DOI={10.1051/0004-6361/201425235},
   journal={\aap},
   publisher={EDP Sciences},
   author={Bialas, D. and Lisker, T. and Olczak, C. and Spurzem, R. and Kotulla, R.},
   year={2015},
   month=apr,
   pages={A103}
}

@ARTICLE{Lokas2020,
       author = {{{\L}okas}, Ewa L.},
        title = "{Tidal evolution of galaxies in the most massive cluster of IllustrisTNG-100}",
      journal = {\aap},
         year = 2020,
        month = jun,
       volume = {638},
          eid = {A133},
        pages = {A133},
          doi = {10.1051/0004-6361/202037643},
archivePrefix = {arXiv},
       eprint = {2002.00610},
 primaryClass = {astro-ph.GA},
       adsurl = {https://ui.adsabs.harvard.edu/abs/2020A&A...638A.133L}
}

@ARTICLE{Biviano2009,
       author = {{Biviano}, A. and {Poggianti}, B.~M.},
        title = "{The orbital velocity anisotropy of cluster galaxies: evolution}",
      journal = {\aap},
         year = 2009,
        month = jul,
       volume = {501},
       number = {2},
        pages = {419-427},
          doi = {10.1051/0004-6361/200911757},
archivePrefix = {arXiv},
       eprint = {0905.2853},
 primaryClass = {astro-ph.CO},
       adsurl = {https://ui.adsabs.harvard.edu/abs/2009A&A...501..419B}
}

@article{Boselli_2022,
   title={Ram pressure stripping in high-density environments},
   volume={30},
   ISSN={1432-0754},
   DOI={10.1007/s00159-022-00140-3},
   number={1},
   journal={\aapr},
   publisher={Springer Science and Business Media LLC},
   author={Boselli, Alessandro and Fossati, Matteo and Sun, Ming},
   year={2022},
   month=may }

@ARTICLE{Xie_2025,
       author = {{Xie}, Lizhi and {De Lucia}, Gabriella and {Fossati}, Matteo and {Fontanot}, Fabio and {Hirschmann}, Michaela},
        title = "{The impact of ram pressure on cluster galaxies, insights from GAEA and TNG}",
      journal = {\aap},
         year = 2025,
        month = jun,
       volume = {698},
          eid = {A73},
        pages = {A73},
          doi = {10.1051/0004-6361/202553915},
archivePrefix = {arXiv},
       eprint = {2504.12863},
 primaryClass = {astro-ph.GA},
       adsurl = {https://ui.adsabs.harvard.edu/abs/2025A&A...698A..73X}
}

@article{Boselli_2006,
   title={Environmental Effects on Late‐Type Galaxies in Nearby Clusters},
   volume={118},
   ISSN={1538-3873},
   DOI={10.1086/500691},
   number={842},
   journal={\pasp},
   publisher={IOP Publishing},
   author={Boselli, Alessandro and Gavazzi, Giuseppe},
   year={2006},
   month=apr,
   pages={517–559}
}

@ARTICLE{Muriel14,
       author = {{Muriel}, H. and {Coenda}, V.},
        title = "{Galaxy properties in clusters. II. Backsplash galaxies}",
      journal = {\aap},
         year = 2014,
        month = apr,
       volume = {564},
          eid = {A85},
        pages = {A85},
          doi = {10.1051/0004-6361/201322033},
archivePrefix = {arXiv},
       eprint = {1402.3594},
 primaryClass = {astro-ph.CO},
       adsurl = {https://ui.adsabs.harvard.edu/abs/2014A&A...564A..85M}
}

@ARTICLE{Wetzel2013,
       author = {{Wetzel}, Andrew R. and {Tinker}, Jeremy L. and {Conroy}, Charlie and {van den Bosch}, Frank C.},
        title = "{Galaxy evolution in groups and clusters: satellite star formation histories and quenching time-scales in a hierarchical Universe}",
      journal = {\mnras},
         year = 2013,
        month = jun,
       volume = {432},
       number = {1},
        pages = {336-358},
          doi = {10.1093/mnras/stt469},
archivePrefix = {arXiv},
       eprint = {1206.3571},
 primaryClass = {astro-ph.CO},
       adsurl = {https://ui.adsabs.harvard.edu/abs/2013MNRAS.432..336W}
}

@article{Morandi_2016,
   title={Probing dark energy via galaxy cluster outskirts},
   volume={457},
   ISSN={1365-2966},
   DOI={10.1093/mnras/stw143},
   number={3},
   journal={\mnras},
   publisher={Oxford University Press (OUP)},
   author={Morandi, Andrea and Sun, Ming},
   year={2016},
   month=feb,
   pages={3266–3284}
}

@article{Dai_2018,
   title={Probing Dark Matter Subhalos in Galaxy Clusters Using Highly Magnified Stars},
   volume={867},
   ISSN={1538-4357},
   DOI={10.3847/1538-4357/aae478},
   number={1},
   journal={\apj},
   publisher={American Astronomical Society},
   author={Dai, Liang and Venumadhav, Tejaswi and Kaurov, Alexander A. and Miralda-Escud, Jordi},
   year={2018},
   month=oct,
   pages={24}
}

@ARTICLE{Zwicky1933,
       author = {{Zwicky}, F.},
        title = "{Die Rotverschiebung von extragalaktischen Nebeln}",
      journal = {Helvetica Physica Acta},
         year = 1933,
        month = jan,
       volume = {6},
        pages = {110-127},
       adsurl = {https://ui.adsabs.harvard.edu/abs/1933AcHPh...6..110Z}
}

@ARTICLE{Bahcall81,
       author = {{Bahcall}, N.~A.},
        title = "{The relation between velocity dispersion and central galaxy density in clusters of galaxies.}",
      journal = {\apj},
         year = 1981,
        month = aug,
       volume = {247},
        pages = {787-791},
          doi = {10.1086/159090},
       adsurl = {https://ui.adsabs.harvard.edu/abs/1981ApJ...247..787B}
}

@ARTICLE{Hubble31,
       author = {{Hubble}, Edwin and {Humason}, Milton L.},
        title = "{The Velocity-Distance Relation among Extra-Galactic Nebulae}",
      journal = {\apj},
         year = 1931,
        month = jul,
       volume = {74},
        pages = {43},
          doi = {10.1086/143323},
       adsurl = {https://ui.adsabs.harvard.edu/abs/1931ApJ....74...43H}
}

@ARTICLE{Zwicky1937,
       author = {{Zwicky}, F.},
        title = "{On the Masses of Nebulae and of Clusters of Nebulae}",
      journal = {\apj},
         year = 1937,
        month = oct,
       volume = {86},
        pages = {217},
          doi = {10.1086/143864},
       adsurl = {https://ui.adsabs.harvard.edu/abs/1937ApJ....86..217Z}
}

@INPROCEEDINGS{Biviano2000,
       author = {{Biviano}, Andrea},
        title = "{From Messier to Abell: 200 Years of Science with Galaxy Clusters}",
    booktitle = {Constructing the Universe with Clusters of Galaxies},
         year = 2000,
       editor = {{Durret}, Florence and {Gerbal}, Daniel},
        month = jan,
          eid = {1},
        pages = {1},
          doi = {10.48550/arXiv.astro-ph/0010409},
archivePrefix = {arXiv},
       eprint = {astro-ph/0010409},
 primaryClass = {astro-ph},
       adsurl = {https://ui.adsabs.harvard.edu/abs/2000cucg.confE...1B}
}

@article{van_der_Marel_2000,
   title="{The Velocity and Mass Distribution of Clusters of Galaxies from the CNOC1 Cluster Redshift Survey}",
   volume={119},
   ISSN={0004-6256},
   DOI={10.1086/301351},
   number={5},
   journal={\aj},
   publisher={American Astronomical Society},
   author={van der Marel, Roeland P. and Magorrian, John and Carlberg, Ray G. and Yee, H. K. C. and Ellingson, E.},
   year={2000},
   month=may,
   pages={2038–2052}
}

@ARTICLE{Cuciti2023,
       author = {{Cuciti}, V. and {Cassano}, R. and {Sereno}, M. and {Brunetti}, G. and {Botteon}, A. and {Shimwell}, T.~W. and {Bruno}, L. and {Gastaldello}, F. and {Rossetti}, M. and {Zhang}, X. and {Simionescu}, A. and {Br{\"u}ggen}, M. and {van Weeren}, R.~J. and {Jones}, A. and {Akamatsu}, H. and {Bonafede}, A. and {De Gasperin}, F. and {Di Gennaro}, G. and {Pasini}, T. and {R{\"o}ttgering}, H.~J.~A.},
        title = "{The Planck clusters in the LOFAR sky. V. LoTSS-DR2: Mass-radio halo power correlation at low frequency}",
      journal = {\aap},
         year = 2023,
        month = dec,
       volume = {680},
          eid = {A30},
        pages = {A30},
          doi = {10.1051/0004-6361/202346755},
archivePrefix = {arXiv},
       eprint = {2305.04564},
 primaryClass = {astro-ph.CO},
       adsurl = {https://ui.adsabs.harvard.edu/abs/2023A&A...680A..30C}
}

@ARTICLE{Grandis2021,
       author = {{Grandis}, Sebastian and {Bocquet}, Sebastian and {Mohr}, Joseph J. and {Klein}, Matthias and {Dolag}, Klaus},
        title = "{Calibration of bias and scatter involved in cluster mass measurements using optical weak gravitational lensing}",
      journal = {\mnras},
         year = 2021,
        month = nov,
       volume = {507},
       number = {4},
        pages = {5671-5689},
          doi = {10.1093/mnras/stab2414},
archivePrefix = {arXiv},
       eprint = {2103.16212},
 primaryClass = {astro-ph.CO},
       adsurl = {https://ui.adsabs.harvard.edu/abs/2021MNRAS.507.5671G}
}

@ARTICLE{Umetsu20,
       author = {{Umetsu}, Keiichi},
        title = "{Cluster-galaxy weak lensing}",
      journal = {\aapr},
         year = 2020,
        month = dec,
       volume = {28},
       number = {1},
          eid = {7},
        pages = {7},
          doi = {10.1007/s00159-020-00129-w},
archivePrefix = {arXiv},
       eprint = {2007.00506},
 primaryClass = {astro-ph.CO},
       adsurl = {https://ui.adsabs.harvard.edu/abs/2020A&ARv..28....7U}
}

@ARTICLE{Giocoli25,
       author = {{Giocoli}, C. and {Despali}, G. and {Meneghetti}, M. and {Rasia}, E. and {Moscardini}, L. and {Borgani}, S. and {Lesci}, G.~F. and {Marulli}, F. and {Cui}, W. and {Yepes}, G.},
        title = "{The Three Hundred project hydrodynamical simulations: Hydrodynamical weak-lensing cluster mass biases and richnesses using different hydro models}",
      journal = {\aap},
         year = 2025,
        month = may,
       volume = {697},
          eid = {A184},
        pages = {A184},
          doi = {10.1051/0004-6361/202553871},
archivePrefix = {arXiv},
       eprint = {2501.14019},
 primaryClass = {astro-ph.CO},
       adsurl = {https://ui.adsabs.harvard.edu/abs/2025A&A...697A.184G}
}

@ARTICLE{Hoekstra2013,
       author = {{Hoekstra}, Henk and {Bartelmann}, Matthias and {Dahle}, H{\r{a}}kon and {Israel}, Holger and {Limousin}, Marceau and {Meneghetti}, Massimo},
        title = "{Masses of Galaxy Clusters from Gravitational Lensing}",
      journal = {\spsr},
         year = 2013,
        month = aug,
       volume = {177},
       number = {1-4},
        pages = {75-118},
          doi = {10.1007/s11214-013-9978-5},
archivePrefix = {arXiv},
       eprint = {1303.3274},
 primaryClass = {astro-ph.CO},
       adsurl = {https://ui.adsabs.harvard.edu/abs/2013SSRv..177...75H}
}

@ARTICLE{Simet2017,
       author = {{Simet}, Melanie and {Battaglia}, Nicholas and {Mandelbaum}, Rachel and {Seljak}, Uro{\v{s}}},
        title = "{Weak lensing calibration of mass bias in the REFLEX+BCS X-ray galaxy cluster catalogue}",
      journal = {\mnras},
         year = 2017,
        month = apr,
       volume = {466},
       number = {3},
        pages = {3663-3673},
          doi = {10.1093/mnras/stw3322},
archivePrefix = {arXiv},
       eprint = {1502.01024},
 primaryClass = {astro-ph.CO},
       adsurl = {https://ui.adsabs.harvard.edu/abs/2017MNRAS.466.3663S}
}

@ARTICLE{McClintock2019,
       author = {{McClintock}, T. and {Varga}, T.~N. and {Gruen}, D. and {Rozo}, E. and {Rykoff}, E.~S. and {Shin}, T. and {Melchior}, P. and {DeRose}, J. and {Seitz}, S. and {Dietrich}, J.~P. and {Sheldon}, E. and {Zhang}, Y. and {von der Linden}, A. and {Jeltema}, T. and {Mantz}, A.~B. and {Romer}, A.~K. and {Allen}, S. and {Becker}, M.~R. and {Bermeo}, A. and {Bhargava}, S. and {Costanzi}, M. and {Everett}, S. and {Farahi}, A. and {Hamaus}, N. and {Hartley}, W.~G. and {Hollowood}, D.~L. and {Hoyle}, B. and {Israel}, H. and {Li}, P. and {MacCrann}, N. and {Morris}, G. and {Palmese}, A. and {Plazas}, A.~A. and {Pollina}, G. and {Rau}, M.~M. and {Simet}, M. and {Soares-Santos}, M. and {Troxel}, M.~A. and {Vergara Cervantes}, C. and {Wechsler}, R.~H. and {Zuntz}, J. and {Abbott}, T.~M.~C. and {Abdalla}, F.~B. and {Allam}, S. and {Annis}, J. and {Avila}, S. and {Bridle}, S.~L. and {Brooks}, D. and {Burke}, D.~L. and {Carnero Rosell}, A. and {Carrasco Kind}, M. and {Carretero}, J. and {Castander}, F.~J. and {Crocce}, M. and {Cunha}, C.~E. and {D'Andrea}, C.~B. and {da Costa}, L.~N. and {Davis}, C. and {De Vicente}, J. and {Diehl}, H.~T. and {Doel}, P. and {Drlica-Wagner}, A. and {Evrard}, A.~E. and {Flaugher}, B. and {Fosalba}, P. and {Frieman}, J. and {Garc{\'\i}a-Bellido}, J. and {Gaztanaga}, E. and {Gerdes}, D.~W. and {Giannantonio}, T. and {Gruendl}, R.~A. and {Gutierrez}, G. and {Honscheid}, K. and {James}, D.~J. and {Kirk}, D. and {Krause}, E. and {Kuehn}, K. and {Lahav}, O. and {Li}, T.~S. and {Lima}, M. and {March}, M. and {Marshall}, J.~L. and {Menanteau}, F. and {Miquel}, R. and {Mohr}, J.~J. and {Nord}, B. and {Ogando}, R.~L.~C. and {Roodman}, A. and {Sanchez}, E. and {Scarpine}, V. and {Schindler}, R. and {Sevilla-Noarbe}, I. and {Smith}, M. and {Smith}, R.~C. and {Sobreira}, F. and {Suchyta}, E. and {Swanson}, M.~E.~C. and {Tarle}, G. and {Tucker}, D.~L. and {Vikram}, V. and {Walker}, A.~R. and {Weller}, J. and {DES Collaboration}},
        title = "{Dark Energy Survey Year 1 results: weak lensing mass calibration of redMaPPer galaxy clusters}",
      journal = {\mnras},
         year = 2019,
        month = jan,
       volume = {482},
       number = {1},
        pages = {1352-1378},
          doi = {10.1093/mnras/sty2711},
archivePrefix = {arXiv},
       eprint = {1805.00039},
 primaryClass = {astro-ph.CO},
       adsurl = {https://ui.adsabs.harvard.edu/abs/2019MNRAS.482.1352M}
}

@ARTICLE{Rozo2010,
       author = {{Rozo}, Eduardo and {Wechsler}, Risa H. and {Rykoff}, Eli S. and {Annis}, James T. and {Becker}, Matthew R. and {Evrard}, August E. and {Frieman}, Joshua A. and {Hansen}, Sarah M. and {Hao}, Jiangang and {Johnston}, David E. and {Koester}, Benjamin P. and {McKay}, Timothy A. and {Sheldon}, Erin S. and {Weinberg}, David H.},
        title = "{Cosmological Constraints from the Sloan Digital Sky Survey maxBCG Cluster Catalog}",
      journal = {\apj},
         year = 2010,
        month = jan,
       volume = {708},
       number = {1},
        pages = {645-660},
          doi = {10.1088/0004-637X/708/1/645},
archivePrefix = {arXiv},
       eprint = {0902.3702},
 primaryClass = {astro-ph.CO},
       adsurl = {https://ui.adsabs.harvard.edu/abs/2010ApJ...708..645R}
}

@ARTICLE{Nelson2014,
       author = {{Nelson}, Kaylea and {Lau}, Erwin T. and {Nagai}, Daisuke},
        title = "{Hydrodynamic Simulation of Non-thermal Pressure Profiles of Galaxy Clusters}",
      journal = {\apj},
         year = 2014,
        month = sep,
       volume = {792},
       number = {1},
          eid = {25},
        pages = {25},
          doi = {10.1088/0004-637X/792/1/25},
archivePrefix = {arXiv},
       eprint = {1404.4636},
 primaryClass = {astro-ph.CO},
       adsurl = {https://ui.adsabs.harvard.edu/abs/2014ApJ...792...25N}
}

@ARTICLE{Salvati2019,
       author = {{Salvati}, Laura and {Douspis}, Marian and {Ritz}, Anna and {Aghanim}, Nabila and {Babul}, Arif},
        title = "{Mass bias evolution in tSZ cluster cosmology}",
      journal = {\aap},
         year = 2019,
        month = jun,
       volume = {626},
          eid = {A27},
        pages = {A27},
          doi = {10.1051/0004-6361/201935041},
archivePrefix = {arXiv},
       eprint = {1901.03096},
 primaryClass = {astro-ph.CO},
       adsurl = {https://ui.adsabs.harvard.edu/abs/2019A&A...626A..27S}
}

@article{Adam_2024,
   title="{The XXL Survey: LI. Pressure profile and $Y_{SZ} - M$ scaling relation in three low-mass galaxy clusters at $z \sim 1$ observed with NIKA2}",
   volume={684},
   ISSN={1432-0746},
   DOI={10.1051/0004-6361/202348049},
   journal={\aap},
   publisher={EDP Sciences},
   author={Adam, R. and Ricci, M. and Eckert, D. and Ade, P. and Ajeddig, H. and Altieri, B. and André, P. and Artis, E. and Aussel, H. and Beelen, A. and Benoist, C. and Benoît, A. and Berta, S. and Bing, L. and Birkinshaw, M. and Bourrion, O. and Boutigny, D. and Bremer, M. and Calvo, M. and Cappi, A. and Catalano, A. and De Petris, M. and Désert, F.-X. and Doyle, S. and Driessen, E. F. C. and Faccioli, L. and Ferrari, C. and Gastaldello, F. and Giles, P. and Gomez, A. and Goupy, J. and Hahn, O. and Hanser, C. and Horellou, C. and Kéruzoré, F. and Koulouridis, E. and Kramer, C. and Ladjelate, B. and Lagache, G. and Leclercq, S. and Lestrade, J.-F. and Macías-Pérez, J. F. and Madden, S. and Maughan, B. and Maurogordato, S. and Maury, A. and Mauskopf, P. and Monfardini, A. and Muñoz-Echeverría, M. and Pacaud, F. and Perotto, L. and Pierre, M. and Pisano, G. and Pompei, E. and Ponthieu, N. and Revéret, V. and Rigby, A. and Ritacco, A. and Romero, C. and Roussel, H. and Ruppin, F. and Sereno, M. and Schuster, K. and Sievers, A. and Tintoré Vidal, G. and Tucker, C. and Zylka, R.},
   year={2024},
   month=mar,
   pages={A18}
}

@ARTICLE{Angulo2012,
       author = {{Angulo}, R.~E. and {Springel}, V. and {White}, S.~D.~M. and {Jenkins}, A. and {Baugh}, C.~M. and {Frenk}, C.~S.},
        title = "{Scaling relations for galaxy clusters in the Millennium-XXL simulation}",
      journal = {\mnras},
         year = 2012,
        month = nov,
       volume = {426},
       number = {3},
        pages = {2046-2062},
          doi = {10.1111/j.1365-2966.2012.21830.x},
archivePrefix = {arXiv},
       eprint = {1203.3216},
 primaryClass = {astro-ph.CO},
       adsurl = {https://ui.adsabs.harvard.edu/abs/2012MNRAS.426.2046A}
}

@ARTICLE{Bocquet2019,
       author = {{Bocquet}, S. and {Dietrich}, J.~P. and {Schrabback}, T. and {Bleem}, L.~E. and {Klein}, M. and {Allen}, S.~W. and {Applegate}, D.~E. and {Ashby}, M.~L.~N. and {Bautz}, M. and {Bayliss}, M. and {Benson}, B.~A. and {Brodwin}, M. and {Bulbul}, E. and {Canning}, R.~E.~A. and {Capasso}, R. and {Carlstrom}, J.~E. and {Chang}, C.~L. and {Chiu}, I. and {Cho}, H.-M. and {Clocchiatti}, A. and {Crawford}, T.~M. and {Crites}, A.~T. and {de Haan}, T. and {Desai}, S. and {Dobbs}, M.~A. and {Foley}, R.~J. and {Forman}, W.~R. and {Garmire}, G.~P. and {George}, E.~M. and {Gladders}, M.~D. and {Gonzalez}, A.~H. and {Grandis}, S. and {Gupta}, N. and {Halverson}, N.~W. and {Hlavacek-Larrondo}, J. and {Hoekstra}, H. and {Holder}, G.~P. and {Holzapfel}, W.~L. and {Hou}, Z. and {Hrubes}, J.~D. and {Huang}, N. and {Jones}, C. and {Khullar}, G. and {Knox}, L. and {Kraft}, R. and {Lee}, A.~T. and {von der Linden}, A. and {Luong-Van}, D. and {Mantz}, A. and {Marrone}, D.~P. and {McDonald}, M. and {McMahon}, J.~J. and {Meyer}, S.~S. and {Mocanu}, L.~M. and {Mohr}, J.~J. and {Morris}, R.~G. and {Padin}, S. and {Patil}, S. and {Pryke}, C. and {Rapetti}, D. and {Reichardt}, C.~L. and {Rest}, A. and {Ruhl}, J.~E. and {Saliwanchik}, B.~R. and {Saro}, A. and {Sayre}, J.~T. and {Schaffer}, K.~K. and {Shirokoff}, E. and {Stalder}, B. and {Stanford}, S.~A. and {Staniszewski}, Z. and {Stark}, A.~A. and {Story}, K.~T. and {Strazzullo}, V. and {Stubbs}, C.~W. and {Vanderlinde}, K. and {Vieira}, J.~D. and {Vikhlinin}, A. and {Williamson}, R. and {Zenteno}, A.},
        title = "{Cluster Cosmology Constraints from the 2500 deg$^{2}$ SPT-SZ Survey: Inclusion of Weak Gravitational Lensing Data from Magellan and the Hubble Space Telescope}",
      journal = {\apj},
         year = 2019,
        month = jun,
       volume = {878},
       number = {1},
          eid = {55},
        pages = {55},
          doi = {10.3847/1538-4357/ab1f10},
archivePrefix = {arXiv},
       eprint = {1812.01679},
 primaryClass = {astro-ph.CO},
       adsurl = {https://ui.adsabs.harvard.edu/abs/2019ApJ...878...55B}
}

@ARTICLE{Pratt2019,
       author = {{Pratt}, G.~W. and {Arnaud}, M. and {Biviano}, A. and {Eckert}, D. and {Ettori}, S. and {Nagai}, D. and {Okabe}, N. and {Reiprich}, T.~H.},
        title = "{The Galaxy Cluster Mass Scale and Its Impact on Cosmological Constraints from the Cluster Population}",
      journal = {SSRv},
         year = 2019,
        month = feb,
       volume = {215},
       number = {2},
          eid = {25},
        pages = {25},
          doi = {10.1007/s11214-019-0591-0},
archivePrefix = {arXiv},
       eprint = {1902.10837},
 primaryClass = {astro-ph.CO},
       adsurl = {https://ui.adsabs.harvard.edu/abs/2019SSRv..215...25P}
}

@ARTICLE{Pizzuti25b,
       author = {{Pizzuti}, Lorenzo and {Barrena}, Rafael and {Sereno}, Mauro and {Streblyanska}, Alina and {Ferragamo}, Antonio and {Maurogordato}, Sophie and {Cappi}, Alberto and {Ettori}, Stefano and {Pratt}, Gabriel W. and {Castignani}, Gianluca and {Donahue}, Megan and {Eckert}, Dominique and {Gastaldello}, Fabio and {Gavazzi}, Raphael and {Haines}, Christopher P. and {Kay}, Scott T. and {Lovisari}, Lorenzo and {Maughan}, Ben J. and {Pointecouteau}, Etienne and {Rasia}, Elena and {Radovich}, Mario and {Sayers}, Jack},
        title = "{CHEX-MATE: Exploring the kinematical properties of Planck galaxy clusters}",
      journal = {\aap},
         year = 2025,
        month = jul,
       volume = {699},
          eid = {A88},
        pages = {A88},
          doi = {10.1051/0004-6361/202555417},
archivePrefix = {arXiv},
       eprint = {2505.03708},
 primaryClass = {astro-ph.CO},
       adsurl = {https://ui.adsabs.harvard.edu/abs/2025A&A...699A..88P}
}

@article{Boumechta24,
  title = {Constraining chameleon screening using galaxy cluster dynamics},
  author = {Boumechta, Yacer and Haridasu, Balakrishna S. and Pizzuti, Lorenzo and Butt, Minahil Adil and Baccigalupi, Carlo and Lapi, Andrea},
  journal = {Phys. Rev. D},
  volume = {108},
  issue = {4},
  pages = {044007},
  numpages = {15},
  year = {2023},
  month = {Aug},
  publisher = {American Physical Society},
  doi = {10.1103/PhysRevD.108.044007}
}

@ARTICLE{Wen2024,
       author = {{Wen}, Z.~L. and {Han}, J.~L. and {Yuan}, Z.~S.},
        title = "{A catalogue of merging clusters of galaxies: cluster partners, merging subclusters, and post-collision clusters}",
      journal = {\mnras},
         year = 2024,
        month = aug,
       volume = {532},
       number = {2},
        pages = {1849-1886},
          doi = {10.1093/mnras/stae1614},
archivePrefix = {arXiv},
       eprint = {2406.00652},
 primaryClass = {astro-ph.CO},
       adsurl = {https://ui.adsabs.harvard.edu/abs/2024MNRAS.532.1849W}
}

@ARTICLE{Ragagnin2024,
       author = {{Ragagnin}, A. and {Meneghetti}, M. and {Calura}, F. and {Despali}, G. and {Dolag}, K. and {Fischer}, M.~S. and {Giocoli}, C. and {Moscardini}, L.},
        title = "{Dianoga SIDM: Galaxy cluster self-interacting dark matter simulations}",
      journal = {\aap},
         year = 2024,
        month = jul,
       volume = {687},
          eid = {A270},
        pages = {A270},
          doi = {10.1051/0004-6361/202449872},
archivePrefix = {arXiv},
       eprint = {2404.01383},
 primaryClass = {astro-ph.CO},
       adsurl = {https://ui.adsabs.harvard.edu/abs/2024A&A...687A.270R}
}

@ARTICLE{Ribeiro2013,
       author = {{Ribeiro}, A.~L.~B. and {Lopes}, P.~A.~A. and {Rembold}, S.~B.},
        title = "{NoSOCS in SDSS. III. The interplay between galaxy evolution and the dynamical state of galaxy clusters}",
      journal = {\aap},
         year = 2013,
        month = aug,
       volume = {556},
          eid = {A74},
        pages = {A74},
          doi = {10.1051/0004-6361/201220801},
archivePrefix = {arXiv},
       eprint = {1306.6935},
 primaryClass = {astro-ph.CO},
       adsurl = {https://ui.adsabs.harvard.edu/abs/2013A&A...556A..74R}
}

@article{Stapelberg_2022,
   title={Triaxiality in galaxy clusters: Mass versus potential reconstructions},
   volume={663},
   ISSN={1432-0746},
   DOI={10.1051/0004-6361/202040238},
   journal={\aap},
   publisher={EDP Sciences},
   author={Stapelberg, S. and Tchernin, C. and Hug, D. and Lau, E. T. and Bartelmann, M.},
   year={2022},
   month=jul,
   pages={A17}
}

@article{Morandi_2011,
   title="{Reconstructing the Triaxiality of The Galaxy Cluster A1689: Solving The X-ray and Strong Lensing Mass discrepancy}",
   volume={729},
   ISSN={1538-4357},
   DOI={10.1088/0004-637x/729/1/37},
   number={1},
   journal={\apj},
   publisher={American Astronomical Society},
   author={Morandi, Andrea and Pedersen, Kristian and Limousin, Marceau},
   year={2011},
   month=Feb,
   pages={37}
}

@article{Limousin_2013,
   title={The Three-Dimensional Shapes of Galaxy Clusters},
   volume={177},
   ISSN={1572-9672},
   DOI={10.1007/s11214-013-9980-y},
   number={1-4},
   journal={\spsr},
   publisher={Springer Science and Business Media LLC},
   author={Limousin, Marceau and Morandi, Andrea and Sereno, Mauro and Meneghetti, Massimo and Ettori, Stefano and Bartelmann, Matthias and Verdugo, Tomas},
   year={2013},
   month=May,
   pages={155–194}
}

@article{Bilton_2020,
   title={The impact of disturbed galaxy clusters on the kinematics of active galactic nuclei},
   volume={499},
   ISSN={1365-2966},
   DOI={10.1093/mnras/staa3154},
   number={3},
   journal={\mnras},
   publisher={Oxford University Press (OUP)},
   author={Bilton, Lawrence E and Pimbblet, Kevin A and Gordon, Yjan A},
   year={2020},
   month=oct,
   pages={3792–3805}
}

@article{Roberts_2019,
   title="{‘Observing’ unrelaxed clusters in dark matter simulations}",
   volume={490},
   ISSN={1365-2966},
   DOI={10.1093/mnras/stz2666},
   number={1},
   journal={\mnras},
   publisher={Oxford University Press (OUP)},
   author={Roberts, Ian D and Parker, Laura C},
   year={2019},
   month=sep,
   pages={773–783}
}

@ARTICLE{Eckert2019,
       author = {{Eckert}, D. and {Ghirardini}, V. and {Ettori}, S. and {Rasia}, E. and {Biffi}, V. and {Pointecouteau}, E. and {Rossetti}, M. and {Molendi}, S. and {Vazza}, F. and {Gastaldello}, F. and {Gaspari}, M. and {De Grandi}, S. and {Ghizzardi}, S. and {Bourdin}, H. and {Tchernin}, C. and {Roncarelli}, M.},
        title = "{Non-thermal pressure support in X-COP galaxy clusters}",
      journal = {\aap},
         year = 2019,
        month = jan,
       volume = {621},
          eid = {A40},
        pages = {A40},
          doi = {10.1051/0004-6361/201833324},
archivePrefix = {arXiv},
       eprint = {1805.00034},
 primaryClass = {astro-ph.CO},
       adsurl = {https://ui.adsabs.harvard.edu/abs/2019A&A...621A..40E}
}

@article{Mitchell_2018,
   title="{A general framework to test gravity using galaxy clusters – I. Modelling the dynamical mass of haloes in $f(R)$ gravity}",
   volume={477},
   ISSN={1365-2966},
   DOI={10.1093/mnras/sty636},
   number={1},
   journal={\mnras},
   publisher={Oxford University Press (OUP)},
   author={Mitchell, Myles A and He, Jian-hua and Arnold, Christian and Li, Baojiu},
   year={2018},
   month=mar,
   pages={1133–1152}
}

@article{Sakstein_2016,
   title={Testing gravity using galaxy clusters: new constraints on beyond Horndeski theories},
   volume={2016},
   ISSN={1475-7516},
   DOI={10.1088/1475-7516/2016/07/019},
   number={07},
   journal={\jcap},
   publisher={IOP Publishing},
   author={Sakstein, Jeremy and Wilcox, Harry and Bacon, David and Koyama, Kazuya and Nichol, Robert C.},
   year={2016},
   month=jul,
   pages={019–019}
}

@ARTICLE{liu21,
       author = {{Liu}, Rayne and {Valogiannis}, Georgios and {Battaglia}, Nicholas and {Bean}, Rachel},
        title = "{Constraints on $f(R)$ and normal-branch Dvali-Gabadadze-Porrati modified gravity model parameters with cluster abundances and galaxy clustering}",
      journal = {\prd},
         year = 2021,
        month = nov,
       volume = {104},
       number = {10},
          eid = {103519},
        pages = {103519},
          doi = {10.1103/PhysRevD.104.103519},
archivePrefix = {arXiv},
       eprint = {2101.08728},
 primaryClass = {astro-ph.CO},
       adsurl = {https://ui.adsabs.harvard.edu/abs/2021PhRvD.104j3519L}
}

@ARTICLE{Hammami17,
       author = {{Hammami}, A. and {Mota}, D.~F.},
        title = "{Probing modified gravity via the mass-temperature relation of galaxy clusters}",
      journal = {\aap},
         year = 2017,
        month = feb,
       volume = {598},
          eid = {A132},
        pages = {A132},
          doi = {10.1051/0004-6361/201629003},
archivePrefix = {arXiv},
       eprint = {1603.08662},
 primaryClass = {astro-ph.CO},
       adsurl = {https://ui.adsabs.harvard.edu/abs/2017A&A...598A.132H}
}

@ARTICLE{Balaguera2014,
       author = {{Balaguera-Antol{\'\i}nez}, Andr{\'e}s},
        title = "{What can the spatial distribution of galaxy clusters tell about their scaling relations?}",
      journal = {\aap},
         year = 2014,
        month = mar,
       volume = {563},
          eid = {A141},
        pages = {A141},
          doi = {10.1051/0004-6361/201322029},
archivePrefix = {arXiv},
       eprint = {1306.1399},
 primaryClass = {astro-ph.CO},
       adsurl = {https://ui.adsabs.harvard.edu/abs/2014A&A...563A.141B}
}

@article{Sartoris_2016,
   title={Next generation cosmology: constraints from the Euclid galaxy cluster survey},
   volume={459},
   ISSN={1365-2966},
   DOI={10.1093/mnras/stw630},
   number={2},
   journal={\mnras},
   publisher={Oxford University Press (OUP)},
   author={Sartoris, B. and Biviano, A. and Fedeli, C. and Bartlett, J. G. and Borgani, S. and Costanzi, M. and Giocoli, C. and Moscardini, L. and Weller, J. and Ascaso, B. and Bardelli, S. and Maurogordato, S. and Viana, P. T. P.},
   year={2016},
   month=mar,
   pages={1764–1780}
}

@article{Cataneo_2018,
   title={Tests of gravity with galaxy clusters},
   volume={27},
   ISSN={1793-6594},
   DOI={10.1142/s0218271818480061},
   number={15},
   journal={International Journal of Modern Physics D},
   publisher={World Scientific Pub Co Pte Ltd},
   author={Cataneo, Matteo and Rapetti, David},
   year={2018},
   month=nov,
   pages={1848006}
}

@article{Biviano_2023,
   title="{CLASH-VLT: The Inner Slope of the MACS J1206.2-0847 Dark Matter Density Profile}",
   volume={958},
   ISSN={1538-4357},
   DOI={10.3847/1538-4357/acf832},
   number={2},
   journal={\apj},
   publisher={American Astronomical Society},
   author={Biviano, Andrea and Pizzuti, Lorenzo and Mercurio, Amata and Sartoris, Barbara and Rosati, Piero and Ettori, Stefano and Girardi, Marisa and Grillo, Claudio and Caminha, Gabriel B. and Nonino, Mario},
   year={2023},
   month=nov,
   pages={148}
}

@article{Walker2019,
  author  = {Walker, S. and Simionescu, A. and Nagai, D. and Okabe, N. and Eckert, D. and Mroczkowski, T. and Akamatsu, H. and Ettori, S. and Ghirardini, V.},
  title   = {The Physics of Galaxy Cluster Outskirts},
  journal = {\spsr},
  volume  = {215},
  number  = {7},
  year    = {2019},
  eprint  = {1810.00890}
}

@article{Miyatake2025,
  author  = {Miyatake, H.},
  title   = {Cosmology with Galaxy Clusters},
  journal = {\pr},
  year    = {2025},
  eprint  = {2505.07697}
}

@ARTICLE{deblok2010,
       author = {{de Blok}, W.~J.~G.},
        title = "{The Core-Cusp Problem}",
      journal = {Adv. in Astr.},
         year = 2010,
        month = jan,
       volume = {2010},
          eid = {789293},
        pages = {789293},
          doi = {10.1155/2010/789293},
archivePrefix = {arXiv},
       eprint = {0910.3538},
 primaryClass = {astro-ph.CO},
       adsurl = {https://ui.adsabs.harvard.edu/abs/2010AdAst2010E...5D}
}

@article{Zhao96,
    author = {Zhao, HongSheng},
    title = {Analytical models for galactic nuclei},
    journal = {\mnras},
    volume = {278},
    number = {2},
    pages = {488-496},
    year = {1996},
    month = {01},
    issn = {0035-8711},
    doi = {10.1093/mnras/278.2.488}
}

@article{Del_Popolo_2021,
   title={Cluster density slopes from dark matter–baryons energy transfer},
   volume={33},
   ISSN={2212-6864},
   DOI={10.1016/j.dark.2021.100847},
   journal={Phys. Dark Universe},
   publisher={Elsevier BV},
   author={Del Popolo, Antonino and Le Delliou, Morgan and Deliyergiyev, Maksym},
   year={2021},
   month=sep,
   pages={100847}
}

@ARTICLE{Spergel2000,
       author = {{Spergel}, David N. and {Steinhardt}, Paul J.},
        title = "{Observational Evidence for Self-Interacting Cold Dark Matter}",
      journal = {\prl},
         year = 2000,
        month = apr,
       volume = {84},
       number = {17},
        pages = {3760-3763},
          doi = {10.1103/PhysRevLett.84.3760},
archivePrefix = {arXiv},
       eprint = {astro-ph/9909386},
 primaryClass = {astro-ph},
       adsurl = {https://ui.adsabs.harvard.edu/abs/2000PhRvL..84.3760S}
}

@article{Tran_2026,
   title={Novel density profile for isothermal cores of dark matter halos},
   volume={113},
   ISSN={2470-0029},
   DOI={10.1103/t8pl-mdhc},
   number={6},
   journal={\prd},
   publisher={American Physical Society (APS)},
   author={Tran, Vinh and Shen, Xuejian and Vogelsberger, Mark and Gilman, Daniel and O’Neil, Stephanie and Roche, Cian and Zier, Oliver and Gao, Jiarun},
   year={2026},
   month=Mar }

@ARTICLE{Stadel2009,
       author = {{Stadel}, J. and {Potter}, D. and {Moore}, B. and {Diemand}, J. and {Madau}, P. and {Zemp}, M. and {Kuhlen}, M. and {Quilis}, V.},
        title = "{Quantifying the heart of darkness with GHALO - a multibillion particle simulation of a galactic halo}",
      journal = {\mnras},
         year = 2009,
        month = sep,
       volume = {398},
       number = {1},
        pages = {L21-L25},
          doi = {10.1111/j.1745-3933.2009.00699.x},
archivePrefix = {arXiv},
       eprint = {0808.2981},
 primaryClass = {astro-ph},
       adsurl = {https://ui.adsabs.harvard.edu/abs/2009MNRAS.398L..21S}
}

@ARTICLE{Rocha2013,
       author = {{Rocha}, Miguel and {Peter}, Annika H.~G. and {Bullock}, James S. and {Kaplinghat}, Manoj and {Garrison-Kimmel}, Shea and {O{\~n}orbe}, Jose and {Moustakas}, Leonidas A.},
        title = "{Cosmological simulations with self-interacting dark matter - I. Constant-density cores and substructure}",
      journal = {\mnras},
         year = 2013,
        month = mar,
       volume = {430},
       number = {1},
        pages = {81-104},
          doi = {10.1093/mnras/sts514},
archivePrefix = {arXiv},
       eprint = {1208.3025},
 primaryClass = {astro-ph.CO},
       adsurl = {https://ui.adsabs.harvard.edu/abs/2013MNRAS.430...81R}
}

@ARTICLE{Kaplinghat2016,
       author = {{Kaplinghat}, Manoj and {Tulin}, Sean and {Yu}, Hai-Bo},
        title = "{Dark Matter Halos as Particle Colliders: Unified Solution to Small-Scale Structure Puzzles from Dwarfs to Clusters}",
      journal = {\prl},
         year = 2016,
        month = jan,
       volume = {116},
       number = {4},
          eid = {041302},
        pages = {041302},
          doi = {10.1103/PhysRevLett.116.041302},
archivePrefix = {arXiv},
       eprint = {1508.03339},
 primaryClass = {astro-ph.CO},
       adsurl = {https://ui.adsabs.harvard.edu/abs/2016PhRvL.116d1302K}
}

@ARTICLE{Tulin2018,
       author = {{Tulin}, Sean and {Yu}, Hai-Bo},
        title = "{Dark matter self-interactions and small scale structure}",
      journal = {\physrep},
         year = 2018,
        month = feb,
       volume = {730},
        pages = {1-57},
          doi = {10.1016/j.physrep.2017.11.004},
archivePrefix = {arXiv},
       eprint = {1705.02358},
 primaryClass = {hep-ph},
       adsurl = {https://ui.adsabs.harvard.edu/abs/2018PhR...730....1T}
}

@ARTICLE{Robertson2018,
       author = {{Robertson}, Andrew and {Massey}, Richard and {Eke}, Vincent and {Tulin}, Sean and {Yu}, Hai-Bo and {Bah{\'e}}, Yannick and {Barnes}, David J. and {Bower}, Richard G. and {Crain}, Robert A. and {Dalla Vecchia}, Claudio and {Kay}, Scott T. and {Schaller}, Matthieu and {Schaye}, Joop},
        title = "{The diverse density profiles of galaxy clusters with self-interacting dark matter plus baryons}",
      journal = {\mnras},
         year = 2018,
        month = may,
       volume = {476},
       number = {1},
        pages = {L20-L24},
          doi = {10.1093/mnrasl/sly024},
archivePrefix = {arXiv},
       eprint = {1711.09096},
 primaryClass = {astro-ph.CO},
       adsurl = {https://ui.adsabs.harvard.edu/abs/2018MNRAS.476L..20R}
}

@ARTICLE{Pozo2025,
       author = {{Pozo}, A. and {Emami}, R. and {Mocz}, P. and {Broadhurst}, T. and {Hernquist}, L. and {Vogelsberger}, M. and {Smith}, R. and {Tremblay}, G. and {Narayan}, R. and {Steiner}, J. and {Grindlay}, J. and {Smoot}, G.},
        title = "{Galaxy formation with wave/fuzzy dark matter: The core-halo structure and the solitonic imprint}",
      journal = {\aap},
         year = 2025,
        month = jul,
       volume = {699},
          eid = {A308},
        pages = {A308},
          doi = {10.1051/0004-6361/202450443},
archivePrefix = {arXiv},
       eprint = {2310.12217},
 primaryClass = {astro-ph.CO},
       adsurl = {https://ui.adsabs.harvard.edu/abs/2025A&A...699A.308P}
}

@article{Maccio_2012,
   title={Cores in warm dark matter haloes: a Catch 22 problem: Cores in warm dark matter haloes},
   volume={424},
   ISSN={0035-8711},
   DOI={10.1111/j.1365-2966.2012.21284.x},
   number={2},
   journal={\mnras},
   publisher={Oxford University Press (OUP)},
   author={Macciò, Andrea V. and Paduroiu, Sinziana and Anderhalden, Donnino and Schneider, Aurel and Moore, Ben},
   year={2012},
   month=jun,
   pages={1105–1112}
}

@ARTICLE{Martizzi2012,
       author = {{Martizzi}, Davide and {Teyssier}, Romain and {Moore}, Ben and {Wentz}, Tina},
        title = "{The effects of baryon physics, black holes and active galactic nucleus feedback on the mass distribution in clusters of galaxies}",
      journal = {\mnras},
         year = 2012,
        month = jun,
       volume = {422},
       number = {4},
        pages = {3081-3091},
          doi = {10.1111/j.1365-2966.2012.20879.x},
archivePrefix = {arXiv},
       eprint = {1112.2752},
 primaryClass = {astro-ph.CO},
       adsurl = {https://ui.adsabs.harvard.edu/abs/2012MNRAS.422.3081M}
}

@ARTICLE{Schaller2015,
       author = {{Schaller}, Matthieu and {Frenk}, Carlos S. and {Bower}, Richard G. and {Theuns}, Tom and {Trayford}, James and {Crain}, Robert A. and {Furlong}, Michelle and {Schaye}, Joop and {Dalla Vecchia}, Claudio and {McCarthy}, I.~G.},
        title = "{The effect of baryons on the inner density profiles of rich clusters}",
      journal = {\mnras},
         year = 2015,
        month = sep,
       volume = {452},
       number = {1},
        pages = {343-355},
          doi = {10.1093/mnras/stv1341},
archivePrefix = {arXiv},
       eprint = {1409.8297},
 primaryClass = {astro-ph.CO},
       adsurl = {https://ui.adsabs.harvard.edu/abs/2015MNRAS.452..343S}
}

@ARTICLE{Governato2010,
       author = {{Governato}, F. and {Brook}, C. and {Mayer}, L. and {Brooks}, A. and {Rhee}, G. and {Wadsley}, J. and {Jonsson}, P. and {Willman}, B. and {Stinson}, G. and {Quinn}, T. and {Madau}, P.},
        title = "{Bulgeless dwarf galaxies and dark matter cores from supernova-driven outflows}",
      journal = {\nat},
         year = 2010,
        month = jan,
       volume = {463},
       number = {7278},
        pages = {203-206},
          doi = {10.1038/nature08640},
archivePrefix = {arXiv},
       eprint = {0911.2237},
 primaryClass = {astro-ph.CO},
       adsurl = {https://ui.adsabs.harvard.edu/abs/2010Natur.463..203G}
}

@ARTICLE{Boldrini2021,
       author = {{Boldrini}, Pierre and {Mohayaee}, Roya and {Silk}, Joe},
        title = "{Flattening of Dark Matter Cusps during Mergers: Model of M31}",
      journal = {\apj},
         year = 2021,
        month = oct,
       volume = {919},
       number = {2},
          eid = {86},
        pages = {86},
          doi = {10.3847/1538-4357/ac12d3},
archivePrefix = {arXiv},
       eprint = {2002.12192},
 primaryClass = {astro-ph.GA},
       adsurl = {https://ui.adsabs.harvard.edu/abs/2021ApJ...919...86B}
}

@ARTICLE{Gnedin2004,
       author = {{Gnedin}, Oleg Y. and {Kravtsov}, Andrey V. and {Klypin}, Anatoly A. and {Nagai}, Daisuke},
        title = "{Response of Dark Matter Halos to Condensation of Baryons: Cosmological Simulations and Improved Adiabatic Contraction Model}",
      journal = {\apj},
         year = 2004,
        month = nov,
       volume = {616},
       number = {1},
        pages = {16-26},
          doi = {10.1086/424914},
archivePrefix = {arXiv},
       eprint = {astro-ph/0406247},
 primaryClass = {astro-ph},
       adsurl = {https://ui.adsabs.harvard.edu/abs/2004ApJ...616...16G}
}

@ARTICLE{Einasto65,
       author = {{Einasto}, J.},
        title = "{On the Construction of a Composite Model for the Galaxy and on the Determination of the System of Galactic Parameters}",
      journal = {Trudy Astr. Inst. Alma-Ata},
         year = 1965,
        month = jan,
       volume = {5},
        pages = {87-100},
       adsurl = {https://ui.adsabs.harvard.edu/abs/1965TrAlm...5...87E}
}

@article{Baes_2022,
   title="{The Einasto model for dark matter haloes}",
   volume={667},
   ISSN={1432-0746},
   DOI={10.1051/0004-6361/202244567},
   journal={\aap},
   publisher={EDP Sciences},
   author={Baes, Maarten},
   year={2022},
   month=Nov,
   pages={A47}
}

@ARTICLE{Wyithe_gNFW_2001,
       author = {{Wyithe}, J.~S.~B. and {Turner}, E.~L. and {Spergel}, D.~N.},
        title = "{Gravitational Lens Statistics for Generalized NFW Profiles: Parameter Degeneracy and Implications for Self-Interacting Cold Dark Matter}",
      journal = {\apj},
         year = 2001,
        month = jul,
       volume = {555},
       number = {1},
        pages = {504-523},
          doi = {10.1086/321437},
archivePrefix = {arXiv},
       eprint = {astro-ph/0007354},
 primaryClass = {astro-ph},
       adsurl = {https://ui.adsabs.harvard.edu/abs/2001ApJ...555..504W}
}

@article{He_2020,
   title={Constraining the inner density slope of massive galaxy clusters},
   volume={496},
   ISSN={1365-2966},
   DOI={10.1093/mnras/staa1769},
   number={4},
   journal={\mnras},
   publisher={Oxford University Press (OUP)},
   author={He, Qiuhan and Li, Hongyu and Li, Ran and Frenk, Carlos S and Schaller, Matthieu and Barnes, David and Bahé, Yannick and Kay, Scott T and Gao, Liang and Dalla Vecchia, Claudio},
   year={2020},
   month=Jun,
   pages={4717–4733}
}

@ARTICLE{Lam2012,
       author = {{Lam}, Tsz Yan and {Nishimichi}, Takahiro and {Schmidt}, Fabian and {Takada}, Masahiro},
        title = "{Testing Gravity with the Stacked Phase Space around Galaxy Clusters}",
      journal = {\prl},
         year = 2012,
        month = aug,
       volume = {109},
       number = {5},
          eid = {051301},
        pages = {051301},
          doi = {10.1103/PhysRevLett.109.051301},
archivePrefix = {arXiv},
       eprint = {1202.4501},
 primaryClass = {astro-ph.CO},
       adsurl = {https://ui.adsabs.harvard.edu/abs/2012PhRvL.109e1301L}
}

@ARTICLE{Wojtak2011,
       author = {{Wojtak}, Rados{\l}aw and {Hansen}, Steen H. and {Hjorth}, Jens},
        title = "{Gravitational redshift of galaxies in clusters as predicted by general relativity}",
      journal = {\nat},
         year = 2011,
        month = sep,
       volume = {477},
       number = {7366},
        pages = {567-569},
          doi = {10.1038/nature10445},
archivePrefix = {arXiv},
       eprint = {1109.6571},
 primaryClass = {astro-ph.CO},
       adsurl = {https://ui.adsabs.harvard.edu/abs/2011Natur.477..567W}
}

@ARTICLE{Milgrom1983,
       author = {{Milgrom}, M.},
        title = "{A modification of the Newtonian dynamics as a possible alternative to the hidden mass hypothesis.}",
      journal = {\apj},
         year = 1983,
        month = jul,
       volume = {270},
        pages = {365-370},
          doi = {10.1086/161130},
       adsurl = {https://ui.adsabs.harvard.edu/abs/1983ApJ...270..365M}
}

@ARTICLE{Moffat2006,
       author = {{Moffat}, J.~W.},
        title = "{Scalar tensor vector gravity theory}",
      journal = {\jcap},
         year = 2006,
        month = mar,
       volume = {2006},
       number = {3},
          eid = {004},
        pages = {004},
          doi = {10.1088/1475-7516/2006/03/004},
archivePrefix = {arXiv},
       eprint = {gr-qc/0506021},
 primaryClass = {gr-qc},
       adsurl = {https://ui.adsabs.harvard.edu/abs/2006JCAP...03..004M}
}

@ARTICLE{Zhang2026,
       author = {{Zhang}, Dong and {Zonoozi}, Akram Hasani and {Kroupa}, Pavel},
        title = "{Revisiting the missing mass problem in MOND for nearby galaxy clusters}",
      journal = {\prd},
         year = 2026,
        month = feb,
       volume = {113},
       number = {4},
          eid = {043027},
        pages = {043027},
          doi = {10.1103/mp3f-q5dc},
archivePrefix = {arXiv},
       eprint = {2602.06082},
 primaryClass = {astro-ph.CO},
       adsurl = {https://ui.adsabs.harvard.edu/abs/2026PhRvD.113d3027Z}
}

@ARTICLE{Sanna2023,
       author = {{Sanna}, Andrea P. and {Matsakos}, Titos and {Diaferio}, Antonaldo},
        title = "{Covariant formulation of refracted gravity}",
      journal = {\aap},
         year = 2023,
        month = jun,
       volume = {674},
          eid = {A209},
        pages = {A209},
          doi = {10.1051/0004-6361/202243553},
archivePrefix = {arXiv},
       eprint = {2109.11217},
 primaryClass = {astro-ph.CO},
       adsurl = {https://ui.adsabs.harvard.edu/abs/2023A&A...674A.209S}
}

@ARTICLE{Cesare2020,
       author = {{Cesare}, V. and {Diaferio}, A. and {Matsakos}, T. and {Angus}, G.},
        title = "{Dynamics of DiskMass Survey galaxies in refracted gravity}",
      journal = {\aap},
         year = 2020,
        month = may,
       volume = {637},
          eid = {A70},
        pages = {A70},
          doi = {10.1051/0004-6361/201935950},
archivePrefix = {arXiv},
       eprint = {2003.07377},
 primaryClass = {astro-ph.GA},
       adsurl = {https://ui.adsabs.harvard.edu/abs/2020A&A...637A..70C}
}

@ARTICLE{Matsakos2016,
       author = {{Matsakos}, Titos and {Diaferio}, Antonaldo},
        title = "{Dynamics of galaxies and clusters in refracted gravity}",
      journal = {arXiv e-prints},
         year = 2016,
        month = mar,
          eid = {arXiv:1603.04943},
        pages = {arXiv:1603.04943},
          doi = {10.48550/arXiv.1603.04943},
archivePrefix = {arXiv},
       eprint = {1603.04943},
 primaryClass = {astro-ph.GA},
       adsurl = {https://ui.adsabs.harvard.edu/abs/2016arXiv160304943M}
}

@ARTICLE{Kelleher2024,
       author = {{Kelleher}, R. and {Lelli}, F.},
        title = "{Galaxy clusters in Milgromian dynamics: Missing matter, hydrostatic bias, and the external field effect}",
      journal = {\aap},
         year = 2024,
        month = aug,
       volume = {688},
          eid = {A78},
        pages = {A78},
          doi = {10.1051/0004-6361/202449968},
archivePrefix = {arXiv},
       eprint = {2405.08557},
 primaryClass = {astro-ph.CO},
       adsurl = {https://ui.adsabs.harvard.edu/abs/2024A&A...688A..78K}
}

@ARTICLE{Brownstein2006,
       author = {{Brownstein}, J.~R. and {Moffat}, J.~W.},
        title = "{Galaxy cluster masses without non-baryonic dark matter}",
      journal = {\mnras},
         year = 2006,
        month = apr,
       volume = {367},
       number = {2},
        pages = {527-540},
          doi = {10.1111/j.1365-2966.2006.09996.x},
archivePrefix = {arXiv},
       eprint = {astro-ph/0507222},
 primaryClass = {astro-ph},
       adsurl = {https://ui.adsabs.harvard.edu/abs/2006MNRAS.367..527B}
}

@ARTICLE{Sanders2003,
       author = {{Sanders}, R.~H.},
        title = "{Clusters of galaxies with modified Newtonian dynamics}",
      journal = {\mnras},
         year = 2003,
        month = jul,
       volume = {342},
       number = {3},
        pages = {901-908},
          doi = {10.1046/j.1365-8711.2003.06596.x},
archivePrefix = {arXiv},
       eprint = {astro-ph/0212293},
 primaryClass = {astro-ph},
       adsurl = {https://ui.adsabs.harvard.edu/abs/2003MNRAS.342..901S}
}

@ARTICLE{Famaey2013,
       author = {{Famaey}, Beno{\^\i}t and {McGaugh}, Stacy S.},
        title = "{Modified Newtonian Dynamics (MOND): Observational Phenomenology and Relativistic Extensions}",
      journal = {Living Rev. Relativ.},
         year = 2012,
        month = dec,
       volume = {15},
          eid = {10},
        pages = {10},
          doi = {10.12942/lrr-2012-10},
archivePrefix = {arXiv},
       eprint = {1112.3960},
 primaryClass = {astro-ph.CO},
       adsurl = {https://ui.adsabs.harvard.edu/abs/2012LRR....15...10F}
}

@ARTICLE{Eckert2017,
       author = {{Eckert}, D. and {Ettori}, S. and {Pointecouteau}, E. and {Molendi}, S. and {Paltani}, S. and {Tchernin}, C.},
        title = "{The XMM cluster outskirts project (X{\ensuremath{-}}COP )}",
      journal = {\an},
         year = 2017,
        month = mar,
       volume = {338},
       number = {293},
        pages = {293-298},
          doi = {10.1002/asna.201713345},
archivePrefix = {arXiv},
       eprint = {1611.05051},
 primaryClass = {astro-ph.CO},
       adsurl = {https://ui.adsabs.harvard.edu/abs/2017AN....338..293E}
}

@article{Dvali_2000,
   title={4D gravity on a brane in 5D Minkowski space},
   volume={485},
   ISSN={0370-2693},
   DOI={10.1016/s0370-2693(00)00669-9},
   number={1-3},
   journal={Physics Letters B},
   publisher={Elsevier BV},
   author={Dvali, Gia and Gabadadze, Gregory and Porrati, Massimo},
   year={2000},
   month=jul,
   pages={208–214}
}

@ARTICLE{Brax2011,
       author = {{Brax}, Philippe and {van de Bruck}, Carsten and {Davis}, Anne-Christine and {Li}, Baojiu and {Schmauch}, Benoit and {Shaw}, Douglas J.},
        title = "{Linear growth of structure in the symmetron model}",
      journal = {\prd},
         year = 2011,
        month = dec,
       volume = {84},
       number = {12},
          eid = {123524},
        pages = {123524},
          doi = {10.1103/PhysRevD.84.123524},
archivePrefix = {arXiv},
       eprint = {1108.3082},
 primaryClass = {astro-ph.CO},
       adsurl = {https://ui.adsabs.harvard.edu/abs/2011PhRvD..84l3524B}
}

@ARTICLE{Hinterbichler2010,
       author = {{Hinterbichler}, Kurt and {Khoury}, Justin},
        title = "{Screening Long-Range Forces through Local Symmetry Restoration}",
      journal = {\prl},
         year = 2010,
        month = jun,
       volume = {104},
       number = {23},
          eid = {231301},
        pages = {231301},
          doi = {10.1103/PhysRevLett.104.231301},
archivePrefix = {arXiv},
       eprint = {1001.4525},
 primaryClass = {hep-th},
       adsurl = {https://ui.adsabs.harvard.edu/abs/2010PhRvL.104w1301H}
}

@ARTICLE{HuSawicki07,
       author = {{Hu}, Wayne and {Sawicki}, Ignacy},
        title = "{Models of $f(R)$ cosmic acceleration that evade solar system tests}",
      journal = {\prd},
         year = 2007,
        month = sep,
       volume = {76},
       number = {6},
          eid = {064004},
        pages = {064004},
          doi = {10.1103/PhysRevD.76.064004},
archivePrefix = {arXiv},
       eprint = {0705.1158},
 primaryClass = {astro-ph},
       adsurl = {https://ui.adsabs.harvard.edu/abs/2007PhRvD..76f4004H}
}

@ARTICLE{Lam2013,
       author = {{Lam}, Tsz Yan and {Schmidt}, Fabian and {Nishimichi}, Takahiro and {Takada}, Masahiro},
        title = "{Modeling the phase-space distribution around massive halos}",
      journal = {\prd},
         year = 2013,
        month = jul,
       volume = {88},
       number = {2},
          eid = {023012},
        pages = {023012},
          doi = {10.1103/PhysRevD.88.023012},
archivePrefix = {arXiv},
       eprint = {1305.5548},
 primaryClass = {astro-ph.CO},
       adsurl = {https://ui.adsabs.harvard.edu/abs/2013PhRvD..88b3012L}
}

@ARTICLE{Zu2014,
       author = {{Zu}, Ying and {Weinberg}, David H. and {Jennings}, Elise and {Li}, Baojiu and {Wyman}, Mark},
        title = "{Galaxy infall kinematics as a test of modified gravity}",
      journal = {\mnras},
         year = 2014,
        month = dec,
       volume = {445},
       number = {2},
        pages = {1885-1897},
          doi = {10.1093/mnras/stu1739},
archivePrefix = {arXiv},
       eprint = {1310.6768},
 primaryClass = {astro-ph.CO},
       adsurl = {https://ui.adsabs.harvard.edu/abs/2014MNRAS.445.1885Z}
}

@article{Peirani_2017,
   title={Density profile of dark matter haloes and galaxies in the horizon–agn simulation: the impact of AGN feedback},
   volume={472},
   ISSN={1365-2966},
   DOI={10.1093/mnras/stx2099},
   number={2},
   journal={\mnras},
   publisher={Oxford University Press (OUP)},
   author={Peirani, Sébastien and Dubois, Yohan and Volonteri, Marta and Devriendt, Julien and Bundy, Kevin and Silk, Joe and Pichon, Christophe and Kaviraj, Sugata and Gavazzi, Raphaël and Habouzit, Mélanie},
   year={2017},
   month=Aug,
   pages={2153–2169}
}

@article{Newman_2013,
   title="{The Density Profiles of Massive, Relaxed Galaxy Clusters. II. Separating Luminous and Dark Matter in Cluster Cores}",
   volume={765},
   ISSN={1538-4357},
   DOI={10.1088/0004-637x/765/1/25},
   number={1},
   journal={\apj},
   publisher={American Astronomical Society},
   author={Newman, Andrew B. and Treu, Tommaso and Ellis, Richard S. and Sand, David J.},
   year={2013},
   month=Feb,
   pages={25}
}

@article{Dehghani_2020,
   title="{Navarro-Frenk-White dark matter profile and the dark halos around disk systems}",
   volume={643},
   ISSN={1432-0746},
   DOI={10.1051/0004-6361/201937079},
   journal={\aap},
   publisher={EDP Sciences},
   author={Dehghani, R. and Salucci, P. and Ghaffarnejad, H.},
   year={2020},
   month=Nov,
   pages={A161}
}

@ARTICLE{Delpopolo09,
       author = {{Del Popolo}, A. and {Kroupa}, P.},
        title = "{Density profiles of dark matter haloes on galactic and cluster scales}",
      journal = {\aap},
         year = 2009,
        month = aug,
       volume = {502},
       number = {3},
        pages = {733-747},
          doi = {10.1051/0004-6361/200811404},
archivePrefix = {arXiv},
       eprint = {0906.1146},
 primaryClass = {astro-ph.CO},
       adsurl = {https://ui.adsabs.harvard.edu/abs/2009A&A...502..733D}
}

@article{Laporte12,
    author = {Laporte, Chervin F. P. and White, Simon D. M. and Naab, Thorsten and Ruszkowski, Mateusz and Springel, Volker},
    title = {Shallow dark matter cusps in galaxy clusters},
    journal = {\mnras},
    volume = {424},
    number = {1},
    pages = {747-753},
    year = {2012},
    month = {07},
    issn = {0035-8711},
    doi = {10.1111/j.1365-2966.2012.21262.x}
}

@ARTICLE{Balmes14,
       author = {{Balm{\`e}s}, I. and {Rasera}, Y. and {Corasaniti}, P.-S. and {Alimi}, J.-M.},
        title = "{Imprints of dark energy on cosmic structure formation - III. Sparsity of dark matter halo profiles}",
      journal = {\mnras},
         year = 2014,
        month = jan,
       volume = {437},
       number = {3},
        pages = {2328-2339},
          doi = {10.1093/mnras/stt2050},
archivePrefix = {arXiv},
       eprint = {1307.2922},
 primaryClass = {astro-ph.CO},
       adsurl = {https://ui.adsabs.harvard.edu/abs/2014MNRAS.437.2328B}
}

@ARTICLE{Umetsu_2016,
       author = {{Umetsu}, Keiichi and {Zitrin}, Adi and {Gruen}, Daniel and {Merten}, Julian and {Donahue}, Megan and {Postman}, Marc},
        title = "{CLASH: Joint Analysis of Strong-lensing, Weak-lensing Shear, and Magnification Data for 20 Galaxy Clusters}",
      journal = {\apj},
         year = 2016,
        month = apr,
       volume = {821},
       number = {2},
          eid = {116},
        pages = {116},
          doi = {10.3847/0004-637X/821/2/116},
archivePrefix = {arXiv},
       eprint = {1507.04385},
 primaryClass = {astro-ph.CO},
       adsurl = {https://ui.adsabs.harvard.edu/abs/2016ApJ...821..116U}
}

@article{Mantz_2016,
   title="{Cosmology and astrophysics from relaxed galaxy clusters – V. Consistency with cold dark matter structure formation}",
   volume={462},
   ISSN={1365-2966},
   DOI={10.1093/mnras/stw1707},
   number={1},
   journal={\mnras},
   publisher={Oxford University Press (OUP)},
   author={Mantz, A. B. and Allen, S. W. and Morris, R. G.},
   year={2016},
   month=jul,
   pages={681–688}
}

@article{Cai2025,
       author = {{Cai}, Yan-Chuan and {Kaiser}, Nick and {Cole}, Shaun and {Frenk}, Carlos},
        title = "{Dynamical analysis of stacked samples of asymmetric non-static self-gravitating systems}",
      journal = {\mnras},
         year = 2025,
        month = aug,
       volume = {541},
       number = {2},
        pages = {1745-1752},
          doi = {10.1093/mnras/staf987},
archivePrefix = {arXiv},
       eprint = {2504.19922},
 primaryClass = {astro-ph.CO},
       adsurl = {https://ui.adsabs.harvard.edu/abs/2025MNRAS.541.1745C}
}

@article{Abdullah2025,
       author = {{Abdullah}, Mohamed H. and {Mabrouk}, Raouf H. and {Ishiyama}, Tomoaki and {Wilson}, Gillian and {Amin}, Magdy Y. and {Khattab}, Elamira Hend and {Abdel Rahman}, H.~I.},
        title = "{Quantifying the Velocity Anisotropy Profile of Galaxy Clusters Using the Uchuu Cosmological Simulation}",
      journal = {\apj},
         year = 2025,
        month = jul,
       volume = {987},
       number = {1},
          eid = {70},
        pages = {70},
          doi = {10.3847/1538-4357/adde4b},
archivePrefix = {arXiv},
       eprint = {2504.04575},
 primaryClass = {astro-ph.CO},
       adsurl = {https://ui.adsabs.harvard.edu/abs/2025ApJ...987...70A}
}

@ARTICLE{Mundow2025,
       author = {{Mundow}, Raeed and {Nusser}, Adi},
        title = "{Estimating Cluster Masses: A Comparative Study between Machine Learning and Maximum Likelihood}",
      journal = {\apj},
         year = 2025,
        month = nov,
       volume = {994},
       number = {1},
          eid = {38},
        pages = {38},
          doi = {10.3847/1538-4357/ae17c0},
archivePrefix = {arXiv},
       eprint = {2507.21876},
 primaryClass = {astro-ph.CO},
       adsurl = {https://ui.adsabs.harvard.edu/abs/2025ApJ...994...38M}
}

@ARTICLE{Biviano_2017,
       author = {{Biviano}, A. and {Moretti}, A. and {Paccagnella}, A. and {Poggianti}, B.~M. and {Bettoni}, D. and {Gullieuszik}, M. and {Vulcani}, B. and {Fasano}, G. and {D'Onofrio}, M. and {Fritz}, J. and {Cava}, A.},
        title = "{The concentration-mass relation of clusters of galaxies from the OmegaWINGS survey}",
      journal = {\aap},
         year = 2017,
        month = nov,
       volume = {607},
          eid = {A81},
        pages = {A81},
          doi = {10.1051/0004-6361/201731289},
archivePrefix = {arXiv},
       eprint = {1708.07349},
 primaryClass = {astro-ph.CO},
       adsurl = {https://ui.adsabs.harvard.edu/abs/2017A&A...607A..81B}
}

@article{Despali_2018,
   title={Modelling the line-of-sight contribution in substructure lensing},
   volume={475},
   ISSN={1365-2966},
   DOI={10.1093/mnras/sty159},
   number={4},
   journal={\mnras},
   publisher={Oxford University Press (OUP)},
   author={Despali, Giulia and Vegetti, Simona and White, Simon D M and Giocoli, Carlo and van den Bosch, Frank C},
   year={2018},
   month=jan,
   pages={5424–5442}
}

@ARTICLE{DeMaio20,
       author = {{DeMaio}, Tahlia and {Gonzalez}, Anthony H. and {Zabludoff}, Ann and {Zaritsky}, Dennis and {Aldering}, Greg and {Brodwin}, Mark and {Connor}, Thomas and {Donahue}, Megan and {Hayden}, Brian and {Mulchaey}, John S. and {Perlmutter}, Saul and {Stanford}, S.~A.},
        title = "{The growth of brightest cluster galaxies and intracluster light over the past 10 billion years}",
      journal = {\mnras},
         year = 2020,
        month = jan,
       volume = {491},
       number = {3},
        pages = {3751-3759},
          doi = {10.1093/mnras/stz3236},
archivePrefix = {arXiv},
       eprint = {1911.07911},
 primaryClass = {astro-ph.GA},
       adsurl = {https://ui.adsabs.harvard.edu/abs/2020MNRAS.491.3751D}
}

@article{Giocoli_2012,
   title={Cosmology in two dimensions: the concentration-mass relation for galaxy clusters},
   volume={426},
   ISSN={0035-8711},
   DOI={10.1111/j.1365-2966.2012.21743.x},
   number={2},
   journal={\mnras},
   publisher={Oxford University Press (OUP)},
   author={Giocoli, Carlo and Meneghetti, Massimo and Ettori, Stefano and Moscardini, Lauro},
   year={2012},
   month=oct,
   pages={1558–1573}
}

@ARTICLE{Gifford2013,
       author = {{Gifford}, Daniel and {Miller}, Christopher and {Kern}, Nicholas},
        title = "{A Systematic Analysis of Caustic Methods for Galaxy Cluster Masses}",
      journal = {\apj},
         year = 2013,
        month = aug,
       volume = {773},
       number = {2},
          eid = {116},
        pages = {116},
          doi = {10.1088/0004-637X/773/2/116},
archivePrefix = {arXiv},
       eprint = {1307.0017},
 primaryClass = {astro-ph.CO},
       adsurl = {https://ui.adsabs.harvard.edu/abs/2013ApJ...773..116G}
}

@ARTICLE{Adhikari2014,
       author = {{Adhikari}, Susmita and {Dalal}, Neal and {Chamberlain}, Robert T.},
        title = "{Splashback in accreting dark matter halos}",
      journal = {\jcap},
         year = 2014,
        month = nov,
       volume = {2014},
       number = {11},
        pages = {019-019},
          doi = {10.1088/1475-7516/2014/11/019},
archivePrefix = {arXiv},
       eprint = {1409.4482},
 primaryClass = {astro-ph.CO},
       adsurl = {https://ui.adsabs.harvard.edu/abs/2014JCAP...11..019A}
}

@article{Abdullah_2020,
doi = {10.3847/1538-4365/ab536e},
year = {2019},
month = {dec},
publisher = {The American Astronomical Society},
volume = {246},
number = {1},
pages = {2},
author = {Abdullah, Mohamed H. and Wilson, Gillian and Klypin, Anatoly and Old, Lyndsay and Praton, Elizabeth and Ali, Gamal B.},
title = "{{GalWeight} Application: A Publicly Available Catalog of Dynamical Parameters of 1800 Galaxy Clusters from SDSS-DR13, (GalWCat19)}",
journal = {\apjs}
}

@article{Adhikari_2016,
   title={Observing dynamical friction in galaxy clusters},
   volume={2016},
   ISSN={1475-7516},
   DOI={10.1088/1475-7516/2016/07/022},
   number={07},
   journal={\jcap},
   publisher={IOP Publishing},
   author={Adhikari, Susmita and Dalal, Neal and Clampitt, Joseph},
   year={2016},
   month=jul,
   pages={022–022}
}

@ARTICLE{Damsted2023,
       author = {{Damsted}, S. and {Finoguenov}, A. and {Clerc}, N. and {Davalgait{\.{e}}}, I. and {Kirkpatrick}, C.~C. and {Mamon}, G.~A. and {Ider Chitham}, J. and {Kiiveri}, K. and {Comparat}, J. and {Collins}, C.},
        title = "{CODEX: Role of velocity substructure in the scaling relations of galaxy clusters}",
      journal = {\aap},
         year = 2023,
        month = aug,
       volume = {676},
          eid = {A127},
        pages = {A127},
          doi = {10.1051/0004-6361/202245308},
archivePrefix = {arXiv},
       eprint = {2307.08749},
 primaryClass = {astro-ph.CO},
       adsurl = {https://ui.adsabs.harvard.edu/abs/2023A&A...676A.127D}
}

@article{Andreon2010b,
    author = {Andreon, S. and Hurn, M. A.},
    title = {The scaling relation between richness and mass of galaxy clusters: a Bayesian approach},
    journal = {\mnras},
    volume = {404},
    number = {4},
    pages = {1922-1937},
    year = {2010},
    month = {05},
    issn = {0035-8711},
    doi = {10.1111/j.1365-2966.2010.16406.x}
}

@ARTICLE{SE15,
       author = {{Sereno}, Mauro and {Ettori}, Stefano},
        title = "{CoMaLit - IV. Evolution and self-similarity of scaling relations with the galaxy cluster mass}",
      journal = {\mnras},
         year = 2015,
        month = jul,
       volume = {450},
       number = {4},
        pages = {3675-3695},
          doi = {10.1093/mnras/stv814},
       adsurl = {https://ui.adsabs.harvard.edu/abs/2015MNRAS.450.3675S}
}

@ARTICLE{Barahona2022,
       author = {{Aguado-Barahona}, A. and {Rubi{\~n}o-Mart{\'\i}n}, J.~A. and {Ferragamo}, A. and {Barrena}, R. and {Streblyanska}, A. and {Tramonte}, D.},
        title = "{Velocity dispersion and dynamical masses for 388 galaxy clusters and groups. Calibrating the M$_{SZ}$ ‒ M$_{dyn}$ scaling relation for the PSZ2 sample}",
      journal = {\aap},
         year = 2022,
        month = mar,
       volume = {659},
          eid = {A126},
        pages = {A126},
          doi = {10.1051/0004-6361/202039980},
archivePrefix = {arXiv},
       eprint = {2111.13071},
 primaryClass = {astro-ph.CO},
       adsurl = {https://ui.adsabs.harvard.edu/abs/2022A&A...659A.126A}
}

@ARTICLE{Borgani04,
       author = {{Borgani}, S. and {Murante}, G. and {Springel}, V. and {Diaferio}, A. and {Dolag}, K. and {Moscardini}, L. and {Tormen}, G. and {Tornatore}, L. and {Tozzi}, P.},
        title = "{X-ray properties of galaxy clusters and groups from a cosmological hydrodynamical simulation}",
      journal = {\mnras},
         year = 2004,
        month = mar,
       volume = {348},
       number = {3},
        pages = {1078-1096},
          doi = {10.1111/j.1365-2966.2004.07431.x},
archivePrefix = {arXiv},
       eprint = {astro-ph/0310794},
 primaryClass = {astro-ph},
       adsurl = {https://ui.adsabs.harvard.edu/abs/2004MNRAS.348.1078B}
}

@ARTICLE{WojtakLokas2007,
       author = {{Wojtak}, Radoslaw and {Lokas}, Ewa L.},
        title = "{Mass modelling of Abell 2634: avoiding the interloper bias}",
      journal = {arXiv e-prints},
         year = 2007,
        month = dec,
          eid = {arXiv:0712.2698},
        pages = {arXiv:0712.2698},
          doi = {10.48550/arXiv.0712.2698},
archivePrefix = {arXiv},
       eprint = {0712.2698},
 primaryClass = {astro-ph},
       adsurl = {https://ui.adsabs.harvard.edu/abs/2007arXiv0712.2698W}
}

@ARTICLE{Old_2015,
       author = {{Old}, L. and {Wojtak}, R. and {Mamon}, G.~A. and {Skibba}, R.~A. and {Pearce}, F.~R. and {Croton}, D. and {Bamford}, S. and {Behroozi}, P. and {de Carvalho}, R. and {Mu{\~n}oz-Cuartas}, J.~C. and {Gifford}, D. and {Gray}, M.~E. and {von der Linden}, A. and {Merrifield}, M.~R. and {Muldrew}, S.~I. and {M{\"u}ller}, V. and {Pearson}, R.~J. and {Ponman}, T.~J. and {Rozo}, E. and {Rykoff}, E. and {Saro}, A. and {Sepp}, T. and {Sif{\'o}n}, C. and {Tempel}, E.},
        title = "{Galaxy cluster mass reconstruction project - II. Quantifying scatter and bias using contrasting mock catalogues}",
      journal = {\mnras},
         year = 2015,
        month = may,
       volume = {449},
       number = {2},
        pages = {1897-1920},
          doi = {10.1093/mnras/stv421},
archivePrefix = {arXiv},
       eprint = {1502.07347},
 primaryClass = {astro-ph.CO},
       adsurl = {https://ui.adsabs.harvard.edu/abs/2015MNRAS.449.1897O}
}

@article{Harko_2007,
   title={Virial theorem and the dynamics of clusters of galaxies in the brane world models},
   volume={76},
   ISSN={1550-2368},
   DOI={10.1103/physrevd.76.044013},
   number={4},
   journal={\prd},
   publisher={American Physical Society (APS)},
   author={Harko, T. and Cheng, K. S.},
   year={2007},
   month=aug }

@ARTICLE{LopezC22,
       author = {{L{\'o}pez-Corredoira}, M. and {Betancort-Rijo}, J.~E. and {Scarpa}, R. and {Chrob{\'a}kov{\'a}}, {\v{Z}}.},
        title = "{Virial theorem in clusters of galaxies with MOND}",
      journal = {\mnras},
         year = 2022,
        month = dec,
       volume = {517},
       number = {4},
        pages = {5734-5743},
          doi = {10.1093/mnras/stac3117},
archivePrefix = {arXiv},
       eprint = {2210.13961},
 primaryClass = {astro-ph.GA},
       adsurl = {https://ui.adsabs.harvard.edu/abs/2022MNRAS.517.5734L}
}

@ARTICLE{Serra2011a,
       author = {{Serra}, Ana Laura and {Dom{\'\i}nguez Romero}, Mariano Javier L.},
        title = "{Measuring the dark matter equation of state}",
      journal = {\mnras},
         year = 2011,
        month = jul,
       volume = {415},
       number = {1},
        pages = {L74-L77},
          doi = {10.1111/j.1745-3933.2011.01082.x},
archivePrefix = {arXiv},
       eprint = {1103.5465},
 primaryClass = {gr-qc},
       adsurl = {https://ui.adsabs.harvard.edu/abs/2011MNRAS.415L..74S}
}

@ARTICLE{TheWhite1986,
       author = {{The}, Lih Sin and {White}, Simon D.~M.},
        title = "{The mass of the Coma cluster}",
      journal = {\aj},
         year = 1986,
        month = dec,
       volume = {92},
        pages = {1248-1253},
          doi = {10.1086/114258},
       adsurl = {https://ui.adsabs.harvard.edu/abs/1986AJ.....92.1248T}
}

@ARTICLE{Girardi98,
       author = {{Girardi}, Marisa and {Giuricin}, Giuliano and {Mardirossian}, Fabio and {Mezzetti}, Marino and {Boschin}, Walter},
        title = "{Optical Mass Estimates of Galaxy Clusters}",
      journal = {\apj},
         year = 1998,
        month = sep,
       volume = {505},
       number = {1},
        pages = {74-95},
          doi = {10.1086/306157},
archivePrefix = {arXiv},
       eprint = {astro-ph/9804187},
 primaryClass = {astro-ph},
       adsurl = {https://ui.adsabs.harvard.edu/abs/1998ApJ...505...74G}
}

@ARTICLE{Limber1960,
       author = {{Limber}, D. Nelson and {Mathews}, William G.},
        title = "{The Dynamical Stability of Stephan's Quintet.}",
      journal = {\apj},
         year = 1960,
        month = sep,
       volume = {132},
        pages = {286},
          doi = {10.1086/146928},
       adsurl = {https://ui.adsabs.harvard.edu/abs/1960ApJ...132..286L}
}

@ARTICLE{Adhikari2018,
       author = {{Adhikari}, Susmita and {Sakstein}, Jeremy and {Jain}, Bhuvnesh and {Dalal}, Neal and {Li}, Baojiu},
        title = "{Splashback in galaxy clusters as a probe of cosmic expansion and gravity}",
      journal = {\jcap},
         year = 2018,
        month = nov,
       volume = {2018},
       number = {11},
          eid = {033},
        pages = {033},
          doi = {10.1088/1475-7516/2018/11/033},
archivePrefix = {arXiv},
       eprint = {1806.04302},
 primaryClass = {astro-ph.CO},
       adsurl = {https://ui.adsabs.harvard.edu/abs/2018JCAP...11..033A}
}

@ARTICLE{Diaferio2005,
       author = {{Diaferio}, Antonaldo and {Geller}, Margaret J. and {Rines}, Kenneth J.},
        title = "{Caustic and Weak-Lensing Estimators of Galaxy Cluster Masses}",
      journal = {\apjl},
         year = 2005,
        month = aug,
       volume = {628},
       number = {2},
        pages = {L97-L100},
          doi = {10.1086/432880},
archivePrefix = {arXiv},
       eprint = {astro-ph/0506560},
 primaryClass = {astro-ph},
       adsurl = {https://ui.adsabs.harvard.edu/abs/2005ApJ...628L..97D}
}

@ARTICLE{Pizzardo23,
       author = {{Pizzardo}, Michele and {Geller}, Margaret J. and {Kenyon}, Scott J. and {Damjanov}, Ivana and {Diaferio}, Antonaldo},
        title = "{Galaxy cluster mass accretion rates from IllustrisTNG}",
      journal = {\aap},
         year = 2023,
        month = dec,
       volume = {680},
          eid = {A48},
        pages = {A48},
          doi = {10.1051/0004-6361/202347470},
archivePrefix = {arXiv},
       eprint = {2307.07398},
 primaryClass = {astro-ph.CO},
       adsurl = {https://ui.adsabs.harvard.edu/abs/2023A&A...680A..48P}
}

@ARTICLE{Umetsu25,
       author = {{Umetsu}, Keiichi and {Pizzardo}, Michele and {Diaferio}, Antonaldo and {Geller}, Margaret J.},
        title = "{Cluster Lensing Mass Inversion (CLUMI+): Combining Dynamics and Weak Lensing around Galaxy Clusters}",
      journal = {\apj},
         year = 2025,
        month = sep,
       volume = {990},
       number = {1},
          eid = {70},
        pages = {70},
          doi = {10.3847/1538-4357/aded91},
archivePrefix = {arXiv},
       eprint = {2505.04694},
 primaryClass = {astro-ph.CO},
       adsurl = {https://ui.adsabs.harvard.edu/abs/2025ApJ...990...70U}
}

@article{Shi_2024,
doi = {10.3847/1538-4357/ad64cf},
year = {2024},
month = {sep},
publisher = {The American Astronomical Society},
volume = {973},
number = {2},
pages = {82},
author = {Shi, Rui and Wang, Wenting and Li, Zhaozhou and Zhu, Ling and Smith, Alexander and Cole, Shaun and Gao, Hongyu and Chen, Xiaokai and Li, Qingyang and Han, Jiaxin},
title = {Inferring the Mass Content of Galaxy Clusters with Satellite Kinematics and {Jeans} Anisotropic Modeling},
journal = {\apj}
}

@ARTICLE{Malavasi2020,
       author = {{Malavasi}, Nicola and {Aghanim}, Nabila and {Tanimura}, Hideki and {Bonjean}, Victor and {Douspis}, Marian},
        title = "{Like a spider in its web: a study of the large-scale structure around the Coma cluster}",
      journal = {\aap},
         year = 2020,
        month = feb,
       volume = {634},
          eid = {A30},
        pages = {A30},
          doi = {10.1051/0004-6361/201936629},
archivePrefix = {arXiv},
       eprint = {1910.11879},
 primaryClass = {astro-ph.CO},
       adsurl = {https://ui.adsabs.harvard.edu/abs/2020A&A...634A..30M}
}

@ARTICLE{Lau2010,
       author = {{Lau}, Erwin T. and {Nagai}, Daisuke and {Kravtsov}, Andrey V.},
        title = "{Effects of Baryon Dissipation on the Dark Matter Virial Scaling Relation}",
      journal = {\apj},
         year = 2010,
        month = jan,
       volume = {708},
       number = {2},
        pages = {1419-1425},
          doi = {10.1088/0004-637X/708/2/1419},
archivePrefix = {arXiv},
       eprint = {0908.2133},
 primaryClass = {astro-ph.CO},
       adsurl = {https://ui.adsabs.harvard.edu/abs/2010ApJ...708.1419L}
}

@article{Palmese_2020,
   title={Stellar mass as a galaxy cluster mass proxy: application to the Dark Energy Survey redMaPPer clusters},
   volume={493},
   ISSN={1365-2966},
   DOI={10.1093/mnras/staa526},
   number={4},
   journal={\mnras},
   publisher={Oxford University Press (OUP)},
   author={Palmese, A and Annis, J and Burgad, J and Farahi, A and Soares-Santos, M and Welch, B and da Silva Pereira, M and Lin, H and Bhargava, S and Hollowood, D L and Wilkinson, R and Giles, P and Jeltema, T and Romer, A K and Evrard, A E and Hilton, M and Vergara Cervantes, C and Bermeo, A and Mayers, J and DeRose, J and Gruen, D and Hartley, W G and Lahav, O and Leistedt, B and McClintock, T and Rozo, E and Rykoff, E S and Varga, T N and Wechsler, R H and Zhang, Y and Avila, S and Brooks, D and Buckley-Geer, E and Burke, D L and Carnero Rosell, A and Carrasco Kind, M and Carretero, J and Castander, F J and Collins, C and da Costa, L N and Desai, S and De Vicente, J and Diehl, H T and Dietrich, J P and Doel, P and Flaugher, B and Fosalba, P and Frieman, J and García-Bellido, J and Gerdes, D W and Gruendl, R A and Gschwend, J and Gutierrez, G and Honscheid, K and James, D J and Krause, E and Kuehn, K and Kuropatkin, N and Liddle, A and Lima, M and Maia, M A G and Mann, R G and Marshall, J L and Menanteau, F and Miquel, R and Ogando, R L C and Plazas, A A and Roodman, A and Rooney, P and Sahlen, M and Sanchez, E and Scarpine, V and Schubnell, M and Serrano, S and Sevilla-Noarbe, I and Sobreira, F and Stott, J and Suchyta, E and Swanson, M E C and Tarle, G and Thomas, D and Tucker, D L and Viana, P T P and Vikram, V and Walker, A R},
   year={2020},
   month=feb,
   pages={4591–4606}
}

@article{Andreon_2010,
   title={The stellar mass fraction and baryon content of galaxy clusters and groups: Stellar mass fraction and baryon content},
   volume={407},
   ISSN={0035-8711},
   DOI={10.1111/j.1365-2966.2010.16856.x},
   number={1},
   journal={\mnras},
   publisher={Oxford University Press (OUP)},
   author={Andreon, S.},
   year={2010},
   month=aug,
   pages={263–276}
}

@ARTICLE{Sartoris_2020,
       author = {{Sartoris}, B. and {Biviano}, A. and {Rosati}, P. and {Mercurio}, A. and {Grillo}, C. and {Ettori}, S. and {Nonino}, M. and {Umetsu}, K. and {Bergamini}, P. and {Caminha}, G.~B. and {Girardi}, M.},
        title = "{CLASH-VLT: a full dynamical reconstruction of the mass profile of Abell S1063 from 1 kpc out to the virial radius}",
      journal = {\aap},
         year = 2020,
        month = may,
       volume = {637},
          eid = {A34},
        pages = {A34},
          doi = {10.1051/0004-6361/202037521},
archivePrefix = {arXiv},
       eprint = {2003.08475},
 primaryClass = {astro-ph.CO},
       adsurl = {https://ui.adsabs.harvard.edu/abs/2020A&A...637A..34S}
}

@ARTICLE{Sereno25b,
       author = {{Sereno}, Mauro},
        title = "{Self-similarity of the mass distribution in rich galaxy clusters up to z{\ensuremath{\sim}}1 tracked with weak lensing}",
      journal = {\aap},
         year = 2025,
        month = apr,
       volume = {696},
          eid = {A227},
        pages = {A227},
          doi = {10.1051/0004-6361/202553989},
archivePrefix = {arXiv},
       eprint = {2503.15605},
 primaryClass = {astro-ph.CO},
       adsurl = {https://ui.adsabs.harvard.edu/abs/2025A&A...696A.227S}
}

@article{Mamon_2013,
   title="{MAMPOSSt: Modelling Anisotropy and Mass Profiles of Observed Spherical Systems – I. Gaussian 3D velocities}",
   volume={429},
   ISSN={0035-8711},
   DOI={10.1093/mnras/sts565},
   number={4},
   journal={\mnras},
   publisher={Oxford University Press (OUP)},
   author={Mamon, Gary A. and Biviano, Andrea and Boué, Gwenaël},
   year={2013},
   month=jan,
   pages={3079–3098}
}

@ARTICLE{Dejonghe1992,
       author = {{Dejonghe}, Herwig and {Merritt}, David},
        title = "{Inferring the Mass of Spherical Stellar Systems from Velocity Moments}",
      journal = {\apj},
         year = 1992,
        month = jun,
       volume = {391},
        pages = {531},
          doi = {10.1086/171368},
       adsurl = {https://ui.adsabs.harvard.edu/abs/1992ApJ...391..531D}
}

@ARTICLE{Solanes90,
       author = {{Solanes}, J.~M. and {Salvador-Sole}, E.},
        title = "{Analytical anisotropic models of clusters of galaxies}",
      journal = {\aap},
         year = 1990,
        month = aug,
       volume = {234},
       number = {1-2},
        pages = {93-98},
       adsurl = {https://ui.adsabs.harvard.edu/abs/1990A&A...234...93S}
}

@ARTICLE{Host2009,
       author = {{Host}, Ole and {Hansen}, Steen H. and {Piffaretti}, Rocco and {Morandi}, Andrea and {Ettori}, Stefano and {Kay}, Scott T. and {Valdarnini}, Riccardo},
        title = "{Measurement of the Dark Matter Velocity Anisotropy in Galaxy Clusters}",
      journal = {\apj},
         year = 2009,
        month = jan,
       volume = {690},
       number = {1},
        pages = {358-366},
          doi = {10.1088/0004-637X/690/1/358},
archivePrefix = {arXiv},
       eprint = {0808.2049},
 primaryClass = {astro-ph},
       adsurl = {https://ui.adsabs.harvard.edu/abs/2009ApJ...690..358H}
}

@ARTICLE{Balestra2016,
       author = {{Balestra}, I. and {Mercurio}, A. and {Sartoris}, B. and {Girardi}, M. and {Grillo}, C. and {Nonino}, M. and {Rosati}, P. and {Biviano}, A. and {Ettori}, S. and {Forman}, W. and {Jones}, C. and {Koekemoer}, A. and {Medezinski}, E. and {Merten}, J. and {Ogrean}, G.~A. and {Tozzi}, P. and {Umetsu}, K. and {Vanzella}, E. and {van Weeren}, R.~J. and {Zitrin}, A. and {Annunziatella}, M. and {Caminha}, G.~B. and {Broadhurst}, T. and {Coe}, D. and {Donahue}, M. and {Fritz}, A. and {Frye}, B. and {Kelson}, D. and {Lombardi}, M. and {Maier}, C. and {Meneghetti}, M. and {Monna}, A. and {Postman}, M. and {Scodeggio}, M. and {Seitz}, S. and {Ziegler}, B.},
        title = "{CLASH-VLT: Dissecting the Frontier Fields Galaxy Cluster MACS J0416.1-2403 with {\ensuremath{\sim}}800 Spectra of Member Galaxies}",
      journal = {\apjs},
         year = 2016,
        month = jun,
       volume = {224},
       number = {2},
          eid = {33},
        pages = {33},
          doi = {10.3847/0067-0049/224/2/33},
archivePrefix = {arXiv},
       eprint = {1511.02522},
 primaryClass = {astro-ph.CO},
       adsurl = {https://ui.adsabs.harvard.edu/abs/2016ApJS..224...33B}
}

@ARTICLE{Read2021,
       author = {{Read}, J.~I. and {Mamon}, G.~A. and {Vasiliev}, E. and {Watkins}, L.~L. and {Walker}, M.~G. and {Pe{\~n}arrubia}, J. and {Wilkinson}, M. and {Dehnen}, W. and {Das}, P.},
        title = "{Breaking beta: a comparison of mass modelling methods for spherical systems}",
      journal = {\mnras},
         year = 2021,
        month = feb,
       volume = {501},
       number = {1},
        pages = {978-993},
          doi = {10.1093/mnras/staa3663},
archivePrefix = {arXiv},
       eprint = {2011.09493},
 primaryClass = {astro-ph.GA},
       adsurl = {https://ui.adsabs.harvard.edu/abs/2021MNRAS.501..978R}
}

@ARTICLE{Cappellari2008,
       author = {{Cappellari}, Michele},
        title = "{Measuring the inclination and mass-to-light ratio of axisymmetric galaxies via anisotropic {Jeans} models of stellar kinematics}",
      journal = {\mnras},
         year = 2008,
        month = oct,
       volume = {390},
       number = {1},
        pages = {71-86},
          doi = {10.1111/j.1365-2966.2008.13754.x},
archivePrefix = {arXiv},
       eprint = {0806.0042},
 primaryClass = {astro-ph},
       adsurl = {https://ui.adsabs.harvard.edu/abs/2008MNRAS.390...71C}
}

@ARTICLE{Valk2025,
       author = {{Valk}, Greique A. and {Rembold}, Sandro B.},
        title = "{Unravelling the orbits of cluster galaxy populations according to their dominant gas ionization source}",
      journal = {\mnras},
         year = 2025,
        month = jan,
       volume = {536},
       number = {3},
        pages = {2730-2748},
          doi = {10.1093/mnras/stae2779},
archivePrefix = {arXiv},
       eprint = {2412.13089},
 primaryClass = {astro-ph.GA},
       adsurl = {https://ui.adsabs.harvard.edu/abs/2025MNRAS.536.2730V}
}

@ARTICLE{Li2023,
       author = {{Li}, Pengfei and {Tian}, Yong and {J{\'u}lio}, Mariana P. and {Pawlowski}, Marcel S. and {Lelli}, Federico and {McGaugh}, Stacy S. and {Schombert}, James M. and {Read}, Justin I. and {Yu}, Po-Chieh and {Ko}, Chung-Ming},
        title = "{Measuring galaxy cluster mass profiles into the low-acceleration regime with galaxy kinematics}",
      journal = {\aap},
         year = 2023,
        month = sep,
       volume = {677},
          eid = {A24},
        pages = {A24},
          doi = {10.1051/0004-6361/202346431},
archivePrefix = {arXiv},
       eprint = {2303.10175},
 primaryClass = {astro-ph.CO},
       adsurl = {https://ui.adsabs.harvard.edu/abs/2023A&A...677A..24L}
}

@ARTICLE{Stark2019,
       author = {{Stark}, Alejo and {Miller}, Christopher J. and {Halenka}, Vitali},
        title = "{Deriving Galaxy Cluster Velocity Anisotropy Profiles from a Joint Analysis of Dynamical and Weak Lensing Data}",
      journal = {\apj},
         year = 2019,
        month = mar,
       volume = {874},
       number = {1},
          eid = {33},
        pages = {33},
          doi = {10.3847/1538-4357/ab06fa},
archivePrefix = {arXiv},
       eprint = {1711.10018},
 primaryClass = {astro-ph.CO},
       adsurl = {https://ui.adsabs.harvard.edu/abs/2019ApJ...874...33S}
}

@article{Springel_2005,
   title={The cosmological simulation code gadget-2},
   volume={364},
   ISSN={1365-2966},
   DOI={10.1111/j.1365-2966.2005.09655.x},
   number={4},
   journal={\mnras},
   publisher={Oxford University Press (OUP)},
   author={Springel, Volker},
   year={2005},
   month=dec,
   pages={1105–1134}
}

@article{Bonafede_2011,
   title={A non-ideal magnetohydrodynamic gadget: simulating massive galaxy clusters},
   volume={418},
   ISSN={0035-8711},
   DOI={10.1111/j.1365-2966.2011.19523.x},
   number={4},
   journal={\mnras},
   publisher={Oxford University Press (OUP)},
   author={Bonafede, A. and Dolag, K. and Stasyszyn, F. and Murante, G. and Borgani, S.},
   year={2011},
   month=nov,
   pages={2234–2250}
}

@article{Hansen_2006,
   title={A universal density slope – Velocity anisotropy relation for relaxed structures},
   volume={11},
   ISSN={1384-1076},
   DOI={10.1016/j.newast.2005.09.001},
   number={5},
   journal={New Astronomy},
   publisher={Elsevier BV},
   author={Hansen, Steen H. and Moore, Ben},
   year={2006},
   month=mar,
   pages={333–338}
}

@ARTICLE{Abdullah2013,
       author = {{Abdullah}, Mohamed H. and {Praton}, Elizabeth A. and {Ali}, Gamal B.},
        title = "{Distortion of infall regions in redshift space-I}",
      journal = {\mnras},
         year = 2013,
        month = sep,
       volume = {434},
       number = {3},
        pages = {1989-2007},
          doi = {10.1093/mnras/stt1145},
archivePrefix = {arXiv},
       eprint = {1408.3710},
 primaryClass = {astro-ph.CO},
       adsurl = {https://ui.adsabs.harvard.edu/abs/2013MNRAS.434.1989A}
}

@ARTICLE{Biviano2004,
       author = {{Biviano}, A. and {Katgert}, P.},
        title = "{The ESO Nearby Abell Cluster Survey. XIII. The orbits of the different types of galaxies in rich clusters}",
      journal = {\aap},
         year = 2004,
        month = sep,
       volume = {424},
        pages = {779-791},
          doi = {10.1051/0004-6361:20041306},
archivePrefix = {arXiv},
       eprint = {astro-ph/0406155},
 primaryClass = {astro-ph},
       adsurl = {https://ui.adsabs.harvard.edu/abs/2004A&A...424..779B}
}

@article{Biviano_2003,
   title="{The Mass Profile of Galaxy Clusters out to $\sim 2 \, r_{200}$}",
   volume={585},
   ISSN={1538-4357},
   DOI={10.1086/345893},
   number={1},
   journal={\apj},
   publisher={American Astronomical Society},
   author={Biviano, Andrea and Girardi, Marisa},
   year={2003},
   month=mar,
   pages={205–214}
}

@article{Geller_1999,
   title="{The Mass Profile of the Coma Galaxy Cluster}",
   volume={517},
   ISSN={0004-637X},
   DOI={10.1086/312024},
   number={1},
   journal={\apj},
   publisher={American Astronomical Society},
   author={Geller, M. J. and Diaferio, Antonaldo and Kurtz, M. J.},
   year={1999},
   month=may,
   pages={L23–L26}
}

@article{Carlberg_1997,
   title={The Average Mass Profile of Galaxy Clusters},
   volume={485},
   ISSN={0004-637X},
   DOI={10.1086/310801},
   number={1},
   journal={\apj},
   publisher={American Astronomical Society},
   author={Carlberg, R. G. and Yee, H. K. C. and Ellingson, E. and Morris, S. L. and Abraham, R. and Gravel, P. and Pritchet, C. J. and Smecker-Hane, T. and Hartwick, F. D. A. and Hesser, J. E. and Hutchings, J. B. and Oke, J. B.},
   year={1997},
   month=aug,
   pages={L13–L16}
}

@ARTICLE{Heisler85,
       author = {{Heisler}, J. and {Tremaine}, S. and {Bahcall}, J.~N.},
        title = "{Estimating the masses of galaxy groups: alternatives to the virial theorem.}",
      journal = {\apj},
         year = 1985,
        month = nov,
       volume = {298},
        pages = {8-17},
          doi = {10.1086/163584},
       adsurl = {https://ui.adsabs.harvard.edu/abs/1985ApJ...298....8H}
}

@article{Smith1936,
  author       = {Smith, S.},
  title        = "{The Mass of the Virgo Cluster}",
  journal      = {\apj},
  volume       = {83},
  pages        = {23},
  year         = {1936},
  doi          = {10.1086/143716}
}

@article{Page1952,
  author       = {Page, T.},
  title        = {The Masses of Clusters of Galaxies},
  journal      = {\apj},
  volume       = {116},
  pages        = {63},
  year         = {1952},
  doi          = {10.1086/145599}
}

@ARTICLE{4most19,
       author = {{de Jong}, R.~S. and {Agertz}, O. and {Berbel}, A.~A. and {Aird}, J. and {Alexander}, D.~A. and {Amarsi}, A. and {Anders}, F. and {Andrae}, R. and {Ansarinejad}, B. and {Ansorge}, W. and {Antilogus}, P. and {Anwand-Heerwart}, H. and {Arentsen}, A. and {Arnadottir}, A. and {Asplund}, M. and {Auger}, M. and {Azais}, N. and {Baade}, D. and {Baker}, G. and {Baker}, S. and {Balbinot}, E. and {Baldry}, I.~K. and {Banerji}, M. and {Barden}, S. and {Barklem}, P. and {Barth{\'e}l{\'e}my-Mazot}, E. and {Battistini}, C. and {Bauer}, S. and {Bell}, C.~P.~M. and {Bellido-Tirado}, O. and {Bellstedt}, S. and {Belokurov}, V. and {Bensby}, T. and {Bergemann}, M. and {Bestenlehner}, J.~M. and {Bielby}, R. and {Bilicki}, M. and {Blake}, C. and {Bland-Hawthorn}, J. and {Boeche}, C. and {Boland}, W. and {Boller}, T. and {Bongard}, S. and {Bongiorno}, A. and {Bonifacio}, P. and {Boudon}, D. and {Brooks}, D. and {Brown}, M.~J.~I. and {Brown}, R. and {Br{\"u}ggen}, M. and {Brynnel}, J. and {Brzeski}, J. and {Buchert}, T. and {Buschkamp}, P. and {Caffau}, E. and {Caillier}, P. and {Carrick}, J. and {Casagrande}, L. and {Case}, S. and {Casey}, A. and {Cesarini}, I. and {Cescutti}, G. and {Chapuis}, D. and {Chiappini}, C. and {Childress}, M. and {Christlieb}, N. and {Church}, R. and {Cioni}, M.-R.~L. and {Cluver}, M. and {Colless}, M. and {Collett}, T. and {Comparat}, J. and {Cooper}, A. and {Couch}, W. and {Courbin}, F. and {Croom}, S. and {Croton}, D. and {Daguis{\'e}}, E. and {Dalton}, G. and {Davies}, L.~J.~M. and {Davis}, T. and {de Laverny}, P. and {Deason}, A. and {Dionies}, F. and {Disseau}, K. and {Doel}, P. and {D{\"o}scher}, D. and {Driver}, S.~P. and {Dwelly}, T. and {Eckert}, D. and {Edge}, A. and {Edvardsson}, B. and {Youssoufi}, D.~E. and {Elhaddad}, A. and {Enke}, H. and {Erfanianfar}, G. and {Farrell}, T. and {Fechner}, T. and {Feiz}, C. and {Feltzing}, S. and {Ferreras}, I. and {Feuerstein}, D. and {Feuillet}, D. and {Finoguenov}, A. and {Ford}, D. and {Fotopoulou}, S. and {Fouesneau}, M. and {Frenk}, C. and {Frey}, S. and {Gaessler}, W. and {Geier}, S. and {Gentile Fusillo}, N. and {Gerhard}, O. and {Giannantonio}, T. and {Giannone}, D. and {Gibson}, B. and {Gillingham}, P. and {Gonz{\'a}lez-Fern{\'a}ndez}, C. and {Gonzalez-Solares}, E. and {Gottloeber}, S. and {Gould}, A. and {Grebel}, E.~K. and {Gueguen}, A. and {Guiglion}, G. and {Haehnelt}, M. and {Hahn}, T. and {Hansen}, C.~J. and {Hartman}, H. and {Hauptner}, K. and {Hawkins}, K. and {Haynes}, D. and {Haynes}, R. and {Heiter}, U. and {Helmi}, A. and {Aguayo}, C.~H. and {Hewett}, P. and {Hinton}, S. and {Hobbs}, D. and {Hoenig}, S. and {Hofman}, D. and {Hook}, I. and {Hopgood}, J. and {Hopkins}, A. and {Hourihane}, A. and {Howes}, L. and {Howlett}, C. and {Huet}, T. and {Irwin}, M. and {Iwert}, O. and {Jablonka}, P. and {Jahn}, T. and {Jahnke}, K. and {Jarno}, A. and {Jin}, S. and {Jofre}, P. and {Johl}, D. and {Jones}, D. and {J{\"o}nsson}, H. and {Jordan}, C. and {Karovicova}, I. and {Khalatyan}, A. and {Kelz}, A. and {Kennicutt}, R. and {King}, D. and {Kitaura}, F. and {Klar}, J. and {Klauser}, U. and {Kneib}, J.-P. and {Koch}, A. and {Koposov}, S. and {Kordopatis}, G. and {Korn}, A. and {Kosmalski}, J. and {Kotak}, R. and {Kovalev}, M. and {Kreckel}, K. and {Kripak}, Y. and {Krumpe}, M. and {Kuijken}, K. and {Kunder}, A. and {Kushniruk}, I. and {Lam}, M.~I. and {Lamer}, G. and {Laurent}, F. and {Lawrence}, J. and {Lehmitz}, M. and {Lemasle}, B. and {Lewis}, J. and {Li}, B. and {Lidman}, C. and {Lind}, K. and {Liske}, J. and {Lizon}, J.-L. and {Loveday}, J. and {Ludwig}, H.-G. and {McDermid}, R.~M. and {Maguire}, K. and {Mainieri}, V. and {Mali}, S. and {Mandel}, H.},
        title = "{4MOST: Project overview and information for the First Call for Proposals}",
      journal = {The Messenger},
         year = 2019,
        month = mar,
       volume = {175},
        pages = {3-11},
          doi = {10.18727/0722-6691/5117},
archivePrefix = {arXiv},
       eprint = {1903.02464},
 primaryClass = {astro-ph.IM},
       adsurl = {https://ui.adsabs.harvard.edu/abs/2019Msngr.175....3D}
}

@ARTICLE{Rosati14,
       author = {{Rosati}, P. and {Balestra}, I. and {Grillo}, C. and {Mercurio}, A. and {Nonino}, M. and {Biviano}, A. and {Girardi}, M. and {Vanzella}, E. and {Clash-VLT Team}},
        title = "{CLASH-VLT: A VIMOS Large Programme to Map the Dark Matter Mass Distribution in Galaxy Clusters and Probe Distant Lensed Galaxies}",
      journal = {The Messenger},
         year = 2014,
        month = dec,
       volume = {158},
        pages = {48-53},
       adsurl = {https://ui.adsabs.harvard.edu/abs/2014Msngr.158...48R}
}

@ARTICLE{Balboni2025,
       author = {{Balboni}, M. and {Ettori}, S. and {Gastaldello}, F. and {Cassano}, R. and {Bonafede}, A. and {Cuciti}, V. and {Botteon}, A. and {Brunetti}, G. and {Bartalucci}, I. and {Gaspari}, M. and {Gavazzi}, R. and {Ghizzardi}, S. and {Gitti}, M. and {Lovisari}, L. and {Maughan}, B.~J. and {Molendi}, S. and {Pointecouteau}, E. and {Pratt}, G.~W. and {Rasia}, E. and {Riva}, G. and {Rossetti}, M. and {Rottgering}, H. and {Sayers}, J. and {van Weeren}, R.~J.},
        title = "{CHEX-MATE: Scaling relations of radio halo profiles for clusters in the LoTSS DR2 area}",
      journal = {\aap},
         year = 2025,
        month = mar,
       volume = {695},
          eid = {A180},
        pages = {A180},
          doi = {10.1051/0004-6361/202453183},
archivePrefix = {arXiv},
       eprint = {2502.18568},
 primaryClass = {astro-ph.CO},
       adsurl = {https://ui.adsabs.harvard.edu/abs/2025A&A...695A.180B}
}

@ARTICLE{Cassano2007,
       author = {{Cassano}, R. and {Brunetti}, G. and {Setti}, G. and {Govoni}, F. and {Dolag}, K.},
        title = "{New scaling relations in cluster radio haloes and the re-acceleration model}",
      journal = {\mnras},
         year = 2007,
        month = jul,
       volume = {378},
       number = {4},
        pages = {1565-1574},
          doi = {10.1111/j.1365-2966.2007.11901.x},
archivePrefix = {arXiv},
       eprint = {0704.3490},
 primaryClass = {astro-ph},
       adsurl = {https://ui.adsabs.harvard.edu/abs/2007MNRAS.378.1565C}
}

@ARTICLE{DESI21,
       author = {{DESI Collaboration} and {Abareshi}, B. and {Aguilar}, J. and {Ahlen}, S. and {Alam}, Shadab and {Alexander}, David M. and {Alfarsy}, R. and {Allen}, L. and {Allende Prieto}, C. and {Alves}, O. and {Ameel}, J. and {Armengaud}, E. and {Asorey}, J. and {Aviles}, Alejandro and {Bailey}, S. and {Balaguera-Antol{\'\i}nez}, A. and {Ballester}, O. and {Baltay}, C. and {Bault}, A. and {Beltran}, S.~F. and {Benavides}, B. and {BenZvi}, S. and {Berti}, A. and {Besuner}, R. and {Beutler}, Florian and {Bianchi}, D. and {Blake}, C. and {Blanc}, P. and {Blum}, R. and {Bolton}, A. and {Bose}, S. and {Bramall}, D. and {Brieden}, S. and {Brodzeller}, A. and {Brooks}, D. and {Brownewell}, C. and {Buckley-Geer}, E. and {Cahn}, R.~N. and {Cai}, Z. and {Canning}, R. and {Capasso}, R. and {Carnero Rosell}, A. and {Carton}, P. and {Casas}, R. and {Castander}, F.~J. and {Cervantes-Cota}, J.~L. and {Chabanier}, S. and {Chaussidon}, E. and {Chuang}, C. and {Circosta}, C. and {Cole}, S. and {Cooper}, A.~P. and {da Costa}, L. and {Cousinou}, M.-C. and {Cuceu}, A. and {Davis}, T.~M. and {Dawson}, K. and {de la Cruz-Noriega}, R. and {de la Macorra}, A. and {de Mattia}, A. and {Della Costa}, J. and {Demmer}, P. and {Derwent}, M. and {Dey}, A. and {Dey}, B. and {Dhungana}, G. and {Ding}, Z. and {Dobson}, C. and {Doel}, P. and {Donald-McCann}, J. and {Donaldson}, J. and {Douglass}, K. and {Duan}, Y. and {Dunlop}, P. and {Edelstein}, J. and {Eftekharzadeh}, S. and {Eisenstein}, D.~J. and {Enriquez-Vargas}, M. and {Escoffier}, S. and {Evatt}, M. and {Fagrelius}, P. and {Fan}, X. and {Fanning}, K. and {Fawcett}, V.~A. and {Ferraro}, S. and {Ereza}, J. and {Flaugher}, B. and {Font-Ribera}, A. and {Forero-Romero}, J.~E. and {Frenk}, C.~S. and {Fromenteau}, S. and {G{\"a}nsicke}, B.~T. and {Garcia-Quintero}, C. and {Garrison}, L. and {Gazta{\~n}aga}, E. and {Gerardi}, F. and {Gil-Mar{\'\i}n}, H. and {Gontcho A Gontcho}, S. and {Gonzalez-Morales}, Alma X. and {Gonzalez-de-Rivera}, G. and {Gonzalez-Perez}, V. and {Gordon}, C. and {Graur}, O. and {Green}, D. and {Grove}, C. and {Gruen}, D. and {Gutierrez}, G. and {Guy}, J. and {Hahn}, C. and {Harris}, S. and {Herrera}, D. and {Herrera-Alcantar}, Hiram K. and {Honscheid}, K. and {Howlett}, C. and {Huterer}, D. and {Ir{\v{s}}i{\v{c}}}, V. and {Ishak}, M. and {Jelinsky}, P. and {Jiang}, L. and {Jimenez}, J. and {Jing}, Y.~P. and {Joyce}, R. and {Jullo}, E. and {Juneau}, S. and {Kara{\c{c}}ayl{\i}}, N.~G. and {Karamanis}, M. and {Karcher}, A. and {Karim}, T. and {Kehoe}, R. and {Kent}, S. and {Kirkby}, D. and {Kisner}, T. and {Kitaura}, F. and {Koposov}, S.~E. and {Kov{\'a}cs}, A. and {Kremin}, A. and {Krolewski}, Alex and {L'Huillier}, B. and {Lahav}, O. and {Lambert}, A. and {Lamman}, C. and {Lan}, Ting-Wen and {Landriau}, M. and {Lane}, S. and {Lang}, D. and {Lange}, J.~U. and {Lasker}, J. and {Le Guillou}, L. and {Leauthaud}, A. and {Le Van Suu}, A. and {Levi}, Michael E. and {Li}, T.~S. and {Magneville}, C. and {Manera}, M. and {Manser}, Christopher J. and {Marshall}, B. and {Martini}, Paul and {McCollam}, W. and {McDonald}, P. and {Meisner}, Aaron M. and {Mena-Fern{\'a}ndez}, J. and {Meneses-Rizo}, J. and {Mezcua}, M. and {Miller}, T. and {Miquel}, R. and {Montero-Camacho}, P. and {Moon}, J. and {Moustakas}, J. and {Mueller}, E. and {Mu{\~n}oz-Guti{\'e}rrez}, Andrea and {Myers}, Adam D. and {Nadathur}, S. and {Najita}, J. and {Napolitano}, L. and {Neilsen}, E. and {Newman}, Jeffrey A. and {Nie}, J.~D. and {Ning}, Y. and {Niz}, G. and {Norberg}, P. and {Noriega}, Hern{\'a}n E. and {O'Brien}, T. and {Obuljen}, A. and {Palanque-Delabrouille}, N. and {Palmese}, A. and {Zhiwei}, P. and {Pappalardo}, D. and {PENG}, X. and {Percival}, W.~J. and {Perruchot}, S. and {Pogge}, R. and {Poppett}, C. and {Porredon}, A. and {Prada}, F. and {Prochaska}, J. and {Pucha}, R. and {P{\'e}rez-Fern{\'a}ndez}, A. and {P{\'e}rez-R{\`a}fols}, I. and {Rabinowitz}, D. and {Raichoor}, A.},
        title = "{Overview of the Instrumentation for the Dark Energy Spectroscopic Instrument}",
      journal = {\aj},
         year = 2022,
        month = nov,
       volume = {164},
       number = {5},
          eid = {207},
        pages = {207},
          doi = {10.3847/1538-3881/ac882b},
archivePrefix = {arXiv},
       eprint = {2205.10939},
 primaryClass = {astro-ph.IM},
       adsurl = {https://ui.adsabs.harvard.edu/abs/2022AJ....164..207D}
}

@article{Short_2010,
   title="{The evolution of galaxy cluster X-ray scaling relations: Evolution of cluster scaling relations}",
   volume={408},
   ISSN={0035-8711},
   DOI={10.1111/j.1365-2966.2010.17267.x},
   number={4},
   journal={\mnras},
   publisher={Oxford University Press (OUP)},
   author={Short, C. J. and Thomas, P. A. and Young, O. E. and Pearce, F. R. and Jenkins, A. and Muanwong, O.},
   year={2010},
   month=sep,
   pages={2213–2233}
}

@ARTICLE{Arnaud2010,
       author = {{Arnaud}, M. and {Pratt}, G.~W. and {Piffaretti}, R. and {B{\"o}hringer}, H. and {Croston}, J.~H. and {Pointecouteau}, E.},
        title = "{The universal galaxy cluster pressure profile from a representative sample of nearby systems (REXCESS) and the Y$_{SZ}$ - M$_{500}$ relation}",
      journal = {\aap},
         year = 2010,
        month = jul,
       volume = {517},
          eid = {A92},
        pages = {A92},
          doi = {10.1051/0004-6361/200913416},
archivePrefix = {arXiv},
       eprint = {0910.1234},
 primaryClass = {astro-ph.CO},
       adsurl = {https://ui.adsabs.harvard.edu/abs/2010A&A...517A..92A}
}

@article{Alam_2021,
   title={Completed SDSS-IV extended Baryon Oscillation Spectroscopic Survey: Cosmological implications from two decades of spectroscopic surveys at the Apache Point Observatory},
   volume={103},
   ISSN={2470-0029},
   DOI={10.1103/physrevd.103.083533},
   number={8},
   journal={\prd},
   publisher={American Physical Society (APS)},
   author={Alam, Shadab and Aubert, Marie and Avila, Santiago and Balland, Christophe and Bautista, Julian E. and Bershady, Matthew A. and Bizyaev, Dmitry and Blanton, Michael R. and Bolton, Adam S. and Bovy, Jo and Brinkmann, Jonathan and Brownstein, Joel R. and Burtin, Etienne and Chabanier, Solène and Chapman, Michael J. and Choi, Peter Doohyun and Chuang, Chia-Hsun and Comparat, Johan and Cousinou, Marie-Claude and Cuceu, Andrei and Dawson, Kyle S. and de la Torre, Sylvain and de Mattia, Arnaud and Agathe, Victoria de Sainte and des Bourboux, Hélion du Mas and Escoffier, Stephanie and Etourneau, Thomas and Farr, James and Font-Ribera, Andreu and Frinchaboy, Peter M. and Fromenteau, Sebastien and Gil-Marín, Héctor and Le Goff, Jean-Marc and Gonzalez-Morales, Alma X. and Gonzalez-Perez, Violeta and Grabowski, Kathleen and Guy, Julien and Hawken, Adam J. and Hou, Jiamin and Kong, Hui and Parker, James and Klaene, Mark and Kneib, Jean-Paul and Lin, Sicheng and Long, Daniel and Lyke, Brad W. and de la Macorra, Axel and Martini, Paul and Masters, Karen and Mohammad, Faizan G. and Moon, Jeongin and Mueller, Eva-Maria and Muñoz-Gutiérrez, Andrea and Myers, Adam D. and Nadathur, Seshadri and Neveux, Richard and Newman, Jeffrey A. and Noterdaeme, Pasquier and Oravetz, Audrey and Oravetz, Daniel and Palanque-Delabrouille, Nathalie and Pan, Kaike and Paviot, Romain and Percival, Will J. and Pérez-Ràfols, Ignasi and Petitjean, Patrick and Pieri, Matthew M. and Prakash, Abhishek and Raichoor, Anand and Ravoux, Corentin and Rezaie, Mehdi and Rich, James and Ross, Ashley J. and Rossi, Graziano and Ruggeri, Rossana and Ruhlmann-Kleider, Vanina and Sánchez, Ariel G. and Sánchez, F. Javier and Sánchez-Gallego, José R. and Sayres, Conor and Schneider, Donald P. and Seo, Hee-Jong and Shafieloo, Arman and Slosar, Anže and Smith, Alex and Stermer, Julianna and Tamone, Amelie and Tinker, Jeremy L. and Tojeiro, Rita and Vargas-Magaña, Mariana and Variu, Andrei and Wang, Yuting and Weaver, Benjamin A. and Weijmans, Anne-Marie and Yèche, Christophe and Zarrouk, Pauline and Zhao, Cheng and Zhao, Gong-Bo and Zheng, Zheng},
   year={2021},
   month=apr }

@article{Hopkins_2013,
   title="{Galaxy And Mass Assembly (GAMA): spectroscopic analysis}",
   volume={430},
   ISSN={0035-8711},
   DOI={10.1093/mnras/stt030},
   number={3},
   journal={\mnras},
   publisher={Oxford University Press (OUP)},
   author={Hopkins, A. M. and Driver, S. P. and Brough, S. and Owers, M. S. and Bauer, A. E. and Gunawardhana, M. L. P. and Cluver, M. E. and Colless, M. and Foster, C. and Lara-López, M. A. and Roseboom, I. and Sharp, R. and Steele, O. and Thomas, D. and Baldry, I. K. and Brown, M. J. I. and Liske, J. and Norberg, P. and Robotham, A. S. G. and Bamford, S. and Bland-Hawthorn, J. and Drinkwater, M. J. and Loveday, J. and Meyer, M. and Peacock, J. A. and Tuffs, R. and Agius, N. and Alpaslan, M. and Andrae, E. and Cameron, E. and Cole, S. and Ching, J. H. Y. and Christodoulou, L. and Conselice, C. and Croom, S. and Cross, N. J. G. and De Propris, R. and Delhaize, J. and Dunne, L. and Eales, S. and Ellis, S. and Frenk, C. S. and Graham, Alister W. and Grootes, M. W. and Häußler, B. and Heymans, C. and Hill, D. and Hoyle, B. and Hudson, M. and Jarvis, M. and Johansson, J. and Jones, D. H. and van Kampen, E. and Kelvin, L. and Kuijken, K. and López-Sánchez, Á. and Maddox, S. and Madore, B. and Maraston, C. and McNaught-Roberts, T. and Nichol, R. C. and Oliver, S. and Parkinson, H. and Penny, S. and Phillipps, S. and Pimbblet, K. A. and Ponman, T. and Popescu, C. C. and Prescott, M. and Proctor, R. and Sadler, E. M. and Sansom, A. E. and Seibert, M. and Staveley-Smith, L. and Sutherland, W. and Taylor, E. and Van Waerbeke, L. and Vázquez-Mata, J. A. and Warren, S. and Wijesinghe, D. B. and Wild, V. and Wilkins, S.},
   year={2013},
   month=feb,
   pages={2047–2066}
}

@article{Kimmig_2023,
   title="{The Hateful Eight: Connecting Massive Substructures in Galaxy Clusters like A2744 to Their Dynamical Assembly State Using the Magneticum Simulations}",
   volume={949},
   ISSN={1538-4357},
   DOI={10.3847/1538-4357/acc740},
   number={2},
   journal={\apj},
   publisher={American Astronomical Society},
   author={Kimmig, Lucas C. and Remus, Rhea-Silvia and Dolag, Klaus and Biffi, Veronica},
   year={2023},
   month=jun,
   pages={92}
}

@article{De_Luca_2021,
   title="{The Three Hundred project: dynamical state of galaxy clusters and morphology from multiwavelength synthetic maps}",
   volume={504},
   ISSN={1365-2966},
   DOI={10.1093/mnras/stab1073},
   number={4},
   journal={\mnras},
   publisher={Oxford University Press (OUP)},
   author={De Luca, Federico and De Petris, Marco and Yepes, Gustavo and Cui, Weiguang and Knebe, Alexander and Rasia, Elena},
   year={2021},
   month=apr,
   pages={5383–5400}
}

@article{Mu_oz_Rodr_guez_2022,
   title={Cosmic evolution of the incidence of active galactic nuclei in massive clusters: simulations versus observations},
   volume={518},
   ISSN={1365-2966},
   DOI={10.1093/mnras/stac3114},
   number={1},
   journal={\mnras},
   publisher={Oxford University Press (OUP)},
   author={Muñoz Rodríguez, Iván and Georgakakis, Antonis and Shankar, Francesco and Allevato, Viola and Bonoli, Silvia and Brusa, Marcella and Lapi, Andrea and Viitanen, Akke},
   year={2022},
   month=oct,
   pages={1041–1056}
}

@article{Sif_n_2025,
   title="{CHANCES, the Chilean Cluster Galaxy Evolution Survey: Selection and initial characterisation of clusters and superclusters}",
   volume={697},
   ISSN={1432-0746},
   DOI={10.1051/0004-6361/202452710},
   journal={\aap},
   publisher={EDP Sciences},
   author={Sifón, Cristóbal and Finoguenov, Alexis and Haines, Christopher P. and Jaffé, Yara and Amrutha, B. M. and Demarco, Ricardo and Lima, E. V. R. and Lima-Dias, Ciria and Méndez-Hernández, Hugo and Merluzzi, Paola and Monachesi, Antonela and Teixeira, Gabriel S. M. and Tejos, Nicolas and Almeida-Fernandes, F. and Araya-Araya, Pablo and Argudo-Fernández, Maria and Baier-Soto, Raúl and Bilton, Lawrence E. and Bom, C. R. and Calderón, Juan Pablo and Cassarà, Letizia P. and Comparat, Johan and Courtois, H. M. and D’Ago, Giuseppe and Dupuy, Alexandra and Fritz, Alexander and Haack, Rodrigo F. and Herpich, Fabio R. and Ibar, E. and Kuchner, Ulrike and Lacerna, Ivan and Lopes, Amanda R. and Lopez, Sebastian and Lösch, Elismar and McGee, Sean and Mendes de Oliveira, C. and Morelli, Lorenzo and Moretti, Alessia and Pallero, Diego and Piraino-Cerda, Franco and Pompei, Emanuela and Rescigno, U. and Smith Castelli, Analía V. and Smith, Rory and Sodré Jr, Laerte and Tempel, Elmo},
   year={2025},
   month=may,
   pages={A92}
}

@ARTICLE{Bertotti2003,
       author = {{Bertotti}, B. and {Iess}, L. and {Tortora}, P.},
        title = "{A test of general relativity using radio links with the Cassini spacecraft}",
      journal = {\nat},
         year = 2003,
        month = sep,
       volume = {425},
       number = {6956},
        pages = {374-376},
          doi = {10.1038/nature01997},
       adsurl = {https://ui.adsabs.harvard.edu/abs/2003Natur.425..374B}
}

@article{Krishnendu_2021,
   title={Testing General Relativity with Gravitational Waves: An Overview},
   volume={7},
   ISSN={2218-1997},
   DOI={10.3390/universe7120497},
   number={12},
   journal={Universe},
   publisher={MDPI AG},
   author={Krishnendu, N. V. and Ohme, Frank},
   year={2021},
   month=Dec,
   pages={497}
}

@article{Ishak_2018,
   title={Testing general relativity in cosmology},
   volume={22},
   ISSN={1433-8351},
   DOI={10.1007/s41114-018-0017-4},
   journal={Living Rev. Relativ.},
   publisher={Springer Science and Business Media LLC},
   author={Ishak, Mustapha},
   year={2018},
   month=Dec }

@article{Rosselli_2023,
   title={Testing general relativity: New measurements of gravitational redshift in galaxy clusters},
   volume={669},
   ISSN={1432-0746},
   DOI={10.1051/0004-6361/202244244},
   journal={\aap},
   publisher={EDP Sciences},
   author={Rosselli, D. and Marulli, F. and Veropalumbo, A. and Cimatti, A. and Moscardini, L.},
   year={2023},
   month=Jan,
   pages={A29}
}

@ARTICLE{Chameleon,
       author = {{Khoury}, Justin and {Weltman}, Amanda},
        title = "{Chameleon cosmology}",
      journal = {\prd},
         year = {2004},
        month = {feb},
       volume = {69},
       number = {4},
          eid = {044026},
        pages = {044026},
          doi = {10.1103/PhysRevD.69.044026},
archivePrefix = {arXiv},
       eprint = {astro-ph/0309411},
 primaryClass = {astro-ph},
       adsurl = {https://ui.adsabs.harvard.edu/abs/2004PhRvD..69d4026K}
}

@ARTICLE{Wilcox2015,
       author = {{Wilcox}, Harry and {Bacon}, David and {Nichol}, Robert C. and {Rooney}, Philip J. and {Terukina}, Ayumu and {Romer}, A. Kathy and {Koyama}, Kazuya and {Zhao}, Gong-Bo and {Hood}, Ross and {Mann}, Robert G. and {Hilton}, Matt and {Manolopoulou}, Maria and {Sahl{\'e}n}, Martin and {Collins}, Chris A. and {Liddle}, Andrew R. and {Mayers}, Julian A. and {Mehrtens}, Nicola and {Miller}, Christopher J. and {Stott}, John P. and {Viana}, Pedro T.~P.},
        title = "{The XMM Cluster Survey: testing chameleon gravity using the profiles of clusters}",
      journal = {\mnras},
         year = 2015,
        month = sep,
       volume = {452},
       number = {2},
        pages = {1171-1183},
          doi = {10.1093/mnras/stv1366},
archivePrefix = {arXiv},
       eprint = {1504.03937},
 primaryClass = {astro-ph.CO},
       adsurl = {https://ui.adsabs.harvard.edu/abs/2015MNRAS.452.1171W}
}

@ARTICLE{Pizzuti26b,
       author = {{Pizzuti}, Lorenzo and {Rivano}, Federico and {Umetsu}, Keiichi and {Biviano}, Andrea},
        title = "{CLASH-VLT: The Fifth Force in Chameleon Gravity from Joint Lensing and Kinematics Cluster Mass Profiles}",
      journal = {Universe},
         year = 2026,
       volume = {12},
       number = {5},
          eid = {124},
        pages = {124},
          doi = {10.3390/universe12050124},
archivePrefix = {arXiv},
       eprint = {2604.23631},
 primaryClass = {astro-ph.CO},
       adsurl = {https://ui.adsabs.harvard.edu/abs/2026arXiv260423631P}
}

@article{Pizzuti2024b,
   title={Mass modeling and kinematics of galaxy clusters in modified gravity},
   volume={2024},
   ISSN={1475-7516},
   DOI={10.1088/1475-7516/2024/11/014},
   number={11},
   journal={\jcap},
   publisher={IOP Publishing},
   author={Pizzuti, Lorenzo and Boumechta, Yacer and Haridasu, Sandeep and Pombo, Alexandre M. and Dossena, Sofia and Butt, Minahil Adil and Benetti, Francesco and Baccigalupi, Carlo and Lapi, Andrea},
   year={2024},
   month=Nov,
   pages={014}
}

@ARTICLE{Koyama2016,
       author = {{Koyama}, Kazuya},
        title = "{Cosmological tests of modified gravity}",
      journal = {\rpp},
         year = 2016,
        month = apr,
       volume = {79},
       number = {4},
          eid = {046902},
        pages = {046902},
          doi = {10.1088/0034-4885/79/4/046902},
archivePrefix = {arXiv},
       eprint = {1504.04623},
 primaryClass = {astro-ph.CO},
       adsurl = {https://ui.adsabs.harvard.edu/abs/2016RPPh...79d6902K}
}

@ARTICLE{Jyoti2019,
       author = {{Jyoti}, Dhrubo and {Mu{\~n}oz}, Julian B. and {Caldwell}, Robert R. and {Kamionkowski}, Marc},
        title = "{Cosmic time slip: Testing gravity on supergalactic scales with strong-lensing time delays}",
      journal = {\prd},
         year = 2019,
        month = aug,
       volume = {100},
       number = {4},
          eid = {043031},
        pages = {043031},
          doi = {10.1103/PhysRevD.100.043031},
archivePrefix = {arXiv},
       eprint = {1906.06324},
 primaryClass = {astro-ph.CO},
       adsurl = {https://ui.adsabs.harvard.edu/abs/2019PhRvD.100d3031J}
}

@article{Ferraro_2011,
       author = {{Ferraro}, Simone and {Schmidt}, Fabian and {Hu}, Wayne},
        title = "{Cluster abundance in $f(R)$ gravity models}",
      journal = {\prd},
         year = 2011,
        month = mar,
       volume = {83},
       number = {6},
          eid = {063503},
        pages = {063503},
          doi = {10.1103/PhysRevD.83.063503},
archivePrefix = {arXiv},
       eprint = {1011.0992},
 primaryClass = {astro-ph.CO},
       adsurl = {https://ui.adsabs.harvard.edu/abs/2011PhRvD..83f3503F}
}

@ARTICLE{Cataneo2015,
       author = {{Cataneo}, Matteo and {Rapetti}, David and {Schmidt}, Fabian and {Mantz}, Adam B. and {Allen}, Steven W. and {Applegate}, Douglas E. and {Kelly}, Patrick L. and {von der Linden}, Anja and {Morris}, R. Glenn},
        title = "{New constraints on $f(R)$ gravity from clusters of galaxies}",
      journal = {\prd},
         year = 2015,
        month = aug,
       volume = {92},
       number = {4},
          eid = {044009},
        pages = {044009},
          doi = {10.1103/PhysRevD.92.044009},
archivePrefix = {arXiv},
       eprint = {1412.0133},
 primaryClass = {astro-ph.CO},
       adsurl = {https://ui.adsabs.harvard.edu/abs/2015PhRvD..92d4009C}
}

@article{Vogt24,
  title = {Constraining $f(R)$ gravity using future galaxy cluster abundance and weak-lensing mass calibration datasets},
  author = {Vogt, Sophie M. L. and Bocquet, Sebastian and Davies, Christopher T. and Mohr, Joseph J. and Schmidt, Fabian},
  journal = {Phys. Rev. D},
  volume = {109},
  issue = {12},
  pages = {123503},
  numpages = {24},
  year = {2024},
  month = {Jun},
  publisher = {American Physical Society},
  doi = {10.1103/PhysRevD.109.123503}
}

@article{Terukina_2014,
   title="{Testing chameleon gravity with the Coma cluster}",
   volume={2014},
   ISSN={1475-7516},
   DOI={10.1088/1475-7516/2014/04/013},
   number={04},
   journal={\jcap},
   publisher={IOP Publishing},
   author={Terukina, Ayumu and Lombriser, Lucas and Yamamoto, Kazuhiro and Bacon, David and Koyama, Kazuya and Nichol, Robert C.},
   year={2014},
   month=Apr,
   pages={013–013}
}

@article{Mitchell_2021,
   title="{A general framework to test gravity using galaxy clusters III: observable-mass scaling relations in $f(R)$ gravity}",
   volume={502},
   ISSN={1365-2966},
   DOI={10.1093/mnras/stab479},
   number={4},
   journal={\mnras},
   publisher={Oxford University Press (OUP)},
   author={Mitchell, Myles A and Arnold, Christian and Li, Baojiu},
   year={2021},
   month=Feb,
   pages={6101–6116}
}

@article{Lombriser_2012,
   title={Cluster density profiles as a test of modified gravity},
   volume={85},
   ISSN={1550-2368},
   DOI={10.1103/physrevd.85.102001},
   number={10},
   journal={\prd},
   publisher={American Physical Society (APS)},
   author={Lombriser, Lucas and Schmidt, Fabian and Baldauf, Tobias and Mandelbaum, Rachel and Seljak, Uroš and Smith, Robert E.},
   year={2012},
   month=May }

@ARTICLE{Shankaranarayanan2022,
       author = {{Shankaranarayanan}, S. and {Johnson}, Joseph P.},
        title = "{Modified theories of gravity: Why, how and what?}",
      journal = {Gen. Relativ. Gravit.},
         year = 2022,
        month = may,
       volume = {54},
       number = {5},
          eid = {44},
        pages = {44},
          doi = {10.1007/s10714-022-02927-2},
archivePrefix = {arXiv},
       eprint = {2204.06533},
 primaryClass = {gr-qc},
       adsurl = {https://ui.adsabs.harvard.edu/abs/2022GReGr..54...44S}
}

@BOOK{Cantata21,
       author = {{Saridakis}, Emmanuel N. and {Lazkoz}, Ruth and {Salzano}, Vincenzo and {Moniz}, Paulo Vargas and {Capozziello}, Salvatore and {Beltr{\'a}n Jim{\'e}nez}, Jose and {De Laurentis}, Mariafelicia and {Olmo}, Gonzalo J.},
        title = "{Modified Gravity and Cosmology: An Update by the {CANTATA} Network}",
         year = 2021,
    publisher = {Springer},
      address = {Cham},
          doi = {10.1007/978-3-030-83715-0},
       adsurl = {https://ui.adsabs.harvard.edu/abs/2021mgcu.book.....S}
}

@ARTICLE{Clifton2012,
       author = {{Clifton}, Timothy and {Ferreira}, Pedro G. and {Padilla}, Antonio and {Skordis}, Constantinos},
        title = "{Modified gravity and cosmology}",
      journal = {\physrep},
         year = 2012,
        month = mar,
       volume = {513},
       number = {1},
        pages = {1-189},
          doi = {10.1016/j.physrep.2012.01.001},
archivePrefix = {arXiv},
       eprint = {1106.2476},
 primaryClass = {astro-ph.CO},
       adsurl = {https://ui.adsabs.harvard.edu/abs/2012PhR...513....1C}
}

@article{YOO_2012,
   title={THEORETICAL MODELS OF DARK ENERGY},
   volume={21},
   ISSN={1793-6594},
   DOI={10.1142/s0218271812300029},
   number={12},
   journal={\ijmp},
   publisher={World Scientific Pub Co Pte Ltd},
   author={Yoo, JAEWON and Watanabe, YUKI},
   year={2012},
   month=Nov,
   pages={1230002}
}

@article{kjpb-r698,
  title = {Scalar-field dark energy models: Current and forecast constraints},
  author = {Shajib, Anowar J. and Frieman, Joshua A.},
  journal = {Phys. Rev. D},
  volume = {112},
  issue = {6},
  pages = {063508},
  numpages = {15},
  year = {2025},
  month = {Sep},
  publisher = {American Physical Society},
  doi = {10.1103/kjpb-r698}
}

@article{Abbott_2022,
   title="{Dark Energy Survey Year 3 results: Cosmological constraints from galaxy clustering and weak lensing}",
   volume={105},
   ISSN={2470-0029},
   DOI={10.1103/physrevd.105.023520},
   number={2},
   journal={\prd},
   publisher={American Physical Society (APS)},
   author={Abbott, T. M. C. and Aguena, M. and Alarcon, A. and Allam, S. and Alves, O. and Amon, A. and Andrade-Oliveira, F. and Annis, J. and Avila, S. and Bacon, D. and Baxter, E. and Bechtol, K. and Becker, M. R. and Bernstein, G. M. and Bhargava, S. and Birrer, S. and Blazek, J. and Brandao-Souza, A. and Bridle, S. L. and Brooks, D. and Buckley-Geer, E. and Burke, D. L. and Camacho, H. and Campos, A. and Carnero Rosell, A. and Carrasco Kind, M. and Carretero, J. and Castander, F. J. and Cawthon, R. and Chang, C. and Chen, A. and Chen, R. and Choi, A. and Conselice, C. and Cordero, J. and Costanzi, M. and Crocce, M. and da Costa, L. N. and da Silva Pereira, M. E. and Davis, C. and Davis, T. M. and De Vicente, J. and DeRose, J. and Desai, S. and Di Valentino, E. and Diehl, H. T. and Dietrich, J. P. and Dodelson, S. and Doel, P. and Doux, C. and Drlica-Wagner, A. and Eckert, K. and Eifler, T. F. and Elsner, F. and Elvin-Poole, J. and Everett, S. and Evrard, A. E. and Fang, X. and Farahi, A. and Fernandez, E. and Ferrero, I. and Ferté, A. and Fosalba, P. and Friedrich, O. and Frieman, J. and García-Bellido, J. and Gatti, M. and Gaztanaga, E. and Gerdes, D. W. and Giannantonio, T. and Giannini, G. and Gruen, D. and Gruendl, R. A. and Gschwend, J. and Gutierrez, G. and Harrison, I. and Hartley, W. G. and Herner, K. and Hinton, S. R. and Hollowood, D. L. and Honscheid, K. and Hoyle, B. and Huff, E. M. and Huterer, D. and Jain, B. and James, D. J. and Jarvis, M. and Jeffrey, N. and Jeltema, T. and Kovacs, A. and Krause, E. and Kron, R. and Kuehn, K. and Kuropatkin, N. and Lahav, O. and Leget, P.-F. and Lemos, P. and Liddle, A. R. and Lidman, C. and Lima, M. and Lin, H. and MacCrann, N. and Maia, M. A. G. and Marshall, J. L. and Martini, P. and McCullough, J. and Melchior, P. and Mena-Fernández, J. and Menanteau, F. and Miquel, R. and Mohr, J. J. and Morgan, R. and Muir, J. and Myles, J. and Nadathur, S. and Navarro-Alsina, A. and Nichol, R. C. and Ogando, R. L. C. and Omori, Y. and Palmese, A. and Pandey, S. and Park, Y. and Paz-Chinchón, F. and Petravick, D. and Pieres, A. and Plazas Malagón, A. A. and Porredon, A. and Prat, J. and Raveri, M. and Rodriguez-Monroy, M. and Rollins, R. P. and Romer, A. K. and Roodman, A. and Rosenfeld, R. and Ross, A. J. and Rykoff, E. S. and Samuroff, S. and Sánchez, C. and Sanchez, E. and Sanchez, J. and Sanchez Cid, D. and Scarpine, V. and Schubnell, M. and Scolnic, D. and Secco, L. F. and Serrano, S. and Sevilla-Noarbe, I. and Sheldon, E. and Shin, T. and Smith, M. and Soares-Santos, M. and Suchyta, E. and Swanson, M. E. C. and Tabbutt, M. and Tarle, G. and Thomas, D. and To, C. and Troja, A. and Troxel, M. A. and Tucker, D. L. and Tutusaus, I. and Varga, T. N. and Walker, A. R. and Weaverdyck, N. and Wechsler, R. and Weller, J. and Yanny, B. and Yin, B. and Zhang, Y. and Zuntz, J.},
   year={2022},
   month=jan }

@article{Adame_2025,
   title="{DESI 2024 VI:  cosmological constraints from the measurements of baryon acoustic oscillations}",
   volume={2025},
   ISSN={1475-7516},
   DOI={10.1088/1475-7516/2025/02/021},
   number={02},
   journal={\jcap},
   publisher={IOP Publishing},
   author={Adame, A.G. and Aguilar, J. and Ahlen, S. and Alam, S. and Alexander, D.M. and Alvarez, M. and Alves, O. and Anand, A. and Andrade, U. and Armengaud, E. and Avila, S. and Aviles, A. and Awan, H. and Bahr-Kalus, B. and Bailey, S. and Baltay, C. and Bault, A. and Behera, J. and BenZvi, S. and Bera, A. and Beutler, F. and Bianchi, D. and Blake, C. and Blum, R. and Brieden, S. and Brodzeller, A. and Brooks, D. and Buckley-Geer, E. and Burtin, E. and Calderon, R. and Canning, R. and Carnero Rosell, A. and Cereskaite, R. and Cervantes-Cota, J.L. and Chabanier, S. and Chaussidon, E. and Chaves-Montero, J. and Chen, S. and Chen, X. and Claybaugh, T. and Cole, S. and Cuceu, A. and Davis, T.M. and Dawson, K. and de la Macorra, A. and de Mattia, A. and Deiosso, N. and Dey, A. and Dey, B. and Ding, Z. and Doel, P. and Edelstein, J. and Eftekharzadeh, S. and Eisenstein, D.J. and Elliott, A. and Fagrelius, P. and Fanning, K. and Ferraro, S. and Ereza, J. and Findlay, N. and Flaugher, B. and Font-Ribera, A. and Forero-Sánchez, D. and Forero-Romero, J.E. and Frenk, C.S. and Garcia-Quintero, C. and Gaztañaga, E. and Gil-Marín, H. and Gontcho, S.Gontcho A. and Gonzalez-Morales, A.X. and Gonzalez-Perez, V. and Gordon, C. and Green, D. and Gruen, D. and Gsponer, R. and Gutierrez, G. and Guy, J. and Hadzhiyska, B. and Hahn, C. and Hanif, M.M.S. and Herrera-Alcantar, H.K. and Honscheid, K. and Howlett, C. and Huterer, D. and Iršič, V. and Ishak, M. and Juneau, S. and Karaçaylı, N.G. and Kehoe, R. and Kent, S. and Kirkby, D. and Kremin, A. and Krolewski, A. and Lai, Y. and Lan, T.-W. and Landriau, M. and Lang, D. and Lasker, J. and Le Goff, J.M. and Le Guillou, L. and Leauthaud, A. and Levi, M.E. and Li, T.S. and Linder, E. and Lodha, K. and Magneville, C. and Manera, M. and Margala, D. and Martini, P. and Maus, M. and McDonald, P. and Medina-Varela, L. and Meisner, A. and Mena-Fernández, J. and Miquel, R. and Moon, J. and Moore, S. and Moustakas, J. and Mueller, E. and Muñoz-Gutiérrez, A. and Myers, A.D. and Nadathur, S. and Napolitano, L. and Neveux, R. and Newman, J.A. and Nguyen, N.M. and Nie, J. and Niz, G. and Noriega, H.E. and Padmanabhan, N. and Paillas, E. and Palanque-Delabrouille, N. and Pan, J. and Penmetsa, S. and Percival, W.J. and Pieri, M.M. and Pinon, M. and Poppett, C. and Porredon, A. and Prada, F. and Pérez-Fernández, A. and Pérez-Ràfols, I. and Rabinowitz, D. and Raichoor, A. and Ramírez-Pérez, C. and Ramirez-Solano, S. and Rashkovetskyi, M. and Ravoux, C. and Rezaie, M. and Rich, J. and Rocher, A. and Rockosi, C. and Roe, N.A. and Rosado-Marin, A. and Ross, A.J. and Rossi, G. and Ruggeri, R. and Ruhlmann-Kleider, V. and Samushia, L. and Sanchez, E. and Saulder, C. and Schlafly, E.F. and Schlegel, D. and Schubnell, M. and Seo, H. and Shafieloo, A. and Sharples, R. and Silber, J. and Slosar, A. and Smith, A. and Sprayberry, D. and Tan, T. and Tarlé, G. and Taylor, P. and Trusov, S. and Ureña-López, L.A. and Vaisakh, R. and Valcin, D. and Valdes, F. and Vargas-Magaña, M. and Verde, L. and Walther, M. and Wang, B. and Wang, M.S. and Weaver, B.A. and Weaverdyck, N. and Wechsler, R.H. and Weinberg, D.H. and White, M. and Yu, J. and Yu, Y. and Yuan, S. and Yèche, C. and Zaborowski, E.A. and Zarrouk, P. and Zhang, H. and Zhao, C. and Zhao, R. and Zhou, R. and Zhuang, T. and Zou, H.},
   year={2025},
   month=feb,
   pages={021}
}

@article{PhysRevD.110.123508,
  title = {Quantifying the ${S}_{8}$ tension and evidence for interacting dark energy from redshift-space distortion measurements},
  author = {Sabogal, Miguel A. and Silva, Emanuelly and Nunes, Rafael C. and Kumar, Suresh and Di Valentino, Eleonora and Giar\`e, William},
  journal = {Phys. Rev. D},
  volume = {110},
  issue = {12},
  pages = {123508},
  numpages = {13},
  year = {2024},
  month = {Dec},
  publisher = {American Physical Society},
  doi = {10.1103/PhysRevD.110.123508}
}

@article{Planck2018,
   title="{Planck 2018 results: VI. Cosmological parameters}",
   volume={641},
   ISSN={1432-0746},
   DOI={10.1051/0004-6361/201833910},
   journal={\aap},
   publisher={EDP Sciences},
   author={Aghanim, N. and Akrami, Y. and Ashdown, M. and Aumont, J. and Baccigalupi, C. and Ballardini, M. and Banday, A. J. and Barreiro, R. B. and Bartolo, N. and Basak, S. and Battye, R. and Benabed, K. and Bernard, J.-P. and Bersanelli, M. and Bielewicz, P. and Bock, J. J. and Bond, J. R. and Borrill, J. and Bouchet, F. R. and Boulanger, F. and Bucher, M. and Burigana, C. and Butler, R. C. and Calabrese, E. and Cardoso, J.-F. and Carron, J. and Challinor, A. and Chiang, H. C. and Chluba, J. and Colombo, L. P. L. and Combet, C. and Contreras, D. and Crill, B. P. and Cuttaia, F. and de Bernardis, P. and de Zotti, G. and Delabrouille, J. and Delouis, J.-M. and Di Valentino, E. and Diego, J. M. and Doré, O. and Douspis, M. and Ducout, A. and Dupac, X. and Dusini, S. and Efstathiou, G. and Elsner, F. and Enßlin, T. A. and Eriksen, H. K. and Fantaye, Y. and Farhang, M. and Fergusson, J. and Fernandez-Cobos, R. and Finelli, F. and Forastieri, F. and Frailis, M. and Fraisse, A. A. and Franceschi, E. and Frolov, A. and Galeotta, S. and Galli, S. and Ganga, K. and Génova-Santos, R. T. and Gerbino, M. and Ghosh, T. and González-Nuevo, J. and Górski, K. M. and Gratton, S. and Gruppuso, A. and Gudmundsson, J. E. and Hamann, J. and Handley, W. and Hansen, F. K. and Herranz, D. and Hildebrandt, S. R. and Hivon, E. and Huang, Z. and Jaffe, A. H. and Jones, W. C. and Karakci, A. and Keihänen, E. and Keskitalo, R. and Kiiveri, K. and Kim, J. and Kisner, T. S. and Knox, L. and Krachmalnicoff, N. and Kunz, M. and Kurki-Suonio, H. and Lagache, G. and Lamarre, J.-M. and Lasenby, A. and Lattanzi, M. and Lawrence, C. R. and Le Jeune, M. and Lemos, P. and Lesgourgues, J. and Levrier, F. and Lewis, A. and Liguori, M. and Lilje, P. B. and Lilley, M. and Lindholm, V. and López-Caniego, M. and Lubin, P. M. and Ma, Y.-Z. and Macías-Pérez, J. F. and Maggio, G. and Maino, D. and Mandolesi, N. and Mangilli, A. and Marcos-Caballero, A. and Maris, M. and Martin, P. G. and Martinelli, M. and Martínez-González, E. and Matarrese, S. and Mauri, N. and McEwen, J. D. and Meinhold, P. R. and Melchiorri, A. and Mennella, A. and Migliaccio, M. and Millea, M. and Mitra, S. and Miville-Deschênes, M.-A. and Molinari, D. and Montier, L. and Morgante, G. and Moss, A. and Natoli, P. and Nørgaard-Nielsen, H. U. and Pagano, L. and Paoletti, D. and Partridge, B. and Patanchon, G. and Peiris, H. V. and Perrotta, F. and Pettorino, V. and Piacentini, F. and Polastri, L. and Polenta, G. and Puget, J.-L. and Rachen, J. P. and Reinecke, M. and Remazeilles, M. and Renzi, A. and Rocha, G. and Rosset, C. and Roudier, G. and Rubiño-Martín, J. A. and Ruiz-Granados, B. and Salvati, L. and Sandri, M. and Savelainen, M. and Scott, D. and Shellard, E. P. S. and Sirignano, C. and Sirri, G. and Spencer, L. D. and Sunyaev, R. and Suur-Uski, A.-S. and Tauber, J. A. and Tavagnacco, D. and Tenti, M. and Toffolatti, L. and Tomasi, M. and Trombetti, T. and Valenziano, L. and Valiviita, J. and Van Tent, B. and Vibert, L. and Vielva, P. and Villa, F. and Vittorio, N. and Wandelt, B. D. and Wehus, I. K. and White, M. and White, S. D. M. and Zacchei, A. and Zonca, A.},
   year={2020},
   month=sep,
   pages={A6}
}

@ARTICLE{Cicoli22,
       author = {{Cicoli}, Michele},
        title = "{Recent progress on inflation and dark energy from string theory}",
      journal = {Gen. Relativ. Gravit.},
         year = 2026,
        month = mar,
       volume = {58},
       number = {4},
          eid = {33},
        pages = {33},
          doi = {10.1007/s10714-026-03538-x},
archivePrefix = {arXiv},
       eprint = {2604.16281},
 primaryClass = {hep-th},
       adsurl = {https://ui.adsabs.harvard.edu/abs/2026GReGr..58...33C}
}

@article{Weinberg89,
  title = {The cosmological constant problem},
  author = {Weinberg, Steven},
  journal = {Rev. Mod. Phys.},
  volume = {61},
  issue = {1},
  pages = {1--23},
  numpages = {0},
  year = {1989},
  month = {Jan},
  publisher = {American Physical Society},
  doi = {10.1103/RevModPhys.61.1}
}

@ARTICLE{Joyce2016,
       author = {{Joyce}, Austin and {Lombriser}, Lucas and {Schmidt}, Fabian},
        title = "{Dark Energy Versus Modified Gravity}",
      journal = {\arnps},
         year = 2016,
        month = oct,
       volume = {66},
       number = {1},
        pages = {95-122},
          doi = {10.1146/annurev-nucl-102115-044553},
archivePrefix = {arXiv},
       eprint = {1601.06133},
 primaryClass = {astro-ph.CO},
       adsurl = {https://ui.adsabs.harvard.edu/abs/2016ARNPS..66...95J}
}

@ARTICLE{DeFelice2010,
       author = {{De Felice}, Antonio and {Tsujikawa}, Shinji},
        title = "{$f(R)$ Theories}",
      journal = {Living Rev. Relativ.},
         year = 2010,
        month = dec,
       volume = {13},
          eid = {3},
        pages = {3},
          doi = {10.12942/lrr-2010-3},
archivePrefix = {arXiv},
       eprint = {1002.4928},
 primaryClass = {gr-qc},
       adsurl = {https://ui.adsabs.harvard.edu/abs/2010LRR....13....3D}
}

@ARTICLE{Carroll2004,
       author = {{Carroll}, Sean M. and {Duvvuri}, Vikram and {Trodden}, Mark and {Turner}, Michael S.},
        title = "{Is cosmic speed-up due to new gravitational physics?}",
      journal = {\prd},
         year = 2004,
        month = aug,
       volume = {70},
       number = {4},
          eid = {043528},
        pages = {043528},
          doi = {10.1103/PhysRevD.70.043528},
archivePrefix = {arXiv},
       eprint = {astro-ph/0306438},
 primaryClass = {astro-ph},
       adsurl = {https://ui.adsabs.harvard.edu/abs/2004PhRvD..70d3528C}
}

@ARTICLE{Perlmutter99,
       author = {{Perlmutter}, S. and {Aldering}, G. and {Goldhaber}, G. and {Knop}, R.~A. and {Nugent}, P. and {Castro}, P.~G. and {Deustua}, S. and {Fabbro}, S. and {Goobar}, A. and {Groom}, D.~E. and {Hook}, I.~M. and {Kim}, A.~G. and {Kim}, M.~Y. and {Lee}, J.~C. and {Nunes}, N.~J. and {Pain}, R. and {Pennypacker}, C.~R. and {Quimby}, R. and {Lidman}, C. and {Ellis}, R.~S. and {Irwin}, M. and {McMahon}, R.~G. and {Ruiz-Lapuente}, P. and {Walton}, N. and {Schaefer}, B. and {Boyle}, B.~J. and {Filippenko}, A.~V. and {Matheson}, T. and {Fruchter}, A.~S. and {Panagia}, N. and {Newberg}, H.~J.~M. and {Couch}, W.~J. and {Project}, The Supernova Cosmology},
        title = "{Measurements of {\ensuremath{\Omega}} and {\ensuremath{\Lambda}} from 42 High-Redshift Supernovae}",
      journal = {\apj},
         year = 1999,
        month = jun,
       volume = {517},
       number = {2},
        pages = {565-586},
          doi = {10.1086/307221},
archivePrefix = {arXiv},
       eprint = {astro-ph/9812133},
 primaryClass = {astro-ph},
       adsurl = {https://ui.adsabs.harvard.edu/abs/1999ApJ...517..565P}
}

@article{Faber_2006,
   title={Combining rotation curves and gravitational lensing: how to measure the equation of state of dark matter in the galactic halo},
   volume={372},
   ISSN={1365-2966},
   DOI={10.1111/j.1365-2966.2006.10845.x},
   number={1},
   journal={\mnras},
   publisher={Oxford University Press (OUP)},
   author={Faber, T. and Visser, M.},
   year={2006},
   month=Oct,
   pages={136–142}
}

@ARTICLE{Riess1998,
       author = {{Riess}, Adam G. and {Filippenko}, Alexei V. and {Challis}, Peter and {Clocchiatti}, Alejandro and {Diercks}, Alan and {Garnavich}, Peter M. and {Gilliland}, Ron L. and {Hogan}, Craig J. and {Jha}, Saurabh and {Kirshner}, Robert P. and {Leibundgut}, B. and {Phillips}, M.~M. and {Reiss}, David and {Schmidt}, Brian P. and {Schommer}, Robert A. and {Smith}, R. Chris and {Spyromilio}, J. and {Stubbs}, Christopher and {Suntzeff}, Nicholas B. and {Tonry}, John},
        title = "{Observational Evidence from Supernovae for an Accelerating Universe and a Cosmological Constant}",
      journal = {\aj},
         year = 1998,
        month = sep,
       volume = {116},
       number = {3},
        pages = {1009-1038},
          doi = {10.1086/300499},
archivePrefix = {arXiv},
       eprint = {astro-ph/9805201},
 primaryClass = {astro-ph},
       adsurl = {https://ui.adsabs.harvard.edu/abs/1998AJ....116.1009R}
}

@ARTICLE{Stelle1977,
       author = {{Stelle}, K.~S.},
        title = "{Renormalization of higher-derivative quantum gravity}",
      journal = {\prd},
         year = 1977,
        month = aug,
       volume = {16},
       number = {4},
        pages = {953-969},
          doi = {10.1103/PhysRevD.16.953},
       adsurl = {https://ui.adsabs.harvard.edu/abs/1977PhRvD..16..953S}
}

@ARTICLE{Eddington1923,
       author = {{Eddington}, A.~S.},
        title = "{Can Gravitation be Explained?}",
      journal = {\jrasc},
         year = 1923,
        month = dec,
       volume = {17},
        pages = {387},
       adsurl = {https://ui.adsabs.harvard.edu/abs/1923JRASC..17..387E}
}

@article{Liu_2022,
doi = {10.3847/1538-4357/ac4c3b},
year = {2022},
month = {mar},
publisher = {The American Astronomical Society},
volume = {927},
number = {1},
pages = {28},
author = {Liu, Xiao-Hui and Li, Zhen-Hua and Qi, Jing-Zhao and Zhang, Xin},
title = {Galaxy-scale Test of General Relativity with Strong Gravitational Lensing},
journal = {\apj}
}

@ARTICLE{Pizzuti22b,
       author = {{Pizzuti}, Lorenzo},
        title = "{Testing Screening Mechanisms with Mass Profiles of Galaxy Clusters}",
      journal = {Universe},
         year = 2022,
        month = mar,
       volume = {8},
       number = {3},
          eid = {157},
        pages = {157},
          doi = {10.3390/universe8030157},
archivePrefix = {arXiv},
       eprint = {2204.04432},
 primaryClass = {astro-ph.CO},
       adsurl = {https://ui.adsabs.harvard.edu/abs/2022Univ....8..157P}
}

@ARTICLE{Pizzuti19,
       author = {{Pizzuti}, Lorenzo and {Saltas}, Ippocratis D. and {Casas}, Santiago and {Amendola}, Luca and {Biviano}, Andrea},
        title = "{Future constraints on the gravitational slip with the mass profiles of galaxy clusters}",
      journal = {\mnras},
         year = 2019,
        month = jun,
       volume = {486},
       number = {1},
        pages = {596-607},
          doi = {10.1093/mnras/stz825},
archivePrefix = {arXiv},
       eprint = {1901.01961},
 primaryClass = {astro-ph.CO},
       adsurl = {https://ui.adsabs.harvard.edu/abs/2019MNRAS.486..596P}
}

@article{Planck2016,
   title="{Planck 2015 results: XIV. Dark energy and modified gravity}",
   volume={594},
   ISSN={1432-0746},
   DOI={10.1051/0004-6361/201525814},
   journal={\aap},
   publisher={EDP Sciences},
   author={Planck Collaboration: Ade, P. A. R. and Aghanim, N. and Arnaud, M. and Ashdown, M. and Aumont, J. and Baccigalupi, C. and Banday, A. J. and Barreiro, R. B. and Bartolo, N. and Battaner, E. and Battye, R. and Benabed, K. and Benoît, A. and Benoit-Lévy, A. and Bernard, J.-P. and Bersanelli, M. and Bielewicz, P. and Bock, J. J. and Bonaldi, A. and Bonavera, L. and Bond, J. R. and Borrill, J. and Bouchet, F. R. and Bucher, M. and Burigana, C. and Butler, R. C. and Calabrese, E. and Cardoso, J.-F. and Catalano, A. and Challinor, A. and Chamballu, A. and Chiang, H. C. and Christensen, P. R. and Church, S. and Clements, D. L. and Colombi, S. and Colombo, L. P. L. and Combet, C. and Couchot, F. and Coulais, A. and Crill, B. P. and Curto, A. and Cuttaia, F. and Danese, L. and Davies, R. D. and Davis, R. J. and de Bernardis, P. and de Rosa, A. and de Zotti, G. and Delabrouille, J. and Désert, F.-X. and Diego, J. M. and Dole, H. and Donzelli, S. and Doré, O. and Douspis, M. and Ducout, A. and Dupac, X. and Efstathiou, G. and Elsner, F. and Enßlin, T. A. and Eriksen, H. K. and Fergusson, J. and Finelli, F. and Forni, O. and Frailis, M. and Fraisse, A. A. and Franceschi, E. and Frejsel, A. and Galeotta, S. and Galli, S. and Ganga, K. and Giard, M. and Giraud-Héraud, Y. and Gjerløw, E. and González-Nuevo, J. and Górski, K. M. and Gratton, S. and Gregorio, A. and Gruppuso, A. and Gudmundsson, J. E. and Hansen, F. K. and Hanson, D. and Harrison, D. L. and Heavens, A. and Helou, G. and Henrot-Versillé, S. and Hernández-Monteagudo, C. and Herranz, D. and Hildebrandt, S. R. and Hivon, E. and Hobson, M. and Holmes, W. A. and Hornstrup, A. and Hovest, W. and Huang, Z. and Huffenberger, K. M. and Hurier, G. and Jaffe, A. H. and Jaffe, T. R. and Jones, W. C. and Juvela, M. and Keihänen, E. and Keskitalo, R. and Kisner, T. S. and Knoche, J. and Kunz, M. and Kurki-Suonio, H. and Lagache, G. and Lähteenmäki, A. and Lamarre, J.-M. and Lasenby, A. and Lattanzi, M. and Lawrence, C. R. and Leonardi, R. and Lesgourgues, J. and Levrier, F. and Lewis, A. and Liguori, M. and Lilje, P. B. and Linden-Vørnle, M. and López-Caniego, M. and Lubin, P. M. and Ma, Y.-Z. and Macías-Pérez, J. F. and Maggio, G. and Maino, D. and Mandolesi, N. and Mangilli, A. and Marchini, A. and Maris, M. and Martin, P. G. and Martinelli, M. and Martínez-González, E. and Masi, S. and Matarrese, S. and McGehee, P. and Meinhold, P. R. and Melchiorri, A. and Mendes, L. and Mennella, A. and Migliaccio, M. and Mitra, S. and Miville-Deschênes, M.-A. and Moneti, A. and Montier, L. and Morgante, G. and Mortlock, D. and Moss, A. and Munshi, D. and Murphy, J. A. and Narimani, A. and Naselsky, P. and Nati, F. and Natoli, P. and Netterfield, C. B. and Nørgaard-Nielsen, H. U. and Noviello, F. and Novikov, D. and Novikov, I. and Oxborrow, C. A. and Paci, F. and Pagano, L. and Pajot, F. and Paoletti, D. and Pasian, F. and Patanchon, G. and Pearson, T. J. and Perdereau, O. and Perotto, L. and Perrotta, F. and Pettorino, V. and Piacentini, F. and Piat, M. and Pierpaoli, E. and Pietrobon, D. and Plaszczynski, S. and Pointecouteau, E. and Polenta, G. and Popa, L. and Pratt, G. W. and Prézeau, G. and Prunet, S. and Puget, J.-L. and Rachen, J. P. and Reach, W. T. and Rebolo, R. and Reinecke, M. and Remazeilles, M. and Renault, C. and Renzi, A. and Ristorcelli, I. and Rocha, G. and Rosset, C. and Rossetti, M. and Roudier, G. and Rowan-Robinson, M. and Rubiño-Martín, J. A. and Rusholme, B. and Salvatelli, V. and Sandri, M. and Santos, D. and Savelainen, M. and Savini, G. and Schaefer, B. M. and Scott, D. and Seiffert, M. D. and Shellard, E. P. S. and Spencer, L. D. and Stolyarov, V. and Stompor, R. and Sudiwala, R. and Sunyaev, R. and Sutton, D. and Suur-Uski, A.-S. and Sygnet, J.-F. and Tauber, J. A. and Terenzi, L. and Toffolatti, L. and Tomasi, M. and Tristram, M. and Tucci, M. and Tuovinen, J. and Valenziano, L. and Valiviita, J. and Van Tent, B. and Viel, M. and Vielva, P. and Villa, F. and Wade, L. A. and Wandelt, B. D. and Wehus, I. K. and White, M. and Yvon, D. and Zacchei, A. and Zonca, A.},
   year={2016},
   month=sep,
   pages={A14}
}

@article{8r4r-qg6t,
  title = {Testing the equivalence principle on cosmological scales using peculiar acceleration power spectra},
  author = {Lu, Guoyuan and Zheng, Yi and Zhang, Le and Li, Xiaodong and Ding, Jiacheng and Chan, Kwan Chuen},
  journal = {Phys. Rev. D},
  volume = {112},
  issue = {10},
  pages = {103542},
  numpages = {11},
  year = {2025},
  month = {Nov},
  publisher = {American Physical Society},
  doi = {10.1103/8r4r-qg6t}
}

@article{Touboul17,
  title = {MICROSCOPE Mission: First Results of a Space Test of the Equivalence Principle},
  author = {Touboul, Pierre and M\'etris, Gilles and Rodrigues, Manuel and Andr\'e, Yves and Baghi, Quentin and Berg\'e, Jo\"el and Boulanger, Damien and Bremer, Stefanie and Carle, Patrice and Chhun, Ratana and Christophe, Bruno and Cipolla, Valerio and Damour, Thibault and Danto, Pascale and Dittus, Hansjoerg and Fayet, Pierre and Foulon, Bernard and Gageant, Claude and Guidotti, Pierre-Yves and Hagedorn, Daniel and Hardy, Emilie and Huynh, Phuong-Anh and Inchauspe, Henri and Kayser, Patrick and Lala, St\'ephanie and L\"ammerzahl, Claus and Lebat, Vincent and Leseur, Pierre and Liorzou, Fran\ifmmode \mbox{\c{c}}\else \c{c}\fi{}oise and List, Meike and L\"offler, Frank and Panet, Isabelle and Pouilloux, Benjamin and Prieur, Pascal and Rebray, Alexandre and Reynaud, Serge and Rievers, Benny and Robert, Alain and Selig, Hanns and Serron, Laura and Sumner, Timothy and Tanguy, Nicolas and Visser, Pieter},
  journal = {Phys. Rev. Lett.},
  volume = {119},
  issue = {23},
  pages = {231101},
  numpages = {7},
  year = {2017},
  month = {Dec},
  publisher = {American Physical Society},
  doi = {10.1103/PhysRevLett.119.231101}
}

@article{Collett_2018,
   title={A precise extragalactic test of General Relativity},
   volume={360},
   ISSN={1095-9203},
   DOI={10.1126/science.aao2469},
   number={6395},
   journal={Science},
   publisher={American Association for the Advancement of Science (AAAS)},
   author={Collett, Thomas E. and Oldham, Lindsay J. and Smith, Russell J. and Auger, Matthew W. and Westfall, Kyle B. and Bacon, David and Nichol, Robert C. and Masters, Karen L. and Koyama, Kazuya and van den Bosch, Remco},
   year={2018},
   month=jun,
   pages={1342–1346}
}

@article{Baker_2015,
   title={LINKING TESTS OF GRAVITY ON ALL SCALES: FROM THE STRONG-FIELD REGIME TO COSMOLOGY},
   volume={802},
   ISSN={1538-4357},
   DOI={10.1088/0004-637x/802/1/63},
   number={1},
   journal={\apj},
   publisher={American Astronomical Society},
   author={Baker, Tessa and Psaltis, Dimitrios and Skordis, Constantinos},
   year={2015},
   month=Mar,
   pages={63}
}

@ARTICLE{EHT_2019,
       author = {{Event Horizon Telescope Collaboration} and {Akiyama}, Kazunori and {Alberdi}, Antxon and {Alef}, Walter and {Asada}, Keiichi and {Azulay}, Rebecca and {Baczko}, Anne-Kathrin and {Ball}, David and {Balokovi{\'c}}, Mislav and {Barrett}, John and {Bintley}, Dan and {Blackburn}, Lindy and {Boland}, Wilfred and {Bouman}, Katherine L. and {Bower}, Geoffrey C. and {Bremer}, Michael and {Brinkerink}, Christiaan D. and {Brissenden}, Roger and {Britzen}, Silke and {Broderick}, Avery E. and {Broguiere}, Dominique and {Bronzwaer}, Thomas and {Byun}, Do-Young and {Carlstrom}, John E. and {Chael}, Andrew and {Chan}, Chi-kwan and {Chatterjee}, Shami and {Chatterjee}, Koushik and {Chen}, Ming-Tang and {Chen}, Yongjun and {Cho}, Ilje and {Christian}, Pierre and {Conway}, John E. and {Cordes}, James M. and {Crew}, Geoffrey B. and {Cui}, Yuzhu and {Davelaar}, Jordy and {De Laurentis}, Mariafelicia and {Deane}, Roger and {Dempsey}, Jessica and {Desvignes}, Gregory and {Dexter}, Jason and {Doeleman}, Sheperd S. and {Eatough}, Ralph P. and {Falcke}, Heino and {Fish}, Vincent L. and {Fomalont}, Ed and {Fraga-Encinas}, Raquel and {Freeman}, William T. and {Friberg}, Per and {Fromm}, Christian M. and {G{\'o}mez}, Jos{\'e} L. and {Galison}, Peter and {Gammie}, Charles F. and {Garc{\'\i}a}, Roberto and {Gentaz}, Olivier and {Georgiev}, Boris and {Goddi}, Ciriaco and {Gold}, Roman and {Gu}, Minfeng and {Gurwell}, Mark and {Hada}, Kazuhiro and {Hecht}, Michael H. and {Hesper}, Ronald and {Ho}, Luis C. and {Ho}, Paul and {Honma}, Mareki and {Huang}, Chih-Wei L. and {Huang}, Lei and {Hughes}, David H. and {Ikeda}, Shiro and {Inoue}, Makoto and {Issaoun}, Sara and {James}, David J. and {Jannuzi}, Buell T. and {Janssen}, Michael and {Jeter}, Britton and {Jiang}, Wu and {Johnson}, Michael D. and {Jorstad}, Svetlana and {Jung}, Taehyun and {Karami}, Mansour and {Karuppusamy}, Ramesh and {Kawashima}, Tomohisa and {Keating}, Garrett K. and {Kettenis}, Mark and {Kim}, Jae-Young and {Kim}, Junhan and {Kim}, Jongsoo and {Kino}, Motoki and {Koay}, Jun Yi and {Koch}, Patrick M. and {Koyama}, Shoko and {Kramer}, Michael and {Kramer}, Carsten and {Krichbaum}, Thomas P. and {Kuo}, Cheng-Yu and {Lauer}, Tod R. and {Lee}, Sang-Sung and {Li}, Yan-Rong and {Li}, Zhiyuan and {Lindqvist}, Michael and {Liu}, Kuo and {Liuzzo}, Elisabetta and {Lo}, Wen-Ping and {Lobanov}, Andrei P. and {Loinard}, Laurent and {Lonsdale}, Colin and {Lu}, Ru-Sen and {MacDonald}, Nicholas R. and {Mao}, Jirong and {Markoff}, Sera and {Marrone}, Daniel P. and {Marscher}, Alan P. and {Mart{\'\i}-Vidal}, Iv{\'a}n and {Matsushita}, Satoki and {Matthews}, Lynn D. and {Medeiros}, Lia and {Menten}, Karl M. and {Mizuno}, Yosuke and {Mizuno}, Izumi and {Moran}, James M. and {Moriyama}, Kotaro and {Moscibrodzka}, Monika and {M{\"u}ller}, Cornelia and {Nagai}, Hiroshi and {Nagar}, Neil M. and {Nakamura}, Masanori and {Narayan}, Ramesh and {Narayanan}, Gopal and {Natarajan}, Iniyan and {Neri}, Roberto and {Ni}, Chunchong and {Noutsos}, Aristeidis and {Okino}, Hiroki and {Olivares}, H{\'e}ctor and {Ortiz-Le{\'o}n}, Gisela N. and {Oyama}, Tomoaki and {{\"O}zel}, Feryal and {Palumbo}, Daniel C.~M. and {Patel}, Nimesh and {Pen}, Ue-Li and {Pesce}, Dominic W. and {Pi{\'e}tu}, Vincent and {Plambeck}, Richard and {PopStefanija}, Aleksandar and {Porth}, Oliver and {Prather}, Ben and {Preciado-L{\'o}pez}, Jorge A. and {Psaltis}, Dimitrios and {Pu}, Hung-Yi and {Ramakrishnan}, Venkatessh and {Rao}, Ramprasad and {Rawlings}, Mark G. and {Raymond}, Alexander W. and {Rezzolla}, Luciano and {Ripperda}, Bart and {Roelofs}, Freek and {Rogers}, Alan and {Ros}, Eduardo and {Rose}, Mel and {Roshanineshat}, Arash and {Rottmann}, Helge and {Roy}, Alan L. and {Ruszczyk}, Chet and {Ryan}, Benjamin R. and {Rygl}, Kazi L.~J. and {S{\'a}nchez}, Salvador and {S{\'a}nchez-Arguelles}, David and {Sasada}, Mahito and {Savolainen}, Tuomas and {Schloerb}, F. Peter and {Schuster}, Karl-Friedrich and {Shao}, Lijing and {Shen}, Zhiqiang and {Small}, Des and {Sohn}, Bong Won and {SooHoo}, Jason and {Tazaki}, Fumie and {Tiede}, Paul and {Tilanus}, Remo P.~J. and {Titus}, Michael and {Toma}, Kenji and {Torne}, Pablo and {Trent}, Tyler and {Trippe}, Sascha and {Tsuda}, Shuichiro and {van Bemmel}, Ilse and {van Langevelde}, Huib Jan and {van Rossum}, Daniel R. and {Wagner}, Jan and {Wardle}, John and {Weintroub}, Jonathan and {Wex}, Norbert and {Wharton}, Robert and {Wielgus}, Maciek and {Wong}, George N. and {Wu}, Qingwen and {Young}, Ken and {Young}, Andr{\'e}},
        title = "{First M87 Event Horizon Telescope Results. I. The Shadow of the Supermassive Black Hole}",
      journal = {\apjl},
         year = 2019,
        month = apr,
       volume = {875},
       number = {1},
          eid = {L1},
        pages = {L1},
          doi = {10.3847/2041-8213/ab0ec7},
archivePrefix = {arXiv},
       eprint = {1906.11238},
 primaryClass = {astro-ph.GA},
       adsurl = {https://ui.adsabs.harvard.edu/abs/2019ApJ...875L...1E}
}

@ARTICLE{Nicolis2009,
       author = {{Nicolis}, Alberto and {Rattazzi}, Riccardo and {Trincherini}, Enrico},
        title = "{Galileon as a local modification of gravity}",
      journal = {\prd},
         year = 2009,
        month = mar,
       volume = {79},
       number = {6},
          eid = {064036},
        pages = {064036},
          doi = {10.1103/PhysRevD.79.064036},
archivePrefix = {arXiv},
       eprint = {0811.2197},
 primaryClass = {hep-th},
       adsurl = {https://ui.adsabs.harvard.edu/abs/2009PhRvD..79f4036N}
}

@ARTICLE{Baker_2017,
       author = {{Baker}, T. and {Bellini}, E. and {Ferreira}, P.~G. and {Lagos}, M. and {Noller}, J. and {Sawicki}, I.},
        title = "{Strong Constraints on Cosmological Gravity from GW170817 and GRB 170817A}",
      journal = {\prl},
         year = 2017,
        month = dec,
       volume = {119},
       number = {25},
          eid = {251301},
        pages = {251301},
          doi = {10.1103/PhysRevLett.119.251301},
archivePrefix = {arXiv},
       eprint = {1710.06394},
 primaryClass = {astro-ph.CO},
       adsurl = {https://ui.adsabs.harvard.edu/abs/2017PhRvL.119y1301B}
}

@ARTICLE{H0DN2026,
       author = {{H0DN Collaboration} and {Casertano}, Stefano and {Anand}, Gagandeep and {Anderson}, Richard I. and {Beaton}, Rachael and {Bhardwaj}, Anupam and {Blakeslee}, John P. and {Boubel}, Paula and {Breuval}, Louise and {Brout}, Dillon and {Cantiello}, Michele and {Cruz Reyes}, Mauricio and {Cs{\"o}rnyei}, Geza and {de Jaeger}, Thomas and {Dhawan}, Suhail and {Di Valentino}, Eleonora and {Galbany}, Llu{\'\i}s and {Gil-Mar{\'\i}n}, H{\'e}ctor and {Graczyk}, Dariusz and {Huang}, Caroline and {Jensen}, Joseph B. and {Kervella}, Pierre and {Leibundgut}, Bruno and {Lengen}, Bastian and {Li}, Siyang and {Macri}, Lucas and {{\"O}z{\"u}lker}, Emre and {Pesce}, Dominic W. and {Riess}, Adam and {Romaniello}, Martino and {Said}, Khaled and {Sch{\"o}neberg}, Nils and {Scolnic}, Dan and {Sicignano}, Teresa and {Skowron}, Dorota M. and {Uddin}, Syed A. and {Verde}, Licia and {Nota}, Antonella},
        title = "{The Local Distance Network: A community consensus report on the measurement of the Hubble constant at {\ensuremath{\sim}}1\% precision}",
      journal = {\aap},
         year = 2026,
        month = apr,
       volume = {708},
          eid = {A166},
        pages = {A166},
          doi = {10.1051/0004-6361/202557993},
archivePrefix = {arXiv},
       eprint = {2510.23823},
 primaryClass = {astro-ph.CO},
       adsurl = {https://ui.adsabs.harvard.edu/abs/2026A&A...708A.166H}
}

@ARTICLE{Wright2025,
       author = {{Wright}, Angus H. and {St{\"o}lzner}, Benjamin and {Asgari}, Marika and {Bilicki}, Maciej and {Giblin}, Benjamin and {Heymans}, Catherine and {Hildebrandt}, Hendrik and {Hoekstra}, Henk and {Joachimi}, Benjamin and {Kuijken}, Konrad and {Li}, Shun-Sheng and {Reischke}, Robert and {von Wietersheim-Kramsta}, Maximilian and {Yoon}, Mijin and {Burger}, Pierre and {Chisari}, Nora Elisa and {de Jong}, Jelte and {Dvornik}, Andrej and {Georgiou}, Christos and {Harnois-D{\'e}raps}, Joachim and {Jalan}, Priyanka and {William}, Anjitha John and {Joudaki}, Shahab and {Lesci}, Giorgio Francesco and {Linke}, Laila and {Loureiro}, Arthur and {Mahony}, Constance and {Maturi}, Matteo and {Miller}, Lance and {Moscardini}, Lauro and {Napolitano}, Nicola R. and {Porth}, Lucas and {Radovich}, Mario and {Schneider}, Peter and {Tr{\"o}ster}, Tilman and {Valentijn}, Edwin and {Wittje}, Anna and {Yan}, Ziang and {Zhang}, Yun-Hao},
        title = "{KiDS-Legacy: Cosmological constraints from cosmic shear with the complete Kilo-Degree Survey}",
      journal = {\aap},
         year = 2025,
        month = nov,
       volume = {703},
          eid = {A158},
        pages = {A158},
          doi = {10.1051/0004-6361/202554908},
archivePrefix = {arXiv},
       eprint = {2503.19441},
 primaryClass = {astro-ph.CO},
       adsurl = {https://ui.adsabs.harvard.edu/abs/2025A&A...703A.158W}
}

@ARTICLE{Falck2015,
       author = {{Falck}, Bridget and {Koyama}, Kazuya and {Zhao}, Gong-Bo},
        title = "{Cosmic web and environmental dependence of screening: Vainshtein vs. chameleon}",
      journal = {\jcap},
         year = 2015,
        month = jul,
       volume = {2015},
       number = {7},
        pages = {049-049},
          doi = {10.1088/1475-7516/2015/07/049},
archivePrefix = {arXiv},
       eprint = {1503.06673},
 primaryClass = {astro-ph.CO},
       adsurl = {https://ui.adsabs.harvard.edu/abs/2015JCAP...07..049F}
}

@ARTICLE{Sagunski2021,
       author = {{Sagunski}, Laura and {Gad-Nasr}, Sophia and {Colquhoun}, Brian and {Robertson}, Andrew and {Tulin}, Sean},
        title = "{Velocity-dependent self-interacting dark matter from groups and clusters of galaxies}",
      journal = {\jcap},
         year = 2021,
        month = jan,
       volume = {2021},
       number = {1},
          eid = {024},
        pages = {024},
          doi = {10.1088/1475-7516/2021/01/024},
archivePrefix = {arXiv},
       eprint = {2006.12515},
 primaryClass = {astro-ph.CO},
       adsurl = {https://ui.adsabs.harvard.edu/abs/2021JCAP...01..024S}
}

@ARTICLE{ODonnell_2026,
       author = {{O'Donnell}, Jackson H. and {Jeltema}, Tesla E. and {Roberts}, M. Grant and {Nightingale}, James and {Flowers}, Abigail and {Aldas}, Dhruv},
        title = "{Constraint on dark matter self-interaction from combined strong lensing and stellar kinematics in MACS J0138-2155}",
      journal = {\prd},
         year = 2026,
        month = mar,
       volume = {113},
       number = {6},
          eid = {063531},
        pages = {063531},
          doi = {10.1103/hfgd-245k},
archivePrefix = {arXiv},
       eprint = {2508.20179},
 primaryClass = {astro-ph.CO},
       adsurl = {https://ui.adsabs.harvard.edu/abs/2026PhRvD.113f3531O}
}

@ARTICLE{Sand2004,
       author = {{Sand}, David J. and {Treu}, Tommaso and {Smith}, Graham P. and {Ellis}, Richard S.},
        title = "{The Dark Matter Distribution in the Central Regions of Galaxy Clusters: Implications for Cold Dark Matter}",
      journal = {\apj},
         year = 2004,
        month = mar,
       volume = {604},
       number = {1},
        pages = {88-107},
          doi = {10.1086/382146},
archivePrefix = {arXiv},
       eprint = {astro-ph/0309465},
 primaryClass = {astro-ph},
       adsurl = {https://ui.adsabs.harvard.edu/abs/2004ApJ...604...88S}
}

@ARTICLE{Rinaldi2026,
       author = {{Rinaldi}, Pierluigi and {Alberts}, Stacey and {Willmer}, Christopher N.~A. and {Carreira}, Courtney and {Williams}, Christina C. and {Noirot}, Ga{\"e}l and {Gilbert}, Carys J.~E. and {Bunker}, Andrew J. and {Baker}, William M. and {Barchiesi}, Luigi and {Ji}, Zhiyuan and {Lyu}, Jianwei and {Tacchella}, Sandro and {Wu}, Zihao and {Zhu}, Yongda},
        title = "{Filling the Gap in Cluster Evolution: JWST's Glimpse into a Young, Star-Forming Cluster at Cosmic Noon}",
      journal = {arXiv e-prints},
         year = 2026,
        month = feb,
          eid = {arXiv:2602.24162},
        pages = {arXiv:2602.24162},
          doi = {10.48550/arXiv.2602.24162},
archivePrefix = {arXiv},
       eprint = {2602.24162},
 primaryClass = {astro-ph.GA},
       adsurl = {https://ui.adsabs.harvard.edu/abs/2026arXiv260224162R}
}

@article{Bilton2018,
    author = {Bilton, Lawrence E and Pimbblet, Kevin A},
    title = {The kinematics of cluster galaxies via velocity dispersion profiles},
    journal = {\mnras},
    volume = {481},
    number = {2},
    pages = {1507-1521},
    year = {2018},
    month = {12},
    issn = {0035-8711},
    doi = {10.1093/mnras/sty2379}
}

@ARTICLE{Gouin2021,
       author = {{Gouin}, C. and {Bonnaire}, T. and {Aghanim}, N.},
        title = "{Shape and connectivity of groups and clusters: Effect of the dynamical state and accretion history}",
      journal = {\aap},
         year = 2021,
        month = jul,
       volume = {651},
          eid = {A56},
        pages = {A56},
          doi = {10.1051/0004-6361/202140327},
archivePrefix = {arXiv},
       eprint = {2101.04686},
 primaryClass = {astro-ph.CO},
       adsurl = {https://ui.adsabs.harvard.edu/abs/2021A&A...651A..56G}
}

@ARTICLE{Musso2021,
       author = {{Musso}, Marcello and {Sheth}, Ravi K.},
        title = "{Excursion set peaks in energy as a model for haloes}",
      journal = {\mnras},
         year = 2021,
        month = dec,
       volume = {508},
       number = {3},
        pages = {3634-3648},
          doi = {10.1093/mnras/stab2640},
archivePrefix = {arXiv},
       eprint = {1907.09147},
 primaryClass = {astro-ph.CO},
       adsurl = {https://ui.adsabs.harvard.edu/abs/2021MNRAS.508.3634M}
}

@ARTICLE{Wojtak2013,
       author = {{Wojtak}, Rados{\l}aw},
        title = "{Phase-space shapes of clusters and rich groups of galaxies}",
      journal = {\aap},
         year = 2013,
        month = nov,
       volume = {559},
          eid = {A89},
        pages = {A89},
          doi = {10.1051/0004-6361/201322509},
archivePrefix = {arXiv},
       eprint = {1310.3624},
 primaryClass = {astro-ph.CO},
       adsurl = {https://ui.adsabs.harvard.edu/abs/2013A&A...559A..89W}
}

@ARTICLE{Zhang_2023,
       author = {{Zhang}, Zhuowen and {Wu}, Hao-Yi and {Zhang}, Yuanyuan and {Frieman}, Joshua and {To}, Chun-Hao and {DeRose}, Joseph and {Costanzi}, Matteo and {Wechsler}, Risa H. and {Adhikari}, Susmita and {Rykoff}, Eli and {Jeltema}, Tesla and {Evrard}, August and {Rozo}, Eduardo},
        title = "{Modelling galaxy cluster triaxiality in stacked cluster weak lensing analyses}",
      journal = {\mnras},
         year = 2023,
        month = aug,
       volume = {523},
       number = {2},
        pages = {1994-2013},
          doi = {10.1093/mnras/stad1404},
archivePrefix = {arXiv},
       eprint = {2202.08211},
 primaryClass = {astro-ph.CO},
       adsurl = {https://ui.adsabs.harvard.edu/abs/2023MNRAS.523.1994Z}
}

@ARTICLE{Kim_2024,
       author = {{Kim}, Junhan and {Sayers}, Jack and {Sereno}, Mauro and {Bartalucci}, Iacopo and {Chappuis}, Loris and {De Grandi}, Sabrina and {De Luca}, Federico and {De Petris}, Marco and {Donahue}, Megan E. and {Eckert}, Dominique and {Ettori}, Stefano and {Gaspari}, Massimo and {Gastaldello}, Fabio and {Gavazzi}, Raphael and {Gavidia}, Adriana and {Ghizzardi}, Simona and {Iqbal}, Asif and {Kay}, Scott T. and {Lovisari}, Lorenzo and {Maughan}, Ben J. and {Mazzotta}, Pasquale and {Okabe}, Nobuhiro and {Pointecouteau}, Etienne and {Pratt}, Gabriel W. and {Rossetti}, Mariachiara and {Umetsu}, Keiichi},
        title = "{CHEX-MATE: CLUster Multi-Probes in Three Dimensions (CLUMP-3D). I. Gas analysis method using X-ray and Sunyaev-Zel'dovich effect data}",
      journal = {\aap},
         year = 2024,
        month = jun,
       volume = {686},
          eid = {A97},
        pages = {A97},
          doi = {10.1051/0004-6361/202347399},
archivePrefix = {arXiv},
       eprint = {2307.04794},
 primaryClass = {astro-ph.CO},
       adsurl = {https://ui.adsabs.harvard.edu/abs/2024A&A...686A..97K}
}

@ARTICLE{Kaiser1986,
       author = {{Kaiser}, N.},
        title = "{Evolution and clustering of rich clusters.}",
      journal = {\mnras},
         year = 1986,
        month = sep,
       volume = {222},
        pages = {323-345},
          doi = {10.1093/mnras/222.2.323},
       adsurl = {https://ui.adsabs.harvard.edu/abs/1986MNRAS.222..323K}
}

@ARTICLE{DeLucia2007,
       author = {{De Lucia}, Gabriella and {Blaizot}, J{\'e}r{\'e}my},
        title = "{The hierarchical formation of the brightest cluster galaxies}",
      journal = {\mnras},
         year = 2007,
        month = feb,
       volume = {375},
       number = {1},
        pages = {2-14},
          doi = {10.1111/j.1365-2966.2006.11287.x},
archivePrefix = {arXiv},
       eprint = {astro-ph/0606519},
 primaryClass = {astro-ph},
       adsurl = {https://ui.adsabs.harvard.edu/abs/2007MNRAS.375....2D}
}

@article{Saro_2013,
doi = {10.1088/0004-637X/772/1/47},
year = {2013},
month = {jul},
publisher = {The American Astronomical Society},
volume = {772},
number = {1},
pages = {47},
author = {Saro, Alex and Mohr, Joseph J. and Bazin, Gurvan and Dolag, Klaus},
title = {TOWARD UNBIASED GALAXY CLUSTER MASSES FROM LINE-OF-SIGHT VELOCITY DISPERSIONS},
journal = {\apj}
}

@ARTICLE{Shirasaki2021,
       author = {{Shirasaki}, Masato and {Egami}, Eiichi and {Okabe}, Nobuhiro and {Miyazaki}, Satoshi},
        title = "{Stacked phase-space density of galaxies around massive clusters: comparison of dynamical and lensing masses}",
      journal = {\mnras},
         year = 2021,
        month = sep,
       volume = {506},
       number = {3},
        pages = {3385-3405},
          doi = {10.1093/mnras/stab1961},
archivePrefix = {arXiv},
       eprint = {2101.01342},
 primaryClass = {astro-ph.GA},
       adsurl = {https://ui.adsabs.harvard.edu/abs/2021MNRAS.506.3385S}
}

@ARTICLE{Pizzuti2020syst,
       author = {{Pizzuti}, L. and {Sartoris}, B. and {Borgani}, S. and {Biviano}, A.},
        title = "{Calibration of systematics in constraining modified gravity models with galaxy cluster mass profiles}",
      journal = {\jcap},
         year = 2020,
        month = apr,
       volume = {2020},
       number = {4},
          eid = {024},
        pages = {024},
          doi = {10.1088/1475-7516/2020/04/024},
archivePrefix = {arXiv},
       eprint = {1912.09096},
 primaryClass = {astro-ph.CO},
       adsurl = {https://ui.adsabs.harvard.edu/abs/2020JCAP...04..024P}
}

\end{document}